\documentclass[11pt]{article}
\pdfoutput=1

\usepackage{my-j}
\usepackage[T1]{fontenc} 

\usepackage{longtable}
\usepackage[vcentermath]{youngtab}
\usepackage{array,multirow}
 \usepackage{multirow}
\usepackage{amssymb}
\usepackage{amsfonts}
\usepackage{amsbsy}
\usepackage{amsmath}
\usepackage{amsthm}
\usepackage{graphicx}
\usepackage{epstopdf}
\usepackage{color}
\usepackage{hyperref}
\usepackage{caption}
\usepackage{subcaption}
\usepackage{wasysym}

\newcommand{\set}[1]{\{ #1\}}

\def \be {\begin{equation}}
\def \ee {\end{equation}}
\def \bsp {\begin{split}}
\def \esp {\end{split}}
\def \bea {\begin{eqnarray}}
\def \eea {\end{eqnarray}}

\def\Z{\mathbb{Z}}
\def\F{\mathbb{F}}
\def\Q{\mathbb{Q}}
\def\C{\mathbb{C}}

\def\R{\mathbb{R}}

\def\P{\mathbb{P}}

\def\n3a{t}

\def\tr{{\mathrm{tr}}}

\def\O{\mathcal{O}}

\def\phit{\phi_0}

\def\ge{{\mathfrak{e}}}
\def\gso{{\mathfrak{so}}}
\def\gsu{{\mathfrak{su}}}
\def\gsp{{\mathfrak{sp}}}
\def\gf{{\mathfrak{f}}}
\def\gg{{\mathfrak{g}}}
\def\gu{{\mathfrak{u}}}

\DeclareMathOperator{\Hom}{Hom}

\newcommand{\ds}{\ensuremath{\Delta}}
\newcommand{\dd}{\ensuremath{\nabla}}

\newcommand{\vb}{\ensuremath{v^{(B)}}}

\newcommand{\eq}[1]{(\ref{#1})}

\title{
Comparing
elliptic and toric hypersurface\\
 Calabi-Yau threefolds at large Hodge numbers}
\author{Yu-Chien Huang,}
\author{Washington Taylor}

\affiliation{Center for Theoretical Physics\\
Department of Physics\\
Massachusetts Institute of Technology\\
77 Massachusetts Avenue\\
Cambridge, MA 02139, USA}

\emailAdd{{\tt yc\_huang} {\rm at} {\tt mit.edu}}
\emailAdd{{\tt wati} {\rm at} {\tt mit.edu}}

\preprint{MIT-CTP-4906}

\abstract{ We compare the sets of Calabi-Yau threefolds with large
Hodge numbers   that are constructed using
  toric hypersurface methods with those can be constructed as elliptic
  fibrations using Weierstrass model techniques motivated by F-theory.
There is a close correspondence between the structure of ``tops'' in
the toric polytope construction and Tate form tunings of Weierstrass
models for elliptic fibrations.
  We find that  all of the 
Hodge number pairs
($h^{1, 1},h^{2, 1}$)
with
$h^{1,1}$ or
  $h^{2, 1}\geq 240$ that are
associated with threefolds in the Kreuzer-Skarke
database can be realized
explicitly by generic or tuned Weierstrass/Tate
models for elliptic fibrations over complex base surfaces.
This includes a relatively small number of somewhat exotic
constructions, including elliptic fibrations over non-toric bases,
models with new Tate tunings that can give rise to
 exotic matter in the 6D F-theory picture, 
tunings of gauge groups over non-toric curves,
tunings with very large 
Hodge number shifts and associated nonabelian gauge groups, and tuned
Mordell-Weil sections associated with U(1) factors in the
corresponding 6D theory. }

\begin{document}
\maketitle

\flushbottom

%--------------------------------

\section{Introduction}

While Calabi-Yau threefolds have played an important role in string
theory since the early days of the subject \cite{chsw}, the set of
these geometries is still relatively poorly understood.  Following the
approach of Batyrev \cite{Batyrev}, in 2000 Kreuzer and Skarke carried
out a complete analysis of all reflexive polytopes in four
dimensions, giving a systematic classification of those Calabi-Yau
(CY)
threefolds that can be realized as hypersurfaces in toric varieties
\cite{Kreuzer:2000xy}.  For many years the resulting database
\cite{database} has represented the bulk of the known set of
Calabi-Yau threefolds, particularly at large Hodge numbers.  More
recently, the study of F-theory \cite{Vafa-F-theory, Morrison-Vafa}
has motivated an alternative method for the systematic construction of
Calabi-Yau threefolds that have the structure of an elliptic fibration
(with section).  By systematically classifying all bases that support
an elliptically fibered CY \cite{clusters, Morrison:2012js, Martini-WT,
  Wang-WT} and then systematically considering all possible
Weierstrass tunings \cite{Johnson-WT, Johnson:2016qar} over each such
base, it is possible in principle to construct all elliptically
fibered Calabi-Yau threefolds. While there are some technical issues
that must still be resolved for a complete classification from this
approach, at large Hodge numbers this method gives a reasonably
complete picture of the set of possibilities.  One perhaps surprising
result that has recently become apparent both from this work and from
other perspectives \cite{Candelas-cs, Gray-hl, Anderson:2015iia, Anderson:2016ler,Anderson:2016cdu,Anderson:2017aux} is that
a very large fraction of the set of Calabi-Yau threefolds that can be
constructed by {\it any} known mechanism are actually elliptically
fibered, particularly at large Hodge numbers.

The goal of this paper is to carry out a direct comparison of the set
of elliptically fibered Calabi-Yau threefolds that can be constructed
using Weierstrass/Tate F-theory based methods with those that arise
through reflexive polytope constructions.  While the general methods
for construction of elliptic Calabi-Yau threefolds can include
non-toric bases \cite{Martini-WT, Wang-WT}, and even over toric bases
there are non-toric Weierstrass tunings \cite{Johnson-WT,
  Johnson:2016qar}, we focus here on the subset of constructions that
have the potential for a toric description through a reflexive
polytope.  
In \S\ref{sec:F-theory}, we review some of the basics of
F-theory and the systematic construction of elliptic Calabi-Yau
threefolds through the geometry of the base and the tuning of
Weierstrass or Tate models from the generic structure over each base.
In \S\ref{sec:reflexive}, we review the Batyrev construction and
reflexive polytopes, and the structure of elliptic fibrations in this
context.  In particular, in \S\ref{sp231} we  describe the precise
correspondence between a particular
fibration structure for a reflexive polytope and Tate form Weierstrass
models. In \S\ref{sec:Tate}, we restrict attention to toric base
surfaces $B_2$ and identify the set of tuned Weierstrass/Tate models
over such bases that naturally correspond to a reflexive polytope in the
Batyrev construction.  This gives us a systematic way of constructing
from the point of view of elliptic fibrations over a chosen base a
large set of elliptic Calabi-Yau threefolds that are expected to be
seen in the Kreuzer-Skarke database with a specific ($\P^{2,3,1}$)
fiber type.  At large Hodge numbers, for reasons discussed further in
\S\ref{bwlhn}, we expect that this should give most or all elliptic
fibrations that arise in the KS database; we find that this is in fact
the case.

The main results of the paper are in
\S\ref{sec:systematic-construction}
and \S\ref{restcases}, where we describe an algorithm to
systematically run through all tuned
Tate models over toric bases and we compare  the results  of running
this algorithm to the  Kreuzer-Skarke database.  The initial result,
described in \S\ref{sec:systematic-construction}, is that
these simply constructed sets match almost perfectly in the large
Hodge number regimes that we study: both at large $h^{2, 1}$ and at
large $h^{1, 1}$ all the models constructed by an appropriate set
of Tate tunings over toric bases appear in the KS database, and
virtually all the Hodge numbers in the database are reproduced by
elliptic Calabi-Yau threefolds produced using this approach.  
There is a small set of large Hodge numbers (18 out of 1,827)
associated with toric hypersurface Calabi-Yaus, however, that are not
reproduced by our initial scan.  By examining these individual cases,
as described in \S\ref{restcases},
we find that all these exceptions also correspond to elliptic
fibrations though with more exotic structure, such as
non-flat fibrations resolved through extra blow-ups
in the base that take the base outside the toric class, and/or force
Mordell-Weil sections on the elliptic fiber.  The upshot is that when
these more exotic constructions are included, {\it all} Hodge number
pairs
with either $h^{1, 1}$ or $h^{2, 1}$ at least 240 are reproduced by an
elliptic Calabi-Yau over some explicitly determined base surface.  We
conclude in \S\ref{sec:conclusions} with a summary of the results and
some related open questions.

Note that in this paper the focus is on understanding in some detail
the connection
between elliptic fibration geometry and polytope geometry  for these
different approaches to  construction of elliptic Calabi-Yau
threefolds.  In a companion paper \cite{Huang-Taylor-fibers} we will describe a more direct 
analysis of the polytopes in the KS database that also shows
explicitly that there is a toric fiber associated with an elliptic fibration for every polytope
in the database at large Hodge numbers.
The principal
class of Tate tunings that we consider in this paper have a
complementary description in the language of ``tops''
\cite{Candelas:1996su}.  
The construction of many polytopes in the KS database through
combining K3 tops and ``bottoms'' was accomplished in
\cite{Candelas-cs}, and
a systematic approach to constructing toric
hypersurface Calabi-Yau threefold with a given base and gauge group
using the language of tops is developed in \cite{Braun-16}, with
particular application to models with gauge group SU(5) as also
studied in e.g.\ \cite{bmpw,Borchmann:2013hta}.  One of
the main results of this paper is the systematic relationship of such
constructions with certain classes of Tate tunings.  This leads in
some cases to the identification of new Tate tunings from observed
polytope structures, and the
observation that some polytopes in the KS database have a more complex
structure that does not admit a direct description in terms of
standard
tops.
On the other hand, new structures of tops are also found through the
construction of polytopes via the correspondence with Tate tunings.

\section{F-theory physics and elliptic Calabi-Yau threefold geometry}
\label{sec:F-theory}

We briefly summarize here how the massless spectrum of a
six-dimensional effective theory from F-theory compactification  is
related to the geometric data of the internal manifold, which is
an elliptically fibered Calabi-Yau threefold (CY3) over a two-dimensional
base $B_2$ (complex dimensions).  
F-theory models can then be systematically studied by first choosing a
base $B_2$ and then specifying an elliptic fibration in Weierstrass
form
over that base.
Further background on F-theory can be found in for
example \cite{Vafa-F-theory, Morrison-Vafa, Morrison-TASI,
  Taylor:2011wt}. 

F-theory compactified on a (possibly singular) elliptically fibered
Calabi-Yau threefold $X$ gives a 6D effective supergravity theory.
Such a compactification of F-theory is equivalent to M-theory on the
resolved Calabi-Yau $\tilde{X}$ in the decompactification limit of
M-theory, where in the F-theory picture the resolved components of the
elliptic fiber are shrunk to zero size.  F-theory can also be thought
of as a nonperturbative formulation of type IIB string theory.  In
this picture the type IIB theory is compactified on the base $B_2$.
In this F-theory description, spacetime filling 7-branes sit at the
codimension-one loci in the base where the fibration degenerates.  The
non-abelian gauge symmetries of the 6D effective theory arise from the
seven-branes and can be inferred from the singularity types of the
elliptic fibers along the codimension-one loci in the base, according
to the Kodaira classification (Table \ref{t:Kodaira}).  At the
intersections of seven-branes there are localized matter fields that
are hypermultiplets in the 6D theory; the representations of the
matter fields can be determined from the detailed form of the
singularities over the codimension-two points in the base (see
e.g. \cite{Katz:1996xe, Bershadsky:1996nh,
  Morrison:2011mb}). Therefore the physics data can be extracted by
studying the singular fibers by means of the Weierstrass models (short
form) or the Tate models (long form) of $X$ that we review in
\S\ref{NHC} and \ref{TT}
\footnote{The short form Weierstrass model is the most
  general form for an elliptic Calabi-Yau threefold. The cases discussed
  in this paper are elliptically fibered Calabi-Yau threefolds that
always have a section and therefore in principle admit a short form Weierstrass form realization.
There can also be genus one fibered Calabi-Yau threefolds (lacking a
global section), which can be related to Weierstrass models of
elliptic Calabi-Yau threefolds through the Jacobian construction
(described from the physics perspective in
\cite{Braun:2011ux, Huang:2013yta}).  The
physics of these threefolds is more subtle, involving discrete gauge
groups
\cite{Braun:2014oya, Morrison:2014era, aggk,
  Mayrhofer:2014opa, Cvetic:2015moa}.
In a few cases we find it useful to use the Jacobian construction even
for cases with a section, giving an explicit transformation to the short
Weierstrass form.}.
  There can also be abelian gauge symmetries, which arise
from additional rational sections of the elliptic fibration
\cite{Morrison-Vafa}. The study of $\gu(1)$ symmetries is more subtle
in that it relates to the global structure of the fibration, as
opposed to non-abelian symmetries where we can just study singular
fibers locally. We will see cases with abelian factors in
\S\ref{restcases}, with a detailed example worked out in
Appendix~\ref{u1}.  In \S\ref{zariski}, we review the Zariski
decomposition, which allows us to determine the order of vanishing and
consequent gauge group of a Weierstrass or Tate form description of an
elliptic fibration, and in \S\ref{sec:Zariski-Tate} we describe how
this method can be applied systematically in the context of Tate
tunings.  In \S\ref{anomaly}, we review the 6D anomaly cancellation
conditions and their connection to the matter content of a 6D theory
and the Hodge numbers of the corresponding Calabi-Yau threefold.  In
\S\ref{rpt} we review the constraints imposed by global symmetry
groups on the set of gauge groups that can be supported on curves
intersecting a given curve, and we conclude the overview of F-theory
in \S\ref{sec:bases} with a summary of the systematic classification
of complex surfaces that can support elliptic Calabi-Yau threefolds
and can be used for F-theory compactification.

\subsection{Hodge numbers and the 6D massless spectrum}
\label{hns}
By going to the 5D Coulomb branch after reduction on a circle,
the F-/M-theory correspondence can be used to relate the geometry of
$\tilde{X}$ to the associated 6D supergravity theory
\cite{Morrison:1996pp, Bonetti:2011mw}.  In particular, the
Hodge
numbers, $h^{1,1}$ and
$h^{2,1}$, of $\tilde{X}$ can be related
to the (massless)
matter content of the 6D theory:
\begin{eqnarray}
\label{h11}
h^{1,1}(\tilde{X}) = r+T+2,
\end{eqnarray}
where $T$ is the number of  tensor multiplets, which is determined
already by the choice of base $B_2$,
\begin{equation}
T=h^{1,1}(B_2)-1,
\label{baseT}
\end{equation}
and $r=r_\text{abelian}+\sum_i r_i$ is the total rank of the gauge group,
\begin{equation}
G=U(1)^{r_\text{abelian}}\times\prod_\text{non-abelian factors $i$} G_i,
\end{equation}
of the 6D effective theory.  We also have
\begin{equation}
\label{h21n}
h^{2,1}(\tilde{X})=H_\text{neutral}-1,
\end{equation}
where $H_\text{neutral}$ is the number of hypermultiplets that  are
neutral under the Cartan subalgebra\footnote{In other words, this
  counts fields that are neutral
  matter fields in the 5D M-theory sense but may transform under the unhiggsed
  non-abelian factors of the 6D F-theory. Often, matter charged under
  the non-abelian factors is still charged under the Cartan subalgebra, but
  for certain representations of some non-abelian groups there can be
charged
 matter
that is neutral under the Cartan
subalgebra. \label{fn6}}
of the
gauge group $G$ of the 6D F-theory.

The spectra of 6D theories are constrained by consistency conditions
associated with the absence of anomalies, which we
describe in further detail
in \S\ref{anomaly}. The
gravitational anomaly cancellation condition
(\ref{eq:anomaly-r4}) gives
$H -V = 273 -29T$,
where $V$ is the dimension of the gauge group $G$, and
$H=H_\text{charged}+H_\text{neutral}$ is the total number of
hypermultiplets  (separated into neutral and charged matter under the Cartan of the gauge group $G$). So we have another expression 
\begin{equation}
h^{2,1}(\tilde{X})= 272+ V-29T-H_\text{charged} \,.
\label{h21}
\end{equation}
This is more useful for some of our purposes than equation
(\ref{h21n}).  In particular, as we discuss in further detail in the
following section,
we are interested in studying various
specializations (tunings) of a generic elliptically fibered CY3 over a given base $B_2$.  
The number of tensors $T$ is fixed for a given
base.  Thus,
if
we start with known Hodge numbers $h^{1,1}$ and $h^{2,1}$
for the generic elliptic fibration over a given (e.g.\ toric
\cite{Morrison:2012js, Hodge}) base,  
and specialize/tune to a model with a larger gauge group and increased
matter content,
then
the Hodge numbers of the tuned model can be simply
calculated by adding to those of the generic models respectively the
shifts
\begin{eqnarray}
&&\Delta h^{1,1}=\Delta r, \label{eq:dh11}\\
&&\Delta h^{2,1}=\Delta V - \Delta H_\text{charged}.
\label{eq:dh21}
\end{eqnarray}
Such a specialization/tuning amounts physically to undoing a Higgsing
transition, and the second of these relations simply expresses the
physical expectation that the number of matter degrees of freedom that
are lost (``eaten'') in a Higgsing transition is equal to the number
of gauge bosons lost to symmetry breaking.  Note that the data on the
right hand sides are associated in general with tuned non-abelian
gauge symmetries but also in some special cases involve abelian
factors.
Note also that the right-hand sides of (\ref{eq:dh11}) and
(\ref{eq:dh21}) are always non-negative and non-positive respectively for any tuning. In most cases, 
the gauge group increases in rank and some of the $h^{2, 1}$
moduli are used to
implement the tuning. In {\it rank-preserving} tunings, however, the Hodge numbers do not change (see Table \ref{t:rpt}) --- $h^{1,1}$ of course does not change in a rank-preserving enhancement; $h^{2,1}$ does not change either in these tunings, as one can check by considering carefully the matter  charged under the Cartan subalgebra (cf. footnote \ref{fn6}.) 

\begin{table}[]
\centering
\begin{tabular}{|c|l|}
\hline
Rank $r$ & Algebras                  \\ \hline
2        & $\gsu(3), \gg_2$          \\ \hline
3        & $\gsu(4), \gso(7)$        \\ \hline
4        & $\gso(8), \gso(9), \gf_4$ \\ \hline
$r\geq$5 & $\gso(r), \gso(r+1)$      \\ \hline
\end{tabular}
\caption{\footnotesize Rank preserving tunings: tunings
of these four classes of gauge algebras do not change $h^{1,1}$  or $h^{2,1}$.}
\label{t:rpt}
\end{table}

\subsection{Generic and tuned
Weierstrass models  for elliptic Calabi-Yau
manifolds}
\label{NHC}

An elliptic fibration with a section over a base $B$ 
can be described by the Weierstrass model
\begin{equation}
y^2=x^3+fx z^4+g z^6 \,.
\label{W}
\end{equation}
The Calabi-Yau condition on the total space $X$ requires that $f, g$ are sections of ${\cal O} (-4K_B)$, ${\cal O} (-6K_B)$, where $K_B$ is the
canonical class of the base.
More abstractly, we take the weighted projective bundle
\begin{equation}
\pi: P=\mathbb{P}_{\mathbb{P}^{2,3,1}}[\mathcal{L}^2\oplus \mathcal{L}^3\oplus \mathcal{O}_B]\rightarrow B,
\label{fs}
\end{equation}
where  $\mathcal{L}=\mathcal{O}(-K_B)$ is required by the Calabi-Yau condition and $x\in \mathcal{O}_P(2)\otimes\pi^*\mathcal{L}^2, y\in
\mathcal{O}_P(3)\otimes\pi^*\mathcal{L}^3, z\in\mathcal{O}_P(1)$ and
$[x:y:z]$ can be viewed as  weighted projective coordinates of  the
$\P^{2,3,1}$, while $f$ and $g$ are sections of, to be more precise, $\pi^*\mathcal{L}^4$ and
$\pi^*\mathcal{L}^6$ respectively.

 Consider an elliptic Calabi-Yau threefold
over a complex two-dimensional base $B_2$, so the divisors in the base
are curves. The elliptic fiber becomes singular over the
codimension-one loci in the base where the discriminant
\begin{equation}
\label{delta}
\Delta=4f^3+27g^2
\end{equation}
vanishes. 
The type of
singular fiber 
at a generic point
along
an irreducible component $\{\sigma=0\}$ of the discriminant locus $\{\Delta=0\}$
is characterized by the
Kodaira singularity type, which is determined by the
 orders of vanishing of $f$, $g$, and
$\Delta$ in an expansion in $\sigma$
  (see Table \ref{t:Kodaira}). The physics interpretation is that
there are seven-branes on which open strings (and junctions) end
located at the
discriminant locus, and the resulting gauge symmetries can be
determined (up to \emph{monodromies}) by the type of the singular
fiber. The gauge algebras that are further determined by
monodromy conditions \cite{Bershadsky:1996nh, Katz:2011qp} are those
of types $I_n, I_0^*, I_n^*, IV, IV^*$, where some factorizability
conditions are imposed on the terms of $f, g, \Delta$ of lowest degrees of
vanishing order along $\{\sigma=0\}$.  We summarize these conditions
in Table $\ref{monocon}$,  in terms of the first
non-vanishing sections $f_i(\zeta), g_j(\zeta), \Delta_k(\zeta)$ in
the local expansions
\begin{eqnarray}
\label{fe}
&&f(\sigma, \zeta)=f_0(\zeta)+f_1(\zeta)\sigma+\cdots,\\
\label{ge}
&&g(\sigma, \zeta)=g_0(\zeta)+g_1(\zeta)\sigma+\cdots,\\
\label{de}
&&\Delta(\sigma, \zeta)=\Delta_0(\zeta)+\Delta_1(\zeta)\sigma+\cdots,
\end{eqnarray}
where $\{\zeta=0\}$ defines a divisor that intersects $\{\sigma=0\}$
transversely so that $\sigma, \zeta$ together serve as local
coordinates on an open patch of
base.

\begin{table}
\begin{center}
\begin{tabular}{|c |c |c |c |c |c |}
\hline
Type &
ord ($f$) &
ord ($g$) &
ord ($\Delta$) &
singularity & nonabelian symmetry algebra\\ \hline \hline
$I_0$&$\geq $ 0 & $\geq $ 0 & 0 & none & none \\
$I_n$ &0 & 0 & $n \geq 2$ & $A_{n-1}$ & $\gsu(n)$  or $\gsp(\lfloor
n/2\rfloor)$\\ 
$II$ & $\geq 1$ & 1 & 2 & none & none \\
$III$ &1 & $\geq 2$ &3 & $A_1$ & $\gsu(2)$ \\
$IV$ & $\geq 2$ & 2 & 4 & $A_2$ & $\gsu(3)$  or $\gsu(2)$\\
$I_0^*$&
$\geq 2$ & $\geq 3$ & $6$ &$D_{4}$ & $\gso(8)$ or $\gso(7)$ or $\gg_2$ \\
$I_n^*$&
2 & 3 & $n \geq 7$ & $D_{n -2}$ & $\gso(2n-4)$  or $\gso(2n -5)$ \\
$IV^*$& $\geq 3$ & 4 & 8 & $E_6$ & $\ge_6$  or $\gf_4$\\
$III^*$&3 & $\geq 5$ & 9 & $E_7$ & $\ge_7$ \\
$II^*$& $\geq 4$ & 5 & 10 & $E_8$ & $\ge_8$ \\
\hline
non-min &$\geq 4$ & $\geq6$ & $\geq12$ & \multicolumn{2}{c|}{ does not occur in 
F-theory } \\ 
\hline
\end{tabular}
\end{center}
\caption[x]{\footnotesize Kodaira classification of singularities in
  the elliptic fiber along codimension one loci in the base
in terms of orders of vanishing of the parameters $f, g$ in the
Weierstrass model (\ref{W}) and the discriminant locus $\Delta$.
}
\label{t:Kodaira}
\end{table}

\begin{table}[]
\centering
\begin{tabular}{|c|c|c|c|c|l|}
\hline
                         & ord($f$)                  & ord($g$)                  & ord($\Delta$)              & algebra                    & monodromy condition                                                                                                                                                                                                                                                                       \\ \hline
\multirow{2}{*}{$I_n$}   & \multirow{2}{*}{$0$}      &
\multirow{2}{*}{$0$}      & \multirow{2}{*}{$n$}       & $\gsu(n)$
& \begin{tabular}[c]{@{}l@{}}since $\Delta_0=0$, locally\\ $f_0(\zeta)=-\frac1{3}u_0^2$ and $g_0(\zeta)=\frac2{27}u_0^3$\\ for some $u_0(\zeta)$, which is a perfect square\end{tabular}                                                                                                    \\ \cline{5-6} 
                         &                           &                           &                            & $\gsp(\lfloor n/2\rfloor)$ & otherwise                                                                                                                                                                                                                                                                                 \\ \hline
\multirow{2}{*}{$IV$}    & \multirow{2}{*}{$\geq 2$} & \multirow{2}{*}{$2$}      & \multirow{2}{*}{$4$}       & $\gsu(3)$                  & $g_2(\zeta)$ is a perfect square                                                                                                                                                                                                                                                          \\ \cline{5-6} 
                         &                           &                           &                            & $\gsu(2)$                  & otherwise                                                                                                                                                                                                                                                                                 \\ \hline
\multirow{3}{*}{$I_0^*$} & \multirow{3}{*}{$\geq 2$} & \multirow{3}{*}{$\geq 3$} & \multirow{3}{*}{$6$}       & $\gso(8)$                  & \begin{tabular}[c]{@{}l@{}}$x^3+f_2(\zeta)x+g_3(\zeta)$ \\ $=(x-a)(x-b)(x+a+b)$\\ for some $a(\zeta), b(\zeta)$\end{tabular}                                                                                                                                                            \\ \cline{5-6} 
                         &                           &
&                            & $\gso(7)$
& \begin{tabular}[c]{@{}l@{}}$x^3+f_2(\zeta)x+g_3(\zeta)$
    \\ $=(x-a)(x^2+ax+b)$\\ for some $a(\zeta), b(\zeta)$ (but not
    $\gso(8)$ condition)\end{tabular}                                                                                                                                                                \\ \cline{5-6} 
                         &                           &                           &                            & $\gg_2$                    & otherwise                                                                                                                                                                                                                                                                                 \\ \hline
\multirow{2}{*}{$I_n^*$} & \multirow{2}{*}{$2$}      &
\multirow{2}{*}{$3$}      & \multirow{2}{*}{$n\geq 7$} & $\gso(2n-4)$
& \begin{tabular}[c]{@{}l@{}}since $\Delta_6=0$,
locally  \\ $f_2(\zeta)=-\frac1{3}u_1^2$ and $g_3(\zeta)=\frac2{27}u_1^3$\\ for some $u_1(\zeta)$;\\ $\frac{\Delta_n(\zeta)}{u_1^3}$ is a perfect square for odd $n$\\ $\frac{\Delta_n(\zeta)}{u_1^2}$ is a perfect square for even $n$\end{tabular} \\ \cline{5-6} 
                         &                           &                           &                            & $\gso(2n-5)$               & otherwise                                                                                                                                                                                                                                                                                 \\ \hline
\multirow{2}{*}{$IV^*$}  & \multirow{2}{*}{$\geq 3$} & \multirow{2}{*}{$4$}      & \multirow{2}{*}{$8$}       & $\ge_6$                    & $g_4(\zeta)$ is a perfect square                                                                                                                                                                                                                                                          \\ \cline{5-6} 
                         &                           &                           &                            & $\gf_4$                    & otherwise                                                                                                                                                                                                                                                                                 \\ \hline
\end{tabular}
\caption{\footnotesize Monodromy conditions for certain algebras to satisfy in additional to the desired orders of vanishing of $f, g, \Delta$:  $f_i(\zeta), g_j(\zeta), \Delta_k(\zeta)$ are coefficients of the expansions in equations (\ref{fe})-(\ref{de}).}
\label{monocon}
\end{table}

\begin{table}[ht]
\begin{center}
\begin{tabular}{|c|c|c|c|c|c|c|c|} \hline
 type & group & $ a_1$ &
$a_2$ & $a_3$ &$ a_4 $& $ a_6$ &$\Delta$ \\ \hline $I_0 $ & --- &$ 0 $ &$ 0
$ &$ 0 $ &$ 0 $ &$ 0$ &$0$ \\ \hline $I_1 $ & --- &$0 $ &$ 0 $ &$ 1 $ &$ 1
$ &$ 1 $ &$1$ \\ \hline $I_2 $ &$SU(2)$ &$ 0 $ &$ 0 $ &$ 1 $ &$ 1 $ &$2$ &$
2 $ \\ \hline $I_{3}^{ns} $ & $Sp(1)$ &$0$ &$0$ &$2$ &$2$ &$3$ &$3$ \\ \hline
$I_{3}^{s}$ & $SU(3)$ &$0$ &$1$ &$1$ &$2$ &$3$ &$3$ \\ \hline
$I_{2n}^{ns}$ &$ Sp(n)$ &$0$ &$0$ &$n$ &$n$ &$2n$ &$2n$ \\ \hline
$I_{2n}^{s}$ &$SU(2n)$ &$0$ &$1$ &$n$ &$n$ &$2n$ &$2n$ \\ \hline
$I_{2n}^{s}$ \scriptsize{(2nd version)} &$SU(2n)^\circ$ &$0$ &$2$ &$n-1$ &$n+1$ &$2n$ &$2n$ \\ \hline
$I_{2n+1}^{ns}$ &$Sp(n)$ & $0$ &$0$ &$n+1$ &$n+1$ &$2n+1$ &$2n+1$
\\ \hline $I_{2n+1}^s$ &$SU(2n+1)$ &$0$ &$1$ &$n$ &$n+1$ &$2n+1$ &$2n+1$
\\ \hline $II$ & --- &$1$ &$1$ &$1$ &$1$ &$1$ &$2$ \\ \hline $III$ &$SU(2)$ &$1$
&$1$ &$1$ &$1$ &$2$ &$3$ \\
\hline $IV^{ns} $ &$Sp(1)$ &$1$ &$1$ &$1$
&$2$ &$2$ &$4$ \\ 
\hline $IV^{s}$ &$SU(3)$ &$1$ &$1$ &$1$ &$2$ &$3 \; \; \; (2)^\star$ &$4$
\\ \hline $I_0^{*\,ns} $ &$G_2$ &$1$ &$1$ &$2$ &$2$ &$3$ &$6$ \\ \hline
$I_0^{*\,ss}$ &$SO(7)$ &$1$ &$1$ &$2$ &$2$ &$4$ &$6$ \\ \hline $I_0^{*\,s}
$ &$SO(8)$ &$1$ &$1$ &$2$ &$2$ &$(4, 3)^\star$ & $6$ \\ 
%\hline $I_{1}^{*\,ns}$
%&$SO(9)$ &$1$ &$1$ &$2$ &$3$ &$4$ &$7$ \\ \hline $I_{1}^{*\,s}$ &$SO(10) $
%&$1$ &$1$ &$2$ &$3$ &$5$ &$7$ \\ \hline $I_{2}^{*\,ns}$ &$SO(11)$ &$1$ &$1$
%&$3$ &$3$ &$5$ &$8$ \\ \hline $I_{2}^{*\,s}$ &$SO(12)$ &$1$ &$1$ &$3$
%&$3$ &$6$&$8$\\ 
\hline 
$I_{2n-3}^{*\,ns}$ &$SO(4n+1)$ &$1$ &$1$ &$n$ &$n+1$
&$2n$ &$2n+3$ \\ \hline $I_{2n-3}^{*\,s}$ &$SO(4n+2)$ &$1$ &$1$ &$n$ &$n+1$
&$2n+1 \; \; \; (2n)^{\star}$ &$2n+3$ \\ \hline $I_{2n-2}^{*\,ns}$ &$SO(4n+3)$ &$1$ &$1$ &$n+1$
&$n+1$ &$2n+1$ &$2n+4$ \\ \hline $I_{2n-2}^{*\,s}$ &$SO(4n+4)$ &$1$ &$1$
&$n+1$ &$n+1$ &$2n+2 \; \; \; (2n + 1)^{\star}$ 
&$2n+4$ \\ \hline $IV^{*\,ns}$ &$F_4 $ &$1$ &$2$ &$2$ &$3$ &$4$
&$8$\\ 
\hline $IV^{*\,s} $ &$E_6$ &$1$ &$2$ &$2$ &$3$ &$5 \; \; \;(4)^{\star}$ & $8$\\ \hline
$III^{*} $ &$E_7$ &$1$ &$2$ &$3$ &$3$ &$5$ & $9$\\ \hline $II^{*} $
&$E_8\,$ &$1$ &$2$ &$3$ &$4$ &$5$ & $10$ \\ \hline
 non-min & --- &$ 1$ &$2$ &$3$ &$4$ &$6$ &$12$ \\ \hline
\end{tabular}
\end{center}
\caption{\footnotesize Tate forms: Extends earlier versions of table by including
  alternative $SU(2n)$ and $SO(2k)$ tunings that can be realized
  purely by orders of vanishing without additional monodromy
  constraints.  In particular, alternate tuning $({}^{\circ})$ of SU(6)
  gives alternate exotic matter content; see text for further details.
  Groups and tunings marked with $^\star$ require additional monodromy
  conditions.}
 \label{tab:tatealg}
\end{table}

A generic Weierstrass model (i.e. with coefficients at a generic point
in the moduli space)
for an elliptically fibered CY$3$
over a given base $B_2$ corresponds
physically to a maximally Higgsed phase.  In the maximally Higgsed
phase over many bases the gauge group and matter content are still 
nontrivial.  The minimal gauge
algebras and matter configuration associated with a given base $B_2$
are carried by  {\em non-Higgsable
clusters} (NHCs) \cite{clusters}, which are isolated 
rational
curves of
self-intersection $m$, $-12\leq m\leq -3$, and clusters 
of multiple rational curves of self-intersection $\leq -2$: $\{-2,
-3\}$, $\{-2, -2, -3\}$, and
$\{-2, -3, -2\}$.  The sections $f, g,
\Delta$  automatically vanish to higher orders
along these curves
in any Weierstrass model over the given base.  This can be understood
geometrically as an effect in which the curvature over the negative
self-intersection curves must be cancelled by 7-branes to maintain the
Calabi-Yau structure of the elliptic fibration. 
The orders of vanishing  and the corresponding minimal
gauge groups on these NHCs are listed in Table \ref{t:dic} in \S\ref{sec:multiple-tops}.

Starting from the generic model over a given base $B$, we can
systematically tune the Weierstrass model coefficients $f$ and $g$ to
increase the order of vanishing over various curves beyond what is
imposed by the NHCs, producing additional or enhanced gauge groups on
some curves in the base.  
Many aspects of such tunings are described in a systematic fashion in
\cite{Johnson:2016qar}.
While over some bases there is a great deal
of freedom to tune many different gauge group factors on various
curves in the Weierstrass model, there are also limitations imposed by
the constraint that there be no codimension one loci over which $f, g$
vanish to orders $(4, 6)$.  In this paper we also avoid cases with
codimension two $(4, 6)$ loci by blowing up such points on the base as
part of the resolution process.  Such singularities can be related
to 6D superconformal field theories; in the geometric picture such
singularities are associated with non-flat fibers\footnote{Resolution
  of non-flat fibers in related cases
of tuned Weierstrass models has recently been considered for example
in \cite{Buchmuller:2017wpe, Dierigl:2018nlv}; the explicit connection
between resolutions giving non-flat fibrations and flat fibrations over a resolved base
through sequences of flops are described explicitly in the papers
\cite{Bhardwaj:2018vuu, Apruzzi:2018nre} that appeared after the
initial appearance of this preprint.} and a resolution
of the singularity can generally
be found by first blowing up the $(4, 6)$ point in
the base, which modifies the geometry of the base $B$, increasing
$h^{1, 1}(B)$ by one.  While in many cases the extent to which
enhanced gauge groups can be tuned in the Weierstrass model over any
given base can be determined by considerations such as the low-energy
anomaly consistency conditions, the precise set of possible tunings is
most clearly described in terms of an explicit description of the
Weierstrass coefficients. In the case of toric bases, the
complete set of monomials in $f, g$ has a simple description (see
e.g. \cite{Morrison:2012js, Johnson:2016qar}) and we have very strong control
over the parameters of the Weierstrass model.

\subsection{Tate form and the Tate algorithm}
\label{TT}
The Tate algorithm is a systematic procedure for determining the
Kodaira singularity type of an elliptic fibration, and provides a
convenient way to study Kodaira singularities
in the context of F-theory
\cite{Bershadsky:1996nh,Katz:2011qp}.  
The associated ``Tate forms'' for the different singularities match up
neatly with the toric construction that we focus on in this paper.
We start with an
equation for an elliptic curve in the general form
\begin{equation}
 y^2+a_1xyz+a_3yz^3=x^3+a_2x^2z^2+a_4xz^4+a_6z^6,
\label{T}
\end{equation}
where for an elliptic fibration $a_n$ are  sections of line
bundles $\mathcal{O}(-nK_B)$.  The general form (\ref{T}) can be
related to the standard Weierstrass form (\ref{W}) by completing the
square in $y$ and shifting $x$, which gives the relations
\begin{eqnarray}
b_2 &= & a_1^2 +4a_2, \label{eq:tw1}\\
b_4 &= & a_1a_3 +2a_4,\\
b_6 &= & a_3^2 + 4a_6,\\
b_8 &= & b_2a_6 - a_1a_3a_4 + a_2a_3^2 - a_4^2,\\
f &= & -\frac{1}{48}(b_2^2 -24b_4),\\
g &= & -\frac{1}{864}(-b_2^3 + 36b_2b_4 -216b_6),\\
\Delta &= & -b_2^2b_8 -8b_4^3 -27b_6^2 + 9b_2b_4b_6.
\label{TtoW}
\end{eqnarray}
An advantage of the general form (\ref{T}) is that by
requiring specific vanishing orders
of the $a_n$'s according to Table \ref{tab:tatealg}, specific desired
vanishing orders of $(f,g,\Delta)$ can be arranged to implement any of
the possible
gauge algebras.  Moreover, the monodromy conditions in Table
\ref{monocon} imposed by some gauge algebras on $f$, $g$, or $\Delta$
are also satisfied automatically by these ``Tate form'' models. 
For example, for tunings of fiber types
$I_m$ or $I_m^{*}$ where $\Delta$ is required to vanish to a
certain order while  ord($f$) and ord($g$) are kept fixed, the
vanishing order of $a_n$'s prescribed by the Tate algorithm 
immediately give the desired ord($\Delta$). 
This makes the Tate form  much more
convenient for constructing these singular fibers by only requiring
the order of vanishing of the $a_n$'s to be specified, in contrast to
the Weierstrass form (\ref{W})
where it is necessary to carefully tune the
coefficients of $f$ and $g$ to arrange for a vanishing of $\Delta$ to
higher order.
The Tate forms described in Table~\ref{tab:tatealg} are also connected
very directly to the geometry of reflexive polytopes.  As we discuss
in the subsequent sections, tuning a Tate form can be described
by simply removing certain monomials from the general form (\ref{T}),
which corresponds geometrically to removing certain points from a
lattice in the toric construction.
We refer to tunings of this type as ``Tate tunings'' in contrast to tunings of the coefficients of $f$ and $g$; when applied to the polytope toric construction, we refer to Tate tunings as ``polytope tunings''.

Note that Table \ref{tab:tatealg} has incorporated some results of the
present study into the Tate table originally described in
the F-theory context in
\cite{Bershadsky:1996nh} and later modified in \cite{Katz:2011qp}.
The most significant new feature is an alternate Tate form for the
algebras $\gsu(2n)$, with $a_2$ vanishing to order 2.  For $n = 3$, in
particular, this Tate form gives a tuning with exotic 3-index
antisymmetric SU(6) matter.  An example of a polytope that realizes
this tuning is described in \S\ref{2ndversion}.  For higher $n$, in
cases where $a_1$ is a constant --- i.e. on curves of
self-intersection $-2$ --- this simply gives an alternate Tate tuning
of SU($2n$).  On any other kind of curve, at the codimension two loci
where $a_1 = 0$ there is a codimension two (4, 6) singularity when $n
> 3$.
This can immediately be seen from the fact that at the locus $a_1 =
0$, each $a_k$ vanishes to order $k$ so that
(\ref{eq:tw1}--\ref{TtoW}) give a vanishing of $(f, g, \Delta)$ to
orders $(4, 6, 12)$.
Resolving this singularity generally involves blowing up a point on
the base, so that the resulting elliptic fibration is naturally
thought of as living on a base with larger $h^{1,1}$, but this kind of
Tate model for SU(8) and higher would be relevant in a complete
analysis of all reflexive polytopes.

We have also identified Tate tunings of $\gso(4n + 4)$, 
like those of
$\gso(4n + 2)$ that do not require an extra monodromy condition  and
only require the vanishing order of $a_i$'s; this arises naturally
in the context of the geometric constructions of polytopes.  We
discuss briefly how these two types of Tate tunings are relevant in
the constructions of this paper.  For $\gso(4n + 4)$, if $a_6$ is of
order $2n + 1$, then the necessary monodromy condition is that
\cite{Katz:2011qp, Grassi-Morrison} $(a_4^2 -4a_2a_6)/z^{2n+ 2}|_{z =
  0}$ is a perfect square.  This condition is clearly automatically
satisfied if $a_6$ is actually of order $2n + 2$, so can be guaranteed
simply by setting certain monomials in the Tate coefficients to vanish
(in a local coordinate system, which can become global in the toric
context used in the later sections of the paper).  On the other hand,
if the leading terms in $a_2, a_4, a_6$ are each constrained to be powers of a single
monomial $m, m^{n + 1}, m^{2n + 1}$, then the monodromy condition will
be automatically satisfied with $a_6$ of order $2n + 1$ without
specifying any particular coefficients for these monomials.  We
encounter both kinds of situation in this paper.  For $\gso(8)$, the
monodromy condition
when $a_6$ is of order 4
 is that $(a_2^2 -4a_4)/z^2|_{z = 0}$ is a perfect
square \cite{Katz:2011qp}.   
\footnote{To relate this to the condition stated in
Table~\ref{monocon}, note that the leading term in the discriminant when that condition is
satisfied becomes $-(a-b)^2 (2a + b)^2 (2b + a)^2$, so that condition
implies the perfect square condition.  Going the other way, when the
perfect square condition is satisfied we can determine $a, b$ 
by noting that $a_2/3$ is one of the roots $a, b, -a-b$ of the cubic
$x^3+ f_2x + g_3$, so without loss of generality we have
$a = a_2/3$, and solving for $b$ gives $b = -a_2/6 + (a_2^2 -4a_4)^{1/2}/2$.}
This can be satisfied if $a_2, a_4$ contain only a single
monomial each $m, m^2$
at leading order, but cannot be imposed by simply setting the
orders of vanishing of each $a_i$.
The situation is similar when $a_6$ is of order 3, though the
monodromy condition is more complicated when $a_2, a_4, a_6$ are not
single monomials $m, m^2, m^3$.
This is the only gauge algebra with
no monodromy-independent Tate tuning except through this kind of
single monomial condition.  Finally, for $\gso(4n + 2)$,
with $a_6$ of order $2n$, the monodromy condition is that $(a_3^2 +
4a_6)/z^{2n}|_{z = 0}$ is a perfect square, satisfied in particular if $a_6$ is
actually of order $2n + 1$ or if
the leading terms in $a_3, a_6$ are each a single monomial
proportional to $m, m^2$.  We explore further, for example, in Section
\ref{so8} for $\gso(12)$ the subtleties in using the Tate tuning \{1,
1, 3, 3, 5\} described in \cite{Bershadsky:1996nh}, which requires an
additional monodromy condition, vs. our alternative tuning \{1, 1, 3,
3, 6\}; In fact, analogous situations occur in tuning all gauge
algebras with monodromies.

\subsection{The Zariski decomposition}
\label{zar}
\label{zariski}
A central feature of the geometry of an F-theory base surface  is
the structure of the intersection form on curves (divisors) in $B_2$.  The
intersection form on $H_2 (B, {\mathbb Z})$ has signature (1, $T$).
Curves of negative self-intersection $C \cdot C < 0$ are rigid.  A
  simple but useful algebraic geometry identity, which follows from
  the Riemann-Roch theorem, is that

\begin{equation}
C \cdot (C + K_B) = 2g-2 \,,
\label{eq:genus-relation}
\end{equation}
for any curve $C$ of genus $g$.  We are primarily interested in
rational (genus 0) curves, for which therefore $C \cdot C = -K_B \cdot
C-2$.  All toric curves on a toric base $B_2$ are rational, and the
intersection product of toric curves has a simple structure that we
review in the following section.

To study the orders of vanishing of $f$, $g$ and $\Delta$ along some
irreducible divisors in the base, aside from looking explicitly at the
sets of monomials of $f$, $g$ and $\Delta$, it is convenient to
consider the more abstract ``Zariski decomposition'', in which an
effective divisor $A$ is decomposed into (minimal) 
multiples of irreducible effective divisors $C_i$ of negative
self-intersection and a residual part $Y$
\begin{equation}
A =\sum_i  q_i C_i+Y, \phantom{0}  q_i  \in \mathbb{Q},
\end{equation}
where $Y$ is effective and satisfies
\begin{equation}
 Y\cdot C_i = 0, \phantom{0} \forall i.
 \label{inequal}
\end{equation}
Then the  order of vanishing
along the curve $C_i$ of a section of the line bundle corresponding to
the divisor $A$  must be at least $c_i=\lceil q_i\rceil$. 
Mathematically, the Zariski decomposition is normally considered over
the rationals, so $q_i \in\Q$.  Here, however, we are simply
interested in the smallest integer coefficient of $C_i$
compatible with the decomposition over the ring of integers.
For example, consider the decomposition   
\begin{equation}
-nK_B=\sum_i c_i C_i + Y
\label{decom}
\end{equation}
The goal is to find the minimal set of integer
values $c_i$ such that the conditions $Y\cdot C_i\geq 0$ are satisfied. 
Taking the intersection product on both sides with $C_j$,  the conditions can be rewritten as the set of inequalities
\begin{equation}
\label{inequality}
%-nK\cdot C_j=\sum_i c_i C_i\cdot C_j + X\cdot C_j
%-nK\cdot C_j-\sum_i c_i C_i\cdot C_j \geq 0.
v_{j,n}-
\sum_{i} M_{ji} c_i \geq 0\,\, ,\,\, \forall j,
\end{equation}
where $M_{ji}\equiv C_j\cdot C_i$ are pairwise intersection
numbers (non-negative for $i\neq j$) and self-intersection numbers
$M_{jj}=C_j\cdot C_j\equiv m_j$, and $v_{j,n}\equiv-nK_B\cdot C_j$.

The Zariski decomposition of $-4K_B$ and $-6K_B$ was used in
\cite{clusters} to analyze the non-Higgsable clusters that can arise
in 6D theories.  More generally, we can use the same approach to
analyze models where we tune a given gauge factor on a specific
divisor beyond the minimal content specified by the non-Higgsable
cluster structure.  In such a situation, we would choose by hand to
take some values of $c_i$ in (\ref{decom}) to be larger than the
minimal possible values; this may in turn force other coefficients
$c_j$ to increase.  As a simple example, consider a pair of $-2$
curves (i.e.\ curves of self-intersection $-2$) $C, D$ that intersect
at a point ($C \cdot D = 1$).  The Zariski decomposition of the
discriminant locus gives simply $ -12K_B = Y$, since $K_B \cdot C = K_B
\cdot D = 0$ from (\ref{eq:genus-relation}), so the discriminant need
not vanish on $C$ or $D$.  If, however, we tune for example an 
$\gsu(4)$
gauge algebra on $D$ so that $\Delta$ vanishes to order 4 on $D$ then we
have the Zariski decomposition $-12K_B-4D = 2C + Y'$, since
$-4D \cdot C = -4$, implying that $\Delta$ must also vanish to order 2
on $C$, so that $C$ must therefore
also carry at least an $\gsu (2)$ gauge
algebra.

\subsection{Zariski decomposition of a Tate tuning}
\label{sec:Zariski-Tate}

A particular application of the Zariski decomposition that we use here
extensively is in the context of a Tate tuning.
In particular, assume that we have an elliptic fibration in
 the Tate form (\ref{T})
over a complex surface base $B$, and we have a set of curves $C_j$ in
the base that includes all curves of negative self-intersection.
The parameter space of the elliptic fibration is given by the five sections
$a_{n}\in \mathcal{O}(-nK), n=1,2,3,4,6$. 
We denote by $c_{j, n}$ the order of vanishing of $a_n$ on $C_j$.
The minimal necessary order of vanishing of each $a_n$ on each curve
$C_j$ can be determined by applying the Zariski decomposition for
$-nK$.  This gives rise to a set of  vanishing orders 
$c_{j, n}$ associated with each non-Higgsable cluster, which we list
in
Table~\ref{c0}.
These are the minimal values $c_{j, n} = c^{{\rm NHC}}_{j, n}$ that satisfy the
inequalities
(\ref{inequality}) for each value of $n \in\{1, 2, 3, 4, 6\}$.
In doing a Tate tuning,  
we impose the additional condition that over certain curves $C_j$, the
vanishing order is  at least some specified value that is
higher than the minimum imposed by the NHCs,
 $c_{j, n} \geq c^{{\rm tuned}}_{j, n} \geq c^{{\rm NHC}}_{j, n}$.  We can then
use the Zariski decomposition to determine the minimum values of the
$c_{j, n}$ compatible with this lower bound that also satisfy the
inequalities (\ref{inequality}).

More concretely, to determine the unique minimum set of values $c_{j,
  n}$ that satisfy the inequalities (\ref{inequality}), we proceed
iteratively, following an algorithm
described in appendix A of \cite{clusters}.
For each $n$,
we begin with an initial assignment of vanishing orders
\begin{equation}
c^{(0)}_{j,n} = c^{{\rm tuned}}_{j, n}
\label{eq:0}
\end{equation}
when we are imposing a given tuning.  When we are computing the
minimal values from NHC's without tuning we simply use the minimal
order of vanishing from the Zariski decomposition on each isolated curve
of self-intersection $m_j = M_{jj}$,
\begin{eqnarray}
&& c_{j,n}^{(0)} = \begin{cases}
 \left\lceil\frac{n(2+m_j)}{m_j}\right\rceil, & m_j\leq -3, \\
  0, & m_j> -3\,.
\label{isoNHC}
\end{cases}
\end{eqnarray}
We can then use the inequalities (\ref{inequality}) to determine the
minimal correction that is needed to each vanishing order (label $n$
dropped for clarity of the notation),
\begin{equation}
\Delta c_j^{(1)}=\text{Max}\left(0,\left\lceil\frac{v_j-
\sum_{i} M_{ji}\left.(c_i^{(0)}\right.)}{m_j}\right\rceil  \right).
\end{equation}
The  second corrections are obtained similarly, replacing $c^{(0)}$ on
the RHS with $c^{(1)} = c^{(0)} + \Delta c_j^{(1)}$.  We continue to repeat this procedure until
the corrections in the $f$-th step all become zero, $\Delta
c_j^{(f)}=0$ for all $j$. 
The final solutions $\{c_j\}$ are obtained iteratively this way  by
adding the non-negative correction values  $\{\Delta c_j^{(k)}\}$:
\begin{eqnarray}
\nonumber
&&c_j=c_j^{(0)}+\Delta c_j^{(1)}+\Delta c_j^{(2)}+\cdots+0,\\
&&\text{where }\Delta
c_j^{(l+1)}=\text{Max}\left(0,\left\lceil\frac{v_j-
\sum_{i}
M_{ji}\left.(c_i^{(0)}+\sum_{k=1}^l \Delta c_i^{(k)}\right.)}{m_j}\right\rceil  \right).
\label{iterat}
\end{eqnarray}
At each step this algorithm clearly increases the orders of vanishing
in a minimal way, so when the algorithm terminates the solution is
clearly a minimal solution of the inequalities (\ref{inequality}).
Note that in some cases, the algorithm leads to a runaway behavior
when there is no acceptable solution without (4, 6) loci.  When this
occurs, or when one of the factors of the gauge algebra exceeds that
desired by the tuning,
we terminate the 
algorithm and do not consider this tuning as a viable possibility.

As an example, consider the set of curves $\{C_j\}$ to be the
NHC $\{-3, -2\}$, so $M_{ji}=\{\{-3, 1\}, \{1, -2\}\}$, and
\begin{eqnarray}
\nonumber
&&\{\{v_1,v_{2}\}\rvert n=1,2,3,4,6\}=\{\{-1,0\},\{-2,0\},\{-3,0\},\{-4,0\},\{-6,0\}\},\\\nonumber
&&\{\{c_{1, n}^{(0)},c_{2, n}^{(0)}\}\rvert n\}=\{\{1,0\},\{1,0\},\{1,0\},\{2,0\},\{2,0\}\}.
\end{eqnarray}
Then the vanishing orders calculated from (\ref{iterat}) are
$\{c_{1,n}\}=\{1, 1, 2, 2, 3\}$ and $\{c_{2,n}\}=\{1, 1, 1, 1, 2\}$,
as shown
in Table \ref{c0}. 

Note that a tuning beyond that shown in Table~\ref{c0} does not
necessarily increase the gauge group on any of the curves.  In
particular, for some gauge groups there are multiple possible Tate
tunings.  Both for the generic gauge group associated with the generic elliptic fibration over a given base
and for constructions with gauge groups that are enhanced through a Tate tuning, this means that there may be distinct Tate
tunings with the same physical properties.  As we will see later,
these distinct Tate tunings can correspond through distinct polytopes
to different Calabi-Yau threefold constructions.
Note also that for the toric bases we are studying here, an
essentially equivalent analysis could be carried out by explicitly
working with the various monomials in the sections $a_n$, which in the
toric context are simply points in a dual lattice, as we discuss in
the next section.  We use the Zariski procedure because it is more
efficient and more general; the results of this analysis should,
however, match an explicit toric computation in each case.

\begin{table}[]
\centering
\begin{tabular}{|c|c|}
\hline
NHC                                  & $\{c^{{\rm NHC}}_{j,n}\}$                                                   \\ \hline
\{-3\}                                   & \{\{1, 1, 1, 2, 2\}\}                                                                \\ \hline
\{-4\}                              & \{\{1, 1, 2, 2, 3\}\}                                                                \\ \hline
\{-5\}                           & \{\{1, 2, 2, 3, 4\}\}                                                                \\ \hline
\{-6\}                         & \{\{1, 2, 2, 3, 4\}\}                                                                \\ \hline
\{-7\}                      & \{\{1, 2, 3, 3, 5\}\}                                                                \\ \hline
\{-8\}                        & \{\{1, 2, 3, 3, 5\}\}                                                                \\ \hline
\{-12\}                      & \{\{1, 2, 3, 4, 5\}\}                                                                \\ \hline
\{-3, -2\}                           & \{\{1, 1, 2, 2, 3\}, \{1, 1, 1, 1, 2\}\}                                         \\ \hline
\multicolumn{1}{|l|}{\{-3, -2, -2\}} & \multicolumn{1}{l|}{\{\{1, 1, 2, 2, 3\}, \{1, 1, 2, 2, 2\}, \{1, 1, 1, 1, 1\}\}} \\ \hline
\multicolumn{1}{|l|}{\{-2, -3, -2\}} & \multicolumn{1}{l|}{\{\{1, 1, 1, 1, 2\}, \{1, 2, 2, 2, 4\}, \{1, 1, 1, 1, 2\}\}} \\ \hline
\end{tabular}
\caption{The minimal vanishing orders of sections  $a_{1,2,3,4,6}$ over NHCs}
\label{c0}
\end{table}

\subsection{Matter content from anomaly constraints in F-theory}
\label{anomaly}

Six-dimensional $\mathcal{N}=(1,0)$ supergravity theories potentially 
suffer from gravitational, gauge, and mixed
gauge-gravitational anomalies.
We focus here primarily on nonabelian gauge anomalies, though similar
considerations hold for abelian gauge factors.
 On the one hand, the anomaly
information can be encoded in an $8$-form $I_8$,  which is
built from the 2-forms characterizing the non-abelian field strength $F$ and the
Riemann tensor $R$, and which has  coefficients
that can be computed in terms of $T, V, H$, and the
explicit numbers of chiral matter
fields in different representations.
On the other hand, the anomalies can be cancelled
through a generalized Green-Schwarz term if $I_8$ factorizes for some
constant coefficients $a^\alpha, b_i^\beta$ in the vector space
$\mathbb{R}^{1,T}$
associated with self-dual and anti self-dual two-forms $B_{\mu \nu}$
in the gravity and tensor multiplets,
\begin{eqnarray}
I_8=\frac1{2}\Omega_{\alpha\beta}X^\alpha_4X_4^\beta,
\end{eqnarray}
where
\begin{equation}
X^\alpha_4=\frac1{2}a^\alpha\tr R^2+\sum_i b_i^\alpha\frac{2}{\lambda_i}\tr F_i^2\,.
\end{equation}
Here $\Omega_{\alpha\beta}$ is a signature $(1,T)$ inner product on
the vector space,
and $\lambda_i$ are normalization constants for
the non-abelian gauge group factors $G_i$. Then, using the notation
and conventions of \cite{KMT-II}, the conditions for
anomaly cancellation are obtained by equating the coefficients of each
term from the two polynomials
\begin{eqnarray}
R^4&:&\phantom{0} H-V=273-29T, \label{eq:anomaly-r4}\\
F^4&:&\phantom{0} 0=B^i_{Adj}-\sum_R x^i_RB^i_R,\\
(R^2)^2&:&\phantom{0} a\cdot a=9-T,\\
\label{f2r2}
F^2R^2&:& \phantom{0} a\cdot b_i =\frac1{6}\lambda_i\left(A^i_{Adj}-\sum_R x^i_R A^i_R\right),\\
\label{f22}
(F^2)^2&:& \phantom{0} b_i \cdot b_i=\frac1{3} \lambda_i^2\left(\sum_R x_R^iC_R^i-C^i_{Adj}\right),\\
\label{f2f2}
F_i^2F_j^2&:&\phantom{0} b_i\cdot b_j=2\sum_{R,S} x_{RS}^{ij}A_R^iA_S^j, \phantom{0} i\neq j,
\end{eqnarray}
where $A_R, B_R, C_R$ are group theory coefficients\footnote{A 
  summary of $A_R, B_R, C_R$ in different representations and
  $\lambda_i$ for different non-abelian gauge groups can be found
  in appendix B in \cite{Johnson:2016qar}.} defined by
\begin{eqnarray}
\tr_R F^2&=&A_R\tr_\text{fund.}F^2,\\
\tr_R F^4&=&B_R \tr_\text{fund.}F^4+C_R(\tr_\text{fund.}F^2)^2 \,,
\end{eqnarray}
%and
$x_R^i$
is the number of matter
fields
\footnote{For each
  representation the matter content contains one complex scalar field
  and a corresponding field in the conjugate representation. For
  special representations like the $\mathbf{2}$ of $SU(2)$, the
  representation is pseudoreal, so that the conjugate need not be
  included; such a field is generally referred to as a ``half-hypermultiplet''.} in the representation $R$
 of the non-abelian factor $G_i$, and
$x_{RS}^{ij}$ is the number of matter
fields in the $(R, S)$-representation of
$G_i\otimes G_j$.

For 6D theories coming from an F-theory compactification, the vectors
$a, b^i$ are related to homology classes in the base $B_2$ through the
relations
\begin{eqnarray}
a & \leftrightarrow & K_B,\\
b_i &\leftrightarrow & C_i,
\end{eqnarray}
where, again, $K_B$ is the canonical class of $B_2$, and $C_i\in
H_2(B_2,\mathbb{Z})$ are irreducible curves in the base supporting the
singular fibers associated with the non-abelian gauge group factors
$G_i$.  With this identification, the Dirac inner products between vectors
in $\mathbb{R}^{1,T}$ are related to intersection products between divisors in
the base.

In principle, the matter content of a 6D theory can be determined by a
careful analysis of the codimension two singularities in the geometry.
In many situations, however, the generic matter content of a
low-energy theory is uniquely determined by the gauge group content
and anomaly cancellation simply from the values of the vectors $a,
b^i$.  For example, a theory with an  SU($N$) gauge factor associated
with a vector $b$ generically has $g$ adjoint matter fields, $(8-N) n
+ 16 (1-g)$ fundamental matter fields, and $(n + 2-2g)$ two-index
antisymmetric matter fields, where $n = b \cdot b$ and $g = 1 + (a
\cdot b + n)/2$ (see e.g. \cite{Johnson:2016qar}); this simplifies in
the $g = 0$ case of primary interest to us here to a spectrum of $n +
2$ two-index antisymmetric matter fields
and $16 + (8-N) n$ fundamental fields.  For most of the 
theories we consider here the matter content follows uniquely in this
way from the values of $a, b^i$.  In some situations,
however, more exotic matter representations can arise; we encounter
some cases of this later in this paper, such as the three-index
antisymmetric representation of SU(6).

In general, the anomaly constraints on 6D theories provide a powerful
set of consistency conditions that we use in many places in this paper
to analyze and check various models that arise through tunings; in
particular, using the anomaly conditions to determine the matter
spectrum gives a direct and simple way in many cases to compute the
Hodge numbers of the associated elliptic Calabi-Yau manifold that can
be matched to the Hodge numbers of a toric hypersurface construction.

\subsection{Global symmetry constraints in F-theory}
\label{rpt}

In many cases, the anomaly cancellation conditions impose constraints
not only on the matter content of the theory but also on what gauge
groups may be combined  on intersecting curves, corresponding to
vectors $b^i$ with non-vanishing inner products in the low-energy
theory.  For example, two gauge factors of 
$\gg_2$ or larger in
the Kodaira classification cannot be associated with vectors $b, b'$
having $b \cdot b' > 0$; in the low-energy supergravity theory this is
  ruled out by the anomaly conditions while in the
F-theory picture this would correspond
  to a configuration with a codimension two (4, 6) point at the
  intersection between the corresponding curves.  In addition to these
  types of constraints, another set of constraints on what combination
  of gauge groups can be tuned on specific
negative self-intersection curves in a base $B_2$ 
can be derived from the low-energy theory by considering the maximum
global symmetry of an SCFT that arises by shrinking a curve $C$ of
self-intersection $n< 0$ that supports a given gauge factor $G_i$
\cite{Bertolini:2015bwa}.  While in most cases these global symmetry
conditions simply match with the expectation from anomaly
cancellation, in some circumstances the global symmetry condition
imposes stronger constraints.  For example the ``$E_8$ rule''
\cite{Heckman:2013pva} 
states
that the maximal global symmetry on a  $-1$ curve 
that does not carry a nontrivial gauge algebra is $\ge_8$; i.e.,  the
direct sum of the gauge algebras carried by the curves intersecting
the $-1$ curve should be a subalgebra of $\ge_8$.
While the global symmetry constraints are completely consistent with
F-theory geometry, they may not be a complete and sufficient set of
constraints; for example a similar constraint appears to hold in
F-theory for the algebras on a set of curves intersecting a 0 curve
\cite{Johnson:2016qar}, though the low-energy explanation for this is
not understood in terms of global constraints from SCFT's.

%rank preserving tunings
%Be careful in the cases of product gauge groups. Trick: use the lowest dimension groups in the calculation.
%[$h^{2,1}$ from monomial counting]

The maximal global symmetry groups realized in 6D F-theory for each
possible algebra on a curve of self-intersection $m\leq -1$ are worked
out in \cite{Bertolini:2015bwa}. We  use their results 
%in Tables 6.1 and 6.2
 in our algorithm to constrain possible gauge algebra tunings.  More
 explicitly, given a gauge algebra on a curve, the maximal global
 symmetry on the curve is determined, so the direct sum of the
 algebras on the curves intersecting it should be a subalgebra of the
 maximal global symmetry algebra. For instance, consider a linear
 chain of three curves $\{C_1, C_2, C_3\}$ carrying gauge algebras
 $\{\gg_1, \gg_2, \gg_3\}$.  These can be either minimal or enhanced
 algebras, but they have to satisfy $\gg_3 \oplus\gg_1\subset
 \gg_{2}^{\text{(glob)}}$, where $\gg_{2}^{\text{(glob)}}$ is the maximal global
 symmetry algebra  given $\gg_2$ on the curve $C_2$, as enumerated in
 the tables in \cite{Bertolini:2015bwa}. This will be
 useful for us to constrain the possible tunings on a curve when the
 gauge symmetries on its neighboring curves are known, making our
 search over possible tunings more efficient. This is also convenient
 sometimes for us to determine the gauge algebras that have monodromy
 conditions without having to figure out the monodromy directly; the
 trick to doing this is described in \S\ref{so8}.  We also
 include the ``$E_8$ rule'' in our algorithm in \S\ref{algo}, corresponding to
 the case where $m = -1$ and $\gg_2$ is trivial.

\subsection{Base surfaces for 6D F-theory models}
\label{sec:bases}

There is a finite set of complex base surfaces that support elliptic
Calabi-Yau threefolds.  It was shown by Grassi \cite{Grassi} that all
such bases can be realized by blowing up a finite set of points on the
minimal bases $\P^2, \F_m$ with $0 \leq m \leq 12$, and the Enriques
surface.  This leads to a systematic constructive approach to
classifying the set of allowed F-theory bases.  The structure of
non-Higgsable clusters limits the configurations of negative
self-intersection curves that can arise on any given base, so we can
in principle construct all allowed bases by blowing up points in all
possible ways and truncating the set of possibilities when a
disallowed configuration such as a curve of self-intersection $-13$ or
below arises.  This was used in \cite{Morrison:2012js} to classify the full set
of toric bases $B_2$ that can support elliptic Calabi-Yau threefolds
(toric geometry is described in more detail in the following
section). While further progress has been made \cite{Martini-WT,
  Wang-WT} in classifying non-toric bases, we focus here primarily on
toric base surfaces, as these are the primary bases that arise in the
toric hypersurface construction of Calabi-Yau threefolds.  
Note,
however, that as we discuss later in the paper, particularly in
e.g.\ \S\ref{multiplicity}, \S\ref{semitoric}, there are cases in the
Kreuzer-Skarke database where a toric polytope corresponds to an
elliptic fibration over a non-toric base.  The primary context in
which this distinction is relevant involves curves of
self-intersection $-9, -10,$ and $-11$.  As discussed in
\cite{clusters}, the Weierstrass model over such curves automatically
has 1, 2, or 3 points on the curve where $f, g$ vanish to degrees $(4,
6)$.  Such points on the base can be blown up for a smooth Calabi-Yau
resolution, so that the actual base supporting the elliptic fibration
is generally a non-toric complex surface.\footnote{More precisely, as
  described in \cite{Braun:2011ux} and \S\ref{NHC},
and discussed in more detail in
  \S\ref{multiplicity}, the original base supports an elliptic
  fibration that is ``non-flat,'' meaning that the fiber becomes two
  dimensional at some points, while the elliptic fibration over the
  blown up base is a flat elliptic fibration.}  In
the simplest cases, such as $\F_{11}$ and $\F_{10}$, the blown up base
 still has a toric description; in other
 simple cases, such as
$\F_{9}$, the resulting surface is a ``semi-toric'' surface
admitting only a single $\C^*$ action
\cite{Martini-WT}, but on surfaces with, for example, multiple curves
of self-intersection $-9, -10, -11$, the blow-up of all (4, 6) points
in the base gives generally a non-toric
base that is neither toric nor admits a
single $\C^*$ action.  Despite this
complication, this blow-up and resolution process is automatically
handled in a natural way in the framework of the toric hypersurface
construction, so that (non-flat)
elliptic fibrations over bases with these types
of curves arise naturally in the Kreuzer-Skarke database.  Thus, we
include toric bases with curves of self-intersection $-9, -10, -11$ in
the set of bases we consider for Tate/Weierstrass constructions.  The
complete set of such bases was enumerated in \cite{Morrison:2012js}, where it
was shown that there are 61,539 toric bases that support elliptic CY3's.  Generic elliptic
Calabi-Yau threefolds over these bases give rise to a range of Hodge
number pairs that fill out the range of known Calabi-Yau Hodge
numbers, in a ``shield'' shape with peaks at (11, 491), (251, 251),
and (491, 11) \cite{Hodge}.
As we see in \S\ref{restcases}, in some cases the base needed for a
tuned Weierstrass model to match a toric hypersurface construction is
even more exotic than those arising from blowing up points on curves
of self-intersection $-9, -10, -11$.  In these more complicated cases
as well,
however, the general story is the same.  The polytope construction
gives rise to a non-flat
elliptic fibration with codimension two (4, 6) points
on the toric
base.  Blowing these curves up
gives rise to another, generically non-toric, base with multiple additional curves.  After these blow-ups, there
is a tuned Weierstrass model giving a
(flat) elliptic fibration over the
new base. While the toric base is what arises most clearly from the
polytope construction, the structure of the blown up base admitting
the flat elliptic fibration is relevant when considering F-theory
models and anomaly cancellation.

In \S\ref{sec:Tate} we consider Tate  tunings over
the toric bases and compare to the toric hypersurface construction
of elliptic Calabi-Yau threefolds, which we now describe in more
detail.

\section{Elliptic fibrations in the toric reflexive polytope construction}
\label{sec:reflexive}

\subsection{Brief review of toric varieties}
\label{toricbasic}
Following \cite{Fulton, mirror-symmetry}, we review some basic
features of toric geometry. An $n$-dimensional toric variety
$X_\Sigma$ can be constructed by defining the \emph{fan} of the toric
variety. A fan $\Sigma$ is a collection of cones\footnote{More
  rigorously, these are  {\em strongly convex cones}, which are
  generated by a finite number of vectors in $N$ and which contain no line through the origin.} in $N_\R$=
$N\otimes\R$, each with the apex at the origin, and
where $N$ is a rank
$n$ lattice, satisfying the conditions that
\begin{itemize}
\item Each face of a cone in $\Sigma$  is also a cone in $\Sigma$.
\item The intersection of two cones in $\Sigma$ is a face of each.
\end{itemize}
Then $X_\Sigma$ can be described by the homogeneous coordinates $z_i$
corresponding to the one-dimensional cones $v_i$ (also called rays) of
$\Sigma$; $X_\Sigma$ may be constructed as the quotient of an open subset
in $\C^k$ ($k$ is the number of rays), by a group $G$,
\begin{equation}
X_\Sigma=\frac{\C^k-Z(\Sigma)}{G},
\end{equation}
where 
\begin{itemize}
\item $Z(\Sigma)\subset \C^k$ is the union of the zero sets of the polynomial sets $\mathcal{S}=\{z_i\}$ associated with the sets of rays \{$v_i$\} that do not span a cone of $\Sigma$.
\item $G\subset (\C^*)^k$ is the kernel of the map
\[\phi: (\C^*)^k\rightarrow (\C^*)^n,
(z_1,..,z_k)\mapsto(\prod_{j=1}^{k} z_j^{v_{j,1}},\ldots,
\prod_{j=1}^{k} z_j^{v_{j,n}}),\] where $v_{j,l}$ specifies the
$l$th component of the ray $v_{j}$ in the coordinate representation in
$\C^n$. 
\end{itemize}
Toric divisors $D_i$ are given by the sets $D_i \equiv\{z_i=0\}$
associated to all the rays $v_i$. The anti-canonical class $-K$ of a
toric variety is given by the sum of toric divisors
\begin{equation}
-K =\sum_i D_i.
\label{anticlass}
\end{equation}
Smooth two-dimensional toric varieties are particularly simple.  The
irreducible effective toric divisors are rational curves with one
intersecting another forming a closed linear chain.  This is easily seen
from the 2D toric fan description, where each ray of the 2D fan
corresponds to an irreducible effective toric divisor. The intersection
products are also easy to read off from the fan diagram, where (including divisors cyclically by setting $D_{k+1}\equiv D_1$, etc.)
\begin{eqnarray}
D_i\cdot D_{i+1}=1  ,
\label{ijin}
\end{eqnarray}
and the self-intersection of each curve is
\begin{equation}
D_i\cdot D_i=m_i,
\label{iiin}
\end{equation}
where $m_i$ is such that
\begin{equation}
\label{self}
-m_iv_i= v_{i-1}+v_{i+1},
\end{equation}
and zero otherwise. We will generally denote the data defining a smooth 2D
toric base by the sequence of self-intersection numbers. (The 2D fan can be
recovered given the intersections, up to lattice automorphisms.)

%Therefore the data of a toric base is simply given by a sequence of the self intersection numbers of the toric divisors connecting one by one in the same order with the divisors corresponding to the two end positions also connecting to each other. In terms of the notation from the last paragraph we have $M_{ji}=m_j$ for $i=j$, $M_{ji}=1$ for adjacent $C_i$, $C_j$, and  $M_{ji}=0$ otherwise. 

%The base will be one of the data from 2D toric bases for elliptically fibered Calab-Yau threefolds that have been classified in \cite{toric}. There are some properties particularly simple in the 2-dimensional smooth toric varieties which are useful to know: 

In the context of this paper, toric varieties play two distinct but related roles. On the one hand, we can use toric geometry to describe many of the bases that support  elliptically fibered Calabi-Yau threefolds. On the other hand, toric geometry can be used to describe ambient fourfolds in which CY threefolds  can be embedded as hypersurfaces, as we describe in the next section.

\subsection{Batyrev's construction of Calabi-Yau manifolds from reflexive polytopes}

Given a \emph{lattice polytope}, which is the convex hull of a finite
set of lattice points (in particular, the vertices are lattice
points), we may define a 
{\em face fan} by taking rays to be the vertices of
the lattice polytope, and the top-dimensional ($n$-dimensional) cones to
correspond to the facets of the polytope.
By including more
lattice points in addition to vertices of the polytope as rays, 
and
thus subdividing (``triangulating'')
the facets of the polytope into multiple smaller top-dimensional
cones, we can refine the fan to impose further properties such as
simpliciality or smoothness.\footnote{A fan is simplicial if all its
  cones are simplicial. A cone is simplicial if its generators are
  linearly independent over $\R$. A fan is smooth if the fan is
  simplicial and for each top-dimensional cone its generators generate
  the lattice $N$. \label{footnote}} 
In this way, a lattice polytope can be associated with a toric
variety.  In general, a given lattice polytope can lead to many
different varieties, depending upon the refinement 
of the face fan. Even for a given set of additional rays added, there
can be many different triangulations of the fan.

We will be interested in
particular in the fans from \emph{reflexive polytopes,} which  are
defined as follows. Let $N$ be a rank $n$ lattice, $N_{\R}\equiv
N\otimes \R$.
A lattice polytope $\nabla\subset N$ containing the origin is  reflexive if its \emph{dual polytope} is also a lattice
polytope. The dual of a polytope $\nabla$ in $N$ is defined to be 
\begin{equation}
\nabla^*=\{u\in M_\R=  M\otimes \R: \langle u,v\rangle\geq-1, \forall v\in \nabla\},
\label{dual}
\end{equation}
where $M = N^*=\Hom(N,\Z)$ is the dual lattice.  If $\nabla$ is reflexive, its dual polytope $\Delta=\nabla^*$ is also reflexive as $(\nabla^*)^*=\nabla$. We call the pair of reflexive
polytopes a mirror pair. Both of them contain the origin as the only
interior lattice point.  Calabi-Yau manifolds in Batyrev's
construction \cite{Batyrev:1994hm} are built out of reflexive
polytopes. Given a mirror pair $\dd\subset N$ and $\ds\subset M$, the
(possibly refined) face fan of $\dd$ describes a toric ambient
variety, in which a Calabi-Yau hypersurface is embedded using the
anti-canonical class of the ambient toric variety, so that the
hypersurface itself has trivial canonical class. Explicitly, a section
of the anti-canonical bundle is given by
\begin{equation}
p=\sum_i^{\text{\# lattice points in $\ds$ }} c_im_i,
\label{p}
\end{equation}
where $c_i$ are generic coefficients taking values in $\C$ and each monomial $m_i$ is given by an associated lattice point $w_i$ in $\ds$
\begin{equation}
m_i=\prod_j z_j^{\langle w_i,v_j\rangle+1},
\label{m}
\end{equation}
where $z_j$ is the homogeneous coordinate associated with the ray
$v_j$ in the fan associated to $\dd$.   The well-definedness of each $m_i$ in terms of the homogeneous coordinates $z_j$
is guaranteed by the linear equivalence relations among the divisors
associated to $v_j$'s, and holomorphicity in the $z_j$s by the reflexivity of
$\dd$. 
We can check that Equation (\ref{p})
indeed defines a section of the anti-canonical class, so that a CY
hypersurface is cut out by $p=0$. 
We can determine the class by looking at any one of the
monomials; we pick the origin since we know it is always an
interior point. Its associated monomial by equation (\ref{m}) is
simply the product of all homogeneous coordinates associated to all
toric divisors $\prod_{j=1}^\text{\# rays} z_j$, which immediately we
see by equation (\ref{anticlass}) is a section in the anti-canonical
class. For the smoothness of the Calabi-Yau, as mentioned
previously, there exists a refinement\footnote{Appropriate subdivisions of the face
  fan of the toric ambient variety by additional lattice points in the facets of the polytope 
  give the resolved description of the embedded Calabi-Yau, where extra coordinates in
  equation (\ref{m}) define the exceptional divisors in the
  resolution of the ambient space.
The added lattice points that do not lie in the interior of the facets also
correspond to exceptional divisors in the resolution of the Calabi-Yau. (Generic hypersurface CYs do not meet the divisors that correspond to interior points of facets.) 
} of the face fan of $\dd$ such that the fan is
simplicial so the ambient toric variety will have at most orbifold
singularities. 
In the case of $n\leq4$, with the anti-canonical embedding, a
hypersurface will generically avoid these singularities and therefore
is generically smooth.

M. Kreuzer and H. Skarke have classified all 473,800,776  four-dimensional
reflexive polytopes for the Batyrev Calabi-Yau construction
\cite{database, palp}. A pair of reflexive polytopes in the KS database
are described in the format:
\begin{center}
M:$\#$ lattice points,  $\#$ vertices (of  $\ds$) N:$\#$ lattice points, $\#$ vertices (of $\dd$) H: $h^{1,1}$, $h^{2,1}$.
\end{center}
We will refer to $\dd$ as the (fa)N polytope and $\ds$ as
the M(onomial) polytope to remind ourselves that $\dd$ gives the fan of
the ambient toric variety for the CY anti-canonical embedding and
$\ds$ determines the monomials of the anti-canonical
hypersurface. In many cases, it is sufficient to specify polytopes
with the information given in the format above, but sometimes there can be
distinct polytopes with identical information of this type, in which case we will
either give further the vertices of the
N polytope to specify the polytope more precisely, or 
indicate its numerical order as it appears among those with identical data in the KS database website (\href{http://hep.itp.tuwien.ac.at/~kreuzer/CY/}{http://hep.itp.tuwien.ac.at/\~{}kreuzer/CY/}) with a
superscript, e.g., M:165 11 N:18 9 H:9,129$^{ [1]}$ or M:165 11 N:18 9
H:9,129$^{ [2]}$.

Note that conversely, we can start from $\ds$ and associate it with
the polytope that defines the fan of the ambient space, and calculate
monomials associated with lattice points in $\dd$. Then the
hypersurface CY is mirror to the previous one. The Hodge numbers of
mirror pairs are
related by $h^{p,q}(CY_{\dd})=h^{d-p,q}(CY_{\ds})$, where
$d=n-1$ is the complex dimension of the CY; in particular, we will
look at 4 dimensional reflexive polytopes for CY threefolds, where the
only non-trivial Hodge numbers are $h^{1,1}$ and $h^{2,1}$, and mirror
CY hypersurfaces have exchanged values for $h^{1,1}$ and
$h^{2,1}$. 
\iffalse
For the CY associated with $\dd$, the Hodge numbers are given by
\begin{eqnarray}
\label{latth21}
&&h^{2,1}=\text{pts}(\ds)-\sum_{\theta\in F^\ds_3}	\text{int}(\theta)+\sum_{\theta\in F^\ds_2} \text{int}(\theta)\text{int}(\tilde{\theta})-5,\\
\label{latth11}
&&h^{1,1}=\text{pts}(\dd)-\sum_{\tilde{\theta}\in F_3^\dd}	\text{int}(\tilde{\theta})+\sum_{\tilde{\theta}\in F_2^\dd} \text{int}(\tilde{\theta})\text{int}(\theta)-5,
\end{eqnarray}
where $\theta$ are faces of $\ds$,  $\tilde{\theta}$ are faces of
$\dd$, $F^{\dd\slash\ds}_l$ denotes the set of  $l$-dimensional faces
of $\dd$ or $\ds$ ($l<n$), and $\text{pts}(\dd\slash\ds):=$ number of
  lattice points of $\dd$ or $\ds$,
  int$(\theta\slash\tilde{\theta}):=$ number of lattice points
  interior to $\theta$ or $\tilde{\theta}$. As $\dd$ and $\ds$ are a
  pair of 4D reflexive polytopes, there is a one-to-one
  correspondence between $l$-dimensional faces $\theta$ of $\ds$ and
  $(4-l)$-dimensional faces $\tilde{\theta}$ of $\dd$ related by the
  dual operation 
\begin{equation}
\theta^*=\{y\in \dd, \langle y, pt\rangle=-1 \rvert \text{ for all $pt$ that are vertices of $\theta$}\},
\end{equation}
which makes the duality between the Hodge number formulas
manifest. \fi 
As $\dd$ and $\ds$ are a
  pair of 4D reflexive polytopes, there is a one-to-one
  correspondence between $l$-dimensional faces $\theta$ of $\ds$ and
  $(4-l)$-dimensional faces $\tilde{\theta}$ of $\dd$ related by the
  dual operation 
\begin{equation}
\theta^*=\{y\in \dd, \langle y, pt\rangle=-1 \rvert \text{ for all $pt$ that are vertices of $\theta$}\} \,.\label{eq:dual-faces}
\end{equation}
For the CY associated with $\dd$, the Hodge numbers are given by
\begin{eqnarray}
\label{latth21}
&&h^{2,1}=\text{pts}(\ds)-\sum_{\theta\in F^\ds_3}	\text{int}(\theta)+\sum_{\theta\in F^\ds_2} \text{int}(\theta)\text{int}({\theta}^*)-5,\\
\label{latth11}
&&h^{1,1}=\text{pts}(\dd)-\sum_{\tilde{\theta}\in F_3^\dd}	\text{int}(\tilde{\theta})+\sum_{\tilde{\theta}\in F_2^\dd} \text{int}(\tilde{\theta})\text{int}(\tilde{\theta}^*)-5,
\end{eqnarray}
where $\theta$ are faces of $\ds$,  $\tilde{\theta}$ are faces of
$\dd$, $F^{\dd\slash\ds}_l$ denotes the set of  $l$-dimensional faces
of $\dd$ or $\ds$ ($l<n$), and $\text{pts}(\dd\slash\ds):=$ number of
  lattice points of $\dd$ or $\ds$,
  int$(\theta\slash\tilde{\theta}):=$ number of lattice points
  interior to $\theta$ or $\tilde{\theta}$. 
The correspondence (\ref{eq:dual-faces}) makes the duality between the Hodge number formulae
manifest.
Note that the Hodge numbers depend only on the polytope and not on the
detailed refinement of the fan.

\subsection{Fibered polytopes in the KS database}
\label{findbase}

For the purpose of studying
(often singular) elliptically fibered Calabi-Yau threefolds
 that arise in the KS database, we will be interested in  4D
reflexive \emph{fibered polytopes} \cite{Avram:1996pj, Kreuzer:1997zg,
  Skarke:1998yk, Braun:2011ux}. A fibered polytope $\dd$ is a 
polytope
in the $N$
lattice that contains a lower-dimensional  subpolytope,
$\dd_2\subset N_2 =\Z^2$, which
passes
through the
origin. 
We are interested in the case where $\dd_2$ is itself a reflexive 2D polytope, containing the origin as an interior point.  
Such
a fibered polytope $\dd$ admits a projection map $\pi: \dd
\rightarrow N_B$ such that $\pi^{-1}(0) = \dd_2$, and $N_B=\Z^2$ is
the quotient of the original lattice $N$ by the projection. We can construct
a set of rays $v_{i}^{(B)}$ in $N_B$ that are the primitive rays with the
property that an integer multiple of $v_{i}^{(B)}$ arises as the image
$\pi (v_i)$ of a ray in $\dd$.  (A \emph{primitive} ray $v \in N$ is one that
cannot be described as an integer multiple $v= nw$ of another ray $w
\in N$, with $n > 1$.)
%The convex hull of these rays defines a base polytope $\dd_{\rm base}$.
We define the base $B_2$ to be the 2D toric variety given by the 2D
fan $\Sigma_B$ with $v_{i}^{(B)}$ taken to be the 1D cones; the 2D
cones are uniquely defined for a 2D variety. Note that in higher
dimensions, the base of the fibration is not uniquely defined as a
toric variety since the cone structure of the base may not be unique.

In the toric geometry language,
a \emph{fan morphism} is a projection $\pi: \Sigma \rightarrow\Sigma_B$
with the property that  for any cone in $\Sigma$
the image is contained in a cone of $\Sigma_B$.  Such a fan morphism
can be
translated to a map between toric varieties  $\pi: X_\Sigma\rightarrow
B_2$. 
Such a map is a
\emph{toric morphism}, which is an equivariant map with respect to the
torus action on the toric varieties that maps the maximal torus in $X_\Sigma$
to the maximal torus in $B_2$. 
As far as the authors are aware,
it is not known whether in general every fibered polytope admits a
triangulation leading to a compatible fan morphism and toric morphism.
Note, however, that the elliptic fiber structure of the polytope does
not depend upon the existence of a triangulation with respect to which there is 
a fan morphism
 $\pi: \Sigma_\dd \rightarrow \Sigma_B$.  Thus, to recognize an elliptic Calabi-Yau
threefold in the KS database, it is only necessary to find a reflexive
subpolytope $\dd_2 \subset\dd$. 
The
Calabi-Yau manifold
defined by an anti-canonical hypersurface in $X_\Sigma$
through the Batyrev construction with reflexive polytopes will then be an elliptically
fibered Calabi-Yau threefold over the base $B_2$
\cite{Kreuzer:1997zg}.
A primary goal of this paper is to relate reflexive polytopes in the
Kreuzer-Skarke database that have this form to
elliptic fibrations of tuned Weierstrass models as described in 
\S \ref{sec:F-theory}.

There are in total 16 2D reflexive polytopes, which give
slightly different realizations of an elliptic curve when an
anti-canonical hypersurface is taken \cite{Braun:2011ux, Braun-16,
  Klevers-16}.  The hypersurface equations $p = 0$, with $p$ given in
 (\ref{p}), of all 16 types
of fibered polytopes can be brought into the Weierstrass form
 (\ref{W}) by the methods described in Appendix A in
\cite{Braun:2011ux}; this gives an equivalent description of the same Calabi-Yau as long as the fibration  has a global section.  The Kreuzer-Skarke database of reflexive
polytopes and associated Calabi-Yau hypersurface constructions
contains a wide range of polytopes that include fibered polytopes with
many different examples of the 16 fiber types.  

For a given base $B_2$
and a given fiber type, there can be a variety of different polytopes
corresponding to configurations with different ``twists'' of the
fibration, associated with different embeddings of the rays $v_i$ defining
the base $B_2$ with respect to the fiber
subpolytope $\dd_2$.  For example, the Hirzebruch surfaces $\F_m$ are
each associated with fibered polytopes with fiber and base $\P^1$,
distinguished by the different twists of the fibration.  For a fibered
polytope $\dd$ with a reflexive subpolytope $\dd_2$, the dual $\Delta$ 
admits a projection to $\ds_2=(\dd_2)^*$.

One of the findings of this paper is that the bulk of KS models with
large Hodge numbers appear to have a description in the form of a
\emph{standard} $\P^{2,3,1}$-fibered type, with a specific form for
the twist of the fiber over the base surface $B_2$; these models can
be connected directly to the Tate form for elliptic fibrations, and in
fact can be constructed from that point of view directly.  On the one
hand, we describe the structure of this type of standard polytope in
\S\ref{sp231}, with the result that the anti-canonical hypersurface
equations from (suitably refined) standard $\P^{2,3,1}$-fibered
polytopes are in the form of equation (\ref{T}).  On the other hand,
we describe the direct construction of polytopes by carrying out Tate
tunings on the effective curves in the toric bases in
\S\ref{sec:Tate}, and develop an algorithm in
\S\ref{sec:systematic-construction} to systematically classify models
of this type that give polytopes and elliptic Calabi-Yau threefolds
with large Hodge numbers; these models are all expected to have a
corresponding standard $\P^{2,3,1}$-fibered polytope, and we compare
the two constructions in the remainder of \S\ref{restcases}.  For a
given base $B_2$ there are generally many distinct polytopes that have
the standard $\P^{2,3,1}$-fibered structure; as we describe in the
following section, these correspond to different Tate tunings over the
same base.

\subsection{Standard $\P^{2,3,1}$-fibered polytopes
and corresponding Tate models}
\label{sp231}

\begin{figure}
\centering
\begin{subfigure}{.45\textwidth}
  \centering
  \includegraphics[width=7cm]{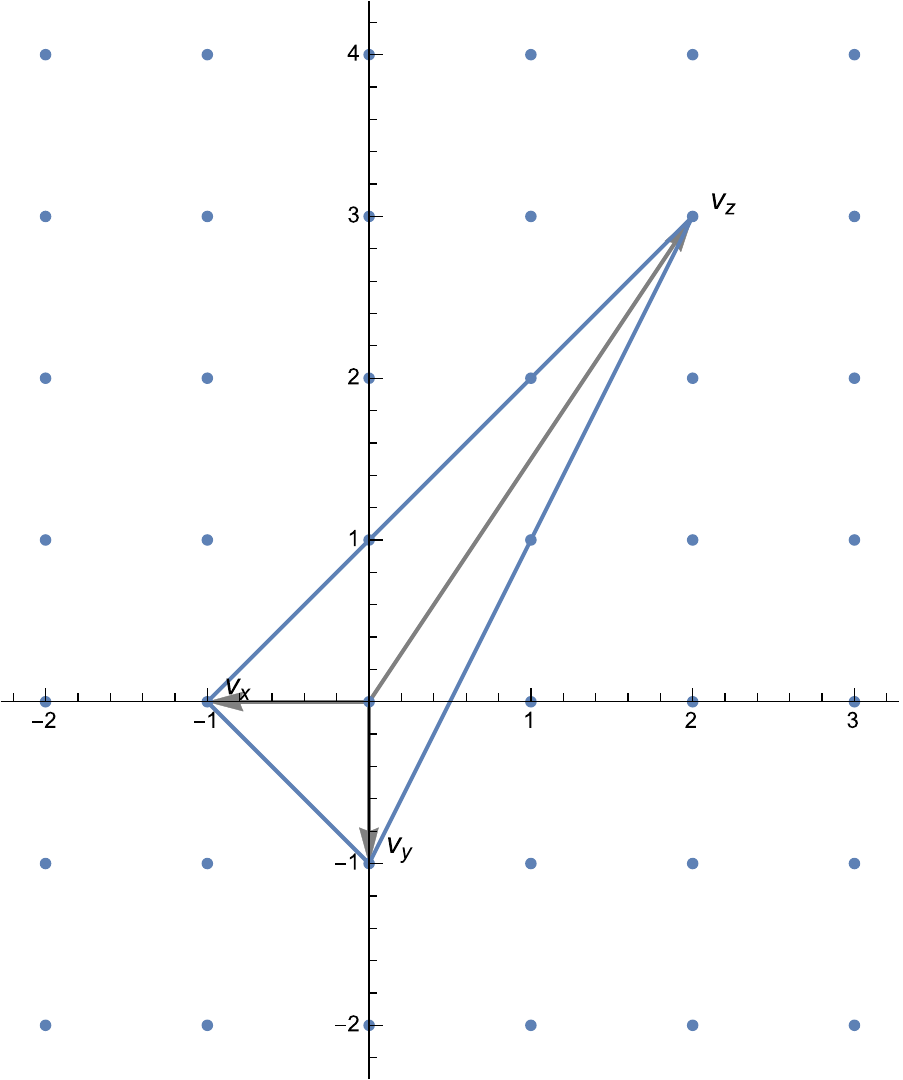}
  \caption{\footnotesize The toric fan for 
$\P^{2,3,1}$.  The convex hull of $\P^{2,3,1}$ plays the role of the
reflexive subpolytope $\dd_2$ for
    \emph{standard $\P^{2,3,1}$-fibered polytopes} $\dd$
in the $N$ lattice. The rays $v_x, v_y, v_z$ are associated with the homogeneous coordinates $x, y, z$, respectively, in the hypersurface equation.}
  \label{fig:sub1}
\end{subfigure}%
\hspace*{0.2in}
\begin{subfigure}{.45\textwidth}
  \centering
  \includegraphics[width=7cm]{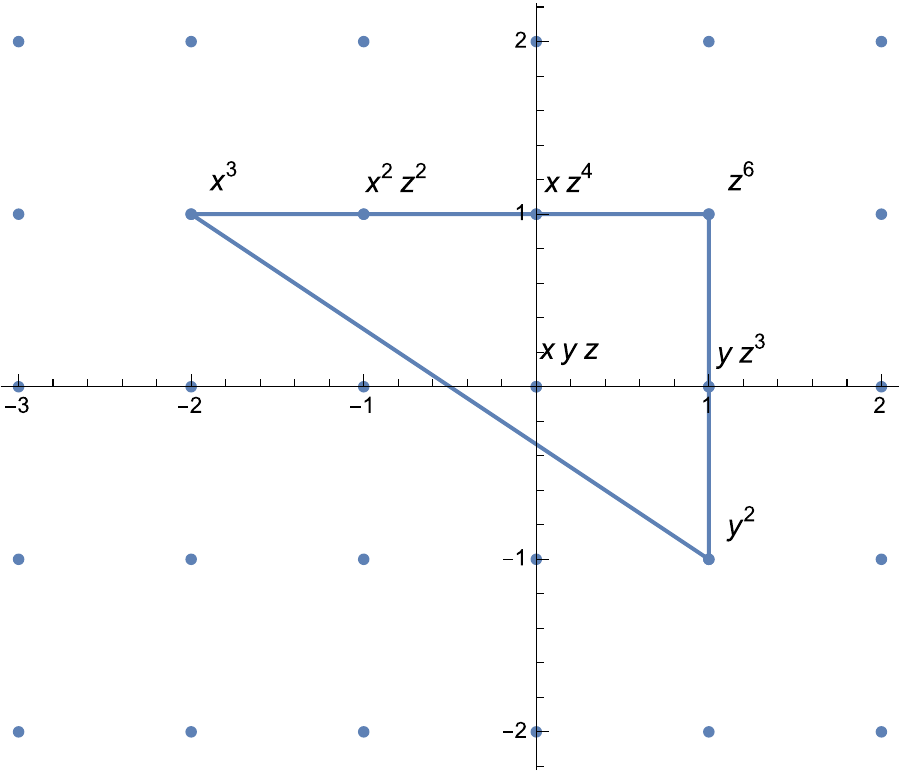}
  \caption{\footnotesize The dual polytope $\ds_2$ to $\dd_2$ in the
    $M_2$ lattice.   Projection onto the $M_2$ plane
    projects the lattice points in $\ds$ into seven lattice points in
    $\ds_2$.  These lattice points correspond to the five sections
    $a_{1, 2, 3, 4, 6}$ in the Tate form of the Weierstrass model,
    indicated in the figure by $xyz, x^2z^2, yz^3, xz^4, z^6$,
    respectively, and to the coefficients of the remaining two terms
    $x^3$, $y^2$ in the hypersurface equation.}
  \label{fig:sub2}
\end{subfigure}
\caption{
The reflexive polytope pair for the $\P^{2,3,1}$ ambient toric fiber.
}
\label{sub}
\end{figure}

The fiber polytope $\dd_2$ that provides a natural
correspondence with the Tate form
models (\ref{T}) is associated with the toric fan giving the weighted
projective space $\P^{2,3,1}$; this is a toric variety given by the rays $v_x
= (-1, 0), v_y = (0, -1), v_z = (2, 3)$
(see Figure~\ref{fig:sub1}).  Given a $\P^{2,3,1}$-fibered
polytope $\dd$ over a  toric base $B_2$, where the fiber is defined
by three rays satisfying $2v_x+3v_y+v_z=0$, we can always perform a
$SL(2,\Z)$ transformation to put the rays in the fiber into the
coordinates
\begin{equation}
v_x=(0,0,-1,0), v_y=(0,0,0,-1), v_z=(0,0,2,3) \,.
\label{fiberrays}
\end{equation}
We can define a {\it standard}\footnote{Because the rays of the base are ``stacked'' in (\ref{23}) over the
vertex $(2, 3)$ of the fiber, we sometimes refer to constructions of
this form as ``stacking'' fibrations.  The ``standard stacking''
we have defined here,
corresponding to the fiber  $\P^{2,3,1}$ and the specific point $(2,
3)$ in the fiber over which the base is stacked plays a special role
in the analysis of this paper.
We describe more general polytopes that
have the form of a ``stacking'' with different fibers and/or different
specified points in the fiber supporting the stacking, which
generalize the specific ``standard stacking'' used here, in the companion
paper \cite{Huang-Taylor-fibers}.}
 $\P^{2,3,1}$-fibered polytope over the base $B_2$ 
as one where there is a coordinate system after an $SL(4,\Z)$
transformation such that the vectors
\begin{equation}
v_{i}^{(a)}=(v_{i, 1}^{(B)}, v_{i, 2}^{(B)}, 2,3)
\label{23}
\end{equation}
are contained within $\dd$
for every ray $v_{i}^{(B)}=(v_{i, 1}^{(B)}, v_{i, 2}^{(B)})$ in
$\Sigma_B$. Note that in fact, these lattice points are all on the
boundary of $\nabla$ since the only interior point of a reflexive
polytope is the origin.
This particular choice of fiber and twist geometry
 represents a very specific class of fibered polytopes that
produce elliptically fibered Calabi-Yau threefolds as hypersurfaces.
These standard  $\P^{2,3,1}$-fibered polytopes play a central role in
the analysis of this paper, and are
a generalization of the well-studied 3D reflexive polytope for a K$3$ surface that
 is an elliptic fibration over a $\P^1$ base
\cite{Skarke:1998yk}.
As mentioned above,
these polytopes appear to be highly prevalent in the Kreuzer Skarke database at
large Hodge numbers.  This seems to occur for several reasons. The
 $\P^{2,3,1}$ fiber is the only one of the 16 reflexive 2D polytopes
that is possible in the presence of curves
of very negative self intersection in the base
(see discussion in \S\ref{bwlhn}).
And the natural correspondence between tuned Tate models and the
particular twist structure defined by
(\ref{23}) makes this twist structure particularly compatible with the
reflexive polytope Calabi-Yau construction.   We do, however, encounter some
specific examples of other twists in later sections.

For a standard  $\P^{2,3,1}$-fibered polytope,
the lattice points of  $\ds\subset M$ in this coordinate system
organize into  the following sets of points:
\begin{equation}
\{(0, 0, 1, -1), (0, 0, -2, 1), (\_,\_,0,0), (\_,\_,-1,1), (\_,\_,1,0), (\_,\_,0,1), (\_,\_,1,1)\}.
\label{m2pts}
\end{equation}
The elliptically fibered CY hypersurface equation $p = 0$ with $p$
from (\ref{p}) then takes precisely the Tate form (\ref{T}).  The sets
of points in (\ref{m2pts}) are associated with the monomials $y^2,
x^3, xy, x^2, y, x, 1$ respectively; $y^2$ and $x^3$ have a single
overall coefficient, and the monomials in the base associated with the
other five sets of points correspond precisely to monomials in the
five sections $\{a_1,a_2,a_3,$ $a_4,a_6\}$ (see figure \ref{sub}.)  In
particular, the condition that $\ds$ is the dual polytope of $\dd$
precisely imposes the condition that $a_n \in{\cal O} (-n K_B)$.  For
example, for $a_6$ we have the condition on the monomial associated
with the point $(m_1, m_2, 1, 1)$ that $ v^{(B)}_1m_1 + v^{(B)}_2m_2 +
2+3 \geq -1$ for each ray $v^{(B)}=\pi((v^{(B)}_1, v^{(B)}_2, 2, 3))$
in the fan of the base $B_2$, so $(m_1, m_2)$ represents a section of
$-6K_{B_2}$, in much the same way that the monomials in (\ref{m})
represent sections of $-K$ of the ambient toric variety.  A similar
computation for each $a_n$ confirms that the corresponding monomials
satisfy $v^{(B)}_1m_1 + v^{(B)}_2m_2 \geq -n$, and the degree $d$ in
the variable $z^{(B)}$ associated with the ray $v^{(B)}$ of a monomial
$(m_1,m_2)$ is given by $v^{(B)}_1m_1 + v^{(B)}_2m_2 = -n+d.$
An analogous computation shows that for the points associated with
$y^2$ and $x^3$ the condition is
$v^{(B)}_1m_1 + v^{(B)}_2m_2 \geq 0$; for any compact base this
implies that $m_1 = m_2 = 0$, so the first two points in
(\ref{m2pts}) are the only points of the form $(m_1, m_2, 1, -1)$
and $(m_1, m_2, -2, 1)$ and are associated with constant functions on
the base.  This matches with the fact that these are sections of
the trivial bundle ${\cal O} $ over the base, and the fact that the
only global holomorphic functions on any compact base are constants.
This proves that for any standard $\P^{2,3,1}$-fibered polytope, the
lattice points in $\ds$ are associated precisely with  the Tate form
of a Weierstrass model over the base, as stated above.

In the simplest cases, all the lattice points of the polytope $\dd$
are simply given by the vectors (\ref{fiberrays}) and the vectors of the form
(\ref{23}).   This corresponds  to the generic elliptic fibration over a
toric base $B_2$ without non-Higgsable clusters.  In other cases,
however, there are lattice points in $\dd$ other than those given by (\ref{fiberrays}) and (\ref{23}). This corresponds to situations with NHCs
or gauge groups tuned over curves in $B_2$ by removing Tate
monomials. 
The set of monomials in $\ds$ completely
span the set of sections of the appropriate line bundles ${\cal O}
(-nK_B)$ for the generic elliptic fibration over a given base.
In the case of NHCs, in particular,
the monomials in $\ds$  span the appropriate set of sections, while in the case of gauge group tunings, some of these monomials are set to zero. From the point of view of the Calabi-Yau geometry, the
lattice points in $\dd$ other than those given by (\ref{fiberrays}) and (\ref{23}) reflect the singular nature of the resulting
Calabi-Yau hypersurface.  
Up to some monodromy
subtleties that we discuss further in \S\ref{sec:Tate}, the 
set of new
 lattice points introduced together with $v_{i}^{(a)}$ in
$\pi^{-1}(v_{i}^{(B)})$ is known as a \emph{top}
\cite{Candelas:1996su,cpr,Perevalov:1997vw}, which forms the extended
Dynkin diagram of the gauge algebra of the singular fiber over the
associated divisor $D_{i}^{(B)}$, with $v_i^{(a)}$ the affine root (this is
the only inverse image when the fiber is smooth).  In 
section
\ref{sec:Tate} we describe in more detail the dictionary between Tate tunings
and toric/polytope geometry for specific gauge groups on particular
local curve configurations in the base geometry.

\subsection{A method for analyzing fibered polytopes: fiber types and 2D toric bases}
\label{fb}

Our primary approach in this paper is to systematically construct Tate
tunings that should have counterparts as reflexive polytopes in the
Kreuzer-Skarke database.  Thus, we start from the F-theory
construction and match the results with the known data in the KS
database.
This gives us something like  a ``sieve'' that leaves behind a set of
special cases of KS data not produced by our algorithm.
After implementing this sieve,
we have then considered separately those few examples in the KS database in
the range of interest that were not found by our F-theory construction.
We have found that there are a few polytopes in the KS database that
can be described in terms of the standard $\P^{2,3,1}$-fibered type; i.e., have Tate
forms, but were nonetheless not found with the initial sieve.  This turns out to be because
they involve such extensive tunings that the starting bases needed are
outside the range we considered. 
There are also data in the KS database that we did not identify in the
original sieve because they are accompanied by more sophisticated constructions involving $\gu(1)$
tunings,  novel  $\gsu(6)$ tunings associated with exotic matter representations, or tunings of generic models over
non-toric bases, which we had not considered. Moreover, we encounter
a type of novel models that did
not arise from our systematic construction because they
are involved with tunings on non-toric curves in the base; they turn
out nonetheless to also be described by reflexive polytopes with toric
fibers associated with
elliptic fibrations.

We had to explicitly study these specific polytopes with Hodge numbers
that we did not immediately identify from
Weierstrass/Tate tunings  to determine whether
these polytopes give hypersurfaces that are
actually elliptically fibered.  We provide here a summary of our
algorithm to analyze reflexive polytopes. We can learn from this
analysis whether one of the 16 reflexive fiber types is a fiber of the
polytope in question; we then define the 2D toric base from the
fibered polytope.  As we describe later in the paper, we can thereby
determine the singularities of the elliptic fiber over the curves in
the base, and then we check that the Hodge numbers of the
inferred tuned model are consistent with those of the polytope model.
Here we briefly summarize the first piece of this analysis: the
algorithm to determine if a given reflexive 2D polytope is a fiber of
a 4D polytope.
There are also software programs like Sage \cite{sage}  with built in
routines to identify the reflexive subpolytopes of a given polytope.

\begin{enumerate}
\item 
We assume that we are interested in a fiber described by the 2D
reflexive polytope $\dd_2$.
To increase the efficiency of the algorithm  in the case that
the number of lattice points in $\dd$ is large (which is true in the
case of large $h^{1,1}$ that we are focusing on), we begin by focusing
on only a subset of these lattice points that can possibly play a role
as the points in a fiber $\dd_2$.  
As mentioned in \S\ref{findbase}, the presence of a fiber
subpolytope $\dd_2 \subset\dd$ implies that there is a projection from
$\ds\rightarrow\ds_2$.  Let us call the maximum value of the inner
product for any pair of vectors in the fiber and its dual
\begin{equation}
M_{\rm max} = {\rm max} \ v\cdot w, \; \;
v\in\dd_2, w \in \ds_2 \,.
\label{eq:Mmax}
\end{equation}
For example, for $\P^{2,3,1}$, $M_\text{max}=5$, and for $\P^{1,1,2}$,
$M_\text{max}=3.$
We can then check for each lattice point $v \in\dd$ whether there
exists a vertex $w$ in $\ds$ with $v\cdot w > M_{\rm max}$.  If there
  is, then $v$ cannot be a ray in a fiber $\dd_2$.
We collect the subset of rays in $\dd$
that are not ruled out by this condition:
\begin{equation}
S=\{v\in\dd:v\cdot w \leq M_{\rm max}\  \forall w\ {\rm vertex\ of}\ 
\ds\} \,.
\label{eq:}
\end{equation}
\item 
We then look for a subset of rays of $S$ that satisfy the necessary
linear relations to be elements of the fiber $\dd_2$.  For example,
for  $\P^{2,3,1}$, we want to find rays $\{v_x,v_y,v_z\}$ that satisfy 
\begin{equation}
2v_x+3v_y+v_z=0.
\label{231c}
\end{equation}
In this case we can look at all pairs of rays $v, v'$ in $S$, and
check to see if $2v+ 3v'$ is also an element of $S$.  If so, we can
then check that the intersection of $\dd$ with the plane spanned by
$v, v'$ precisely contains the 7 points in the polytope $\dd_2$ shown
in Figure~\ref{fig:sub1}.  If this is the case than $\dd$ has a fiber
$\dd_2$.  The other fiber types can be checked in a similar fashion.
\end{enumerate}

By equations (\ref{m}), (\ref{eq:Mmax}) and the projection
$\ds\rightarrow\ds_2$, the maximum exponent of all monomials in the
variables associated with the rays in the fiber should be $M_{\rm max}
+ 1$, and the monomials can be grouped according to the powers of the
fiber coordinates into sets
that are in one-to-one
correspondence with the lattice points in $\ds_2$. For example, for
$\P^{2,3,1}$-fibered polytopes (see Figure \ref{fig:sub2}), we have
the maximum exponent in $z$ among all fiber coordinates;  $M_{\rm
  max}+1 = 6$, and the lattice points in $\ds_2$ are in one-to-one
correspondence with the sections
\begin{equation}
\{y^2, xyz, yz^3, x^3, x^2 z^2, xz^4, z^6\}.
\end{equation}
Note that, following the definition of a standard $\P^{2,3,1}$-fibered
polytope from \S\ref{sp231}, the lattice points in
$\ds_2$ are in one-to-one correspondence with the sections of the line
bundles $\O(-nK_B)$, and the monomials
 $x^3$ and $y^2$  are the
only two independent of the base coordinates.

Similarly, the
 sections of
the $\P^{1,1,2}$-fibered polytope (see Figure~\ref{p112})
are 
 \begin{equation}
 \{y^2, yz^2, xyz, x^2y, z^4, xz^3, x^2z^2, x^3z, x^4\}
\end{equation}
 when the associated rays are such that 
 \begin{equation}
 v_x+2v_y+v_z=0 \,,
 \label{112c}
\end{equation}
and $M_{\rm max}+1 = 4$.  The first step in the algorithm
above is only used to speed up the algorithm, but particularly when
the number of lattice points in $\dd$ is large, this speedup is
significant.  For example, for the polytope associated with the
Calabi-Yau with Hodge numbers H:$491, 11$, the number of lattice
points in $\dd$ is 680, while the number in $S$ is only 9.  Since the
second step of the algorithm is quadratic in the number of lattice
points considered, this represents a speedup by a factor of hundreds
or thousands of times in many cases.  While in this paper we are only
considering a few examples, such a speedup is useful when considering
larger datasets.  In   the companion paper \cite{Huang-Taylor-fibers} we will describe the systematic
application of this algorithm to all elements of the KS database with
large Hodge numbers.

Once we have determined the fiber,
we can then compute the base $B_2$ of the fibration.
We define the set of rays of the fan describing $B_2$ to be 
 \begin{equation}
\{v_{i}^{(B)}/\text{GCD}(v_{i, 1}^{(B)}, v_{i, 2}^{(B)}), \forall v_i \in \dd \},
\label{b2base}
\end{equation}
where $v_{i}^{(B)}\equiv\pi(v_i)=(v_{i, 1}^{(B)}, v_{i, 2}^{(B)})$ and
$\pi$ is the projection along the fiber subpolytope ($\pi(\dd_2)=0$).
The division by $\text{GCD}(v_{i, 1}^{(B)}, v_{i, 2}^{(B)})$ is done
to restrict to primitive rays in the image, as discussed in
\S\ref{findbase}.  Given the rays $v^{(B)}_i$, we associate a 2D cone
with each pair of adjacent rays, giving a unique toric structure to
the base geometry $B_2$.  \footnote{Note that in higher dimensions,
  the cone structure of the fan is not uniquely determined by the
  rays.}  Note that the base defined this way gives a flat toric
fibration, but not necessarily a flat elliptic fibration
\cite{Braun:2011ux}. We discuss this point in more detail in later
sections.

In the regions of the Hodge numbers that we study in this paper, we
also encounter polytopes that have no standard $\P^{2,3,1}$ fiber.  These
polytopes can be described using two different types of models.  One
of these other types of model that we encounter is very similar to the
standard $\P^{2,3,1}$-fibered polytopes, but has a fiber that is a
single blowup of $\P^{2,3,1}$.  This Bl$_{[0,0,1]}\P^{2,3,1}$ fiber,
which is one of the other 16 reflexive 2D fiber types, is shown in
Figure \ref{11}.  The corresponding fiber subpolytope $\ds_2$ is
identical to that for the $\P^{2,3,1}$ fiber except that it has an
additional vertex at ($-1, -1$), so that the number of lattice points
in the plane of the fiber subpolytope is 8 rather than 7.  From the
Tate point of view, such a fiber occurs when all the monomials in the
coefficient $a_6$ are taken to vanish.  This vanishing of $a_6$ forces
a global $\gu(1)$ symmetry that we mentioned earlier
\cite{Braun-16, Mayrhofer:2014opa}. We describe an explicit example of this type
of model in Appendix~\ref{u1}.  Models with this fiber can be treated
in essentially the same fashion as standard $\P^{2,3,1}$-fibered
polytopes.

The other unusual kind of fibration that we encounter in a few models
is a fibered polytope with fiber $\Delta_2$ given by the usual
$\P^{2,3,1}$ polytope, but with a different ``twist'' to the
$\P^{2,3,1}$ bundle over the base.  In other words, while there is a
projection of $\ds$ to the dual polytope $\ds_2$ of the 
$\P^{2,3,1}$ fiber, the base rays in $\dd$ do not all lie in a
plane that contains the vector $v_z$; i.e., the
base of the polytope defined in (\ref{b2base}) can not be constituted by a set of rays all in the form  (\ref{23}).  The consequence of this is that
the hypersurface equations (\ref{p}) for these Calabi-Yau threefolds
do not take on the Tate form (\ref{T}).  In particular, there is generically more than one lattice point projected to the points
in $\ds_2$ associated with $y^2$ and/or $x^3$.  To determine the Weierstrass
form (\ref{W}) for the models of this type that we found and analyze their structure, we found
that it was useful to view them as essentially ``$\P^{1,1,2}$-fibered
polytopes'' (or more precisely, $\P^{1,1,2}$ with two more blowups)
rather than the standard $\P^{2,3,1}$-fibered polytopes (see figure
\ref{p112} for comparison).  This allows us to follow the method for
analyzing $\P^{1,1,2}$-fibered models described in Appendix A of
\cite{Braun:2011ux} to bring them into Weierstrass form. This type of
novel model gives rise to an enhancement over
non-toric curves as we mentioned earlier. We refer to
this type of models as \emph{non-standard $\P^{2,3,1}$-fibered}
polytopes, and describe their analysis in more detail in section
\ref{non}.  The treatment of non-standard $\P^{2,3,1}$ models in terms
of models with a blow-up of $\P^{1,1,2}$ as a fiber is closely
analogous to the analysis of models with a Bl$_{[0,0,1]}\P^{2,3,1}$
fiber as special cases of $\P^{2,3,1}$ Weierstrass/Tate models.

\section{Tate tunings and the Kreuzer-Skarke database}
\label{sec:Tate}

\label{polytune}

We want to understand how the set of  Calabi-Yau threefolds
produced by toric hypersurface constructions through reflexive polytopes
in the Kreuzer-Skarke database can be related to the general
construction of elliptic Calabi-Yau threefolds through tuned
Weierstrass models.
The approach we take is to identify a specific subclass of tuned
models that match with toric hypersurface constructions.  In
particular, we begin with the set of toric bases identified in
\cite{Morrison:2012js} and consider Tate tunings over these bases.

In principle, to find all the elliptically fibered threefolds in the
Kreuzer-Skarke database we might want to consider a variety of tunings
and singularity structures that correspond to all 16 of the toric
fiber types mentioned in \S\ref{findbase}.  To simplify the set of
possibilities, however, we focus on a region of Hodge numbers where we
expect a single toric fiber type to dominate.  A generic Tate-form
elliptic fibration over a given toric base can always be constructed
starting from the ``standard stacking'' procedure as we will describe in
\S\ref{sec:reflexive-smooth} and \S\ref{tops}; this procedure uses the
$\P^{2,3,1}$ fiber type.  
Tuning
the resulting generic Tate model by removing monomials  in the
dual polytope then leads to a set of possible tunings corresponding to
further reflexive polytopes that can appear in the database; we
describe this process in \S\ref{tatetunepoly} and \S\ref{sec:combining}, and
 give an
 example in  Appendix \ref{polytopetuning}.  Such a construction can be carried out
for any base.  The gauge symmetries associated with the tunings can be
read off from the tops
\cite{Candelas:1996su,Skarke:1998yk,cpr,Perevalov:1997vw} of the
polytopes.  We review polytope tops in \S\ref{tp}, 
and
we address some subtle issues about the
multiple tops of a gauge algebra in \S\ref{sec:multiple-tops}, which
are related to the monodromy choices of the Tate tunings of the algebra
that we have discussed in \S\ref{TT}.

The other 15 fiber types, however, implicitly constrain
the Weierstrass model associated with an elliptic fibration. We
explain in \S\ref{bwlhn} some constraints on  the other 15
fiber types, which are related to the structure of the base.  Based on
these constraints, we expect that when we
confine the range of Hodge numbers to relatively large values, as we do
in section
\ref{sec:systematic-construction}, the simplest $\P^{2, 3, 1}$ fiber
type will dominate the set of polytopes.  
\footnote{That this expectation is correctly borne out is also
  verified explicitly with a systematic analysis of the KS database in
  the companion paper \cite{Huang-Taylor-fibers}.}  By focusing on
this simple class of constructions, therefore, we realize almost all
the Hodge numbers in the range of interest with a single class of
Tate-tuned elliptic fibration models.  Although we will not deal with
the matching of the multiplicity in KS database of a given Hodge pair
with our systematic tuning construction, we will explore a bit more
this aspect in sections \ref{so8} and \ref{multiplicity}.

\subsection{Reflexive polytopes from elliptic fibrations without singular fibers}
\label{sec:reflexive-smooth}

In \S\ref{sp231} we defined a standard $\P^{2,3,1}$-fibered
polytope, and showed that there is always a corresponding Tate
model. Now we are trying to do the converse --- given a toric base and
a corresponding Tate model, we wish to construct a corresponding
reflexive polytope.  
As alluded to earlier,
the recipe for the construction of a 4D standard
$\P^{2,3,1}$-fibered polytope for an elliptically fibered threefold is
the natural generalization of the 3D reflexive polytope for a K$3$ surface that
 is an elliptic fibration over a $\P^1$ base as described in e.g.
\cite{Skarke:1998yk}.

To construct a 4D $\P^{2,3,1}$-fibered polytope, we start with the 2D
$\P^{2,3,1}$ fiber and a 2D base, and we construct the polytope in a
straightforward way to have the desired fibration structure over the
base. We denote the toric fan associated with the base $B$ by
$\Sigma_B$, with the set of rays being $\{v_i^{(B)}\}$.  
Taking the fan
of $\P^{2,3,1}$ to be the ambient space of the elliptic fiber,  we
can embed this in the 4D coordinates such that the three rays are
$\{v_x,v_y, v_z\}=\{(0,0,-1,0),(0,0,0,-1),(0,0,2,3)\} $.  Since in
  the Weierstrass or Tate model framework of equation (\ref{fs}) the
  fiber coordinate $z$ is %a section of the tensor product with the
associated with the trivial bundle over the base, the
    lattice point associated with the ray $v_z=(0,0,2,3)$ should be in
    the plane of the base.  Thus, we define a polytope
$\tilde{\dd}$ to be the convex hull of the set
\begin{equation}
\{(v_{i, 1}^{(B)},v_{i, 2}^{(B)},2,3)\rvert v_{i}^{(B)} \text{ rays in } \Sigma_B \}\cup \{(0,0,-1,0),(0,0,0,-1)\},
\label{pile}
\end{equation}
where  $v_{i, 1}^{(B)},v_{i, 2}^{(B)}$ are the first and the second
components of  the 1D ray $v_{i}^{(B)}$ in the smooth 2D toric
base $B$.
From the definition in the previous section, this is a standard
$\P^{2,3,1}$-fibered polytope; we refer sometimes to this construction
as the ``standard stacking'' approach to construction of a  polytope.  
Note that the
4D rays $v_{i}=(v_{i, 1}^{(B)},v_{i, 2}^{(B)},2,3)$
can be vertices of $\tilde{\dd}$
only if $v_{i}^{(B)}$ are associated with curves of self-intersection $D_i\cdot D_i >
  -2$ (see Table \ref{t:nonconvex}).  
We now wish to check
that $\tilde{\dd}$
is reflexive, so it can be used as the reflexive polytope $\dd$ in
Batyrev's construction of a Calabi-Yau threefold.  In some
cases $\tilde{\dd}$ is immediately reflexive, and in other more
complicated cases it must be modified to make it reflexive.

\begin{table}[]
\centering
\begin{tabular}{|c|cccc|}
\hline
NHC &\{-3\}&\{-2, -3\}&\{-2, -2, -3\}&\{-2, -3, -2\} \\
\hline
Fan &  $\begin{array}{l}\includegraphics[width=2cm]{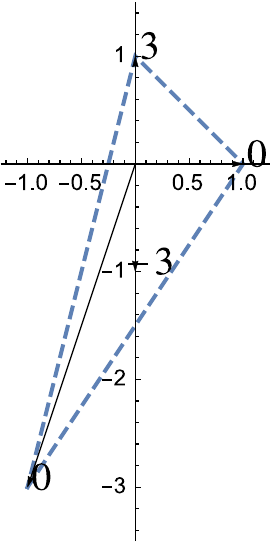}\end{array}$  & $\begin{array}{l}\includegraphics[width=3cm]{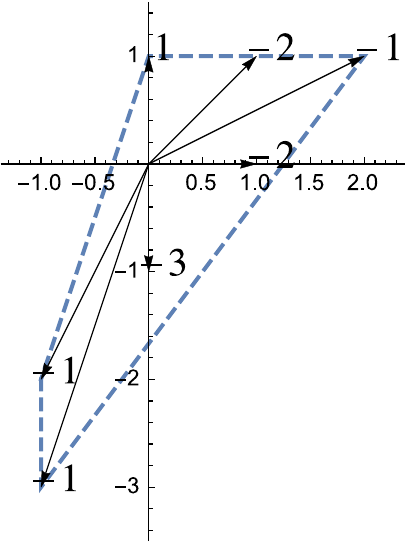}\end{array}$       &  $\begin{array}{l}\includegraphics[width=3cm]{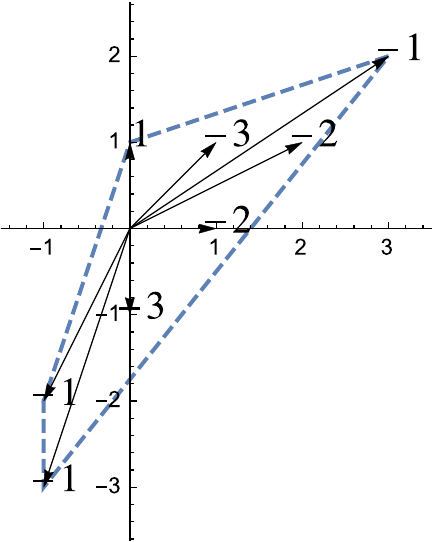}\end{array}$          &           $\begin{array}{l}\includegraphics[width=3cm]{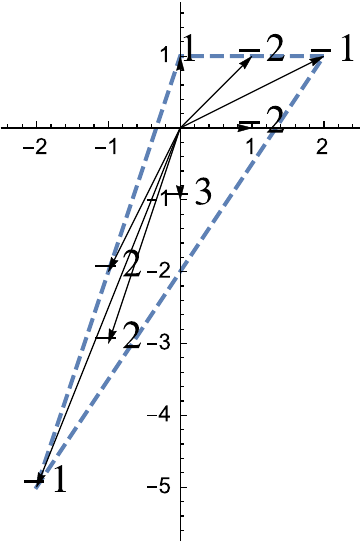}\end{array}$\\
\hline
\end{tabular}
\caption{\footnotesize Non-convexity of NHCs: The rays corresponding
  to an NHC cannot be a vertices; hence, the vertex contribution from
  the base can only come from curves of self-intersection $\geq
  -1$ (isolated -2 curves will be on a 1D face,  and also cannot be vertices).}
\label{t:nonconvex}
\end{table}

We start with the simplest case, in which we have a
generic elliptically fibered Calabi-Yau over a toric base $B$ that
contains no non-Higgsable clusters (i.e., no curves with self-intersection less than $-2$). 
In this case, the Weierstrass/Tate model of the Calabi-Yau is smooth and there is no gauge group in
the 6D supergravity theory.
In this context, lattice points  associated to curves of
   self-intersection $-2$  lie on the 1D faces  of $\tilde{\dd}$ that
   are boundaries of the 2D face $\theta_B$, which is the 2D face
   associated with the base; and there are no interior points in
   $\theta_B$ other than $(0,0,2,3)$.
We can now check directly
that in these simple cases $\tilde{\dd}$
is reflexive without further modification.
The vertices of the polytope dual to the convex hull of the set of
vertices (\ref{pile}), in any case,  are
\begin{equation}
\left\{\left(\left.\frac{6 (v_{i, 2}^{(B)}-v^{(B)}_{j, 2})}{\text{Det}[v_{i}^{(B)},v^{(B)}_j]},-\frac{6 (v_{i, 1}^{(B)}-v^{(B)}_{j, 1})}{\text{Det}[v_{i}^{(B)},v^{(B)}_j]},1,1\right)\right.\right\}\cup \{(0,0,-2,1),(0,0,1,-1)\},
\label{dualpile}
\end{equation}
where $(i,j)$ are taken to run over all
pairs of labels of base rays that correspond to
adjacent vertices of  $\theta_B$.  
The vertices in (\ref{dualpile})
will lie on the $M$ lattice only when the denominators
$\text{Det}[v_{i}^{(B)},v^{(B)}_j]$ are cleared so that all entries
are integers. 
For a smooth 2D base fan,
$\text{Det}[v_{i}^{(B)},v_{i+1}^{(B)}]=1$, so we have a lattice point whenever
$j=i+1$ (including the boundary case $j = 1, i = n$); i.e., 
we get lattice points as long as there are
no non-convex base rays, which would be skipped.  
We also get a lattice point as long as $\vb_i$ and $\vb_j$ are
separated only by some number $k$ of $-2$ curves.  In this case
$\vb_i-\vb_j = k w$, where $w$ is a primitive vector, and
${\rm Det}[\vb_i,\vb_j] = k$, so we again have a cancellation
and the vertex of the dual polyhedra is an integral lattice point.
Thus, as long as the base $B$ contains no non-Higgsable clusters, the
set of vertices (\ref{pile})  immediately provides a reflexive
polytope. \footnote{As we will discuss in \S\ref{imm}, the set
  (\ref{pile}) still gives a reflexive polytope in certain cases when
  the base contain NHCs, but those lattice points corresponding to the
  curves in the base that carry the NHCs
  are not vertices.}

Simple examples of polytopes realized in this way are the elliptically
fibered Calabi-Yau threefolds over the toric bases $\P^2,
\F_{n=0,1,2}$, whose vertex sets of the $M$ polytopes $\ds$ are (\ref{dualpile}),
with the first set of vertices respectively being $\{(-6, -6, 1, 1),
(12, -6, 1, 1), (-6, 12, 1, 1)\}$, $\{(-6, -6, 1, 1), (6(1+n), -6, 1,
1), (6(-n+1), 6, 1, 1), (-6, 6, 1, 1)\}$, given the respective base
rays $\{(1,0),(0,1),(-1,-1)\}$, $\{(1,0),(0,1),(-1,-n),(0,-1)\}$. The
$\P^2$ model gives the only polytope (up to lattice automorphism) with
Hodge numbers H:2,272 in the KS database and the $\F_{n=0,1,2}$ models
give exactly the three data points with Hodge numbers H:3,243.

The bases described by toric varieties with no curves of
self-intersection less than $-2$ are {\it weak Fano}
varieties, and correspond to reflexive 2D polytopes, as we have just
verified explicitly.
We now want to describe the generalization of this construction to
situations where there is a gauge group arising either from a
non-Higgsable cluster in the base or a Tate tuning.  The realization
of reflexive 4D polytopes in these cases arises from a general
relationship between Tate tunings and  ``tops'' in the toric language.

\subsection{Tate tuning and polytope tops}
\label{tp}
We saw in \S\ref{sp231} that for a standard 
$\P^{2,3,1}$-fibered polytope,
the lattice points of $\ds$
that project to each of the different  lattice points of 
$\ds_2$ (figure \ref{sub})
correspond precisely to the sets of
monomials in the coefficients of the Tate form
(\ref{T}).
The lattice points of $\ds$ are thus
divided into 5 groups corresponding to the 5 sections
 $a_n \in{\cal O} (-nK_B)$   and another 2
points corresponding to the constant coefficients of $y^2$ and $x^3$. 
In the previous subsection we described generic elliptic fibrations
over
 weak Fano bases, where the ``standard stacking'' procedure immediately
 gives a reflexive 4D polytope, and no additional rays are needed in
 $\dd$, corresponding to a physics model with no nonabelian gauge
 group.  We now wish to consider how this story changes when there is
 a nontrivial nonabelian gauge group due either to an NHC in the base
 or a Tate tuning of the monomials in the Tate form.

\begin{figure}
  \centering
  \includegraphics[width=7cm]{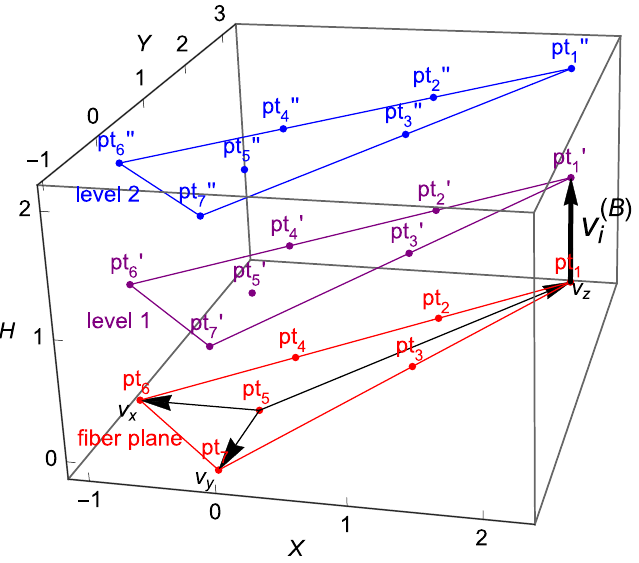}
  \caption{\footnotesize A 3D visualization of the lattice points that
    appear in a top over $v_{i}^{(B)}$: in standard $\P^{2,3,1}$
    models, a top over a ray in the base $v_{i}^{(B)}$ (in the
    direction $H$) is a set of lattice points stacked over the 7
    lattice points of the fiber subpolytope $\P^{2,3,1}$ (in the X-Y
    plane). The level (the multiple of $v_{i}^{(B)}$) where points
are located is indicated by the number of primes. When the gauge
    algebra is trivial over the associated divisor $D_{i}^{(B)}$,
    $pt_1'$ (equation (\ref{23})) is the only point in the top; while
    otherwise there are additional points (cf. Table \ref{t:dic})
    forming the extended Dynkin diagram of the gauge algebra with
    $pt_1'$ the affine node.}
  \label{fig:levels}
\end{figure}

The presence of an NHC in the base or an explicit Tate tuning can 
force some of the coefficients in the $a_n$s to vanish to some specified order along a particular base divisor
$D_{i}^{(B)}$. This absence of monomials in $\ds$  gives rise to a corresponding enlargement of $\dd$ from the standard stacking.
The additional lattice points in the fan polytope $\dd$ correspond to
the exceptional divisors that resolve the singularities of the associated
fibered geometry. These additional lattice points form the
``top'' \cite{Candelas:1996su,Skarke:1998yk,cpr,Perevalov:1997vw} of the
enhanced gauge symmetries over $D_{i}^{(B)}$.  In coordinate
representation, a lattice point in the top of $D_{i}^{(B)}$ is of the
form
\begin{equation}
((l v_{i}^{(B)})_{1, 2}, (pt_{\text{1, 2, 3, 4, 5, 6, or 7}})_{3, 4}),
\end{equation}
where 
\begin{equation}
pt_{1,2,3,4,5,6, 7}=(2,3), (1,2), (1,1), (0,1), (0,0), (-1,0),
(0,-1)
\label{pts}
\end{equation}
are the 7 lattice points in the 2D reflexive fiber subpolytope
$\P^{2,3,1}$, $v_{i}^{(B)}$ is the associated 2D ray, and
$l\in\mathbb{N}$ specifies the ``level'' of the point away from the
fiber plane (see figure \ref{fig:levels}). We adopt the shorthand
notation $pt_j^{(l)}$ or $pt_j^{'\cdots'}$, where the number of primes
specifies the level parameter $l$.
When we denote a top, the points with fewer than the maximal number
of primes over each point are omitted and implied by the point of most
primes with the same index; e.g. $\{ pt_1''', pt_2'', pt_3',
pt_4'\}=\{pt_1', pt_1'', pt_1''', pt_2', pt_2'', pt_3',
pt_4'\}$. 
The tops of the various
gauge algebras have been worked out in the previous literature.  Tops
for gauge algebras of rank no greater than eight
that arise in reflexive polytopes can be looked up for example in Table 3.2 in
  \cite{Candelas:1996su}. 
  We have explicitly calculated
  a few more cases, including the tops of $\gso(n)$ and $\gsu(n)$ gauge algebras
  to rank $12$ in both cases and list the results in tables Table
  \ref{sutop} and Table \ref{sotop}, respectively. 
In \cite{Bouchard:2003bu}, Vincent Bouchard and Harald Skarke
generalized the notion of tops  (including those which may not have a completion to
reflexive polytopes) to include all fiber types,
and they classified all such ``tops in the dual space''  (i.e., the
$M$ lattice space), including higher rank $\gso(n)$
and $\gsu(n)$ tops. The tops in Table
  \ref{sutop} and Table \ref{sotop} were explicitly obtained from
  reflexive polytope constructed from
successive Tate tunings, and we have cross-checked the $\gso(n)$ cases
with the  results of \cite{Bouchard:2003bu} in the dual space, which
agree up to a $GL(2,\Z)$ transformation. 
Note that
for higher rank $\gso(n)$ and $\gsu(n)$ algebras, the $\dd$ polytope grows in the fiber subpolytope direction (as opposed to the level direction), and more $pt$s projecting to the fiber plane are involved. We list the ones we need in Table \ref{sutop} and Table \ref{sotop}:
\begin{eqnarray}
\label{extendedp231}
%\lefteqn
{pt_{8,9,10,11,12,13,14,15,16,17}}%\nonumber\\\nonumber
&=&(-1, -1), (-2, -1), (-3, -2), (-2, -2), (-4, -3), (-5, 
-4), (-3, -3),  \nonumber\\
& & \hspace*{0.3in}(-6, -5), (-7, -6), (-4,-4)\,. %\nonumber
\end{eqnarray}

\begin{table}[]
\centering
\begin{tabular}{|c|l|l|}
\hline
$n$ & Tate form          & Top/Affine Dynkin nodes                                                                                                    \\ \hline
7   & \{0, 1, 3, 4, 7\}  & \{$pt_1'$,$pt_2'$,$pt_3'$,$pt_4'$,$pt_5'$,$pt_6'$,$pt_8'$\}                                                                \\ \hline
8   & \{0, 1, 4, 4, 8\}  & \{$pt_1'$,$pt_2'$,$pt_3'$,$pt_4'$,$pt_5'$,$pt_6'$,$pt_8'$,$pt_9'$\}                                                        \\ \hline
9   & \{0, 1, 4, 5, 9\}  & \{$pt_1'$,$pt_2'$,$pt_3'$,$pt_4'$,$pt_5'$,$pt_6'$,$pt_8'$,$pt_9'$,$pt_{11}'$\}                                             \\ \hline
10  & \{0, 1, 5, 5, 10\} & \{$pt_1'$,$pt_2'$,$pt_3'$,$pt_4'$,$pt_5'$,$pt_6'$,$pt_8'$,$pt_9'$,$pt_{10}'$,$pt_{11}'$\}                                  \\ \hline
11  & \{0, 1, 5, 6, 11\} & \{$pt_1'$,$pt_2'$,$pt_3'$,$pt_4'$,$pt_5'$,$pt_6'$,$pt_8'$,$pt_9'$,$pt_{10}'$,$pt_{11}'$,$pt_{14}'$\}                       \\ \hline
12  & \{0, 1, 6, 6, 12\} & \{$pt_1'$,$pt_2'$,$pt_3'$,$pt_4'$,$pt_5'$,$pt_6'$,$pt_8'$,$pt_9'$,$pt_{10}'$,$pt_{11}'$,$pt_{12}'$,$pt_{14}'$\}            \\ \hline
13  & \{0, 1, 6, 7, 13\} & \{$pt_1'$,$pt_2'$,$pt_3'$,$pt_4'$,$pt_5'$,$pt_6'$,$pt_8'$,$pt_9'$,$pt_{10}'$,$pt_{11}'$,$pt_{12}'$,$pt_{14}'$,$pt_{17}'$\} \\ \hline
\end{tabular}
\caption{\footnotesize The tops of $\gsu(n)$ algebras. The
coordinates of
  the points $pt_{1,2,3,4,5,6,7,8,9,10,11,12,13,14,15,16,17}$ are
  given in
  equations (\ref{pts}) and (\ref{extendedp231}). All lattice points
  in these
tops are of level one, and correspond to affine Dynkin nodes. The
  rank of each algebra is the number of the nodes minus one.}
\label{sutop}
\end{table}

\begin{table}[]
\centering
\begin{tabular}{|c|c|l|l|}
\hline
$n$ & Tate form      & Top                                                                                                                                                                                                   & Affine Dynkin nodes                                                                                                                                                                \\ \hline
13  & \{$1,1,3,4,6$\}  & \{$pt_1''$,$pt_2''$,$pt_3'$,$pt_4''$,$pt_5'$,$pt_6''$\}                                                                                                                                               & \{$pt_1'$,$pt_1''$,$pt_2''$,$pt_3'$,$pt_4''$,$pt_6'$,$pt_6''$\}                                                                                                                    \\ \hline
14  & \{$1,1,3,4,7$\}  & \{$pt_1''$,$pt_2''$,$pt_3'$,$pt_4''$,$pt_5'$,$pt_6''$,$pt_8'$\}                                                                                                                                       & \{$pt_1'$,$pt_1''$,$pt_2''$,$pt_3'$,$pt_4''$,$pt_6'$,$pt_6''$,$pt_8'$\}                                                                                                            \\ \hline
15  & \{$1,1,4,4,7$\}  & \{$pt_1''$,$pt_2''$,$pt_3'$,$pt_4''$,$pt_5'$,$pt_6''$,$pt_8'$,$pt_9''$\}                                                                                                                              & \{$pt_1'$,$pt_1''$,$pt_2''$,$pt_3'$,$pt_4''$,$pt_6''$,$pt_8'$,$pt_9''$\}                                                                                                           \\ \hline
16  & \{$1,1,4,4,8$\}  & \{$pt_1''$,$pt_2''$,$pt_3'$,$pt_4''$,$pt_5'$,$pt_6''$,$pt_8'$,$pt_9''$\}                                                                                                                              & \{$pt_1'$,$pt_1''$,$pt_2''$,$pt_3'$,$pt_4''$,$pt_6''$,$pt_8'$,$pt_9'$,$pt_9''$\}                                                                                                   \\ \hline
17  & \{$1,1,4,5,8$\}  & \{$pt_1''$,$pt_2''$,$pt_3'$,$pt_4''$,$pt_5'$,$pt_6''$,$pt_8'$,$pt_9''$,$pt_{10}''$\}                                                                                                                  & \{$pt_1'$,$pt_1''$,$pt_2''$,$pt_3'$,$pt_4''$,$pt_6''$,$pt_9'$,$pt_9''$,$pt_{10}''$\}                                                                                               \\ \hline
18  & \{$1,1,4,5,9$\}  & \begin{tabular}[c]{@{}l@{}}\{$pt_1''$,$pt_2''$,$pt_3'$,$pt_4''$,$pt_5'$,$pt_6''$,$pt_8'$,$pt_9''$,$pt_{10}'',$\\ $pt_{11}'$\}\end{tabular}                                                            & \begin{tabular}[c]{@{}l@{}}\{$pt_1'$,$pt_1''$,$pt_2''$,$pt_3'$,$pt_4''$,$pt_6''$,$pt_9'$,$pt_9''$,\\ $pt_{10}''$,$pt_{11}'$\}\end{tabular}                                         \\ \hline
19  & \{$1,1,5,5,9$\}  & \begin{tabular}[c]{@{}l@{}}\{$pt_1''$,$pt_2''$,$pt_3'$,$pt_4''$,$pt_5'$,$pt_6''$,$pt_8'$,$pt_9''$,$pt_{10}''$,\\ $pt_{11}'$,$pt_{12}''$\}\end{tabular}                                                & \begin{tabular}[c]{@{}l@{}}\{$pt_1'$,$pt_1''$,$pt_2''$,$pt_3'$,$pt_4''$,$pt_6''$,$pt_9''$,$pt_{10}''$,\\ $pt_{11}'$,$pt_{12}''$\}\end{tabular}                                     \\ \hline
20  & \{$1,1,5,5,10$\} & \begin{tabular}[c]{@{}l@{}}\{$pt_1''$,$pt_2''$,$pt_3'$,$pt_4''$,$pt_5'$,$pt_6''$,$pt_8'$,$pt_9''$,$pt_{10}''$,\\ $pt_{11}'$,$pt_{12}''$\}\end{tabular}                                                & \begin{tabular}[c]{@{}l@{}}\{$pt_1'$,$pt_1''$,$pt_2''$,$pt_3'$,$pt_4''$,$pt_6''$,$pt_9''$,$pt_{10}'$,\\ $pt_{10}''$,$pt_{11}'$,$pt_{12}''$\}\end{tabular}                          \\ \hline
21  & \{$1,1,5,6,10$\} & \begin{tabular}[c]{@{}l@{}}\{$pt_1''$,$pt_2''$,$pt_3'$,$pt_4''$,$pt_5'$,$pt_6''$,$pt_8'$,$pt_9''$,$pt_{10}''$,\\ $pt_{11}'$,$pt_{12}''$,$pt_{13}''$\}\end{tabular}                                    & \begin{tabular}[c]{@{}l@{}}\{$pt_1'$,$pt_1''$,$pt_2''$,$pt_3'$,$pt_4''$,$pt_6''$,$pt_9''$,$pt_{10}'$,\\ $pt_{10}''$,$pt_{12}''$,$pt_{13}''$\}\end{tabular}                         \\ \hline
22  & \{$1,1,5,6,11$\} & \begin{tabular}[c]{@{}l@{}}\{$pt_1''$,$pt_2''$,$pt_3'$,$pt_4''$,$pt_5'$,$pt_6''$,$pt_8'$,$pt_9''$,$pt_{10}''$,\\ $pt_{11}'$,$pt_{12}''$,$pt_{13}''$,$pt_{14}'$\}\end{tabular}                         & \begin{tabular}[c]{@{}l@{}}\{$pt_1'$,$pt_1''$,$pt_2''$,$pt_3'$,$pt_4''$,$pt_6''$,$pt_9''$,$pt_{10}'$,\\ $pt_{10}''$,$pt_{12}''$,$pt_{13}''$,$pt_{14}'$\}\end{tabular}              \\ \hline
23  & \{$1,1,6,6,11$\} & \begin{tabular}[c]{@{}l@{}}\{$pt_1''$,$pt_2''$,$pt_3'$,$pt_4''$,$pt_5'$,$pt_6''$,$pt_8'$,$pt_9''$,$pt_{10}''$,\\ $pt_{11}'$,$pt_{12}''$,$pt_{13}''$,$pt_{14}'$,$pt_{15}''$\}\end{tabular}             & \begin{tabular}[c]{@{}l@{}}\{$pt_1'$,$pt_1''$,$pt_2''$,$pt_3'$,$pt_4''$,$pt_6''$,$pt_9''$,$pt_{10}''$,\\ $pt_{12}''$,$pt_{13}''$,$pt_{14}'$,$pt_{15}''$\}\end{tabular}             \\ \hline
24  & \{$1,1,6,6,12$\} & \begin{tabular}[c]{@{}l@{}}\{$pt_1''$,$pt_2''$,$pt_3'$,$pt_4''$,$pt_5'$,$pt_6''$,$pt_8'$,$pt_9''$,$pt_{10}''$,\\ $pt_{11}'$,$pt_{12}''$,$pt_{13}''$,$pt_{14}'$,$pt_{15}''$\}\end{tabular}             & \begin{tabular}[c]{@{}l@{}}\{$pt_1'$,$pt_1''$,$pt_2''$,$pt_3'$,$pt_4''$,$pt_6''$,$pt_9''$,$pt_{10}''$,\\ $pt_{12}'$,$pt_{12}''$,$pt_{13}''$,$pt_{14}'$,$pt_{15}''$\}\end{tabular}  \\ \hline
25  & \{$1,1,6,7,12$\} & \begin{tabular}[c]{@{}l@{}}\{$pt_1''$,$pt_2''$,$pt_3'$,$pt_4''$,$pt_5'$,$pt_6''$,$pt_8'$,$pt_9''$,$pt_{10}''$,\\ $pt_{11}'$,$pt_{12}''$,$pt_{13}''$,$pt_{14}'$,$pt_{15}''$,$pt_{16}''$\}\end{tabular} & \begin{tabular}[c]{@{}l@{}}\{$pt_1'$,$pt_1''$,$pt_2''$,$pt_3'$,$pt_4''$,$pt_6''$,$pt_9''$,$pt_{10}''$,\\ $pt_{12}'$,$pt_{12}''$,$pt_{13}''$,$pt_{15}''$,$pt_{16}''$\}\end{tabular} \\ \hline
\end{tabular}
\caption{\footnotesize The tops of $\gso(n)$ algebras. 
The
coordinates of
  the points $pt_{1,2,3,4,5,6,7,8,9,10,11,12,13,14,15,16}$ are given
in
  equations (\ref{pts}) and (\ref{extendedp231}). (Only the highest
  level point for each $pt$ is listed in each top, and the lattice points
  of the lower levels are implied.) $\gso(4n-1)$ and $\gso(4n)$ in the
  table have the same top but different (numbers of) affine Dynkin
  nodes as the ranks
(which differ from the number of the nodes by one) are
  different.
These tops match those found in \cite{Bouchard:2003bu} after an
appropriate coordinate transformation.
}
\label{sotop}
\end{table}
 
There is a simple and precise correspondence between tunings of the
Tate form and tops.  This correspondence holds independent of whether
the Tate form corresponds to an NHC or an explicit tuning.
Consider for example a situation where the standard $\P^{2,3,1}$-fibered polytope $\dd$ contains the
lattice point $pt_2' = (v^{(B)}_{1, 2}, 1,2)$.
Recall that the lattice point $pt_1'=(v^{(B)}_{1, 2}, 2, 3)$ imposes
the conditions on the dual lattice points $(m_{1, 2}, 1, 1)$
associated with monomials
in $a_6$ that $v^{(B)} \cdot m + 5 \geq -1 \Rightarrow
v^{(B)} \cdot m \geq -6$ as expected for a section of ${\cal O}
(-6K)$.  The point $pt_2'$ imposes the stronger condition
 $v^{(B)} \cdot m + 3 \geq -1 \Rightarrow
v^{(B)} \cdot m \geq - 6+2$, corresponding to the condition that $a_6$
vanish to order 2 over the corresponding $D^{(B)}$.  
% {\color{red}(a)} 
A similar
calculation shows that
$(a_1, a_2, a_3, a_4, a_6)$ vanish to orders at least $(0,0, 1, 1, 2)$
respectively when the point $pt_2'$ is present in $\dd$.
Indeed, this goes both ways:
% {\color{red}(b)}
 only when the $a_n$s vanish at least to  orders
$(0,0, 1, 1, 2)$, associated with the absence of a certain set of
lattice points in $\ds$,  can the point $pt'_2$ appear in
$\dd$, and indeed if all the $a_n$s vanish to these orders then the
point $pt'_2$ must appear in the polytope $\dd$ dual to $\ds$.  
 Thus, there is a precise local correspondence between Tate
tunings of the $a_n$ coefficients over a certain ray in the base,
associated with lattice points absent from $\ds$, and
the toric top in $\dd$ over that ray.  We tabulate 
a few examples of this correspondence in Table~\ref{t:top-Tate}.
Note that just as multiple Tate tunings can correspond to the same
gauge algebra, the corresponding multiple tops also correspond to the
same gauge algebra.  The multiplicity of constructions
 for a given gauge
algebra was studied from the point of view of tops in \cite{Braun-16}.
One particular situation in which multiple tops are possible for a
fixed gauge algebra corresponds to monodromy-dependent Tate tuning
configurations, which we discuss further in \S\ref{sec:multiple-tops}.

\begin{table}
\begin{center}
\begin{tabular}{|c|c|c|l|}
\hline
point    & ord$(a_1,a_2,a_3,a_4,a_6)$ & group & type        \\ \hline
$pt_2'$  & $(0,0,1,1,2)$                & SU(2) & $I_2$       \\
$pt_3'$  & $(0,1,1,2,3)     $           & SU(3) & $I_3^s$     \\
$pt_1''$ & $(1,1,2,2,3)     $           & G$_2$ & $I_0^{*ns}$ \\
$pt_4'$  & $(0,0,2,2,4)$                & Sp(2) & $I_4^{ns}$  \\ \hline
\end{tabular}
\end{center}
\caption[x]{\footnotesize  Some examples of the
correspondence between additional lattice
  points in $\dd$ associated with a ray $v^{(B)}$ in the base and the
  associated tuning of the Tate coefficients
$(a_1, a_2, a_3, a_4, a_6)$ over the associated divisor.
}
\label{t:top-Tate}
\end{table}

This correspondence leads to a natural association of reflexive
polytopes with elliptic fibrations over toric bases that have Tate
forms.  Over a given base, various gauge groups can arise from a
combination of non-Higgsable clusters and Tate tunings. The interplay
between extra vertices in $\dd$ over nearby divisors and the absence
of monomials in $\ds$ leads to local interactions between the sets of
lattice points in the polytope that are affected by adjacent rays in
the base.  We  consider more explicitly
in the following section how this leads to consistent reflexive
polytopes in both the NHC and Tate tuning cases.

%\subsection{Construction of generic elliptically fibered CY polytope
%models and NHCs Tops}
\subsection{Reflexive polytopes for NHCs and Tate tunings}
\label{tops}

In this subsection
we describe the construction of reflexive polytopes from
elliptic fibrations corresponding to F-theory models with gauge groups
from non-Higgsable clusters or tuning.  We give the construction of
generic models over bases with NHCs in \S\ref{imm} and
\S\ref{sec:dual-dual}, and constructions with tunings in \S\ref{tatetunepoly}.

\subsubsection{NHCs with immediately reflexive polytopes}
\label{imm}
Now consider models  where the base has a non-Higgsable cluster. 
We begin with the simplest cases, where the NHC contains a single
curve of negative self-intersection $-m$, and $m | 12$.  In these
cases, the standard stacking construction described in section
\ref{sec:reflexive-smooth} leads directly to a reflexive polytope.
This can be understood from several points of view.
Due to the factor 6 in the numerators of the first two coordinates in  (\ref{dualpile}),
those cases where a ray is skipped and
$\text{Det}[v_{i}^{(B)},v^{(B)}_j]=3,$ or $6$  also give
lattice points; i.e., when the skipped rays are NHCs $-3$ and $-6$;
furthermore, the NHCs $-4$ and $-12$ are fine as well because of  extra
factors of 2 that arise from the difference terms in the numerators. Therefore the
set (\ref{dualpile}) should also be sufficient to give the $\ds$ polytopes
of the models with the NHCs $-3, -4, -6, -12$, so that the standard
stacking polytope $\dd$ defined through (\ref{pile}) is reflexive.
The values of $m$
compatible with the standard stacking can
also be understood from the bounds on the set of monomials in $a_6$
controlled by the $-m$ curve.  Other than the vertices from $x^3,
y^2$, all vertices of $\ds$
come from lattice points associated with
monomials in $a_6$.
Choosing local toric coordinates
for a set of adjacent rays $v^{(B)}_1, v^{(B)}_2, v^{(B)}_3$ in the
base $B$ so that the ray $v_2$ corresponds to the $-m$ curve,
\begin{equation}
v^{(B)}_1 = (1, 0), v^{(B)}_{2} = (0, -1), v^{(B)}_{ 3} = (-1, -m)\,,
\label{eq:local-toric}
\end{equation} the monomials $(m_1, m_2)$
in $a_6
\in{\cal O} (-6K_B)$ are then bounded by  $ m_1 \geq -6, m_2\leq 6,$ 
 and $6-mm_2\geq m_1$.  The first and the third constraints  intersect at an integral point precisely when $m |
12$.  This intersection point is a vertex of $\ds$, so $\ds$ can only
be a lattice polytope when $m | 12$.  
Note that $6-12/m$ is the order
of vanishing of $a_6$ over the divisor
associated with $v_2^{(B)}$
 since there are no
points in the dual lattice with $m_2 > 12/m$.

As an example, the
reflexive polytope model for the generic elliptically fibered CY over
the base
$\F_{12}$ has $\{v_{i}^{(B)}\}= \{(1,0),(0,-1),(-1,-12),(0,1)\}$
(the self-intersection numbers of the toric
divisors are $\{0,-12,0,12\}$); the vertices of the 2D convex polygon
are $i = 1, 3, 4$, and the dual vertices arise from the pairs
$\{(i,j)\}=\{(1,3),(3,4),(4,1)\}$, so with these pairs, (\ref{dualpile}) gives the
vertices of the 
dual polytope $\ds$, which is a lattice polytope.  Indeed, this
polytope has vertices
\begin{equation}
\{(-6,1,1,1),(78,-6,1,1),(-6,-6,1,1),(0,0,-2,1),(0,0,1,-1)\},
\end{equation}
and is the
only reflexive polytope in the $M$ lattice (up to lattice automorphism)
associated with the Hodge pair H:11,491 in the KS database.

We can understand the reflexive polytopes formed in this way in terms
of the dual Tate tunings and tops described in the previous
subsection.  For example, consider the case of the $-3$ curve NHC.  
 Using again the local toric coordinates (\ref{eq:local-toric}) with
 $m = 3$, the polytope $\dd$ has vertices from (\ref{pile}),
$(1, 0, 2, 3)$ and $(-1, -3, 2, 3)$.  Considering a 3D slice of $\dd$
 that contains the fiber polytope $\dd_2$ and the ray $v^{(B)}_{2}
 = (0, -1)$, we have a picture like Figure~\ref{fig:levels},
where $v_i^{(B)}$ is identified with $v_2^{(B)}$.
The
 boundary of the polytope $\dd$ intersects the vertical line 
$\set{X=2, Y=3}$, which is perpendicular  to the $\set{H=0}$ plane,
 at $(X,Y,H)=(2, 3, 3/2)$; this corresponds in the polytope to
 the midpoint $(0, -3/2, 2, 3)$ of the line
 between the two vertices $(1, 0, 2, 3)$ and $(-1, -3, 2, 3)$.  The
 boundary of the polytope in the 3D slice is therefore the 2-plane
 passing through the points $(2, 3, 3/2), (0, -1, 0), (-1, 0, 0)$.
 This plane passes through the point $pt_2'$ 
($(X,Y,H)=(1,2,1)$ in the Figure), so the reflexive
 polytope associated with a standard stacking from a base with a $-3$
 curve automatically has the point $pt_2'=(0,-1,1,2)$ in the top
in $\dd$. 
Using the same methodology
 as in the $n = 6$ example above, we see that the orders of vanishing
 of the $a_n$s in the dual polytope 
are (1, 1, 1, 2, 2).  From Table~\ref{tab:tatealg}, we see that this
is a type $IV$ singularity; in this case the monodromy condition for
the gauge algebra $\gsu(3)$ is automatically satisfied, so this 
actually corresponds to an $\gsu(3)$ top, as indicated in the first
line of Table~\ref{t:dic}.

\subsubsection{Other NHCs: reflexive polytopes from the dual of the dual}
 \label{sec:dual-dual}

The rest of the NHCs have the issue that there are fractions in the
vertices of the dual polytope described by $(\ref{dualpile})$.  Let us
 denote the convex hull of the set of vertices defined by
 (\ref{pile})
by  $\tilde{\dd}$, and its dual  by  $\tilde{\ds}$.  If
$\tilde{\ds}$ is not
 a lattice polytope then
$\tilde{\dd}$ is not a reflexive polytope.  We have to supply $\tilde{\dd}$ with
additional lattice points to make it into a reflexive polytope $\dd$
so that $\ds=\dd^*$ is a lattice polytope. 

We can turn $\tilde{\dd}$ into a reflexive polytope in a minimal
fashion by taking the ``dual of the dual''.  We begin by defining the
lattice polytope $\ds^\circ = \text{convex\ hull} (\tilde{\dd}^*\cap
M)$ to be the polytope defined by the convex hull of the set of
integral points of $\tilde{\ds}$; the polytope $\ds^\circ$ then has
itself a dual $\dd = (\ds^\circ)^*$.  This gives us the minimal
reflexive polytope $\dd \supset \tilde{\dd}$ in the $N$ lattice that
we are looking for; for any base with NHCs, as we have confirmed by
explicit computation in each case, the resulting $\dd$ 
indeed has a dual
$\ds=\dd^*$ that is a lattice polytope.

This ``dual of the dual'' procedure adds points in the $N$ lattice  that are
needed to complete the tops associated with the
gauge symmetries coming from the NHCs in all cases other than those of a single
curve with self-intersection $n | 12$. 
For example, take the generic
model over $\F_5$ described by the set of rays
$\{(1,0),(0,1),(-1,-5),(0,-1)\}$; if we took just (\ref{pile}) as the
set of vertices, we would have $\{(i,j)\}=\{(1,2),(2,3),(3,1)\}$ in
(\ref{dualpile}) and there would be a non-lattice point
vertex $(-6,12/5,1,1)$ from
$(i,j)=(3,1)$. 
This problem can be seen as arising from the absence of a sufficient
set of lattice points in
$\tilde{\dd}$ over the NHC $-5$-curve $v^{(B)}_4$ to form a complete $\gf_4$
top. 
While the top in $\tilde{\dd}$ (the convex hull of the standard stacking
polytope)
over $v^{(B)}_4$ is $\{pt_1',pt_1'',pt_2',pt_3'\}$, it is $\{pt_1',
pt_1'', pt_1''', pt_2', pt_2'', pt_3', pt_4'\}$ in $\dd$; the latter is
exactly the $\gf_4$ top as described in earlier literature,  which is
obtained explicitly via the
$\ds^\circ$ construction we just described above.

For each of the NHC's, Table~\ref{t:dic} describes the tops that arise
over the divisors supporting the NHC, the corresponding Tate forms,
and the vanishing orders of $f, g, \Delta$ along with the resulting
gauge algebra.  The minimal top associated with the $\ds^\circ$
construction of $\dd$ as the dual of the dual is in each case the
first top listed.  In a number of cases there are other higher Tate
tunings that give different tops but the same gauge algebra, as
discussed further in \S\ref{sec:multiple-tops}.  The global models
describing generic elliptic fibrations over the Hirzebruch surfaces
that incorporate each of the single-curve NHC's are also described
explicitly in Table~\ref{t:fn}, showing how this construction works in
the context of the global polytopes.  While in this paper we focus on
the systematic construction of polytopes through tuning of Tate forms
(corresponding to the structure of $\ds$), one could also construct
general polytopes by considering the different tops over each base and
thus classifying polytopes $\dd$; in Table~\ref{t:dic} we also list
the possible new vertices that may arise in the polytope $\dd$ for
each top.

\subsubsection{Reflexive polytopes from Tate tunings}
\label{tatetunepoly}

We can understand Tate tunings in the polytope in a similar fashion.
Consider starting with the reflexive polytope
$\dd$ associated with the generic elliptic
fibration over a given toric base $B$, constructed as above using the
standard stacking procedure and the dual of the dual if needed for
NHC's.
We take $\ds$ to be the dual polytope of $\dd$, which is also a
lattice polytope.  We can produce an additional gauge group beyond the
minimum imposed from the NHC's by performing a  tuning in the Tate
description of the model, which corresponds to removing certain
vertices from the polytope $\ds$.  Using a Tate tuning from
Table~\ref{tab:tatealg} gives us the set of lattice points that should be
removed from $\ds$ associated with certain coefficients in the $a_n$s
over the divisor(s) in $B$.  Calling the new $M$ polytope that results
from removing these lattice points $\hat{\ds}$, we get an enlarged $N$
polytope $\hat{\dd} = (\hat{\ds})^*$, which has extra lattice points.  In general, each Tate tuning in $\ds$
gives a corresponding top in $\dd$, giving a new reflexive polytope
$\hat{\dd}$.  This gives a large class of constructions for Tate
tunings that should have reflexive polytope analogues in the KS
database.

As a simple example, consider the polytope $\dd$ associated with the
generic elliptic fibration over $\F_2$.  As discussed in \S\ref{sec:reflexive-smooth}
this polytope follows from the standard stacking procedure and has
vertices given by
\begin{equation}
\dd=\text{Conv}\{(1,0,2,3), (-1,-m,2,3), (0,1,2,3),(0,0,-1,0), (0,0,0,-1)\}
\label{eq:dd-2}
\end{equation}
with $m = 2$.  This is a reflexive polytope, identified in the
Kreuzer-Skarke database as M:335 5 N:11 5 H:3,243.
The dual polytope $\ds$ has vertices
\begin{equation}
 \{(-6,-6,1,1),(0,0,-2,1),(18,-6,1,1),(0,0,1,-1),(-6,6,1,1)\}.
\end{equation}
Now consider a Tate tuning of the algebra $\gsu(2)$ over the $-2$
curve $C$ in the base, which corresponds to the 2D toric vector $(0, -1)$.  This is achieved by setting $a_1, a_2, a_3, a_4,
a_6$ to vanish to orders $\{0, 0, 1, 1, 2\}$ in the coordinate
associated with $C$, which is the second coordinate in $\ds$. The set
of the lattice points that have to be removed from $\ds$ to achieve
the required vanishing orders is
$\{(-6,5,1,1),(-6,6,1,1),(-5,5,1,1),(-4,4,0,1),(-4,5,1,1),(-3,3,1,0)\}$. The
resulting new  $M$ polytope after the reduction is
\begin{eqnarray}
\hat{\ds}  & = & \text{Conv}
 \{(-6,-6,1,1), (-6,4,1,1), (-2,2,-1,1), (-2,4,1,1), (0,0,-2,1), (0,0,1,-1), \nonumber\\
& & \hspace*{0.5in}
 (18,-6,1,1)\}.
\end{eqnarray}
This gives the reflexive polytope $\hat{\dd}$ given by augmenting
$\dd$ from (\ref{eq:dd-2}) with the additional lattice point $(0,
  -1, 1, 2)$, which gives the $\gsu(2)$ top over $C,$ as described in
Table~\ref{t:top-Tate}. The resulting polytope
 corresponds to the example M:329 7 N:12
  6 H:4,238 in the KS database. The Hodge numbers from the polytope
  data are consistent with those from the
anomaly cancellation calculation
  in equations (\ref{eq:dh11}) and (\ref{eq:dh21}) with a tuning of $\gsu(2)$
  on the isolated $-2$ curve: $\Delta h^{1,1}=\text{rank}(\gsu(2))= 1,
  \Delta h^{2,1}=\text{dim}(\gsu(2))-4\times {\bf 2}=3-8=-5.$

In general, we find that the correspondence described in the last few
subsections between Tate tunings and tops immediately provides
reflexive polytopes for most Tate tuning constructions.
There are several subtleties in this construction, which we elaborate
in the remainder of this chapter.
 
\subsection{Multiple tops}
\label{sec:multiple-tops}

One  thing that we have found in considering the variety
of Tate tunings and the corresponding models in the KS database is
that for many gauge algebras there are multiple distinct tops that can
arise in the $N$-polytope $\dd$.  
This multiplicity of tops was also discussed in \cite{Braun-16}.
These different tops correspond to distinct Tate tunings of the same gauge algebra.
In many cases these arise in
situations where the gauge algebra in the Weierstrass model depends
upon some monodromy condition, which may be satisfied automatically in
certain cases by the Tate tuning.

As an example of this phenomenon, 
consider
the generic model over the $\F_{m=3}$ base, 
\begin{equation}
\dd=\text{Conv}\{(1,0,2,3), (-1,-3,2,3), (0,1,2,3),(0,0,-1,0),
(0,0,0,-1)\} \,.
\label{Delta}
\end{equation}
This is already a reflexive polytope, M:348 5 N:12 5 H:5,251,
 with the top over the $-3$-curve
$\{pt_1', pt_2'\}$ that we found at the end of \S \ref{imm}.  Naively from Table~\ref{t:top-Tate}, this
might appear to be an
``$\gsu(2)$'' top; however
looking explicitly at the Tate form  associated to
the polytope $\ds$, the vanishing orders along the $-3$-curve are
\{1,1,1,2,2\} in terms of the five sections $a_n$, and $\{2,2,4\}$ in terms
of $\{f, g, \Delta\}$, and the $\gsu(3)$ monodromy condition is
satisfied - hence the gauge algebra is indeed $\gsu(3)$ (indeed, we
know from the presence of the $-3$
 NHC that $\gsu(2)$
is not possible in this geometry.)  
In \S$\ref{TT}$ (see
in particular Table \ref{tab:tatealg}), we described two distinct Tate
tunings for $\gsu(3)$.  In this case, 
the geometry matches the alternate Tate
form for $IV^s$ associated with vanishing of $a_6$ to order 2 and an
additional monodromy condition, and the ``top'' is a non-standard
$\gsu(3)$ top.  There also exists a polytope model 
with  the ``usual'' $\gsu(3)$ top: adding $pt_3'$
($(0,-1,1,1)$) to the top gives another polytope model  M:347 7 N:13 6
H:5,251, which has the standard
$\gsu(3)$ top; on the Tate side this model
can be obtained by
the reduction of the $M$ polytope such that the vanishing orders along the
$-3$-curve become $\{1,1,1,2,3\}$ - the standard Tate form for
$IV^s$. 
Analogous situations arise for the NHCs $-4$ and $-6$ as well: in
these cases, as discussed above,
$\dd$ in equation (\ref{Delta}) is already a reflexive polytope model
of the generic CY over $\F_{m=4,6}$. 
 The tops over the $-m$ curves in these cases look like
 those appearing in the literature for gauge algebras $\gg_2, \gf_4$
respectively, and the
vanishing orders along the $-m$-curves are
$\{1,1,2,2,3\}, \{1,2,2,3,4\}$ for $m=4,6$, which would naively be
tunings for $\gg_2, \gf_4$.  In these cases, however, the gauge algebras are actually
$\gso(8), \ge_6$ with monodromy conditions satisfied. Just like the
case for $\F_3$, there are also generic polytope models over $\F_{4,6}$
that have the usual $\gso(8), \ge_6$ tops and Tate vanishing orders of
$\gso(8), \ge_6$.  The extra lattice points in the tops of these $\dd$ 
polytopes precisely correspond to the reduction in Tate monomials of
the $M$ polytope $\ds$.

In addition to multiple tops associated with monodromy conditions in
Tate tunings, there are also other Tate tuning/top combinations that
can arise for certain gauge groups.  We have not attempted a
systematic analysis of all possibilities, but we have encountered a
range of possibilities simply in analyzing the polytopes of the KS
database with fixed Hodge numbers and associated Tate tunings for the
dual polytopes.  To give a sense of the possibilities that arise, we
list the structures of the polytopes in the KS database that have the
Hodge numbers of generic elliptically fibered CYs over $\F_{m}$ bases
for $0\le m\le 12$ in Table \ref{t:fn}.  The details of the
corresponding Tate forms for the $-m$ NHCs are given in
Table~\ref{t:dic}.  Note in particular, that in addition to those
models mentioned above, there is a third polytope model associated
with the Hodge numbers of the generic elliptic fibration over $\F_6$
in addition to the monodromy construction and the standard
construction discussed above.  This third possibility involves a
further specialization of the vanishing orders of the standard
construction along the $-6$-curve, giving a further reduced $M$
polytope $\ds$.  Another interesting case of multiple tops that arises
in these tables is the possibility of a second type of Tate tuning/top
for $\ge_7$ on a $-8$ curve.  In this case there is no monodromy
issue\footnote{However, note that the same
  Tate vanishing orders $\set{1,2,3,3,5}$ may also give the $\ge_7$
  algebra
over
  $-7$ curves where there  is also charged matter.}, but a second
Tate tuning where the degree of vanishing of $a_6$ is enhanced,
associated with a second $\ge_7$ top and corresponding reflexive
polytopes.

In the analysis in the remainder of this paper we focus on classifying
the possible elliptic fibrations constructions through the set of Tate
tunings.  One could also, however, imagine classifying different
reflexive polytopes by considering all ways of augmenting the set of
vertices (\ref{pile}) associated with the ``standard stacking'' with all
possible tops.  Proceeding in this fashion would require a systematic
way of identifying the complete set of tops for each possible tuned
gauge group.

\begin{table}[]
\centering
\begin{tabular}{|c|c|c|c|l|}
\hline
Hodge pair                              & Mult.~in KS                          & $\F_n$ base      & Gauge symmetry           & Top over the $-n$-curve                               \\ \hline
\multirow{3}{*}{(3,243)}                & \multirow{3}{*}{3}                     & $\F_2$    & trivial  &  \{$pt_1'$\} (affine node)\\ \cline{3-5} 
                                        &                                        & $\F_1$    & trivial&   \{$pt_1'$\} (affine node) \\ \cline{3-5} 
                                        &                                        & $\F_0$    &trivial   &\{$pt_1'$\} (affine node)     \\ \hline
\multirow{3}{*}{(5,251)}                & \multirow{3}{*}{3}                     & $\F_3$    & $\gsu(3)$  & $\{pt_1', pt_2'\}$ ("$\gsu(2)$") \\ \cline{3-5} 
                                        &                                        & $\F_3$    &  $\gsu(3)$  &$\{pt_1', pt_2', pt_3'\}$    \\ \cline{3-5} 
                                        &                                        & $\F_3$    & $\gg_2$ enhanced on -3   &$\{pt_1'', pt_2', pt_3'\}$                \\ \hline
\multirow{4}{*}{(7,271)}                & \multirow{4}{*}{4}                     & $\F_4$    & $\gso(8)$  &$\{pt_1'', pt_2', pt_3'\}$ ("$\gg_2$")   \\ \cline{3-5} 
                                        &                                        & $\F_4$    & $\gso(8)$  & $\{pt_1'', pt_2', pt_3',pt_4'\}$ ("$\gso(7)$") \\ \cline{3-5} 
                                        &                                        & $\F_4$    & $\gf_4$ enhanced on -4   & $\{pt_1''', pt_2'', pt_3',pt_4'\}$               \\ \cline{3-5} 
                                        &                                        & $\F_4$    & $\gso(9)$ enhanced on -4 & $\{pt_1'', pt_2'', pt_3', pt_4'\}$               \\ \hline
(7,295)                                 & 1                                      & $\F_5$    &$\gf_4$    & $\{pt_1''', pt_2'', pt_3',pt_4'\}$                      \\ \hline
\multirow{3}{*}{(9,321)}                & \multirow{3}{*}{3}                     & $\F_6$    &  $\ge_6$    & $\{pt_1''', pt_2'', pt_3',pt_4'\}$ ("$\gf_4$")   \\ \cline{3-5} 
                                        &                                        & $\F_6$    &  $\ge_6$    & $\{pt_1''', pt_2'', pt_3'',pt_4',pt_5',pt_7'\}$                   \\ \cline{3-5} 
                                        &                                        & $\F_6$    & $\ge_6$    & $\{pt_1''', pt_2'', pt_3'',pt_4',pt_5'\}$       \\ \hline
(10,348)                                & 1                                      & $\F_7$  & $\ge_7$  (w/ matter \bf{$\frac1{2}56$})    &$\{pt_1'''', pt_2''', pt_3'',pt_4'',pt_5'\}$                     \\ \hline
\multirow{2}{*}{(10,376)}               & \multirow{2}{*}{2}                     & $\F_8$    & $\ge_7$ w/o matter    & $\{pt_1'''', pt_2''', pt_3'',pt_4'',pt_5',pt_6'\}$                \\ \cline{3-5} 
                                        &                                        & $\F_8$    & $\ge_7$ w/o matter  & $\{pt_1'''', pt_2''', pt_3'',pt_4'',pt_5'\}$      \\ \hline
(11,491)                                & 1                                      & $\F_{12}$ & NHC -12 curve: $\ge_8$   & $\{pt_1^{(6)}, pt_2'''', pt_3''',pt_4'',pt_5'\}$                     \\ \hline
\end{tabular}
\caption{\footnotesize Polytope models associated with generic
  elliptic fibrations over the Hirzebruch surfaces
  $\F_{0,1,\ldots,8,12}$, as well as all other models with the same Hodge
  numbers.  Alternate constructions include multiple tops, some due to
monodromy conditions in Tate tunings, as well as rank-preserving
tunings
(\S\ref{sec:multiple-tops}).}
\label{t:fn}
\end{table}

\begin{table}
\begin{center}
\begin{tabular}{|c |c |c |c|c|c|}
\hline
NHC&
Tops &
Possible vertices&
Tate form&
$(f,g,\Delta)$&
$G$
\\ \hline \hline
-3 &$\{pt_1', pt_2'\}$ & none &$\{1,1,1,2,2\}$&{\multirow{2}{*}{$(2,2,4)$}}&{\multirow{2}{*}{$\gsu(3)$}}\\ 
 &$\{pt_1', pt_2', pt_3'\}$ & $pt_3'$&$\{1,1,1,2,3\}$&& \\
 \hline
-4  &$\{pt_1'', pt_2', pt_3'\}$ & none&$\{1,1,2,2,3\}$&{\multirow{2}{*}{$(2,3,6)$}}& {\multirow{2}{*}{$\gso(8)$}}\\
  &$\{pt_1'', pt_2', pt_3',pt_4'\}$ & $pt_4'$&$\{1,1,2,2,4\}$&& \\
  \hline
-5 &$\{pt_1''', pt_2'', pt_3',pt_4'\}$ & $pt_1'''$&$\{1,2,2,3,4\}$&$(3,4,8)$&$\gf_4$ \\
\hline
-6 &$\{pt_1''', pt_2'', pt_3',pt_4'\}$ & none&$\{1,2,2,3,4\}$&&\\
&$\{pt_1''', pt_2'', pt_3'',pt_4',pt_5'\}$ & $pt_3'', pt_5'$&$\{1,2,2,3,5\}$&$(3,4,8)$&$\ge_6$ \\
&$\{pt_1''', pt_2'', pt_3'',pt_4',pt_5',pt_7'\}$ & $pt_7'$&$\{1,2,2,4,6\}$&&\\
\hline
-7 &$\{pt_1'''', pt_2''', pt_3'',pt_4'',pt_5'\}$ & $pt_1'''', pt_4''$&$\{1,2,3,3,5\}$&$(3,5,9)$&$\ge7$\\
\hline
-8 &$\{pt_1'''', pt_2''', pt_3'',pt_4'',pt_5'\}$ & $pt_4''$&$\{1,2,3,3,5\}$&{\multirow{2}{*}{$(3,5,9)$}}&{\multirow{2}{*}{$\ge_7$}}\\
&$\{pt_1'''', pt_2''', pt_3'',pt_4'',pt_5',pt_6'\}$ & $pt_6'$&$\{1,2,3,3,6\}$&&\\
\hline
-12 &$\{pt_1^{(6)}, pt_2'''', pt_3''',pt_4'',pt_5'\}$ & none&$\{1,2,3,4,5\}$&$(4,5,10)$&$\ge8$\\
\hline\hline
-2, &$\{pt_1', pt_2'\}$, & $\{pt_2'\}$,&$\{1, 1, 1, 1, 2\}$, &$(1,2,3)$,&$\gsu(2)$\\
 -3&$\{pt_1'', pt_2', pt_3'\}$ & $\{pt_1''\}$&$\{1, 1, 2, 2, 3\}$&$(2,3,6)$& $\oplus\gg_2$\\
\hline
-2,  &$\{pt_1'\}$, & none,& $\{1,1,1,1,1\}$,&$(1,1,2)$,&\\
  -2, &$\{pt_1', pt_2'\}$, & none,&$\{1,1,2,2,2\}$, &$(2,2,4)$,&$\gsu(2)$\\
-3&$\{pt_1'', pt_2', pt_3'\}$ & $\{pt_1''\}$&$\{1,1,2,2,3\}$&$(2,3,6)$&$\oplus\gg_2 $\\
\hline
-2, &$\{pt_1',pt_2'\}$, & none,&$\{1, 1, 1, 1, 2\}$, &$(1,2,3)$,&$\gsu(2) $\\
 -3, &$\{pt_1'', pt_2', pt_3',pt_4'\}$, & $\{pt_4'\}$,&$\{1, 2, 2, 2, 4\}$,&$(2,4,6)$,&$\oplus\gso(7) $\\
-2  &$\{pt_1', pt_2'\}$&none& $\{1, 1, 1, 1, 2\}$&$(1,2,3)$&$\oplus\gsu(2)$\\
\hline
\end{tabular}
\begin{tabular}{|c |c |c |c|c|c|}
\hline
-9 blown up at 3pts&$\{pt_1^{(6)}, pt_2'''', pt_3''',pt_4'',pt_5'\}$ & $pt_1^{(6)}$&$\{1,2,3,4,5\}$&$(4,5,10)$&$\ge8$\\
\hline
-10 blown up at 2pts &$\{pt_1^{(6)}, pt_2'''', pt_3''',pt_4'',pt_5'\}$ & $pt_1^{(6)}$&$\{1,2,3,4,5\}$&$(4,5,10)$&$\ge8$\\
\hline
 -11 blown up at 1pt&$\{pt_1^{(6)}, pt_2'''', pt_3''',pt_4'',pt_5'\}$ & $pt_1^{(6)}$&$\{1,2,3,4,5\}$&$(4,5,10)$&$\ge8$\\
\hline
\end{tabular}

\end{center}
\caption[x]{\footnotesize Tops over NHCs and the corresponding Tate
  vanishing orders.  
In each case the first example is the top and associated minimal Tate
tuning associated with the ``dual of the dual'' construction described
in \S\ref{sec:dual-dual}.} 
\label{t:dic}
\end{table}

\begin{figure}
\begin{center}
\includegraphics[width=10cm]{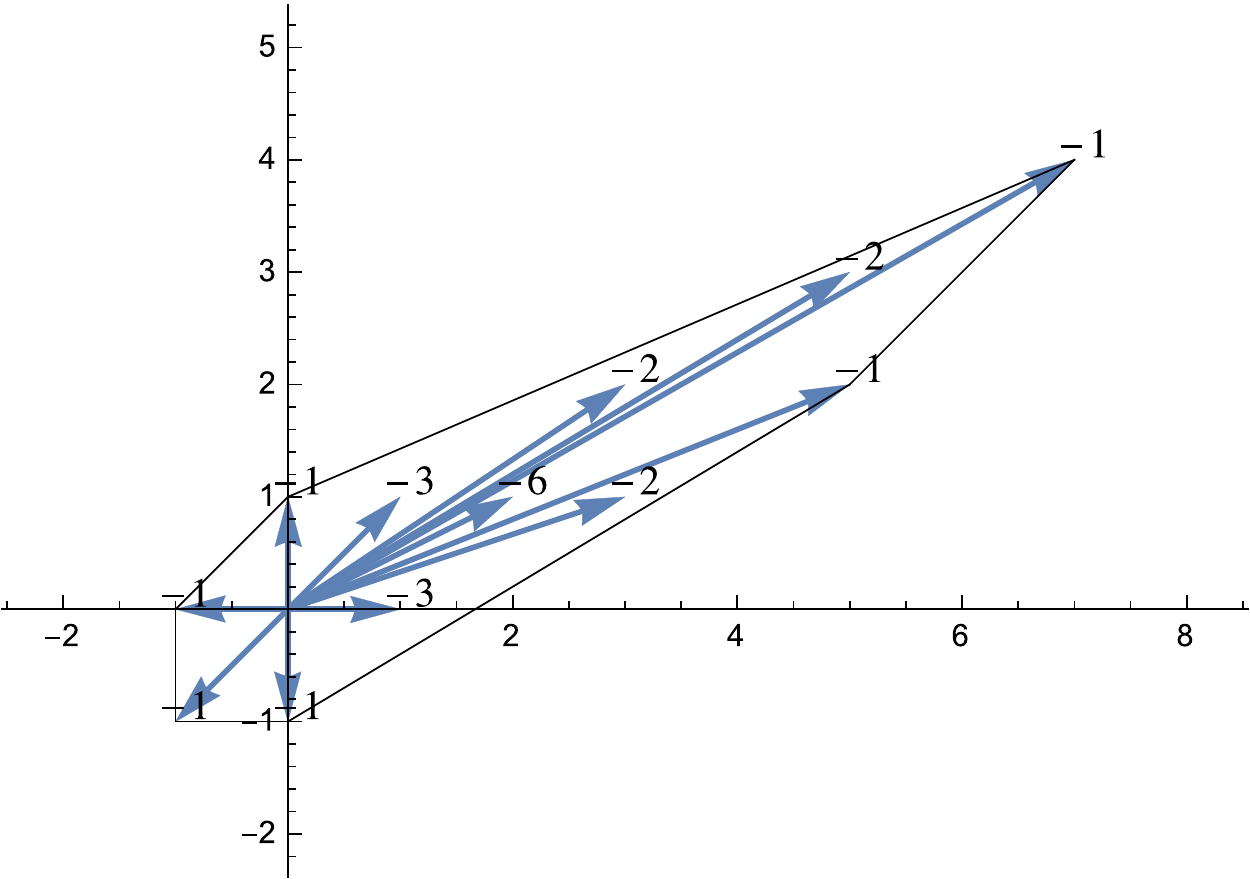}
\end{center}
\caption[x]{\footnotesize  The toric fan of the base of a generic model with small $h^{1,1}$:  \{23, 107, \{-3, -2, -2, -1, -6, -1, -2, -3, -1, -1, -1, -1\}\}. Each $-1$-curve in the base corresponds to a vertex of $\dd$.} 
\label{examplesmall}
\end{figure}

We will not deal systematically  with the explicit triangulation
of $\dd$, corresponding to the resolution of the Calabi-Yau threefold, but make some comments here on the relationship
between extra rays in $\dd$ and the
resolution of the singular fiber associated with a tuned or
non-Higgsable gauge group.
Many of the details of this correspondence were worked out in
\cite{cpr, Perevalov:1997vw}.
  When the gauge algebra is non-trivial over
a divisor $D^{(B)}$, there are lattice points in the top over
$v^{(B)}$ in addition to just $pt_1'$. Specifically, in the cases
where there are no lattice points in the top lying in the interior of
the 2-dimensional faces of $\dd$, the lattice points in the top that
do not lie in the interior of the 3-dimensional faces of $\dd$ form
the Dynkin diagram of the gauge algebra.  These correspond to the
exceptional divisors that arise in the resolution of the corresponding
singularities. However, when there are lattice points lying in the
interior of the 3-dimensional and  the 2-dimensional faces of $\dd$, they contribute to the
second and
third terms, respectively, in Batyrev's $h^{1,1}$ formula
(\ref{latth11}).
The second term corresponds to components that miss the hypersurface,
and  contributions to the third term 
arise when the singularity is not resolved by a toric divisor but
rather by a non-toric deformation, so the Dynkin diagram is not
fully visible from the top. This happens exactly in those gauge
algebras with an additional monodromy condition
that is automatically satisfied. 

In summary, $\dd$ models are divided into two types according to whether there
is a nonzero third term in the $h^{1,1}$ formula (\ref{latth11}): (1) 
Trivial third term: There is no lattice point lying in the interior of any
two-dimensional face. Gauge algebras can be read off directly from
tops (the nodes of the Dynkin diagram are given by the lattice points
in the top that do not lie in interior of facets), which are those in
the literature. The Tate forms are those with
no additional monodromy
condition, which again match those in the literature. The nodes also
correspond to exceptional divisors resolving the singular fiber. (2)
Non-vanishing third term: There are lattice points lying in the
interior of two-dimensional faces.  These cases give rise to the
additional Tate forms we have described. For example, in the gauge algebras involved
with monodromy conditions, there are Tate forms of lower degrees,
which achieve the gauge algebras by satisfying the additional
monodromy conditions automatically. The singular fiber is (partially)
resolved by deformation.  Therefore, there are fewer exceptional
divisors in the top, in which the ``Dynkin diagram'' would seem to be
the lower rank gauge algebra counterpart.

Finally, recall from Table~\ref{t:rpt} that there are  \emph{rank-preserving}
tunings of certain gauge algebras that leave the Hodge numbers of an
elliptic Calabi-Yau unchanged.  These are also associated with further
Tate tunings on $\ds$ and additional tops in $\dd$ that do not change
the Hodge numbers from the generic elliptic fibration over a given
base.  The polytopes listed in Table~\ref{t:fn} include
rank-preserving tunings of
$\gg_2$ over the $-3$ curve in $\F_3$, and $\gf_4, \gso(9)$ over the $-4$ curve in
$\F_4$.  A detailed example of the polytopes associated with different
tunings of $\gsu_3$ and $\gg_2$ over $\F_2$ is given in
Appendix~\ref{2dfiber}.

\subsection{Combining tunings}
\label{sec:combining}

A final important issue that we must consider in attempting to
systematically construct global models associated with polytopes is
whether given a generic model over a given base, all combinations of
Tate tunings that are each possible locally can be combined into an
allowed global model.  This depends on the global structure of the
base and can be tested by the
Tate-Zariski decomposition discussed in
\S\ref{sec:Zariski-Tate}.  As discussed there, we can perform a
Zariski decomposition, with the initial values of $\{c_{j,n}\rvert
n\}$ over each curve set to be the initial values we want in Table
\ref{t:dic}.  We then carry out the Tate-Zariski iteration procedure
and if the Zariski decomposition with the desired vanishing values and
corresponding gauge groups does not exist, there will not be a
corresponding polytope model.  In general, if the Zariski
decomposition works out, there is a corresponding polytope.  We do not
have a proof of this in general but as we see later, at least the
Hodge numbers of every elliptic Calabi-Yau threefold constructed in
this way arise from a polytope in the KS database.  This analysis of
combined tunings through Tate-Zariski is the essential analysis we
carry out in our systematic enumeration of Tate tunings that should
have corresponding polytopes.  To illustrate the issues that can arise
we give a couple of simple examples here, where one but not all of the
possible Tate tunings over a given curve in the base are consistent
with a global model.

Let us consider first as a concrete example the
generic model over the base with toric curves of self-intersection numbers $\{-3, -2, -2, -1, -6, -1, -2, -3, -1, -1,
-1, -1\}$, for which the toric rays take coordinates
$\{v_{i}^{(B)}\}$=$\{(1,1)$, $(3,2)$, $(5,3)$, $(7,4)$, $(2,1)$, $(5,2)$, $(3,1)$, $(1,0)$, $(0,-1)$, $(-1,-1)$, $(-1,0)$, $(0,1)\}$
(figure \ref{examplesmall}). 
We consider Tate tunings that keep the gauge group the same as in
the generic model, determined by the NHCs.
From Table~\ref{t:dic}
and the discussion in the preceding subsection, we see that there are
three different possible Tate tunings over the $-6$ curve:
$\{1, 2, 2, 3, 4\}$,
$\{1, 2, 2, 3, 5\}$,
$\{1, 2, 2, 4, 6\}$.  We wish to know which of these three tunings
leads to a consistent Tate-Zariski decomposition, and which
corresponding polytopes exist.

For the polytope $\dd$ in each of these three cases,
we have the vertices from the fiber
\begin{equation}
\{(0,0,-1,0), (0,0,0,-1)\},
\end{equation}
the vertices from the base, which come from the $-1$'s:
\begin{equation}
 \{(7,4,2,3), (5,2,2,3), (0,-1,2,3), (-1,-1,2,3), (-1,0,2,3), (0,1,2,3)\},
\end{equation}
 and vertices from the tops of the NHCs
\begin{itemize}
\item $-3, -2, -2$: \{(2, 2, 2, 3)\},
\item $-2, -3$: \{(3, 1, 1, 2), (2, 0, 2, 3)\},
\item $-6$ with three choices of different possible top vertices.
\end{itemize}

We now consider each of the tunings in turn over the $-6$ curve:
\begin{enumerate}
\item Minimal tuning $\{1, 2, 2, 3, 4\}$, corresponding to no
  additional top vertex from Table~\ref{t:dic}.
This construction leads to a consistent Zariski decomposition, which gives rise
to the generic polytope model M:148 11 N:33 11 H:23,107$^{[2]}$: we
start with the initial configuration
%the first top realization $\{1,2,2,3,4\}$ for the $-6$ and minimal Tate vanishing order for the other curves:
\begin{eqnarray}
\nonumber
\{\{1,1,2,2,3\},\{1,1,2,2,2\},\{1,1,1,1,1\},\{0,0,0,0,0\},{\color{blue}\{1,2,2,3,4\}},\{0,0,0,0,0\},\\\nonumber
\{1,1,1,1,2\},\{1,1,2,2,3\},\{0,0,0,0,0\},\{0,0,0,0,0\},\{0,0,0,0,0\},\{0,0,0,0,0\}\}.
\\\label{-60}
\end{eqnarray}
After the iteration procedure, the configuration becomes
\begin{eqnarray}
\nonumber
\{\{1,1,2,2,3\},\{1,1,2,2,2\},\{1,1,1,1,1\},\{1,1,0,0,0\},{\color{blue}\{1,2,2,3,4\}},\{1,1,0,0,0\},\\\nonumber
\{1,1,1,1,2\},\{1,1,2,2,3\},\{0,0,0,0,0\},\{0,0,0,0,0\},\{0,0,0,0,0\},\{0,0,0,0,0\}\},\nonumber
\label{-61}
\end{eqnarray}
where each curve still has their suitable Tate vanishing orders, which
persist as $\{1,2,2,3,4\}$ on $-6$. 
\item 
Tate tuning $\{1, 2, 2, 3, 5\}$, corresponding to the additional top vertices $\{pt_3'',
pt_5'\}=\{(2, 1, 0, 0), (4, 2, 1, 1)\}$ over the $-6$ curve. This
works as well and gives the generic polytope model M:147 12 N:35 13
H:23,107$^{[1]}$: we start with the initial configuration in (\ref{-60})
but with the vanishing orders along $-6$ replaced by
\{1,2,2,3,5\}. The configuration after iteration becomes
\begin{eqnarray}
\nonumber
\{\{1,1,2,2,3\},\{1,1,2,2,2\},\{1,1,1,1,1\},\{1,1,0,0,0\},{\color{blue}\{1,2,2,3,5\}},\{1,1,0,0,1\},\\\nonumber
\{1,1,1,1,2\},\{1,1,2,2,3\},\{0,0,0,0,0\},\{0,0,0,0,0\},\{0,0,0,0,0\},\{0,0,0,0,0\}\},\nonumber
\end{eqnarray}
where each curve still has their suitable Tate vanishing orders,
which persist as $\{1,2,2,3,5\}$ on $-6$.
\item 
Tate tuning $\{1, 2, 2, 4, 6\}$, which would correspond to the
additional top vertex
$\{pt_7'\}=\{(2, 1, 0, -1)\}$. This does not give a
consistent polytope.  The iteration of  the initial configuration 
\begin{eqnarray}
\nonumber
\{\{1,1,2,2,3\},\{1,1,2,2,2\},\{1,1,1,1,1\},\{0,0,0,0,0\},{\color{blue}\{1,2,2,4,6\}},\{0,0,0,0,0\},\\\nonumber
\{1,1,1,1,2\},\{1,1,2,2,3\},\{0,0,0,0,0\},\{0,0,0,0,0\},\{0,0,0,0,0\},\{0,0,0,0,0\}\}\nonumber
\end{eqnarray}
becomes
\begin{eqnarray}
\nonumber
\{\{1,1,2,2,3\},\{1,1,2,2,3\},\{1,1,1,2,3\},\{1,1,0,2,3\},\{1,2,2,4,6\},\{1,1,0,2,3\},\\\nonumber\{1,1,1,2,3\},\{1,1,2,2,3\},\{0,0,0,0,0\},\{0,0,0,0,0\},\{0,0,0,0,0\},\{0,0,0,0,0\}\},
\end{eqnarray}
where the vanishing orders over the NHC $-2, -2, -3$ are
disturbed. Hence, unlike the case of the $\F_6$ base where there is a
generic polytope model of vanishing order $\{1,2,2,4,6\}$ on $-6$, the
third Tate tuning and corresponding top realization does not exist for
this base.
\end{enumerate}

%Not only does Tate Zarisk tell the existence of polytope models, more significantly, the results of the Tate-Zariski decomposition equation (\ref{-61}) and (\ref{-62}) give exactly the same results calculated directly from the respective polytope data M:148 11 N:33 11 H:23,107 and M:147 12 N:35 13 H:23,107 of the vanishing order along each curves in the base, which indicates the equivalence of the Zariski decomposition and reflexive polytope construction for the elliptic fibration CY models.

\begin{figure}
\begin{center}
\includegraphics[width=15cm]{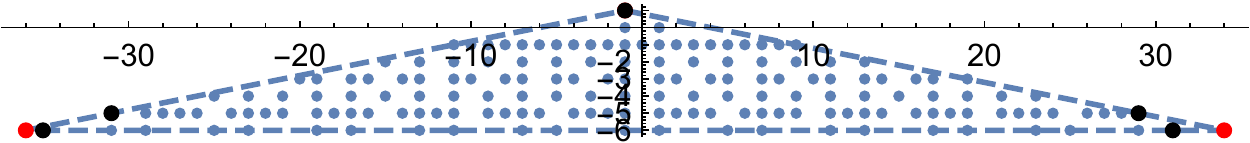}
\end{center}
\caption[x]{\footnotesize  The toric fan (arrows indicating rays are
  simplified to points for clarity) of the base of a generic model
  with large $h^{1,1}$
  \{416,14,\{-12,-1,-2,-2,-3,-1,-5,-1,-3,-2,-2,$\mathbf{-1}$,-11,$\mathbf{-1}$,-2,-2,-3,-1,-5,-1,-3,-2,-2,-1,-12//-12//-12//-12//-12//-12//-12//-12//-12//-12//-12,$\mathbf{-1}$,-2,-2,-3,-1,-5,-1,-3,-2,$\mathbf{-1}$,-8,-1,-2,-3,-2,-1,-8,$\mathbf{0}$\},
  where the five curves corresponding to vertices of the base are in
  boldface, and are denoted by black dots in the fan diagram. The
  point at the top $(-1,1)$ corresponds to the zero curve, which is
  also a vertex of $\dd$. Two red dots in the fan diagram correspond
  to points in the tops: $pt_1^{(6)}$ of $D_{B13}$ and $pt_1''$ of
  $D_{B153}$, respectively; these points thus do not correspond to
  rays of the base fan.}
\label{examplelarge}
\end{figure}

As another illustrative example, consider
the polytopes associated with
Hodge numbers H:416,14, which match those of the generic elliptic
Calabi-Yau threefold
over the base 
$\{416,14,\{-12//-11//-12//-12//-12//-12//-12//-12//-12//-12//-12//-12//-12,-1,-2,-2,-3,-1,-5,-1,-3,-2,-1,-8,-1,-2,-3,-2,-1,-8,0\}\}$ 
(see figure \ref{examplelarge}, by $//$ we denote the sequence of
curves $-1, -2, -2, -3, -1, -5, -1, -3, -2, -2, -1$; there are in total 163 toric curves in
the base, with curves 153 and 162 being the $-8$ curves). 
There are only two polytope models in the
KS database with
H:416,14, and both give polytope models of the CY with generic gauge
group over the given base, with different
detailed Tate tuning/top structure.
Naively one might expect four models, since there are two different
$\ge_7$ tunings possible over each $-8$ curve.  Analyzing the
structure of the polytopes, however, we find:

\begin{enumerate}
\item M:26 6 N:576 {\color{blue}6} H:416,14
\begin{itemize}
\item A vertex from the $0$-curve in the base. In particular,
note that all $-1$ curves in // do not contribute to vertices.
\item Vertices from NHC tops 
\begin{enumerate}
\item $D_{B13}$ ([-11]): $pt_1^{(6)}$
\item $D_{B153}$ (-3 in [-3 -2]): $pt_1''$
\item $D_{B162}$ ([-8]): $pt_6'$
\end{enumerate}
\item and vertices $v_x, v_y$.
\end{itemize}

\item M:29 7 N:575 {\color{blue}7} H:416,14
\begin{itemize}
\item Vertex contributions from the base and the fiber are the same as the first case.
\item Vertices from NHC tops
\begin{enumerate}
\item $D_{B13}$ ([-11]): $pt_1^{(6)}$
\item $D_{B153}$ (-3 in [-3 -2]): $pt_1''$
\item $D_{B156}$ ([-8]): $pt_4''$
\item $D_{B162}$ ([-8]): $pt_4''$
\end{enumerate}
\end{itemize}

\end{enumerate}
In the first model, the top over the first $-8$-curve ($D_{B156}$) is
$\{pt_1'''', pt_2''', pt_3'',pt_4'',pt_5'\}$ while over the second
($D_{B162}$) is $\{pt_1'''', pt_2''', pt_3'',pt_4'',pt_5',pt_6'\}$; in
the second model, it is $\{pt_1'''', pt_2''', pt_3'',pt_4'',pt_5'\}$
over both $-8$-curves; however there is no model of top $\{pt_1'''',
pt_2''', pt_3'',pt_4'',pt_5',pt_6'\}$ over $D_{B156}$.  This matches
with the observation that there is
no corresponding Zariski decomposition --- the vanishing orders can not
be $\{1,2,3,3,6\}$ along $D_{B156}$.

Note that these models also illustrate another point: a vertex of the
base can only come from curves with self-intersection number $m$
greater than $-2$, but all curves with $m>-2$ will not necessarily be
  vertices. Though this generally is the case for small $h^{1,1}$,
  exceptions increase as $h^{1,1}$ increases, since additional rays
  can expand the convex hull of the base polytope. Also, a vertex
  associated with a top can only come from those possibilities listed
  in the third column of Table \ref{t:dic}, but the entries in that
  column are not always vertices, though they are always lattice
  points in the $N$ polytope $\dd$.  This fact can be seen in the
  first model in the second example: $pt_4''$ over $D_{B156}$ ([-8])
  is not a vertex point.

\subsection{Tate tunings and polytope models of $\gso(n)$ gauge algebras}
\label{so8}

\begin{table}[]
\centering
\begin{tabular}{|l|l|}
\hline
Generic models NHCs&  \begin{tabular}[c]{@{}l@{}}$\cdot\oplus\gso(8)\oplus\cdot\oplus\gso(8)\oplus\cdot\oplus\gso(8)\oplus\cdot$\end{tabular}            \\ \hline
M:41 5 N:457 5 H:335,23  & \begin{tabular}[c]{@{}l@{}}\{\{1,0,1,0,0\},\{1,1,2,2,3\},\{1,0,1,0,0\} ,\{1,1,2,2,3\},\\\{1,0,1,0,0\},\{1,1,2,2,3\},\{0,0,0,0,0\}\}\end{tabular}                         \\ \hline
M:40 7 N:460 7 H:335,23  & \begin{tabular}[c]{@{}l@{}}\{\{1,0,1,0,1\},\{1,1,2,2,4\},\{1,0,1,0,2\} ,\{1,1,2,2,4\},\\ \{1,0,1,0,2\},\{1,1,2,2,4\},\{0,0,0,0,0\}\}\end{tabular}                         \\ \hline\hline
Tuned symmetries  & \begin{tabular}[c]{@{}l@{}}$\cdot\oplus\gso(9)\oplus\gsp(1)\oplus{\gso(10)}\oplus\gsp(1)\oplus\gso(9)\oplus\cdot$\end{tabular} \\ \hline
M:39 7 N:465 7 H:338,22  & \begin{tabular}[c]{@{}l@{}}\{\{1,0,1,2,1\},\{1,1,2,3,4\},\{1,0,1,2,2\},{\{1,1,2,3,4\}},\\ \{1,0,1,2,2\},\{1,1,2,3,4\},\{0,0,0,0,0\}\}\end{tabular}           \\ \hline
M:38 8 N:466 8 H:338,22  & \begin{tabular}[c]{@{}l@{}}\{\{1,0,1,2,1\},\{1,1,2,3,4\},\{1,0,1,2,3\},{\{1,1,2,3,5\}},\\ \{1,0,1,2,3\},\{1,1,2,3,4\},\{0,0,0,0,0\}\}\end{tabular}           \\ \hline
Tuned symmetries  & \begin{tabular}[c]{@{}l@{}}$\cdot\oplus\gso(9)\oplus\gsp(1)\oplus\oplus{\gso(11)}\oplus\gsp(1)\oplus\gso(9)\oplus\cdot$\end{tabular} \\ \hline
M:37 7 N:467 7 H:338,22  & \begin{tabular}[c]{@{}l@{}}\{\{1,0,3,2,1\},\{1,1,3,3,4\},\{1,0,3,2,3\},{\{1,1,3,3,5\}},\\\{1,0,2,2,3\},\{1,1,2,3,4\},\{0,0,0,0,0\}\}\end{tabular}           \\ \hline
\hline
Tuned symmetries  & \begin{tabular}[c]{@{}l@{}}$\cdot\oplus\gso(9)\oplus\gsp(1)\oplus\gso(10)\oplus\gsp(1)\oplus\gso(10)\oplus\cdot$\end{tabular} \\ \hline
M:36 9 N:467 9 H:339,21  & \begin{tabular}[c]{@{}l@{}}\{\{1,0,1,2,1\},\{1,1,2,3,4\},\{1,0,1,2,3\},\{1,1,2,3,5\},\\\{1,0,1,2,4\},\{1,1,2,3,5\},\{0,0,0,0,0\}\}\end{tabular}           \\ \hline
\end{tabular}
\caption{\footnotesize An example contrasting  the absence and the existence of multiple realizations:    successive Tate tunings of generic
  CYs over the toric base
  $\{-12//-11//-12//-12//-12//-12//-12//-12//-12//-12//-12,-1,-2,-2,-3,-1,-4,-1,-4,-1,-4,0\}$.  The Tate vanishing orders on  the last seven curves
  $\{-1,-4,-1,-4,-1,-4,0\}$ are indicated.                
       All polytope models in the KS database with the Hodge pairs $\set{225,23}, \set{338,22}, \set{339,21}$ are listed in each of
the three blocks. In models with Hodge pair $\set{338,22}$, both the weaker and the stronger versions of the tuning of $\gso(10)$ on the middle $-4$-curve exist - the weaker version can not correspond to $\gso(9)$ by the global symmetry constraint on the $-4$-curve. On the other hand, there is only one model with Hodge pair $\set{339,21}$, the weaker version of the tuning of $\gso(10)$ on the last $-4$-curve does not exist in the KS database -  the same Tate tuning gives $\gso(9)$ on the last $-4$-curve in the model M:38 8 N:466 8 H:338,22. 
%the correspondence betweenpolytope data in the KS database and
% Several features to be noticed: the
%  $\gso(9)$ vs.~the two realizations of $\gso(10)$ tunings, the
%  correlation between the existence of M:38 8 N:466 8 H:338,22 and
%  M:36 9 N:467 9 H:339,21 as the only polytope with the latter
%Hodge numbers;
%  and the rank preserving tuning $\gso(10)$ to $\gso(11)$ in the
%  middle block.
} 
\label{t:so10}
\end{table}

\begin{table}[]
\centering
\begin{tabular}{|l|l|}

 \hline
 Generic model, in KS & \begin{tabular}[c]{@{}l@{}}$\cdot\oplus\gsu(3)\oplus\cdot\oplus\gso(8)\oplus\cdot\oplus\cdot$\end{tabular}                                                        
 \\ \hline
M:85 6 N:379 6 H:274,58  & \begin{tabular}[c]{@{}l@{}}$\{\{1, 1, 0, 1, 0\}, \{1, 1, 1, 2, 2\}, \{1, 0,0, 0, 0\}, $\\$\{1, 1, 2, 2,   3\}, \{0, 0, 0, 0, 0\}, \{0, 0, 0, 0, 0\}\}$\end{tabular}
%Generic model in KS & \begin{tabular}[c]{@{}l@{}}M:85 6 N:379 6 H:274,58\end{tabular}                                                        
 \\ \hline\hline
 Tuned model, in KS & \begin{tabular}[c]{@{}l@{}}$\cdot\oplus\gg_2\oplus\gsp(1)\oplus{\gso(12)}\oplus\gsp(3)\oplus\cdot$\end{tabular}                                                        
 \\ \hline
M:35 7 N:387 7 H:280,22  & \begin{tabular}[c]{@{}l@{}}$\{\{1,1,3,1,1\},\{1,1,3,2,3\},\{1,0,3,1,2\},$\\ ${\{1,1,3,3,5\}},\{0,0,3,3,6\},\{0,0,0,0,0\}\}$\end{tabular}
%Generic model in KS & \begin{tabular}[c]{@{}l@{}}M:85 6 N:379 6 H:274,58\end{tabular}                                                        
\\ \hline
\end{tabular}
\caption{\footnotesize An example of the non-existence of the stronger
  version of the Tate form: a tuning of  a generic model over the
  base
  $\{-12//-11//-12//-12//-12//-12//-12//-12//-12,-1,-2,-2,-3,-1,-5,-1,-3,-1,-4,-1,0\}$
  on the last $-4$-curve with a $\gso(12)$ gauge algebra (which forces  gauge
  algebras on nearby curves). The Tate vanishing orders on  the last six curves
  $\{-1,-3,-1,-4,-1,0\}$ are indicated. While the weaker version of the Tate form   $\{1,1,3,3,5\}$ exists in the KS database, the stronger version $\{1,1,3,3,6\}$ does not give rise to a Tate-Zariski decomposition with the desired gauge algebras.}
\label{t:so12}

\end{table}

As described in \S\ref{sec:multiple-tops}, for some gauge
algebras such as $\gsu_3$ and $\ge_6$ there are multiple tops
associated with distinct Tate tunings, where one tuning involves an
additional monodromy condition.  This also occurs for the gauge
algebras $\gso(n)$.
We discuss in this subsection some particular aspects of $\gso(n)$
tunings and the associated reflexive polytopes, which have
some unique features.

As can be seen from Table~\ref{tab:tatealg}, for each of the $\gso(n)$
gauge algebras with $n$ even, starting with $n = 8$, there are two
distinct Tate tunings that realize the algebra, with one (both in the
case of $\gso(8)$) involving a monodromy condition.  
(Note that these forms in the table expand on earlier versions of the
table appearing in the literature, which did not include all these possibilities.)
As discussed in
\S\ref{sec:Tate}, the monodromy condition for the weaker Tate
tuning can be realized automatically when the leading terms in certain
$a_i$s are powers of a single monomial, corresponding in the polytope
language to a condition that the associated set of lattice points
contain only a single element with appropriate multiplicity
properties.

As for $\gsu_3, \ge_6$, we find that both kinds of Tate tunings of the
$\gso(2n)$ gauge algebras can arise in corresponding polytopes in the
KS database, corresponding to the usual condition that a global
Tate-Zariski decomposition is possible.  We also note, however, that
when the algebra $\gso(2n -1)$ can be realized on one polytope over a
given curve, then the monodromy realization of $\gso(2n)$ is generally
not possible, though the higher Tate tuning generally is.  This
basically corresponds geometrically to the question of whether the
minimally tuned Tate model with the weaker vanishing condition has the
appropriate single monomials in the $a_i$s, or not.  By the same
token, the gauge algebra $\gso(8)$, which has only monodromy
realizations, can only be realized when neither $\gg_2$ or $\gso(7)$
is possible over a given curve, which essentially reduces the
appearance of this algebra to the NHC structure of $-4$ curves.

To illustrate these points we give a few examples:

For a first example, consider a chain of curves
$\{-1,-4,{-1},-4,{-1},-4,0\}$; by requiring
Tate vanishing orders $\{0,0,1,1,2\}$ ($\gsp(1)$ gauge algebra) on
$D_{B3}$ and $D_{B5}$, the Tate vanishing orders on each of the
curves become
$\{\{0,0,0,0,0\},\{1,1,2,3,4\},\{1,0,1,2,2\},{\{1,1,2,3,4\}},$
$\{1,0,1,2,2\},$ $\{1,1,2,3,4\},\{0,0,0,0,0\}\}$.
Without taking into account the monodromy conditions, it would appear
in this case that
the enhanced algebras were
$\{\cdot\oplus\gso(9)\oplus\gsp(1)\oplus\gso(9)\oplus\gsp(1)\oplus\gso(9)\oplus\cdot\}$;
 explicitly analysis of the  monomials, however, shows that  while $D_{B2}$
and $D_{B6}$ are indeed $\gso(9)$ algebras, 
there is really a $\gso(10)$ algebra
on  $D_{B4}$, since the $\gso(10)$ monodromy condition is automatically
satisfied. 
This can also be understood from the perspective of
global
symmetry constraints \cite{Bertolini:2015bwa}; when the gauge algebra
is $\gso(9)$  on a $-4$-curve, the maximal global symmetry algebra is
$\gsp(1)$, so it is not possible for $\gso(9)$
to appear on $D_{B4}$ next
to two $\gsp(1)$'s.  Thus,
 $D_{B4}$
indeed must carry the gauge algebra  $\gso(10)$, for which the maximal
global symmetry algebra is $\gsp(2)\supset\gsp(1)\oplus\gsp(1)$.

For a similar example, for tunings of $\gso(4k+3)$ and $\gso(4k+4)$
%\footnote{In the
%  original Tate table, $\gso(4k+3)$ and $\gso(4k+4)$ have the same
%  Tate form $\{1, 1, k + 1, k + 1, 2k + 1\}$ with $\gso(4k+4)$ tuning
%  subject to further monodromy condition. We noticed that the Tate
%  form $\{1, 1, k + 1, k + 1, 2k + 2\}$ give rise to $(f,g,\Delta)$
%  that satisfies $\gso(4k+4)$ monodromy automatically (in analogous to
%  that $\{1, 1, k, k + 1, 2k\}$ generically leads to $\gso(4k+1)$
%  while $\{1, 1, k, k + 1, 2k+1\}$ leads to $\gso(4k+2)$), which would
%  be the natural choice to be put in the Tate table. The difference is
%  exactly the focus we are now discussing.}: 
consider the sequence of curves $\{-1, -3, -1,
-4, -1, 0\}$; by requiring vanishing orders of $\{1,1,3,3,5\}$ on
$D_{B4}$ and $\{0, 0, 3, 3, 6\}$ on $D_{B5}$,  the other
vanishing orders are
forced to
$\{\{0,0,0,0,0\}$, $\{1,1,2,2,3\}$, $\{1,0,2,1,2\}$, $\{1,1,3,3,5\}$, $\{0,0,3,3,6\},
\{0,0,0,0,2\}\}$, which gives the gauge algebras
$\{\cdot\oplus\gg_2\oplus\gsp(1)\oplus\gso(12)\oplus\gsp(3)\oplus\cdot\}$;
the algebra $\gso(11)$ is not possible on $D_{B4}$ by global symmetry
constraints.  Examples of these tunings in the context of global
constructions
are given in
Tables \ref{t:so10} and \ref{t:so12}.

In the examples just given, on certain curves the $\gso(2n -1)$ gauge
algebra cannot arise, and the lower Tate tuning with the monodromy
condition is realized.  As mentioned above, when the $\gso(2n -1)$
tuning is allowed, there is not generally a polytope in the KS
database with the same Tate tuning and the monodromy condition
automatically satisfied, and one has to use the higher Tate tuning to
  guarantee the condition.
These facts can be seen in
  contrasting the polytope models, for example, of $\gso(9)$ and the
  two realizations of $\gso(10)$ in table \ref{t:so10}. There is only
  one model in the KS database with the Hodge pair \{339,21\}, M:36 9
  N:467 9 H:339,21, which corresponds to tuning of the generic model
  $\{ 335,23,
  \{-12//-11//-12//-12//-12//-12//-12//-12//-12//-12//-12,-1,-2,-2,-3,{-1,-4,-1,-4,-1,-4,0}\}\}$
  on $\{-1,-4,-1,-4,-1,-4,0\}$ to gauge algebras
  $\{\cdot\oplus\gso(9)\oplus\gsp(1)\oplus\gso(10)\oplus\gsp(1)\oplus{\gso(10)}\oplus\cdot\}$. The
  Tate tuning along the last $-4$-curve is $\{1, 1, 2, 3, 5\}$. There
  is not a second polytope with the same Hodge numbers corresponding to
  the weaker Tate realization $\set{1,1,2,3,4}$ of the gauge algebra
  $\gso(10)$ along the last $-4$-curve.  This matches with the
  observation that the absence of multiple data in the KS database for
  a given tuning is due to
  the existence of the same Tate tuning appearing in the lower rank
  gauge algebras: There is already the  case M:38 8 N:466 8 H:338,22,
  corresponding to tuning of the same generic model to gauge algebras
  $\{\cdot\oplus\gso(9)\oplus\gsp(1)\oplus\gso(10)\oplus\gsp(1)\oplus{\gso(9)}\oplus\cdot\}$,
  and the Tate tuning along the last $-4$-curve is $\{1, 1, 2, 3, 4\}$
  giving an $\gso(9)$ there. On the other hand, there are two models
  with H:338,22, M:39 7 N:465 7 and M:38 8 N:466 8, corresponding to
  the tuning
  $\{\cdot\oplus\gso(9)\oplus\gsp(1)\oplus\gso(10)\oplus\gsp(1)\oplus{\gso(9)}\oplus\cdot\}$
  giving the two different Tate realizations of the $\gso(10)$. In
  this case, the weaker tuning satisfies
the monodromy condition
  automatically, which is expected as
  $\{\cdot\oplus\gso(9)\oplus\gsp(1)\oplus\gso(9)\oplus\gsp(1)\oplus{\gso(9)}\oplus\cdot\}$
  is not allowed as mentioned.

 There is a
similar story between $\gso(11)$ and $\gso(12)$.  For example, we can
tune an $\gso(11)$ on the $-3$-curve of the generic model over $\F_3$
by requiring Tate vanishing orders of $\{1,1,3,3,5\}$, which gives
rise to M:328 8 N:18 7 H:8,242 in KS database. Then to get a polytope
corresponding to a tuning of $\gso(12)$, we need to use
$\{1,1,3,3,6\}$, which has a good Zariski decomposition, and therefore
a corresponding reflexive polytope exists, M:318 10 N:19 8
H:9,233. The Hodge numbers of all these examples are consistent with
calculations from anomalies.

As we have mentioned, there is
 a special situation for the
$\gso(8)$ algebra and related polytopes in the
KS database: all realizations of $\gso(8)$ involve monodromy
constraints.  Thus,
there are no polytopes where there is a Tate tuning of
the algebra $\gso(8)$, and this algebra only arises over the NHC
$-4$. In the case of the NHC $-4$, $\gso(8)$ is the
minimal gauge algebra, so either vanishing orders $\{1,1,2,2,3\}$ or
$\{1,1,2,2,4\}$ will automatically satisfy the $\gso(8)$ monodromy
condition in any Tate tuning over a base with a $-4$
curve.  
%The reason that enhancement of $\gso(8)$ on curves with greater
%selfintersection is not allowed is that first of all, unlike the
%$\gso(4k+2)$ or $\gso(4k+4)$ for $k\in \mathbb{N}_{\geq 2}$ that
%there are Tate forms satisfying the monodromy condition
%automatically, no such a Tate form for $\gso(8)$ exists. 
This unique aspect of $\gso(8)$ matches with the observation that a
tuned $\gso(7)$ cannot be ruled out through the global symmetry group
since the global symmetry group on a tuned $\gso(7)$ curve contains
that on a tuned $\gso(8)$ curve.  Thus, any Tate tuning of
$\{1,1,2,2,3\}$ or $\{1,1,2,2,4\}$ over a curve with self-intersection
greater than $-4$ will lead to a model with, if not $\gg_2$, $\gso(7)$
enhancement.

% The Tate vanishing order of $\gso(n)$ tunings in KS database are summarized in table \ref{t:so}.

\subsection{Multiplicity in the KS database}
\label{multiplicity}

\begin{table}[]
\centering
\begin{tabular}{|c|c|c|c|c|}
\hline
-1                                                            & 0                                                                       & 1                                                             & 0                                                                 & KS data                 \\ \hline
\begin{tabular}[c]{@{}c@{}}\{0,0,0,1,1\}\\ $I_0$\end{tabular} & \begin{tabular}[c]{@{}c@{}}\{0,1,1,2,3\}\\ $I_3$ $\gsu(3)$\end{tabular} & \begin{tabular}[c]{@{}c@{}}\{0,0,0,0,0\}\\ $I_0$\end{tabular} & \begin{tabular}[c]{@{}c@{}}\{1,1,2,3,4\}\\ $\gso(9)$\end{tabular} & M:165 11 N:18 9 H:9,129$^{[1]}$ \\ \hline
\begin{tabular}[c]{@{}c@{}}\{0,1,1,1,1\}\\ $I_1$\end{tabular} & \begin{tabular}[c]{@{}c@{}}\{0,1,1,2,3\}\\ $I_3$ $\gsu(3)$\end{tabular} & \begin{tabular}[c]{@{}c@{}}\{0,0,0,0,0\}\\ $I_0$\end{tabular} & \begin{tabular}[c]{@{}c@{}}\{1,2,2,3,4\}\\ $\gf_4$\end{tabular}   & M:160 9 N:19 8 H:9,129  \\ \hline
\begin{tabular}[c]{@{}c@{}}\{1,0,1,1,1\}\\ $I_1$\end{tabular} & \begin{tabular}[c]{@{}c@{}}\{1,1,2,2,3\}\\ $\gg_2$\end{tabular}         & \begin{tabular}[c]{@{}c@{}}\{0,0,0,0,0\}\\ $I_0$\end{tabular} & \begin{tabular}[c]{@{}c@{}}\{1,1,2,3,4\}\\ $\gso(9)$\end{tabular} & M:155 7 N:19 6 H:9,129  \\ \hline
\begin{tabular}[c]{@{}c@{}}\{1,1,1,1,1\}\\ $II$\end{tabular}  & \begin{tabular}[c]{@{}c@{}}\{1,1,2,2,3\}\\ $\gg_2$\end{tabular}         & \begin{tabular}[c]{@{}c@{}}\{0,0,0,0,0\}\\ $I_0$\end{tabular} & \begin{tabular}[c]{@{}c@{}}\{1,2,2,3,4\}\\ $\gf_4$\end{tabular}   & M:150 5 N:20 5 H:9,129  \\ \hline\hline
\begin{tabular}[c]{@{}c@{}}\{1,1,1,1,1\}\\ $II$\end{tabular}  & \begin{tabular}[c]{@{}c@{}}\{1,1,1,2,3\}\\ $IV^s \gsu(3)$\end{tabular}          & \begin{tabular}[c]{@{}c@{}}\{0,0,0,0,0\}\\ $I_0$\end{tabular} & \begin{tabular}[c]{@{}c@{}}\{1,2,2,3,4\}\\ $\gf_4$\end{tabular}   & no corresponding KS data          \\ \hline
\end{tabular}
\caption{\footnotesize Some rank-preserving tunings over the $\F_1$
  base. Notice that the Tate vanishing orders of the trivial algebra
  on the $-1$-curve change in the Tate-Zariski decomposition as the
  vanishing orders of the two $0$-curves get higher. The last row
  gives an example of a general observation that when the gauge
  algebra tuning is only a further specialization of an existing gauge
  algebra tuning (but not the case of gauge algebras realized by
  different monodromy tunings listed in Table \ref{tab:tatealg} with
  $\star$, which involves with the requirements of additional
  conditions), there would not be a corresponding polytope in the KS
  database even if the Tate-Zariski configuration is stable. The
  example illustrates that since there is the $\gsu(3)$ model in the
  second row realized by $I_3$, there is no model realized by $IV^s$.}
\label{t:f1}
\end{table}

Given a pair of Hodge numbers $h^{1,1}, h^{2,1}$, there are in general
many distinct polytopes in the KS database.  There are many ways in
which such a multiplicity may arise.  Of course, generic or tuned
elliptic fibrations over distinct bases may coincidentally give the
same Hodge numbers.  As discussed above, however, there are also many
closely related constructions that give identical Hodge numbers.
Different realizations of the same gauge algebra through different
Tate tunings may contribute, often related to monodromy tunings as
discussed in the preceding subsections.  There are also
rank-preserving tunings that change the gauge algebra but not the
Hodge numbers.  And in some cases there are non-toric deformations
that can give additional multiplicity.  A complete analysis of the KS
database that accounts for these multiplicities exactly would require
a complete and systematic tracking of all distinct possible Tate
tunings for each gauge algebra combination and a clear and systematic
analysis of the non-toric deformation possibilities.  We have not
attempted such a systematic analysis here.  Rather, in the analysis in
the remainder of the paper we focus on constructing distinct possible
gauge groups through Tate tunings and identifying the distinct Hodge
numbers that can arise for reflexive polytopes in this way. In this
section we discuss in a bit more detail some aspects of the
multiplicity question.

To systematically analyze multiplicities of different Tate tunings of
the same algebra, we would need to consider all combinations of
monodromy and non-monodromy tunings of algebras like $\gsu_3, \ge_6,
\gso(n)$ etc.  Over bases with many curves allowing such tunings this
could give a large combinatorial multiplicity.  For example, consider
the two polytope models in the first block of Table \ref{t:so10}.  We
start with the minimal $\{1,1,2,2,3\}$ Tate vanishing orders for all
three $-4$ curves, which together do have a corresponding Tate-Zariski
decomposition, so there is a corresponding polytope construction.
Then we tune the vanishing orders on the middle $-4$-curve alone to be
$\{1,1,2,2,4\}$.  After iteration, the other two $-4$-curves are
forced to also have $\{1,1,2,2,4\}$ vanishing, giving the second
generic model with all $-4$ curves reaching the second realization.
This exhausts the possibilities.  So from what might appear to in
principle be 8 possible combinations of tunings, only two are actually
consistent.  It can also happen that only the lower-order realization
exists, while the higher-order realization does not have an acceptable
Zariski decomposition and there is no corresponding polytope, as we
have seen for example in the failure to realize the third model of
H:23,107 with the generic gauge group over a $-6$ curve in section
\ref{sec:combining}.  In general, the realization of any given combination must
be checked by performing a global Tate-Zariski decomposition, as local
information may not be completely adequate to rule in or out a
possible tuning.  An example is given by the models in
Table~\ref{t:so12}, where there is no global Zariski decomposition of
the $\{1,1,3,3,6\}$ realization of $\gso(12)$, and the reflexive
polytope model does not exist over the given global base, though it
would seem to be fine if we were to analyze the tuning pattern with
the focus on the local sequence $\{-1,-3,-1,-4,-1,0\}$ only.

Note also that further Tate tunings of a given algebra
may not give rise to a new reflexive polytope, even if the
higher vanishing orders still have a valid Zariski decomposition. We
describe briefly
several examples here: There is only one polytope in the
KS database with
H:4,226, which corresponds to the type $I_2$ $\gsu(2)$ tuning
$\{0,0,1,1,2\}$ on the $-2$-curve of the $\F_2$ base, but there is no
polytope that corresponds to the type $III$ $\gsu(2)$ $\{1,1,1,1,2\}$. It is
even more interesting to compare the
H:5,233 models discussed in Appendix
\ref{polytopetuning}
and H:5,251 in table \ref{t:fn}: there is no $IV$ $\gsu(3)$
for the former since it is just a specialization of the type $I_3$
$\gsu(3)$ tuning, while there are two different $IV$ $\gsu(3)$
realizations for the latter; and both of these sets have the rank-preserving
tuning $\gg_2$ model. Similarly, type $I^{ns}_{2n}$ and type
$I^{ns}_{2n+1}$ $ \gsp(n)$ tunings do not give rise to different
polytopes. Also for three different types $I_0, I_1, II$ of the
trivial algebra, only the one with the lowest vanishing orders that
has a Tate-Zariski decomposition has a reflexive polytope
construction. 
An amusing exercise is illustrated
in Table \ref{t:f1}, where we can
see the changes in three different types of trivial algebra under
various tunings.

Another source of multiplicity comes from tunings of rank-preserving
type as described in the end of \S\ref{hns}. We have seen
several examples in global models: H:7,271 of rank 4 $\gso(8)$,
$\gso(9)$, and $\gf_4$ tunings in table \ref{t:fn}, and H:338,22 of
rank 5 $\gso(10)$ and $\gso(11)$ tunings in table \ref{t:so10}, and
H:9,129 of different combinations of rank preserving tunings in table
\ref{t:f1}. Notice that it is not always true that tuning gauge
algebras with the same rank will lead to the same $h^{2,1}$ shift. For
example, $\gsu(7)$ and $\ge_6$ are not subalgebras of each other, and
the tunings give different $h^{2,1}$s.

Lastly, multiplicity can come from situations where the elliptic
fibration over a toric base has (4, 6) points that must be blown up.
As discussed in \S\ref{sec:bases},
over toric bases
  containing curves with self-intersection number $-9, -10, -11$ 
the generic elliptic fibration is non-flat and the base
must
  be blown up at the $(4,6)$  points 
to give $-12$-curves, over which there is a flat elliptic fibration.
In general the base resulting from these blow-ups will be non-toric, and the blowups give extra tensor multiplets contributing to
  anomaly cancellation \cite{Morrison:2012js, Martini-WT}. 
In some cases, however, the base is still toric after blowing up one
or more of the $(4, 6)$ points; in such cases there will be multiple
entries in the KS database associated with these distinct bases.   In
general we expect that these all represent  smooth Calabi-Yau
threefolds that can be viewed as  non-flat elliptic
fibrations over toric bases or flat elliptic
fibrations over the non-toric bases resolved at the non-toric $(4, 6)$ points, though we have not checked explicitly
that this is true in all cases.
 Examples of some non-flat elliptic fibrations of this type are
analyzed in \cite{Braun:2011ux, Buchmuller:2017wpe, Dierigl:2018nlv}.
To illustrate this structure,
in Appendix \ref{2dfiber} we analyze the non-flat elliptic fibration
structure of the toric hypersurfaces associated with
(flat) toric fibrations
  of the reflexive fibered polytopes over the Hirzebruch surfaces
  $\F_9, \F_{10}, \F_{11}$.  In these cases,
we see explicitly that the fiber over the
  $(4,6)$ points in the $-9,-10,-11$-curves contains extra irreducible
components that may naturally be associated with divisors in the blown
up space.
%singular Weierstrass models with $(4,6)$ points are resolved into irreducible components of the two-dimensional toric fiber . We show explicitly 
%We illustrate the 2D fiber resolution of the elliptic fibration over the toric bases Hirzebruch surfaces $\F_9, \F_{10}$, and $\F_{11}$ in Appendix \ref{2dfiber}. 
%Moreover, each one of the flat elliptic fibration models over the
%semi-toric bases resolved in various ways has a corresponding flat
%toric fibration polytope model with two-dimensional resolved fiber in
%the KS database, so the combination of the non-toric deformations in
%the base counts this type of multiplicity of Hodge pairs. 
The
multiplicity with which the Hodge pairs for the generic elliptic
fibration models over the suitably blown up
Hirzebruch surfaces $\F_{9/10/11}$ 
are listed in Table \ref{t:f9},  the tops over the
-9/-10/-11-curves are listed in the second block in Table \ref{t:dic}.
This illustrates the way in which the same smooth Calabi-Yau threefold
can be realized as a non-flat elliptic fibration over one
or more toric bases as well as sometimes a flat elliptic fibration
over another toric base, with each fibration structure realized in a
different polytope in the KS database.
For example, as illustrated in the table there are 6 distinct
polytopes at Hodge numbers H:14,404, which correspond to toric
realizations of elliptic fibrations over
different ``semi-toric'' bases that admit only a single $\C^*$
action (including various limits
in which $-2$ curves arise).

\begin{longtable}{|c|c|cc|}
\hline
Hodge pair                & Mult. in KS        & \multicolumn{2}{c|}{Bases}                                     \\ \hline
\multirow{6}{*}{(14,404)} & \multirow{6}{*}{6} & \includegraphics[width=3cm]{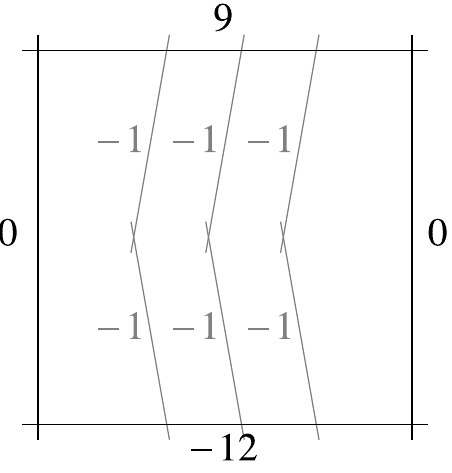}                             & \includegraphics[width=3cm]{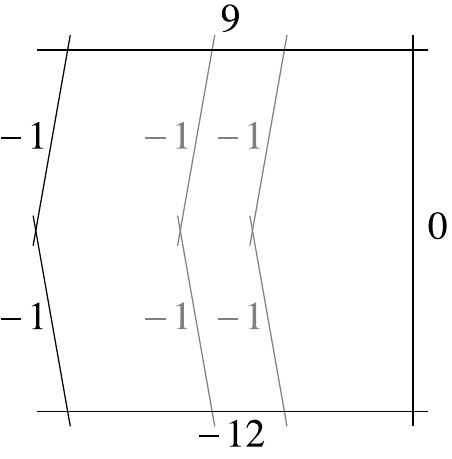}                            \\
                          &                    & \{0, -9, 0, 9\}                & \{-1, -1, -10, 0, 9\}         \\
                          &                    & \includegraphics[width=3cm]{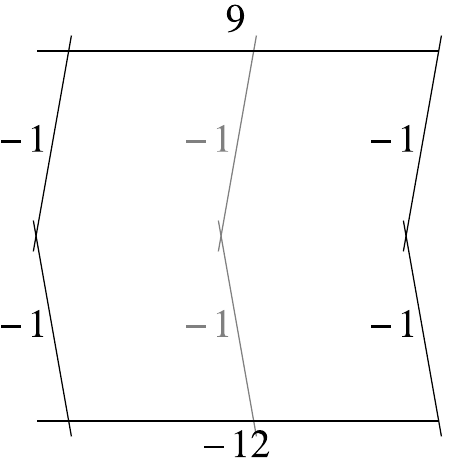}                             & \includegraphics[width=3cm]{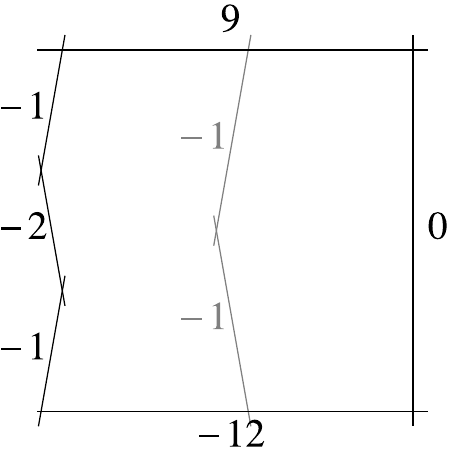}                            \\
                          &                    & \{-1, -1, -11, -1, -1, 9\}     & \{-1, -2, -1, -11, 0, 9\}     \\
                          &                    & \includegraphics[width=3cm]{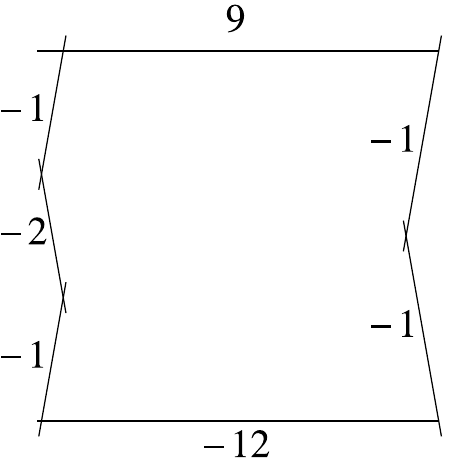}                             & \includegraphics[width=3cm]{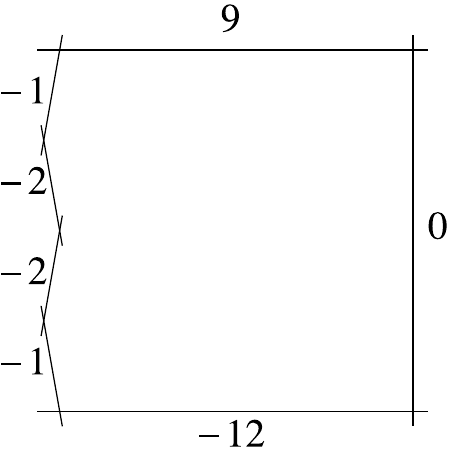}                           \\
                          &                    & \{-1, -2, -1, -12, -1, -1, 9\} & \{-1, -2, -2, -1, -12, 0, 9\} \\ \cline{1-2} \hline
\multirow{4}{*}{(13,433)} & \multirow{4}{*}{4} & \includegraphics[width=3cm]{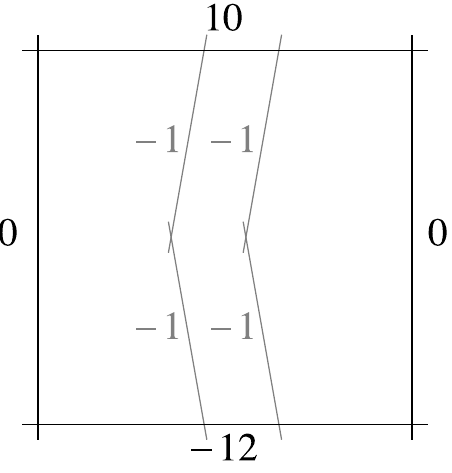}                             & \includegraphics[width=3cm]{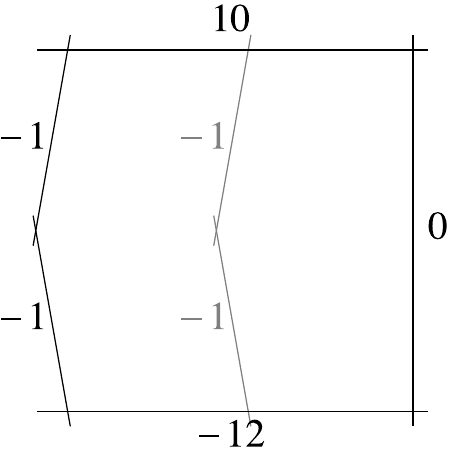}                            \\
                          &                    & \{0, -10, 0, 10\}              & \{-1, -1, -11, 0, 10\}        \\
                          &                    & \includegraphics[width=3cm]{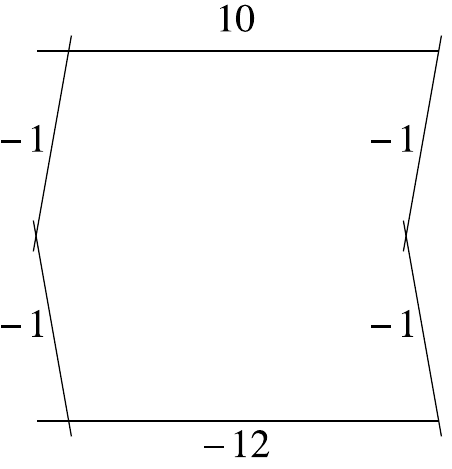}                             &      \includegraphics[width=3cm]{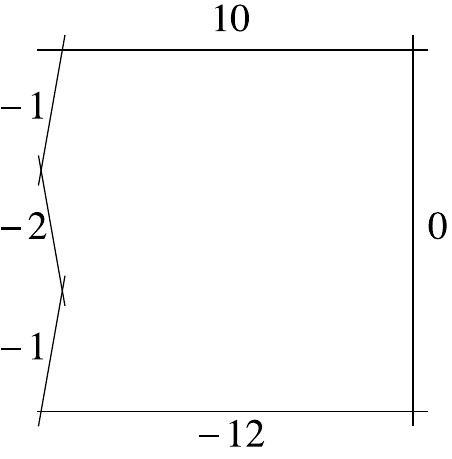}                        \\
                          &                    & \{-1, -1, -12, -1, -1, 10\}    & \{-1, -2, -1, -12, 0, 10\}    \\ \cline{1-2} \hline
\multirow{2}{*}{(12,462)} & \multirow{2}{*}{2} & \includegraphics[width=3cm]{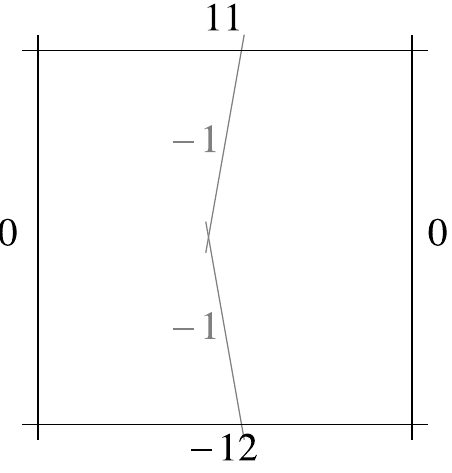}                             & \includegraphics[width=3cm]{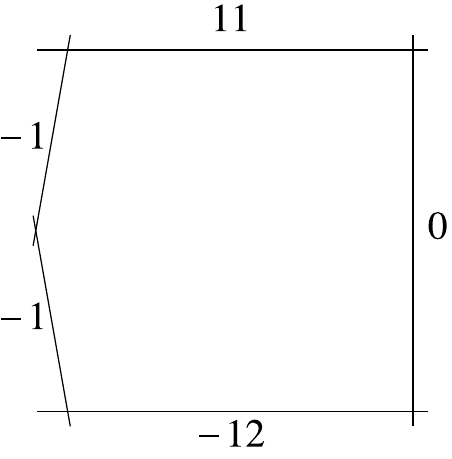}                             \\
                          &                    & \{0, -11, 0, 11\}              & \{-1, -1, -12, 0, 11\}        \\ \cline{1-2}
                          \hline
\caption{\footnotesize 
%The semi-toric bases of flat elliptic fibration models are drawn, and
%  corresponding toric bases of non-flat elliptic fibration models in
%  the polytope construction are indicated below.
A variety of polytope models arise for the Hodge pairs associated with
  the generic elliptic fibrations over the Hirzebruch
  surfaces  $\F_{9/10/11}$. The possibilities are enumerated in this table.
The first graph for each Hodge pair is the generic model, where the
(4, 6) singularities on the $-9, -10,$ or $-11$ curve are at non-toric
points and the elliptic fibration is non-flat.
In these cases
 the blow-ups are handled automatically by the resolution of the
toric geometry, giving a resolved model corresponding to
 a flat elliptic fibration over a ``semi-toric''
base.  There are also toric bases that arise by blowing up one or more
of the (4, 6) points at toric points, giving polytopes with toric
fibrations over blow-ups of the Hirzebruch surfaces.
When multiple (4, 6) points coincide this corresponds to a limit with
a $-2$ curve in the base.  For each polytope the base of the fibration
is a toric surface given by the curves on the outside of the diagram,
with self-intersections as labeled.
} 
\label{t:f9}
\end{longtable}

\subsection{Bases with large Hodge numbers}
\label{bwlhn}
In this work we have confined our study to the simplest
 $\P^{2,3,1}$ fiber type
  polytopes.  In part this is because the standard
 fiber type matches with the Tate structure of the
  Weierstrass model as discussed previously.  Also, however, this
  fiber type dominates the structure at large Hodge numbers.  In
  particular, we can explicitly identify constraints on the bases that
  can be used for
the other 15 fiber types.
These constraints are such that the other fiber types
all lead to problematic codimension one (4, 6) singularities on some
divisor in the base when the base contains curves of sufficiently
negative self-intersection.  In particular, none of the other 15
fibers can be supported over any base that contains a curve of
self-intersection less than $-8$.  This immediately constrains the set
of constructions at large Hodge number, since the generic elliptic
fibrations with the largest Hodge numbers almost always involve $-12$ curves
in the base
(though there are notable exceptions to this general principle,
including the other one of the two fibrations of the H:491:11 polytope).

We leave a more detailed analysis of the constraints on different
fiber types for future work, but briefly
outline the issue that arises for other fiber types
besides the  $\P^{2,3,1}$ fiber.  Consider for
example the $\P^2$ fiber type.  Carrying out the analogue of the
standard stacking procedure for a $\P^2$ fiber, we find that there are 10 dual
monomials analogous to the coefficients $a_1, \ldots, a_6$. These 10 monomials are
sections of line bundles $\mathcal{O}(-K), \mathcal{O}(-2K)$ and $\mathcal{O}(-3K)$.  Any section of a line
bundle $-nK$ must vanish over a $-12$ curve to at least degree $n$
when $n < 5$ by the Zariski decomposition.  This immediately leads to
  the presence of a codimension one (4, 6) singularity over any $-12$
  curve in the base.  Similar issues arise for the other fiber types.

Considering the toric bases, we can simply consider the complete
enumeration carried out in \cite{Morrison:2012js} and identify the bases with
largest Hodge numbers that have curves of self-intersection  no
smaller than $-8$.  The base of this type with the largest $h^{2, 1}$
for the generic elliptic fibration is $\F_8$, over which the generic
elliptic fibration has Hodge numbers $(10, 376)$.  Even over $\F_8$,
the largest $h^{2, 1}$ value that can be achieved for a tuning with
any fiber other than $\P^{2, 3, 1}$ is quite restricted; over this base,
for example, there are 5 other fiber types including $\P^{1, 1, 2}$
that are possible; the generic fibration with each of these fiber
types gives an elliptic Calabi-Yau threefold with Hodge numbers (11,
227).  Any other fibration with these or any other fibers other than
 $\P^{2, 3, 1}$ over any base would seem to give a Calabi-Yau
threefold with an even smaller value of $h^{2, 1}$.  Thus, by restricting to
Hodge numbers above $h^{2, 1}\geq 240$, we can expect that the threefolds
in the KS database that admit elliptic fibrations should be all or
almost all described by the $\P^{2,3,1}$ fiber type.  

Similarly, the largest value of $h^{1,1}$ that can arise for a base
with no curves of self-intersection below $-8$ is 224.  The
corresponding base has a set of toric curves of self-intersection $(0,
-8//-7//-8//-8//-8//-8//-8//-8//-8//-8//-8//-7//-8)$, where $//$
denotes the sequence $-1, -2, -3, -2, -1$ associated with $E_7$ chains
(see e.g.\ \cite{Morrison:2012js}), and a generic elliptic fibration with Hodge
numbers (224,18). There is nothing that can be tuned over this base
without producing a curve of self-intersection below $-8$ so it seems
that confining attention to threefolds with $h^{1,1}\geq 240$ should
again restrict us to primarily $\P^{2,3,1}$ fiber types.  As we see in
\S\ref{restcases}, however, there are a few unusual cases in which bases that
have generic elliptic fibrations with rather small values of $h^{1,1}$
admit extreme tunings that dramatically increase the value of
$h^{1,1}$ without producing curves of highly negative
self-intersection.  In a companion paper \cite{Huang-Taylor-fibers},
we study the fibration structure of the hypersurface models in the KS
database more directly, and confirm both the prevalence of
$\P^{2,3,1}$ fibers at large Hodge numbers and the existence of
exceptions involving extreme tunings.

\section{Systematic construction of Tate-tuned models in
the KS database}
\label{sec:systematic-construction}

Kreuzer and Skarke have classified all 473,800,776 4D reflexive
polytope models, which give 30,108 distinct Hodge pairs. It was
found in \cite{Hodge} that the set of Hodge pairs \{$h^{1,1}$,
$h^{2,1}$\} of all generic elliptically fibered CYs over toric bases
is a subset of all the Hodge pairs in the KS database.  We gave in
\S\ref{sec:dual-dual} a general construction of reflexive
polytope models of these generic elliptic fibrations over toric bases
with NHCs, and we expect that all generic elliptic fibration models
over toric bases have these corresponding reflexive polytope models in
the KS database. We wish to carry out a more comprehensive comparison
by matching tuned Weierstrass models of CYs over 2D toric bases with
4D reflexive polytope models of Calabi-Yau hypersurfaces at large
Hodge numbers.

Of the 30,108 distinct Hodge pairs in the KS database, 1,827 have
either $h^{1,1}\geq 240$ or $h^{2,1}\geq 240$
(only  the Hodge pair $\{251, 251\}$ satisfies both inequalities).  To compare the two
constructions at large Hodge numbers, the next step would be to
construct roughly this number of distinct Weierstrass models of tuned
CYs in these regions.  Not all Weierstrass models correspond to
reflexive polytope constructions, however.  Nonetheless, as discussed
in \S\ref{polytune}, there is a close relation between Tate-tuned
models and $\P^{2,3,1}$-fibered polytopes, which dominate at large
Hodge numbers as argued in \S\ref{bwlhn}.  Therefore, our approach is
to construct systematically all Tate-tuned models via tunings of
generic Tate models over 2D toric bases.  As a preliminary to this
analysis, however, we begin with a simpler systematic analysis of
which gauge group tunings may be possible based on more general
Weierstrass tunings, and then we refine the analysis to Tate tunings.
We describe the logic of this analysis in more detail in \S\ref{algo}.

All Hodge pairs of the Tate-tuned models from this algorithm fall
within those in the KS database. However, there are certain Hodge
numbers in the KS database in the regions of interest at which our
initial analysis identified no matching Tate-tuned model. We therefore
have analyzed directly, via the method described in \S\ref{fb}, the
polytope models with the Hodge numbers that were not found in our
systematic tuning construction; the analysis of these cases is
described in \S\ref{restcases}.  It turns out that all these remaining
polytope models can be described as somewhat more exotic Weierstrass
or Tate tunings over bases that are either toric bases or blow-ups
thereof.  This completes the comparison of the two constructions at
the level of Hodge numbers.  At a basic level the result of this
analysis is that in the regions of interest all the Hodge pairs in the
KS database are realized through generic or tuned elliptic fibrations.
This matches with the results through a direct analysis of the
fibration structure in the companion paper \cite{Huang-Taylor-fibers}.
The more detailed analysis we carry out here, however, gives much more
insight into the structure of these fibrations and the complex variety
of Weierstrass tunings and geometries that are realized through simple
toric hypersurface constructions.

We also discuss briefly the limits of Tate tuning in section
\ref{tatevsKS}, where we collect some results on tunings that are
compatible with the global symmetry constraints but can't not be
realized by Tate tunings. These tunings may be realized by Weierstrass
models and in such cases give new Hodge pairs outside the KS database.

%We look for all Tate-tuned elliptic Clabi-Yau models over toric bases in the large $h^{1,1}$ and large $h^{2,1}$ regions. Given a base, possible tunings are confined by global symmetry constraints and a real tuned model is carried out by Zariski decomposition. We saw in \S\ref{polytune} that there is close relation between Tate-tuned models and $\P^{2,3,1}$ fibered polytopes. Kreuzer and Skarke have classified all 473,800,776 4D reflexive polytope models, and there are 30,108 dinstinct Hodge data. It is true that all Hodge data we got from Tate-tuned models are in KS Hodge data. Conversely, we would expect all $\P^{2,3,1}$ fibered polytopes correspond to some Tate-tuned models and we would rediscover KS Hodge data of them.  The KS Hodge data in the large $h^{2,1}$ region we got were realized by some Tate-tuned models, but it seems to not be the case in the large $h^{1,1}$ region and needs further investigation.

%Not all Weierstrass tuned models can be realized by Tate tuning. We first consider Zariski decomposition of tunings in terms of sections $f$, $g$, and $\Delta$, and then restrict to the subset which can be tuned by Tate algorithm; i.e., has valid Zariski decomposition  of sections $a_{1,2,3,4,6}$. Those tunings can't be realized by Tate may still have a Weierstrass tuning and give rise to a new Hodge data not in the KS database. We listed some examples of Tate tuning swampland we found in this way in table \ref{t:swamp}. We also studied a case where Tate tuning is not allowed but still has Weierstrass tuning. 

\subsection{Algorithm: Global symmetries and Zariski decomposition
for Weierstrass models}
\label{algo}

We give an algorithm in this section to systematically construct all
tunings of enhanced gauge groups
 over
a given 2D toric base, 
starting with the generic model.  Our goal is to construct all Tate-tuned models
over toric bases that give elliptic Calabi-Yau threefolds with Hodge
numbers
in the
regions $h^{1,1}\geq 240$ or $h^{2,1}\geq 240$. 
As we
saw in \S\ref{rpt}, global symmetry constraints on each curve put
upper bounds on the gauge algebras that can be tuned on intersecting
curves. On the other hand, as discussed in \S\ref{sec:combining},
there is an issue of whether local tunings on subsets of curves can be
combined into a global model over some toric base $B_2$. This can be
tested by the Zariski decomposition. More specifically,  our goal is
to
 carry out
explicitly arbitrary combinations of
 the Tate tunings from \S\ref{TT} on the curves in the base,
applying the variant
of the
Zariski decomposition described in \S\ref{sec:Zariski-Tate}
to determine which combinations are globally compatible.
While in principle we could simply iterate over all possible gauge
algebra combinations, using the global symmetry constraints on what
gauge algebras can arise on the curves intersecting a curve of
negative self intersection helps prune the tree and make the algorithm
more efficient.
Global symmetries are also helpful in limiting the set of possible
monodromy-dependent gauge groups 
that can arise on sequences of intersecting curves in
ways that are not apparent at the level of the Zariski decomposition.

Although ultimately we wish to analyze Tate tunings, we perform an
initial analysis of Weierstrass tunings using global symmetry
constraints and the Zariski decomposition.  This gives us a set of
possible tunings that we expect may be possible at the level of the
gauge algebras.  Not all these constructions, however, are compatible
with Tate tunings and with polytopes.  We begin the discussion by
focusing on the Weierstrass tunings and then in  \S\ref{tatevsKS} we
use the results of the Weierstrass tunings to check which Tate tunings
are possible.

Given a 2D toric base $B$, which is represented by a set of
$K$ irreducible
toric curves $\{C_j, j = 1, 2, \ldots, K\}$ intersecting each other in a linear chain, we
first obtain for the generic model over the base
$B$ the orders of
vanishing
$\{c_{j,4}, c_{j,6}, c_{j,12}\}$ of $f$, $g$, and $\Delta$ along each curve. The sets of values $\set{c_{j,4}}$,
$\set{c_{j,6}}$, and $\set{c_{j,12}}$ can be determined by the Zariski
decomposition via the procedure described in equations
(\ref{eq:0})-(\ref{iterat}) with $n=4$, $n=6$, and $n=12$,
respectively, or can be directly read off from the non-Higgsable
cluster structure of the curves $\{C_j\}$.

Now let us consider all possible (Weierstrass) tunings of the generic
model. We describe a procedure to determine an allowed pattern
$\set{\gg_j}$ of tuned algebras $\gg_j$ on each curve $C_j$ in the
base.
Note that in this algorithm we assume that there are no toric $(4, 6)$
points in the base, even after the tuning; such a point would be blown
up to form a different toric base, and the tunings over the blown up
base would be found directly by tunings over that base.  We do allow
non-toric $(4, 6)$ points in the case where the base contains $-9,
-10$ or $-11$ curves; in these cases we essentially treat the curve as
a $-12$ curve supporting an $\ge_8$, understanding that the polytope
hypersurface construction will automatically resolve these
singularities and effectively blow up the non-toric points in the
base, in accord with the discussions in \S\ref{sec:bases}
and \S\ref{multiplicity}.

\subsection{Main structure of the algorithm: 
bases with a non-Higgsable $\ge_7$ or $\ge_8$}
\label{mainalgo}

We consider first the simplest cases, where there is at least one
curve in the toric base of self-intersection $m \leq -9$; such a curve
necessarily carries a non-Higgsable $\ge_8$ gauge algebra.  We start
the procedure by choosing a specific curve with a non-Higgsable
$\ge_8$ and first considering the possible tunings on one of the
adjacent curves.  Let us label the curve with the $\ge_8$ using the
index $j=1$, the curve we attempt the first tuning on by $j=2$, the
subsequent curve by $j=3$, etc. This choice of the initial
configuration is convenient to serve as the starting point of a
branching algorithm because an $\ge_8$ algebra cannot be further
enhanced; moreover, nothing can be tuned next to an $\ge_8$, without
producing a (4, 6) singularity at a toric point, which we are assuming
does not happen as discussed above. Therefore, the gauge algebras on
$C_1$ and $C_2$ are fixed: $\gg_1=\ge_8$ and $\gg_2$ has to be a
trivial algebra.

We then pass to tunings $\gg_3$ on $C_3$. The possible tunings on
$C_3$ are constrained by the global symmetry group
$\gg_2^{(\text{glob})}$ on $C_2$, which is determined by the
self-intersection number of $C_2$ and the gauge algebra $\gg_2$ on
$C_2$.
  Let the set $\set{\gg_{3,\alpha}}$ be the set of algebras that
satisfy the constraint $\gg_1\oplus \gg_{3,\alpha}=\ge_8\oplus
\gg_{3,\alpha} \subset \gg_2^{(\text{glob})}.$
For the global symmetries,  we used the results in  Table 6.1 and Table 6.2 in \cite{Bertolini:2015bwa} for  the maximal global symmetry group $\gg_j^{(\text{glob})}$ on  a curve $C_j$ of negative self-intersection $m$ carrying a gauge algebra $\gg_j$. Additionally, the curves of negative self-intersection that do not support an NHC can carry trivial gauge algebras, of types $I_0, I_1, II$; therefore in such cases when $\gg_j=\cdot$, we use $\gg_j^{(\text{glob})}=\ge_8$ and $\gg_j^{(\text{glob})}=\gsu(2)$ for $m=-1$ and $m=-2$, respectively.
We used the results tabulated in \cite{slansky} for the subgroups of a global symmetry group $\gg_{j-1}^{(\text{glob})}$  to obtain the restricted set of algebras $\set{\gg_{j,\alpha}}$ satisfying the constraint 
$\gg_{j-2}\oplus \gg_{j,\alpha} \subset \gg_{j-1}^{(\text{glob})}$.

 We attempt tunings
one-by-one for each $\gg_{3,\alpha}$.  For each possible algebra we
replace the original orders of vanishing
$\set{c_{3,4},c_{3,6},c_{3,12}}$ with the desired orders of vanishing
corresponding to $\gg_{3,\alpha}$ using Table \ref{t:Kodaira}. We then
perform the Zariski iteration procedure on all curves with
the new $\set{c_{3,4},c_{3,6}, c_{3,12}}$ for $n=4, 6$ and $12$,
respectively. If all the gauge algebras on the curves prior to and
including $C_3$ stay unchanged after the iteration, tuning $\gg_{3,\alpha}$
is not ruled out.  If any of the gauge algebras on the curves prior to
$C_3$ have changed, or the vanishing orders
$\set{c_{3,4},c_{3,6},c_{3,12}}$ do not produce the desired gauge
algebra $\gg_{3,\alpha}$ in the new configuration after the iteration,
tuning $\gg_{3,\alpha}$ is not allowed on $C_3$; in such cases we terminate
the procedure with this $\gg_{3,\alpha}$ branch, and attempt the next
tuning $\gg_{3,\alpha+1}$ on $C_3$. Note, however, that the fact that the
gauge algebras stay unchanged does not mean that the set of values
$\set{c_{j\leq3,4}}$, $\set{c_{j\leq3,6}}$, $\set{c_{j\leq3,12}}$ stay
unchanged under the iterations. Indeed, often it is the case that the
orders of vanishing on curves near $C_3$ are increased, but without
modifying the gauge algebra on $C_2$. In other words, in this case
$\gg_2$ should be the trivial algebra, but it may be type $I_0, I_1,$
or $II$ (cf. also examples in Table \ref{t:f1}.)

Note that the vanishing orders $\set{c_{j> 3,4},c_{j> 3,6}, c_{j> 3,12}}$ can
obtain new values after the initial set of iterations just described. 
If these increase beyond those determined by the initial NHC
configuration, we use the larger vanishing orders as the starting
points in further iterations of the tuning.
We denote by $\gg_{j[i]}$ the gauge algebra on curve $j$ after the
iteration procedure associated with curve $C_i$.
For $i = j$, $\gg_{j[j]}$ denotes a choice of gauge algebra  in a branch, $\gg_{j[j]}\in \{\gg_{j, \alpha}\}$, and $\gg_{j[j]}\supseteq \gg_{j[j-1]}$. Note that we must have $\gg_{j[k]}=\gg_{j[j]} $
for all $k > j$ as we require in the branch that the gauge algebra on $C_j$ stays unchanged in tuning gauge algebras on $C_{k>j}$, but the orders of vanishing realizing the gauge algebra may be different. We can proceed with the new
configuration to the next step of tuning
algebras on $C_4$, as long as $\gg_{4[3]}\oplus \gg_{2[{3}]} \subset
\gg_{3[3]}^{(\text{glob})}$ is satisfied, where now $\gg_{4[3]}$ is the
gauge algebra on $C_4$ in the new configuration and
$\gg_{3[3]}^{(\text{glob})}$ is determined by the self-intersection of
$C_3$ and the gauge algebra $\gg_{3[3]} \in \set{\gg_{3,\alpha}}$.  For example, let us start with $\gg_{3[3]}=\gg_{3,1}$.  We terminate the procedure
with the $\gg_{3,1}$ branch and attempt the next branch of tuning
$\gg_{3,2}$ on $C_3$ if $\gg_{4[3]}\oplus \gg_{2[{3}]} \subset
\gg_{3,1}^{(\text{glob})}$ is violated in the new configuration.

% we regard the new configuration as a tuning pattern as long as $g$
  %But we simply proceed with the new configuration; i.e., we proceed to perform tunings on $C_4$ with possibly minimal gauge algebras on $C_{j\geq 4}$ differing from that in the generic model setting. 

Assume $\gg_{3,1}$ passes the tests above. We then continue the procedure
similarly to tune the curve $C_4$ in the $\gg_{3,1}$ branch with the
new configuration: The set of possible tunings $\set{\gg_{4,\beta}}$
we attempt on $C_4$ is constrained by $\gg_{4,\beta}\oplus \gg_{2[{3}]}
\subset\gg_{3[3]}^{(\text{glob})}$. The branch $\gg_{4,1}$ can be
continued only if $\gg_{4,1}$ passes the two tests (1) the set of
gauge algebras $\set{\gg_{j\leq
    4}}=\set{\ge_8,\cdot,\gg_{3[{4}]},\gg_{4,1}}$ stays unchanged after
performing Zariski iterations on $\set{c_{j,4}}$, $\set{c_{j,6}}$, and
$\set{c_{j,12}}$ with the desired degrees of vanishing $\set{c_{4,4},
  c_{4,6}, c_{4,12}}$ of the tuned gauge algebra plugged into the
configuration, and (2) $\gg_{5[4]}\oplus \gg_{3[{4}]} \subset
\gg_{4,1}^{(\text{glob})}$ is satisfied, where $\gg_{5[4]}$ is the gauge
algebra on $C_5$ \emph{after} the iterations in the newest updated
configuration, and $\gg_{4,1}^{(\text{glob})}$ is again determined by
the self-intersection of $C_4$ and $\gg_{4,1}$.

The procedure continues similarly until the second to the last curve
$C_{K-1}$ is met. As the last curve $C_{K}$ is connected back to the
first curve $C_1$, we need to consider also the global symmetry
constraint on $C_K$ to close the tuning pattern. The set of possible
tunings $\set{\gg_{K-1,\gamma}}$ on $C_{K-1}$ is  constrained by
$\gg_{K-1,\gamma}\oplus \gg_{K-3[K-{2}]}
\subset\gg_{K-2[K-2]}^{(\text{glob})}$. First, the usual two
conditions have to be satisfied for $\gg_{K-1,\gamma}$ to be allowed:
in the new configuration after the iterations associated with the tuning of $\gg_{K-1,\gamma}$ (1) the prior gauge algebras are held fixed, and (2) the global
symmetry constraint on $C_{K-1}$ is satisfied.  Moreover, there is the
additional third condition: (3) the global symmetry constraint on the
curve $C_{K}$ has to be satisfied; i.e.,
$\gg_{K-1{,\gamma}}\oplus\gg_{1}\subset \gg_{K[K-1]}^{(\text{glob})}$,
where $\gg_1$ is held fixed and is $\ge_8$ in the simplest cases, and
$\gg_{K[K -1]}^{(\text{glob})}$ is determined by the self-intersection
of the curve $C_{K}$ and the gauge algebra $\gg_{K}$ after the Zariski
interations for the tuning $\gg_{K-1,\gamma}$. In fact, $\gg_{K}$ is
only allowed to be a trivial algebra in the simplest cases as $C_1$
carries an $\ge_8$ algebra, so no tuning is allowed on $C_{K}$. Hence,
if the global symmetry constraint on $C_{K}$ is satisfied, we are
basically done to this point in the procedure searching for a tuning
pattern. In this case, we obtain a tuning pattern
$\set{\ge_8,\cdot,\gg_{3[K{-1}]},\gg_{4[K{-1}]},\ldots,
  \gg_{K-3[K{-1}]},\gg_{K-2[K{-1}]},\gg_{K-1[K-1]},\cdot}$.

We check all $\gg_{K-1,\gamma}$'s in order similarly to complete the
scan through all possible tuning patterns compatible with the initial
viable possibility for
$\gg_{3[3]}, \ldots, \gg_{K-2[K-2]}$. After all
$\gg_{K-1,\gamma}$'s are processed, 
we proceed iteratively with a nested loop, continuing with the next
possible value of $\gg_{K-2}$, etc. so that all possible combinations
of gauge group tunings are considered.

All tunings increase $h^{1,1}$ and decrease $h^{2,1}$
with respect to the generic model over a given base.  
Thus, to classify all
tuned models of $h^{2,1}\geq 240$, 
we need only consider toric bases for which the generic elliptic
fibration has $h^{2, 1}\geq 240$.  In our initial scan, we also
restricted to bases that have
generic
models with $h^{1,1} \geq 220$.  As we describe in more detail in the
following section, this misses a few cases where there is a large
amount of tuning that significantly changes $h^{1, 1}$.  On the other hand,
as bases associated with
generic models  having
$h^{1,1}>224$ always
contain at least one curve carrying an $\ge_8$ algebra, the algorithm
as described above is quite effective in
dealing with tunings in the
  large $h^{1,1}$ region
as we always have a simple starting point for the iteration. 
In fact,  the algorithm
actually can work in the same way
for
  tunings of generic models with a curve carrying $\ge_7$ in the base;
  i.e., generic models with a curve of self-intersection $m\leq -7$ in
  the base. This is because $\ge_7$ algebras
also cannot be further
  enhanced without modifying the base --- an enhancement to an
  $\ge_8$ algebra  would give additional $(4, 6)$ points that must be
  blown up.
And no non-trivial algebra can be tuned next to an $\ge_7$ algebra.
Thus, in this case we similarly can
  make the convenient choice that the initial configuration  is
  fixed to be $\gg_1=\ge_7, \gg_2=\cdot.$

We make some final comments on two technical issues in the tuning
procedure.  
As mentioned above, in tuning the curve $C_j$, not only do the orders
of vanishing on $C_{j+1}$ (and in some cases on further curves $C_{j +
2}, \ldots)$
also change in general, but the new
vanishing orders
$\set{c_{j+1,4}, c_{j+1,6}, c_{j+1,12}}$ can in some cases correspond to a
different gauge algebra. 
However, because the three Zariski iterations
were performed independently, sometimes these vanishing orders
 do not correspond to any algebras in the Kodaira classification. 
We encountered a few cases of this type, for example where
$\set{c_{j+1,4}, c_{j+1,6}, c_{j+1,12}}=\{1, 2, 4\}$; this can happen
for example 
if a previous $\gsu(n)$ tuning ($\set{0,0,n}$) pushes up the order of
vanishing of $\Delta$ more significantly than  $f, g$ (where some
required orders of $f, g$ already imposed on the curve); however, note
that, this can never happen in a real $\Delta$ as calculated in a
complete model from $f$ and $g$ in equation (\ref{delta}).
In such situations, we modify the orders of vanishing on $C_{j+1}$ to
fit with
those that correspond to the gauge algebra that arises by increasing the values
$c_{j+1,4}, c_{j+1,6}$ minimally.
 Then we perform the
iteration again after the modification, and use the resulting
configuration to test the conditions (1) and (2).  

Another detail to take care is the tuning of algebras only
distinguished by monodromy conditions.  For those cases where there
are distinct algebras associated with different monodromy conditions,
we retain all the possibilities allowed by global symmetries; in the
list of possible tunings we attach an additional label to the orders
of vanishing using a fourth entry $\{c_{j,4}, c_{j,6}, c_{j,12},
\text{algebra}\}$ to ensure that all possible tunings are considered.

\subsection{Special cases: bases lacking curves of self-intersection
$m \leq -7$ and/or having curves of non-negative self intersection}
\label{modif}

The algorithm described in the preceding subsection relies on the
presence of a curve of self-intersection $m \leq -7$ in the base,
where we can begin the iteration process in a simple fashion
 as the gauge algebra on the initial curve is
fixed.  In the regions we are considering, there are very few bases
that lack such curves; we describe here briefly
how
these cases are handled.  Of course, one could simply use a brute
force algorithm of choosing an arbitrary starting point and looping
over all tunings on the initial curve $C_1$.  In principle, however,
for efficiency we would like to choose the curves $C_1, C_2$ such that
there are fewer allowed combinations $\set{\gg_1,\gg_2}$. For example,
for the generic model $\set{11,263,\set{-1,-1,-6,-1,-1,4}}$, we may
choose to rotate the sequence of the curves to be $
\set{-6,-1,-1,4,-1,-1}$, so that there are only two initial
configurations on the $-6$ curve $C_1$, which are the generic gauge
algebra $\set{\ge_6,\cdot}$ and an enhancement on $C_1$
$\set{\ge_7,\cdot}$. Note that in this case there cannot be any
enhancement on $C_2$ as the global symmetry algebra is always the
trivial algebra on $-6$-curves without an further enhancement to
$\ge_7$, so no tunings are allowed on any intersecting curves.
In fact, in the Hodge number regions we are considering, there are
very few cases that lack non-Higgsable $\ge_7$ or $\ge_8$ gauge
algebras. Every base with a generic elliptic fibration having
$h^{1,1}\geq 220$ has a curve of self-intersection $-7$ or below.  In
the region $h^{2,1}\geq 240$, there are 14 generic models that contain
no curve in the base carrying an $\ge_7$ or $\ge_8$ algebra; the
generic models over bases $\F_{0\leq m\leq 6}$ and $\P^2$ compose 9 of
these 14 models, and are discussed further below. In the remaining
cases, there is no choice of the initial configuration that uniquely
determines the initial configuration, and we have to enumerate and
specify different initial configurations $\set{\gg_1,\gg_2}$ 
over a curve of minimal self-intersection
to
perform the algorithm.

A further issue arises for bases that have curves of non-negative self
intersection.
On such curves, there is no
global constraint on the adjacent algebras from the SCFT point of
view.  While there are some analogous constraints in the case of
curves of non-negative self intersection \cite{Johnson:2016qar}, the
constraints are weaker and less completely understood.  So we do not
impose global constraints in these cases.  In principle this can be
handled by simply iterating over all gauge groups, however in practice
the number of cases where this issue is relevant is rather limited and
can be handled efficiently using more specific  methods.

From the minimal model point of view we can fairly easily classify the
types and configurations of non-negative self intersection curves that
can arise.  The minimal model bases $\P^2$ and $\F_{m}$ have three
consecutive curves of non-negative self intersection.  Any blow-up of
one of these bases has either only one such curve or two adjacent such
curves, since blow-ups reduce the self-intersection of curves
containing the blow-up point and do not introduce new curves of
non-negative self intersection.   Blow-ups of the resulting bases again have at most two
curves of non-negative self intersection and when there are two they
are always adjacent.  So the possibilities are actually quite limited.

In general, 
the way we deal with the cases having one or two
non-negative curves for bases with large $h^{2, 1}$ is by performing the algorithm separately in both
opposite directions from a good starting point (curve of maximally
negative self intersection) to get two ``half-patterns'' of tunings, and
connect them appropriately.  In other words,
we start from a chosen curve $C_1$, run
the algorithm in both directions, and stop the tuning procedures when
the first non-negative curve is met in both directions.
We do not impose any
global conditions for the curves of non-negative self intersection.
The combination of the two sets of the half-patterns connected in
this way gives all tuning patterns of a generic model with one or two
non-negative curves in the base. 
For bases with large $h^{1, 1}$, there is generally at most one
non-negative self intersection curve and the nearby gauge group is
generally constrained by global symmetries and nearby
large negative self-interactions; in some of these cases we have used
simpler heuristics to complete the analysis in the presence of
non-negative self-intersection curves.

For the cases $\P^2$ and $\F_{0\leq m\leq 12}$ that have three
non-negative curves, most tunings in fact decrease
the
Hodge number $h^{2, 1}$ below the value 240 of interest.  For example
\cite{Johnson:2016qar}, tuning an $\gsu(2)$ on a $+1$ curve of $\P^2$
changes the Hodge numbers from $(2, 272)$ to $(3, 231)$.  There are
some exceptions: for example tuning an $\gsu(2)$ on the $+12$ curve of
$\F_{12}$ gives a Calabi-Yau with Hodge numbers $(12, 318)$.  But it
turns out
(as we see explicitly from the analysis of the following section)
 that all these cases with $h^{2, 1}\geq 240$ are also
realized in other ways by generic or tuned models over other toric
bases.  So we do not need to explicitly include these in our analysis
since we are not trying to reproduce the precise multiplicity of
models at each Hodge number pair.

Although we have only focused on tuning models in the large Hodge
number regions, one can in principle classify all allowed tuning
patterns of non-abelian gauge algebras on any toric base with the
algorithm described here; though slightly different methods are needed
for tunings over the bases $\P^2$ and $\F_{0\leq m\leq 12}$, an
exhaustive search is straightforward in these cases as there are only
a few curves in these bases (three curves in $\P^2$ and four curves in
$\F_m$).

\subsection{Tate-tuned models}
\label{tatevsKS}

The analysis  described so far in terms of Weierstrass models gives a
large collection of possible gauge algebra tunings over each toric
base.  Not all of these gauge algebra combinations correspond to
hypersurfaces in reflexive polytopes.  There are several reasons for
this.  First, not every Weierstrass tuning can be realized through a
Tate form, so some of these tunings on toric curves will not have standard
$\P^{2,3,1}$-fibered polytope constructions.
Further, some of the combinations of gauge groups that are allowed by
the Zariski analysis and global constraints still cannot be realized
in practice in a global model
---
we
alluded for example at the end of \S\ref{mainalgo} to the fact
that monodromy conditions
are not really taken care of properly in the Zariski decompositions of
$n=4, 6, 12$.
Indeed, an explicit check shows that
not all the Hodge pairs calculated via equations (\ref{eq:dh11})
and (\ref{eq:dh21}) from the Weierstrass tuning patterns we got from
\S\ref{mainalgo} and \S\ref{modif} lie in the KS database. 

We are interested in constraining to a subset of tuning constructions
for which we expect direct polytope constructions.
Hence, for each gauge algebra tuning combination that satisfies the
Weierstrass Zariski analysis and global conditions, we attempt to
construct an explicit Tate-type model by specifying Tate orders of
vanishing
according to
Table \ref{tab:tatealg} for each tuning in a tuning pattern.  We then
perform the Zariski decomposition of the Tate tunings described in
\S\ref{sec:Zariski-Tate}. A tuning pattern gives a genuine Tate-tuned
model if it has the Zariski decompositions of Tate tunings. 
In performing this analysis,
we used in our systematic analysis only the stronger
 version of the Tate
forms
 for the algebras with multiple realizations and/or
monodromy conditions.
In particular, we did not use any of the tunings marked with $\circ$
or $\star$ in Table~\ref{tab:tatealg}.
  The second version of the
$I_{2n}^s$ Tate tuning (marked with $\circ$) was in fact previously
not known and was identified through the analysis of the next
section.  For the $\gso$ algebras, some of the alternate monodromy
tunings were not previously known (for example, the non-$\star$ version of $\gso(4n+4)$ algebras); also, we wished to restrict
attention to cases where the algebra is guaranteed simply by the order
of vanishing of the Tate coefficients.  In general, as we have noted
in the examples in \S\ref{sec:combining} and \S\ref{so8}, the polytope
constructions do not satisfy the monodromy conditions for the higher
rank gauge algebras in these cases.

These principles give us a set of gauge group and Tate tunings over
each toric base that we believe should have direct correspondents in
the KS database through standard $\P^{2,3,1}$-fibered polytopes, given
the correspondence that we established in \S\ref{sec:Tate}. 
We have carried out an explicit comparison of these two sets, and
indeed the Hodge numbers of this more limited class of Tate-tuned
gauge groups all correspond to values that appear in the KS database.
Furthermore, the Hodge pairs from the original Weierstrass analysis
 that are not in the KS database are
exactly those of the tuning patterns that can not be realized by Tate
tuning.  In fact, 
given this restricted set of tunings
we reproduce almost all of the 1,827 distinct Hodge pairs
in the range $h^{1,1}\geq 240$ or $h^{2,1}\geq 240$.  Only 18
of the Hodge pairs in  this range were not found by a ``sieve'' using
the Tate constructions described above.
In the next section we consider the analysis of these 18 outlying
polytope constructions.

A question that we do not explore further here, but which is relevant
to the more general problem of understanding the full set of
Calabi-Yau threefolds and the classification of 6D F-theory models, is
the extent to which tunings are possible that look like they should be
allowed from the Weierstrass Zariski analysis and anomaly cancellation
conditions, but do not correspond to Tate constructions.  Various
aspects of this ``Tate tuning swampland'' were analyzed in
\cite{Johnson:2016qar}.  In the context of this project, we did a
local analysis of the Weierstrass tuning patterns that are not Tate
tuning patterns.  We reproduced some parts of the known Tate tuning
swampland and also found new obstructions. Some examples of the
problematic tunings in the Tate construction are listed in Table
\ref{t:swamp}. An interesting question for further research is which
of these can be realized through good global Weierstrass models when
the indicated sequence of curves arises as part of a toric (or
non-toric) base.

\begin{table}[]
\centering
\begin{tabular}{|l|l|}
\hline
\multicolumn{2}{|l|}{$\ge8$ Tate swamp}                            \\ \hline
\multicolumn{2}{|l|}{$\gsu(3)\oplus\gsp(3)$, $\gsu(3)\oplus\gsp(4)$,}    \\
\multicolumn{2}{|l|}{$\gg_2\oplus\gso(10)$,}                                       \\
\multicolumn{2}{|l|}{$\gso(9)\oplus\gsu(4)$, $\gso(9)\oplus\gsp(2)$, $\gso(10)\oplus\gsu(4)$, $\gso(10)\oplus\gsp(2)$,}\\
\multicolumn{2}{|l|}{$\gso(11)\oplus\gsu(3)$, $\gso(11)\oplus\gsp(2)$,}\\
\multicolumn{2}{|l|}{$\gso(13)\oplus\gsp(1)$, $\gso(13)\oplus\gsu(2)$}\\
\hline\hline
\multicolumn{2}{|l|}{Miscellaneous Tate swamp (some examples)} \\ \hline
Gauge groups                             & Local geometry                            \\ \hline
$\gso(7)\oplus\gsu(2)\oplus\cdot$&   -3, -2, -2\\
$\cdot\oplus\gsu(2)\oplus\gsp(2)$(or $\gsu(4)$)&-2, -2, -1/0\\
$\cdot\oplus\gsu(2)\oplus\gg_2\oplus\gsp(3)$&-2, -2, -2, -1/0\\
 \hline
\end{tabular}
\caption[]{\footnotesize
Tate tuning swamp: We list all subalgebras allowed by the ``$E_8$ rule'' that however can not be realized by Tate tunings. We also give  some examples of the tuning patterns we found that do not violate global symmetry constraints but that can not be realized by Tate tunings (i.e. violate Tate-Zariski decomposition). }
 \label{t:swamp}
\end{table}

\section{Polytope analysis for  cases missing from the
simple tuning construction, and other exotic constructions}
\label{restcases}

As discussed above, there are only 18 Hodge pairs in the
regions $h^{1,1}\geq 240$ or $h^{2,1}\geq 240$ in the KS database
that are not produced by our Tate tuning algorithm. One
of these missing 18 Hodge pairs is in the large $h^{2,1}$ region,
$\set{45,261}$, and the other 17 (see Table \ref{restt}) are in the
large $h^{1,1}$ region.
In this section we analyze the polytopes in the Kreuzer-Skarke
database associated with these 18 Hodge number pairs.

By studying these 18 classes of Calabi-Yau manifolds, we have
identified new tuning constructions that we had not known previously;
the KS database provides us with global models utilizing these
constructions that we did not expect a priori in our original
analysis.  We study the fibration structure of the 18 outstanding
classes by analyzing the polytopes in the way described in section
\ref{fb}.  All the polytopes associated with these 18 Hodge pairs have
a $\P^{2,3,1}$ fibered polytope structure (though in some cases it is
really the more specialized Bl$_{[0,0,1]}\P^{2,3,1}$ fiber that
occurs), but not all of them are the standard $\P^{2,3,1}$-fibered
polytopes that we have defined in \S\ref{sp231}.  In particular, the
CY hypersurface of a standard $\P^{2,3,1}$-fibered polytope (or
Bl$_{[0,0,1]}\P^{2,3,1}$-fibered polytope) has a Tate form, while this
is not the case for other fibration structures that use the same fiber but a
different ``twist''. We analyze the two different types of polytopes
arising from the 18 Hodge pairs separately.
In  \S\ref{withTate} we analyze the standard $\P^{2,3,1}$-fibered
polytopes in the KS database that we have not obtained in our
systematic construction of Tate-tuned models.  In 
\S\ref{non} we analyze the polytopes that do not have the standard
$\P^{2,3,1}$-fibered structure.  We also include in \S\ref{non} some
further examples in the KS database that are outside the range of
focus of this paper but that illustrate some further interesting
exotic structure associated with gauge groups on non-toric curves in
the base.

\begin{table}
\centering
\begin{tabular}{| l l l  |}
 \hline
Standard $\P^{2,3,1}$- & huge tuning &\{$240, 48$\}$, $\{$244, 10$\}$, $\{$250, 10$\}$,  $\{$261, 9$\}\\ \cline{2-3}
fibered polytopes &non-toric base&  $\{$258, 60$\}$ (``$\ge_8$-tuning'')\\\cline{2-2}
 
 \hline
 Bl$_{[0,0,1]}\P^{2,3,1}$-&global $\gu(1)$ tuning and &\{$263, 32$\}$, $\{$251, 35$\}$, $\{$247, 35$\}$, $\{$240, 37$\}$ $\\
 fibered polytopes& non-toric base &(``$\gso(n\geq 13)$-tuning'' on a $-3$-curve) \\
  \hline \hline
Non-standard  $\P^{2,3,1}$-&tuning on non-&\{$261, 51$\}$,$\{$261, 45$\}$, $\{$260, 62$\}$, $\{$260, 54$\}$, $\\
 fibered polytopes&toric  curve &\{$259, 55$\}$, $\{$258, 84$\}$, $\{$254, 56$\}$, $\{$245, 57$\} \\
\hline
\end{tabular}
\caption{\footnotesize 
The Hodge number pairs in the KS database at large $h^{1,1}$ that we
did not obtain from straightforward Tate-tuned models. However, all
these can be reproduced by some flat elliptic fibrations that we
discuss in this section: The standard $\P^{2,3,1}$ models, which have
a Tate form, are studied in \S\ref{withTate}, and the non-standard
$\P^{2,3,1}$ models, which  involve tunings on non-toric
curves in the base, are studied in \S\ref{non}.}
\label{restt} 
\end{table}

\subsection{Fibered polytope models with Tate forms}
\label{withTate}

Of the 18 missing Hodge pairs, there are $1+9$ Hodge pairs in the
large $h^{2, 1}, h^{1,1}$ regions respectively in which there is a
standard $\P^{2,3,1}$-fibered polytope (or a standard
Bl$_{[0,0,1]}\P^{2,3,1}$-fibered polytope), which has a Tate
form. Therefore, we analyze the Tate models explicitly from these
polytopes to learn about the Tate tunings that we missed in our
initial construction.

The Hodge pair in the large $h^{2,1}$ region, $\set{45,261}$, has only
one polytope.  This polytope reveals a second tuning of the type
$I^s_{2n}$ singular fiber that is not just a specialization of the
known Tate tuning. 
We also find that applying
this novel Tate tuning $\gsu(6)$  on a $m\geq-1$-curve gives models
with
the three-index antisymmetric
representation as opposed to the generic fundamental and two-index
antisymmetry representations. We describe this analysis in detail in
\S\ref{2ndversion}.  The polytopes of the nine missing Hodge pairs at
large $h^{1,1}$ with the standard fibration structure are either
extremely tuned models, with bases having generic elliptic fibrations
with $h^{1,1}< 220$ (described in \S\ref{largesh}), or are non-flat
  elliptic fibration models over a toric base (described in
  \S\ref{semitoric}).  In the non-flat elliptic fibration cases, as we
  have discussed at the end of \S\ref{multiplicity}, the CY resolution
  of (4,6) singularities in terms of the polytope model produces
  irreducible components of the ambient toric fiber (as the
  hypersurface equation restricted to the components is trivially
  satisfied over the $(4,6)$ points). Therefore, at these points the
  dimension of the fiber jumps to two giving the non-flat elliptic
  fibration structure.  Associating the additional divisors with
  blow-ups in the base allows us to describe the resulting Calabi-Yau
  threefolds alternatively as flat elliptic fibrations over the blown
  up base.  The resulting models in the cases found here give rise to
  $\ge_8$ tunings or $\gso(n\geq 13)$ tunings on $-3$-curves, and are
  also involved with tuned Mordell-Weil sections, which are associated
  with U(1) factors and U(1)-charged hypermultiplets.

% discussion of the polytopes of the nine Hodge pairs of large $h^{1,1}$ are \S\ref{largesh} and \S\ref{semitoric}: (1) large shifts of Hodge numbers (2) tunings of models over semi-toric bases (where there are examples of $\ge_8$-tunings.), and (3) models with an U(1) factor and additional charged matter representations under the U(1)  (where there are examples of $\gso(n\geq 13)$-tunings on $-3$-curves.)

%class of semi-toric bases

%4,6, fiber dimension jump but  we have been able to identify resolution of the base such that these models are still  ---  we can interpret these reflexive polytope models as elliptic fiberation over semi-toric bases, and the semi-toric bases are somehow encoded in terms of the (4,6) points.\footnote{This is essentially the situation of those generic fibration models over the toric bases involved with curves of self-intersection $-9, -10, -11$ that we have seen in table \ref{t:f9}.}   

\subsubsection{Type $I^s_{2n}$ Tate tunings  and exotic matter}

\label{2ndversion}

%Let's first discuss the large $h^{2,1}$ region.  We got all Hodge pairs  for $h^{2,1} \geq 240$ except \{45, 261\}. From the analysis of the polytope data of this model, we found the second Tate tuning of type $I^s_{2n}$ gauge algebras. After including this second tuning in our construction, all Hodge pairs in the region are realized by at least a Tate-tuned model.  More interestingly, this second realization automatically gives rise to the $\gsu(6)^*$ exotic matter tuning on $-1$-curves \cite{Bertolini:2015bwa}.
%There is only one data with Hodge numbers $\{45, 261\}$ in the KS database. We didn't get a CY  with the  Hodge numbers  from our constuction in the first place with the original Tate algorithm.

The polytope model M:357 8 N:65 8 H:45,261 is a standard
$\P^{2,3,1}$-fibered polytope, and is a Tate-tuned model of the
generic model
\begin{equation}
\nonumber
\{38, 290, \{-2, -2, -1, -6, -1, -3, -1, -5, -1, -3, -2, -2, -1, -12, 0, 6\}\}.
% \{38, 290, \{-12, -1, -2, -2, -3, -1, -5, -1, -3, -1, -6, -1, {\color{blue}-2, -2}, 6,  0\}\},
\end{equation}
The data $\set{a_1,a_2,a_3,a_4,a_6}$ of the Tate form show the  orders of vanishing along each curve
\begin{eqnarray}
\nonumber
&&\set{\set{0,2,2,4,6},\set{0,2,1,4,5},\set{0,2,0,4,4},\set{1,2,2,4,5},\set{1,2,0,4,2},\\&&\nonumber\set{1,2,1,4,3},\set{1,2,0,4,1},\set{1,2,2,4,4},\set{1,2,1,4,1},\set{1,2,2,4,3},\\&&\set{1,2,2,4,2},\set{1,2,2,4,1},\set{1,2,2,4,0},\set{1,2,3,4,5},\set{0,0,0,0,0},\set{0,0,0,0,0}}.
\label{exo}
\end{eqnarray}
In terms of $\{f, g, \Delta\}$ (equations (\ref{eq:tw1})-(\ref{TtoW})), the orders of vanishing are
\begin{eqnarray}
\nonumber
&&\set{\set{0, 0, 6}, \set{0, 0, 3}, \set{0, 0, 0}, \set{3, 4, 8}, \set{1, 0, 0}, \set{2, 2, 4}, \set{1, 0, 0}, \set{3, 4, 8},\\\nonumber&& \set{2, 1, 2}, \set{3, 3, 6}, \set{3, 2, 4}, \set{3, 1, 2}, \set{3, 0, 0}, \set{4, 5, 10}, \set{0, 0, 0}, \set{0, 0, 0}},
\end{eqnarray}
which shows that there is an $\gsu(6)$ enhanced on the first
$-2$-curve, $D_1\equiv\set{b_1=0}$, and an $\gsu(3)$ on the second
$-2$ curve. However, the corresponding Tate tuning is not just a
specialization of the $\gsu(6)$ Tate tuning $\set{0,1,3,3,6}$ in the
literature. Via this example, we found the second version of the
$\gsu(2n)$ tuning, which we have included in the Tate tunings listed
in Table \ref{tab:tatealg}, indicated by $\gsu(2n)^\circ$.

As this is the only polytope associated with the Hodge pair
$\set{45,261}$, it seems that
the traditional $\gsu(2n)$ tuning is somehow not
allowed in this configuration. We checked explicitly by performing a
tuning where we substitute in the vanishing order
$\set{0,1,3,3,6}$ over $D_1$, and perform the Tate-Zariski
decomposition. The vanishing orders after iteration become
\begin{eqnarray}
\nonumber
&&\{\{0,1,3,3,6\}, \{0,1,3,2,5\}, \{0,1,3,1,4\}, \{1,2,3,3,5\}, \{1,2,3,1,2\}, \\\nonumber
&&\{1,2,3,2,3\}, \{1,2,3,1,1\}, \{1,2,3,3,4\}, \{1,2,3,2,1\}, \{1,2,3,3,3\}, \\\nonumber
&&\{1,2,3,3,2\}, \{1,2,3,3,1\}, \{1,2,3,3,0\}, \{1,2,3,4,5\}, \{0,0,0,0,0\}, \{0,0,0,0,0\}\};
\end{eqnarray}
or in terms of $\{f, g, \Delta\}$,
\begin{eqnarray}
\nonumber
&&\{\{0,0,6\}, \{0,0,4\}, \{0,0,2\}, \{3,5,9\}, \{1,2,3\}, \{2,3,6\}, \{1,1,2\}, \{3,4,8\}, \\\nonumber
&&\{2,1,2\}, \{3,3,6\}, \{3,2,4\}, \{3,1,2\}, \{3,0,0\}, \{4,5,10\}, \{0,0,0\}, \{0,0,0\}\},
\end{eqnarray}
which is problematic as the global symmetry constraint on the
$-6$-curve $D_4$ is violated. This confirms again that there has to be
a 
Tate-tuned pattern that is consistent under the Tate-Zariski
decomposition
 for a corresponding polytope to exist. And  we cannot
obtain a polytope of these Hodge numbers using the standard tuning
methods because the
$\gsu(2n)^\circ$ tunings, $\{0, 2, n - 1, n + 1, 2 n\}$, are not
specializations of the standard $\gsu(2n)$ tunings,$\{0, 1, n, n, 2 n\}$.
\vspace*{0.1in}

%Neither of the two Tate tuning realizations is specialization of each other,  so we will not get this Tate-tuned model with the usual Tate tuning. We include both of them into the Tate algoritm in our construction; more generally,  we include both degrees of vanishing order  $\{0, 1, n, n, 2 n\}$ and $\{0, 2, n - 1, n + 1, 2 n\}$ for the $I^s_{2n}$ type $\gsu(2n)$ gauge groups. One can check via equations (\ref{eq:tw1}--\ref{TtoW}) that this second version Tate tuning indeed gives the  vanishing order of $\{f, g, \Delta\}$ to be $\{0,0,2n\}$ as well.

% which we would never get by initially setting the degrees to
% $\{0,1,3,3,6\}$ where while the second entry is less in the later
% case, the third entry is greater. Not one of them is a
% specialization of the other, so we must include both of them in the
% Tate table \ref{} (used for checking  the existence as a Tate model)
% to get the $\{45,261\}$. 

In the case of a $-2$ curve, as in the example encountered at large
$h^{2, 1}$, the matter content associated with the physics of the
exotic $\gsu(6)^\circ$ tuning is equivalent to that of a standard
$\gsu(6)$ tuning over a $-2$ curve.  After incorporating these
alternative $\gsu(2n)^\circ$-tunings into our algorithm, however, we
discovered that this second Tate realization of $\{f, g,
\Delta\}=\{0,0,2n\}$ gives rise to the non-generic three-index
antisymmetric $({\bf 20})$ representation of $\gsu(6)$ when the tuning
is performed on curves of self-intersection $m\geq-1$.  We describe an
example of this explicitly, in the context of a global model that lies
outside the regions of primary interest $h^{1,1},h^{2, 1} \geq 240$.

The polytope model M:280
11 N:28 9 H:18,206 is a Tate-tuned model of
\begin{equation}
%\{11, 263, \{-6, -1, {\color{blue}-2, -1}, 4, 0\}\}
\{11, 263, \{{-1, -2}, -1, -6, 0, 4\}\}.
\label{eq:exotic-matter-base}
\end{equation}
There is an $\gsu(6)^\circ$ tuned on the $-1$-curve $D_1$ and an
$\gsu(3)$ tuned on
the $-2$-curve $D_2$. Interestingly, by explicit
analysis, we find that the $f, g$ from the polytope data automatically satisfy
the conditions for the codimension-two singularity on $D_1$ to support the
three-index
antisymmetric representation of $\gsu(6)$, as described in
\cite{Morrison:2011mb}.  To see this, we fix
the complex structure moduli of $f, g$ to some general enough $\Z$
values to avoid accidental cancellations,
expand $f$ and $g$ in terms of $\sigma\equiv b_1$ where the
coefficients are in terms of a second local coordinate that we choose
to be $b_2$
\begin{eqnarray}
&&f(\sigma, b_2)=f_0(b_2)+f_1(b_2)\sigma+f_2(b_2)\sigma^2+\cdots, \\
&&g(\sigma, b_2)=g_0(b_2)+g_1(b_2)\sigma+g_2(b_2)\sigma^2+\cdots;
\end{eqnarray}
then we find (following the notation in \cite{Morrison:2011mb})
\begin{itemize}
\item $\Delta_0=0: f_0  \sim   -\frac1{48} \phit^4$ and $g_0  \sim
  \frac1{864} \phit^6 $; we choose to set $ \phi_0= 57 + 46 b_2.$ 
\item $\Delta_1=0: g_1 = -\frac{1}{12}\phi_0^2f_1.$
\item $\Delta_2=0: f_1 \sim\frac1{2}\phi_0\psi_1\Rightarrow \psi_1=-(1/6) b_2 (37 + 62 b_2)\phi_0^2$ and $g_2 = \frac1{4}\psi_1^2 - \frac1{12}\phi_0^2f_2.$
\item $\Delta_3=0: \psi_1\sim -\frac1{3}\phi_0\phi_1\Rightarrow \phi_1=(1/2) b_2 (37 + 62 b_2)\phi_0$
and $g_3=-\frac1{12} \phi_0^2 f_3 - \frac1{3} \phi_1 f_2 - \frac1{27} \phi_1^3$.
\item $\Delta_4=0: f_2 + \frac1{3} \phi_1^2=\frac1{2} \phi_0 \psi_2\Rightarrow \psi_2=-(1/12) b_2 (-972 + 321867 b_2 + 818194 b_2^2 + 770316 b_2^3 + 
   257716 b_2^4) $ and $g_4=\frac1{4} \psi_2^2 - \frac1{12} \phi_0^2 f_4 - \frac1{3} \phi_1 f_3$.
\item $\Delta_5=0: \alpha=\text{GCD}[\phi_0, \psi_2]=1\Rightarrow \beta=\phi_0,  \phi_2=-3\psi_2, \nu= (1/2) b_2 (37 + 62 b_2).$
 $f_3= - \frac1{3} \nu \phi_2 -3 \lambda\beta\Rightarrow \lambda= (1/72) b_2^3 (358621+1496554 b_2+1733688 b_2^2+656328 b_2^3)$ and  $g_5=-\frac1{12} \phi_0^2 f_5 - \frac1{3} \phi_1 f_4 + \phi_2 \lambda$. 
\end{itemize}
Hence, $\alpha\neq 0$ and $\beta=0$ over the codimension-two point
$\sigma=\phi_0=0$, which gives rise to a 3-index antisymmetric matter
field.  Indeed, we have to use the representations
$15\times\mathbf{6}+1/2\times\mathbf{20}$, as opposed to the ordinary
$14\times\mathbf{6} + 1 \times\mathbf{15}$ of $\gsu(6) $ on
$-1$-curves, to obtain the correct shifts of the Hodge number
$h^{2,1}$ from anomaly cancellation: $\Delta h^{1,1}= 2 + 5 = 7$, and
$\Delta h^{2,1}= (8+35)+ (6\times \mathbf{3}+
15\times\mathbf{6}+1/2\times\mathbf{20}- \mathbf{3}\times\mathbf{6}
\text{ (shared)}) = -57$.

The conclusion that the $\gsu(6)^\circ$-tuning on the $-1$-curve leads
to this exotic matter representations is not particular to this
 specific global model. Following the same steps, we performed a local analysis on
an isolated $-1$ curve; when we tune the Tate form $\{0,2,2,4,6\}$ on the
curve, we see that
$\alpha\neq 0$ but $\beta=0$ over a point on the curve,
while the Tate form $\{0,1,3,3,6\}$ leads to $\alpha= 0$ over a point
but $\beta\neq0$. Although there is no corresponding polytope model with
ordinary $\gsu(6)$ matter in case of
the global model studied above (there is no polytope in the KS database that
gives a Calabi-Yau with
Hodge numbers $\{18,207\}$, and the tuning $\{0,1,3,3,6\}$ over the
base
(\ref{eq:exotic-matter-base})
does not lead
to a good global Zariski decomposition), we can contrast the two tunings
of $\gsu(6)$  on a $-1$-curve
in
polytopes that describe tunings of the generic model over
the $\F_1$ base M:335 6 N:11 6
H:3,243. Both models exist in the KS database: the
$\gsu(6)$-tuning gives the model M:242 12 N:16 9 H:8,179 and the
$\gsu(6)^\circ$-tuning gives the model M:236 10 N:16 8 H:8,178.

The two different Tate forms of $\gsu(6)$ automatically give
different representations on all curves with self-intersection
$m\geq-1$ (there is only matter in the fundamental representation on
$-2$-curves). For example, consider  tuning the generic model
over $\F_1$ now with $\gsu(6)$ and $\gsu(6)^\circ$
respectively 
on a $0$-curve. The $\gsu(6)$-tuning gives the model with ordinary
matter M:204 11 N:16 9 H:8,152 while the $\gsu(6)$-tuning gives the
model with exotic matter (two half-hypermultiplets in the ${\bf 20}$
representation) M:197 9 N:16 8 H:8,150. The Hodge numbers
from the polytope data are consistent with the calculation from anomaly
cancellation with the respective matter representations (see table
\ref{t:6reps}).

\begin{table}
\centering
\begin{tabular}{|l|l|l|}
\hline
                & Ordinary matter                              & Exotic matter                                        \\ \hline
Tate form       & \{0, 1, 3, 3, 6\}                            & \{0, 2, 2, 4, 6\}                                    \\ \hline
Representations & $(16 + 2m )\mathbf{6} + (m + 2)\mathbf{15}$ &  $(16+3m+2)\mathbf{6}+(m+2)\frac1{2}\mathbf{20}$\\ \hline
\end{tabular}
\caption{
\footnotesize
Representations of $\gsu(6)$ and $\gsu(6)^{\circ}$-tuning on curves of self-intersection $m\geq -2.$}
\label{t:6reps}
\end{table}

\subsubsection{Large Hodge number shifts}
\label{largesh}

Four of the ``extra'' Hodge number pairs in the region $h^{1,1}\geq
240$ turn out to come from standard Tate tunings of generic models
that have $h^{1,1}<220,$ outside the region we considered for starting
  points.  These are listed as ``huge tunings'' in Table~\ref{restt}.
These models 
each contain a chain of
$\{-1, -4\}$s, which allows $\gso(n)$ with $n$ very large to be enhanced
  on the $-4$-curves, producing huge shifts of the Hodge numbers. 
While there are only four specific models of this type among the 18
Hodge pairs in the region of interest not found by Tate tunings, it
seems that this large tuning structure on chains of $-1, -4$ curves is
a common feature and there are many other examples of this in the
database, both increasing multiplicities at large Hodge numbers in
cases that also have Tate tuned realizations, and also occurring at
Hodge numbers outside the range of interest here.

We work out one example here in detail; the others have similar structure.
The example with the largest $h^{1,1}$ (from the four ``extra'' models
of this type)
is
  the polytope M:20 6 N:352 7 H:261,9, which is a Tate-tuned model of the
  generic polytope model
\begin{eqnarray}
\nonumber
&&\{135, 15, \{-12, -1, -2, -2, -3, -1, -4, -1, -4, -1, -4, -1, -4, -1, \\\nonumber
&&-4, -1, -4, -1, -4, -1, -4, -1, -4, -1, -4, -1, -4, -1, -4, -1, -4, \\
&&-1, -4, -1, -4, -1, -4, -1, -4, -1, -3, -2, -2, -1, -12, 0\}\} \,,
\end{eqnarray}
as can be determined by explicitly computing the base polytope of the
toric fibration.
Therefore, the enhanced tunings should give   $\set{\Delta h^{1,1},\Delta h^{2,1}}=\{126, -6\}$. Explicit analysis of the polytope gives the data $\set{m,\set{a_1,a_2,a_3,a_4,a_6},\{f, g, \Delta\}}$ of each $m$-curve 
\begin{eqnarray}
\nonumber
&&\{\{-12, \{1,2,3,4,5\}, \{4,5,10\}\}, \{-1, \{1,1,5,5,0\}, \{2,0,0\}\}, \{-2, \{1,1,5,5,1\}, \{2,1,2\}\}, \\ \nonumber
&&\{-2, \{1,1,5,5,2\}, \{2,2,4\}\}, \{-3, \{1,1,5,5,3\}, \{2,3,6\}\}, \{-1, \{1,0,7,6,1\}, \{0,0,1\}\}, \\ \nonumber
&&\{-4, \{1,1,5,5,4\}, \{2,3,7\}\}, \{-1, \{1,0,7,6,3\}, \{0,0,3\}\}, \{-4, \{1,1,5,5,5\}, \{2,3,8\}\}, \\ \nonumber
&&\{-1, \{1,0,7,6,5\}, \{0,0,5\}\}, \{-4, \{1,1,5,5,6\}, \{2,3,9\}\}, \{-1, \{1,0,7,6,7\}, \{0,0,7\}\}, \\ \nonumber
&&\{-4, \{1,1,5,5,7\}, \{2,3,10\}\}, \{-1, \{1,0,7,6,9\}, \{0,0,9\}\}, \{-4, \{1,1,5,5,8\}, \{2,3,11\}\}, \\ \nonumber
&&\{-1, \{1,0,7,6,11\}, \{0,0,11\}\}, \{-4, \{1,1,5,5,9\}, \{2,3,12\}\}, \{-1, \{1,0,7,6,12\}, \{0,0,12\}\}, \\ \nonumber
&&\{-4, \{1,1,5,5,9\}, \{2,3,12\}\}, \{-1, \{1,0,7,6,12\}, \{0,0,12\}\}, \{-4, \{1,1,5,5,9\}, \{2,3,12\}\}, \\ \nonumber
&&\{-1, \{1,0,7,6,12\}, \{0,0,12\}\}, \{-4, \{1,1,5,5,9\}, \{2,3,12\}\}, \{-1, \{1,0,7,6,12\}, \{0,0,12\}\}, \\ \nonumber
&&\{-4, \{1,1,5,5,9\}, \{2,3,12\}\}, \{-1, \{1,0,7,6,12\}, \{0,0,12\}\}, \{-4, \{1,1,5,5,9\}, \{2,3,12\}\}, \\ \nonumber
&&\{-1, \{1,0,7,6,12\}, \{0,0,12\}\}, \{-4, \{1,1,5,5,9\}, \{2,3,12\}\}, \{-1, \{1,0,7,6,11\}, \{0,0,11\}\}, \\ \nonumber
&&\{-4, \{1,1,5,5,8\}, \{2,3,11\}\}, \{-1, \{1,0,7,6,9\}, \{0,0,9\}\}, \{-4, \{1,1,5,5,7\}, \{2,3,10\}\}, \\ \nonumber
&&\{-1, \{1,0,7,6,7\}, \{0,0,7\}\}, \{-4, \{1,1,5,5,6\}, \{2,3,9\}\}, \{-1, \{1,0,7,6,5\}, \{0,0,5\}\}, \\ \nonumber
&&\{-4, \{1,1,5,5,5\}, \{2,3,8\}\}, \{-1, \{1,0,7,6,3\}, \{0,0,3\}\}, \{-4, \{1,1,5,5,4\}, \{2,3,7\}\}, \\ \nonumber
&&\{-1, \{1,0,7,6,1\}, \{0,0,1\}\},\{-3, \{1,1,5,5,3\}, \{2,3,6\}\}, \{-2, \{1,1,5,5,2\}, \{2,2,4\}\},  \\ \nonumber
&&\{-2, \{1,1,5,5,1\}, \{2,1,2\}\},\{-1, \{1,1,5,5,0\}, \{2,0,0\}\}, \{-12, \{1,2,3,4,5\}, \{4,5,10\}\},  \\ \nonumber
&&\{0, \{0,0,0,0,0\}, \{0,0,0\}\}\}.
\end{eqnarray}
The gauge algebras on $-4$ and $-1$ curves are only determined from
this analysis
up to
monodromies. We can, however, determine the algebras without explicitly
analyzing monomials. First, from the anomaly constraint analyzed in
\cite{Johnson:2016qar}, $\gsu$ cannot be adjacent to $\gso$, so the
algebras on $-1$ curves have to be $\gsp$. The choice $\gso(2n-5)$ or
$\gso(2n-4)$ on $-4$ is determined from global symmetry
constraints. For example, it has to be $\gso(20)$ rather than
$\gso(19)$ between two $\gsp(6)$
algebras for the global symmetry on the $-4$ curve
to be satisfied; while the lower rank $\gso(17), \gso(19)$ has to be
chosen for two $-4$'s connecting to $\gsp(5)$ for the global symmetry constraint
on the $-1$
curve to be satisfied. Hence, the corresponding gauge algebras
are
\begin{eqnarray}
\nonumber
&&\{\{-12,\ge_8\},\{-1,\cdot\},\{-2,\cdot\},\{-2,\gsu(2)\},\{-3,\gg_2\},\{-1,\cdot\},\{-4,\gso(9)\},\{-1,\gsp(1)\},\\ \nonumber
&&\{-4,\gso(11)\},\{-1,\gsp(2)\},\{-4,\gso(13)\},\{-1,\gsp(3)\},\{-4,\gso(15)\},\{-1,\gsp(4)\},\{-4,\gso(17)\},\\ \nonumber
&&\{-1,\gsp(5)\},\{-4,\gso(19)\},\{-1,\gsp(6)\},\{-4,\gso(20)\},\{-1,\gsp(6)\},\{-4,\gso(20)\},\{-1,\gsp(6)\},\\ \nonumber
&&\{-4,\gso(20)\},\{-1,\gsp(6)\},\{-4,\gso(20)\},\{-1,\gsp(6)\},\{-4,\gso(20)\},\{-1,\gsp(6)\},\{-4,\gso(19)\},\\ \nonumber
&&\{-1,\gsp(5)\},\{-4,\gso(17)\},\{-1,\gsp(4)\},\{-4,\gso(15)\},\{-1,\gsp(3)\},\{-4,\gso(13)\},\{-1,\gsp(2)\},\\ \nonumber
&&\{-4,\gso(11)\},\{-1,\gsp(1)\},\{-4,\gso(9)\},\{-1,\cdot\},\{-3,\gg_2\},\{-2,\gsu(2)\},\{-2,\cdot\},\\ \nonumber
&&\{-1,\cdot\},\{-12,\ge_8\},\{0,\cdot\}\},\end{eqnarray} 
which give
the correct Hodge number shifts (in particular,
one can quickly check that according to
the rank of the gauge algebras $\Delta h^{1,1}=126$ as expected above).

\subsubsection{Tate-tuned models corresponding to non-toric bases}
\label{semitoric}

We have not considered tuning an $\ge_8$ algebra on any curve of
self-intersection $m \geq -8$, as it leads to a violation of the
anomaly conditions that corresponds to the appearance of a $(4,6)$
singularity.  Similarly, tunings of $\gso(n\geq13)$ on $-3$-curves are
also ruled out by anomaly cancellation.  Nonetheless, there are
polytope models in the KS database that appear to contain these
tunings, which give rise to Hodge pairs that we have not obtained in
Tate tunings of Kodaira type.  This set of tunings can be understood
as more complicated generalizations of the non-flat structure we have
already described for fibrations over $-9, -10$ and $-11$ curves.  As
we discussed already in that context, over $(4,6)$ points the
resolved fiber in the polytope model is two-dimensional, but we can
understand the Calabi-Yau geometry by resolving the base at these
points to obtain a corresponding flat elliptic fibration model over a
blown up base that is generally non-toric.  In this section we
describe models that involve $\ge_8$ algebras tuned on $-8$ curves and
models involving tunings of $\gso(n\geq13)$ on $-3$-curves.  In the
latter case, the ``extra'' models in the KS database in our region of
interest involve a further complication in which a nontrivial
Mordell-Weil group is generated associated with an abelian U(1) factor
in the F-theory gauge group; a detailed example with that additional
structure is relegated to Appendix~\ref{26332cal}.
\vspace*{0.1in}

We begin with an example of a tuned $\ge_8$ on a $-8$-curve.  This
occurs in the model M:88 8 N:356 8 H:258,60.  The $\dd$ polytope has 
vertices 
$\set{(0, 0, 0, -1), (7, 6, 2, 3), (-1, -1, 2, 3)$, $(-1, -1,
  1, 2), (0, 6, 2, 3),$ $(0, 0, -1, 0), (-42, -36, 2, 3),(-15, -13, 2,
  3)}$. It describes a non-flat  Tate-tuning of the generic
elliptic fibration
\begin{eqnarray}
\nonumber
&&\{252, 78,\{-12//-11//-12//-12//-12//-12//-12//-12//{-8},-1,-2,{-1},0\}\}\nonumber
\end{eqnarray}
where $//$ stands for $\{-1,-2,-2,-3,-1,-5,-1,-3,-2,-2,-1\}$, and
there are in total $101$ curves $D_i$ in the base. There is an
$\ge_8$ tuned on $D_{97}$ and an $\gsu(2)$ tuned on $D_{100}$, where the
orders of vanishing are enhanced to $\set{1,2,3,4,5}$ and
$\set{0,0,1,1,2}$, respectively. As it needs four blowups for a
$-8$-curve to become a $-12$-curve, which carries the
$\ge_8$ gauge
algebra without $(4,6)$ points, we expect that
there are four $(4,6)$
points on the $D_{97}$ over which the resolved fiber become
two-dimensional.

The $(4,6)$ points and the 2D fiber can be understood by an explicit
analysis of the hypersurface $p$ in equation (\ref{p}) restricting to
each irreducible component, which corresponds to a lattice point in
the $\ge_8$ top in equation (\ref{top}) of the non-generic toric fiber
over $D_{97}$.  Analogous to the models over Hirzebruch surfaces
$\F_9/\F_{10}/\F_{11}$ in Appendix \ref{2dfiber}, we find in this case
that over a generic point on the $-8$-curve, $p$ intersects the 9
components in equation (\ref{boundary}) that are the boundary of the
3-dimensional face in a locus comprising nine $\P^1$'s, which form the
$\ge_8$ extended Dynkin diagram, but over four distinct $(4,6)$ points
on $D_{97}$, $p$ intersects also the whole irreducible component
corresponding to $((v_{97}^{(B)})_{1,2}, 0, 0)$ ($pt_5'$) in the top;
i.e., $p\rvert_{D_{97}}=0$ is trivially satisfied over these four
points, and the elliptic fiber over the toric base contains this
irreducible component, which is two-dimensional, at these four points.

% The degrees of vanishing along each curves indicate there is gauge group $\ge8$ tuned on the ``$-8$'' -curve (again actually $-12$ curve) and $\gsu(2)$ tuned on the second to the last curve (the $-1$ curve). 

%Tuning $\ge8$ on the $-8$, there will be four codimension two $(4,6)$ singularities on the curve, blow up the four points into four exceptional $-1$ curves intersecting the $-12$-curve. 
%proper transformation of the curve, which is now a curve with self-intersection number $-12$ (see Fig.~\ref{}). 

The corresponding flat elliptic fibration model has a non-toric base
where the four points on $D_{97}$ are blown up and the proper
transform $-12$-curve intersects with the four exceptional divisor
$-1$-curves. Now we can calculate the Hodge number shifts of the flat
elliptic fibration model via anomaly cancellation: $\Delta T=4$ (each
blowup contributes one additional tensor multiplet), $\Delta
r=(8-7)+1$, $\Delta V=(248-133)+3$, and $\Delta H_c=10\times
\mathbf{2}$; therefore, by equations (\ref{eq:dh11}) and
(\ref{eq:dh21}), $\Delta h^{1,1}=6$ and $\Delta h^{2,1}=-18$, which
gives $\set{252, 78}+\set{6,-18}=\set{258,60}$, as needed.

\vspace*{0.1in}

%Analogous to the ``$\ge_8$-tuning'' on curves resolved at non-toric points to $-12$-curves,  we can do any $\gso(n)$ tunings on curves with self-intersection $\geq-4$ resolved at non-toric points to $-4$-curves.
\begin{table}[]
\centering
\begin{tabular}{|c|c|l|l|}
\hline
$n$     & Tate form         & polytope model & top over the $-3$-curve \\ \hline
$7$  & $\set{1,1,2,2,4}$ &     M:342 8 N:15 7 H:6,248           &   $\set{pt_1'',pt_2',pt_3',pt_4'}$                      \\ \hline
$9$  & $\set{1,1,2,3,4}$ &    M:339 8 N:16 7 H:7,247            &  $\set{pt_1'',pt_2'',pt_3',pt_4'}$                         \\ \hline
$10$ & $\set{1,1,2,3,5}$ &   M:332 10 N:17 8 H:8,242           & $\set{pt_1'',pt_2'',pt_3',pt_4',pt_5'}$                        \\ \hline
$11$ & $\set{1,1,3,3,5}$ &    M:328 8 N:18 7 H:8,242            &$\set{pt_1'',pt_2'',pt_3',pt_4'',pt_5'}$                         \\ \hline
$12$ & $\set{1,1,3,3,6}$ &   M:318 10 N:19 8 H:9,233             &      $\set{pt_1'',pt_2'',pt_3',pt_4'',pt_5',pt_6'}$                \\ \hline
\end{tabular}
\caption{\footnotesize Polytope tunings of M:348 5 N:12 5 H:5,251
  (generic model over $\F_3$): $\gso(n)$-tunings on the $-3$-curve
  with $n < 13$. These are flat elliptic fibration models, where the
    Hodge numbers can be directly calculated from the anomaly
    cancellation conditions.}
\label{so(n)tunings}
\end{table}

\begin{table}[]
\centering
\begin{tabular}{|c|l|l|}
\hline
$n$ & polytope model            & \{2D component, $(4,6)$ point\}                                                                                                                                        \\ \hline
13  & M:312 8 N:20 7 H:10,232  & \{\{$pt_5'$, $c_3 b_1+c_4 b_3$\}\}                                                                                                                                      \\ \hline
14  & M:299 10 N:21 8 H:11,221 & \{\{$pt_5'$, $c_4 b_1+c_5 b_3$\}\}                                                                                                                                      \\ \hline
15  & M:292 8 N:22 7 H:11,221  & \{\{$pt_5'$, $c_3 b_1+c_4 b_3$\}\}                                                                                                                                      \\ \hline
16  & M:276 10 N:23 8 H:12,206 & \{\{$pt_5'$, $c_4 b_1+c_5 b_3$\}\}                                                                                                                                      \\ \hline
17  & M:267 8 N:24 7 H:13,205  & \{\{$pt_5'$, $c_3 b_1+c_4 b_3$\},\{$pt_8'$, $c_3 b_1+c_4 b_3$\}\}                                                                                                          \\ \hline
18  & M:248 10 N:25 8 H:14,188 & \{\{$pt_5'$, $c_4 b_1+c_5 b_3$\},\{$pt_8'$, $c_4 b_1+c_5 b_3$\}\}                                                                                                          \\ \hline
19  & M:238 8 N:26 7 H:14,188  & \{\{$pt_5'$, $c_3 b_1+c_4 b_3$\},\{$pt_8'$, $c_3 b_1+c_4 b_3$\}\}                                                                                                          \\ \hline
20  & M:216 10 N:27 8 H:15,167 & \{\{$pt_5'$, $c_4 b_1+c_5 b_3$\},\{$pt_8'$, $c_4 b_1+c_5 b_3$\}\}                                                                                                          \\ \hline
21  & M:204 8 N:28 7 H:16,166  & \begin{tabular}[c]{@{}l@{}}\{\{$pt_5'$, $c_3 b_1+c_4 b_3$\},\{$pt_8'$, $c_3 b_1+c_4 b_3$\},\\ \{$pt_{11}'$, $c_3 b_1+c_4 b_3$\}\}\end{tabular}                                \\ \hline
22  & M:179 10 N:29 8 H:17,143 & \begin{tabular}[c]{@{}l@{}}\{\{$pt_5'$, $c_4 b_1+c_5 b_3$\},\{$pt_8'$, $c_4 b_1+c_5 b_3$\},\\ \{$pt_{11}'$, $c_4 b_1+c_5 b_3$\}\}\end{tabular}                                \\ \hline
23  & M:166 8 N:30 7 H:17,143  & \begin{tabular}[c]{@{}l@{}}\{\{$pt_5'$, $c_3 b_1+c_4 b_3$\},\{$pt_8'$, $c_3 b_1+c_4 b_3$\},\\ \{$pt_{11}'$, $c_3 b_1+c_4 b_3$\}\}\end{tabular}                                \\ \hline
24  & M:138 8 N:31 7 H:18,116  & \begin{tabular}[c]{@{}l@{}}\{\{$pt_5'$, $c_3 b_1+c_4 b_3$\},\{$pt_8'$, $c_3 b_1+c_4 b_3$\},\\ \{$pt_{11}'$, $c_3 b_1+c_4 b_3$\}\}\end{tabular}                                \\ \hline
25  & M:123 6 N:32 6 H:19,115  & \begin{tabular}[c]{@{}l@{}}\{\{$pt_5'$, $c_2 b_1+c_3 b_3$\},\{$pt_8'$, $c_2 b_1+c_3 b_3$\},\\ \{$pt_{11}'$, $c_2 b_1+c_3 b_3$\},\{$pt_{14}'$, $c_2 b_1+c_3 b_3$\}\}\end{tabular} \\ \hline\end{tabular}
\caption{\footnotesize Polytope tunings of M:348 5 N:12 5 H:5,251
  (generic model over $\F_3$): $\gso(n)$-tunings on the $-3$-curve
  with $13\leq n < 26$. These are non-flat elliptic fibration
    models. The last column gives the $(4,6)$ points and the
    corresponding 2D toric fiber components contained in the
    hypersurface CY  (see Table \ref{sotop} for $pt$ in tops). The
    Hodge numbers can be calculated from the associated flat elliptic
    fibration model over the non-toric base where the $(4,6)$-point
    are blown up.}
\label{so(n)tunings2}
\end{table}

The remaining four Hodge pairs corresponding to standard
$\P^{2,3,1}$-fibered polytopes at large $h^{1,1}$ that were missed in
our Tate tuning set have a combination of two novel features: they
have apparent $\gso(n\geq 13)$ tunings on $-3$ curves, and also have
extra sections associated with a nontrivial Mordell-Weil rank and
corresponding U(1) factors in the F-theory physics.  For clarity, we
delegate a complete example of one of the ``extra'' models of this
type to Appendix \ref{26332cal}, and focus in the rest of this section
on the issue of $\gso(n\geq13)$-tunings on $-3$-curves in the context
of simpler models with relatively small $h^{1,1}$ that do not also
involve the U(1) issue.

As mentioned above, $\gso$($n$)-tunings on $-3$-curves give rise to
$(4,6)$ singularities and two-dimensional resolved fibers when $n\geq
13$. While the anomaly conditions impose an upper bound of $n = 12$
for $\gso(n)$-tunings over $-3$-curves, there is no bound on
$-4$-curves from anomaly conditions \cite{Johnson:2016qar}. Therefore,
in these cases the corresponding flat elliptic fibration models can be
obtained by resolving the $-3$-curves to $-4$-curves that support
$\gso(n\geq13)$ without suffering from (4,6) points.

We start with a generic polytope model over the Hirzebruch
surface $\F_3$, M:348 5 N:12 5 H:5,251, and perform successive
 tunings of
$\gso(n)$ on the $-3$-curve.  For $7\leq n\leq 12$, all these polytope
tunings, except $\gso(8)$\footnote{We do not expect tuned $\gso(8)$ in
  reflexive polytope models; see \S\ref{so8} for discussion.},
give a model in the KS database as expected, and the Hodge numbers of
these polytope models agree with the Hodge numbers calculated from
anomalies. We list these polytope models in Table
\ref{so(n)tunings}.
Note that the tuning from $\gso(10)$ to $\gso(11)$ is a
rank-preserving tuning (see Table~\ref{t:rpt}), so the Hodge numbers
for these cases are identical.

Consider now the $\gso(13)$ polytope tuning on the $-3$-curve.  This
also gives a reflexive polytope, M:312 8 N:20 7 H:10,232, which is
still of the
standard $\P^{2,3,1}$-fibered form over the $\F_3$ base. But this is
a non-flat elliptic fibration.  In fact, we know immediately from the
Hodge numbers that there is some additional subtlety in this tuning.  Naively, $\gso(12)$ to $\gso(13)$ would be a
rank-preserving tuning, and, as for the $\gso(10)$ to $\gso(11)$ tuning,
in the absence of other issues these should have the same Hodge
numbers, but they clearly do not.
An explicit
analysis shows that over a generic point on the $-3$-curve, the
hypersurface equation intersects with seven components associated with
the seven lattice points
$\set{pt_1',pt_1'',pt_2'',pt_3',pt_4'',pt_6',pt_6''}$ in the
$\gso(13)$ top in a locus containing $\P^1$'s which form the $\gso(13)$ extended
Dynkin diagram, and there is a $(4,6)$ point on the $-3$-curve, over
which the  fiber contains the whole irreducible component
associated with the lattice point $pt_5'$ in the top.

Again, we can calculate the Hodge numbers by considering the
corresponding flat elliptic fibration model over the base where the
$(4,6)$ point on the $-3$-curve is blown up.  This blow-up produces an
exceptional $-1$-curve that intersects the proper transform
$-4$-curve, and which can support any $\gso(n)$ tunings without
producing $(4,6)$ points. Therefore, $\Delta h^{1,1}=\Delta T+\Delta
r=1+(6-2)=5$ and $\Delta h^{2,1}=\Delta V -29\Delta T -\Delta H_c=
(78-8)-29-5\times \mathbf{(13-1)}=-19$,\footnote{Note that although
  the representations of $\gso(13)$ tuning on an $-4$-curve are
  $5\times\mathbf{13}$, the components that are charged under the Cartan
  are $5\times(\mathbf{13-1})$ (the Cartan subgroup of $SO(2N+1)$ is
  the same as $SO(2N)$). As $\gso(13)$ is a rank-preserving tuning of
  $\gso(12)$, we can also do the calculation as if it were a $\gso(12)$
  tuning, in which case $\Delta h^{1,2}=\Delta V -29\Delta T -\Delta
  H_c=(66-8)-29-4\times \mathbf{12}=-19$. The two Hodge number shifts
  are the same.} which agrees with $\set{10,232}-\set{5,251}$.

 %be $\F_3$ blown up at a non-toric point on the $-3$ curve to become a $-4$-curve carrying $\gso(13)$ gauge algebra. Let's check the consistency of Hodge numbers from anomaly calculation: $\gso(13)$ is rank preserving tuning to $\gso(12)$ which is also an allowed tuning, so it is easier to treat the calculation as if it were an $\gso(12)$. Then we have $\Delta T=1$ from the non-toric blowup, $\Delta r=6-2$, $\Delta V=66-8$, and  $\Delta H_c=48$ charged matters from $\gso$(13) tuning on $-4$ curve, and therefore $\Delta h^{1,1}=5$ and $\Delta h^{2,1}=-19$ as desired. Again, we can use Zariski decomposition to analyze tunings on a configuration of a $-4$-curve intersecting a $-1$-curve.
 
In the flat elliptic fibration model over the resolved base, as we keep
increasing the $\gso(n)$ tuning, an additional gauge factor $\gsp(m)$
is forced to arise on
the exceptional $-1$-curve starting at $n=17$: a
simple local analysis
shows that tuning $\gso(n)$ on a $-4$-curve forces $\gsp(\lceil
n/4\rceil-4)$ on an intersecting $-1$-curve.  The forced
$\gsp(m)$ is not apparent in the $(f,g,\Delta)$ of the polytope model,
which is the non-flat model over the original $\F_3$ base  where the
$-1$-curve does not exist.  But we have
to carefully consider this forced gauge algebra on the exceptional
$-1$-curve in computing the Hodge numbers from the anomaly equations
(\ref{h11}) and (\ref{h21}).  For example, tuning $\gso(22)$ on the
$-3$-curve gives rise to the model M:179 10 N:29 8 H:17,143. The
corresponding flat fibration model has a $-4$-curve intersecting an
exceptional $-1$-curve replacing the $-3$-curve, and the $\gso(22)$ is
tuned on the $-4$-curve, which forces an $\gsp(2)$ on the
$-1$-curve. Therefore, the shifts of the Hodge numbers are $\Delta
h^{1,1}=\Delta T+\Delta r=1+((11-2)+2)=12$ and $\Delta h^{2,1}=\Delta
V -29\Delta T -\Delta H_c=
((231-8)+10)-29-(14\times\textbf{22}+12\times\textbf{4}-1/2\times22\times4\text{
  (shared)})=-108$, which agree with the Hodge numbers from the
polytope. The Hodge numbers of the polytope models from the successive
tunings can be calculated this way up to $\gso(26)$, at which point all
monomials in $a_6$ are tuned off, and a U(1) global factor
comes into play. See Appendix \ref{26332cal} for an explicit analysis
of one of the models associated with the
missing Hodge pairs where such a U(1)
becomes relevant. We list the non-flat polytope models
of tuning $\gso(n)$, $13\leq n<26$, over the $-3$ curve of $\F_3$
in  Table
  \ref{so(n)tunings2}.

%(notice there are $1/2\times22\times4$ charged matters shared between the $14\times\textbf{22}$ of $\gso(22)$ and $12\times\textbf{4}$ of $\gsp(2)$).
%The correct interpretation of the data would be a smooth CY over a semi-toric base drawn in Fig.~\ref{} with guage group $\gso(22)$ on the $-4$ curve and $\gsp(2)$ on the $-1$ curve. 
 %We wouldn't get these correct numbers if fail to consider the forced $\gsp(2)$.

%In the ranges we consider,  $\ge8$ tunings appear only on  ``-8'' ($-12$ intersecting four $-1$'s), but we can generalize this to $\ge8$ tunings on other curves with self-intersection numbers greater than $-12$ and find a corresponding polytope construction which has dealt with the non-toric resolution for us in the KS database: ``$-11$'' ($-12$ intersecting one $-1$), ``$-10$'' ($-12$ intersecting two $-1$'s),..., ($-7$ intersecting five $-1$'s), etc. Similarly we can also consider $\gso(\text{some big } n)$ tunings on curves with self-intersection numbers greater than $-4$, which becomes the tunings on a $-4$ intersecting appropriate numbers of non-toric exceptional $-1$'s, though only models of $\gso(n>12)$ on ``$-3$'' ($-4$ intersecting a $-1$) occur in the ranges we consider. Since there is additional $U(1)$ factor to consider,  we discuss these models in the next section.

%\subsubsection{Tunings with a forced global $\gu(1)$}
%\label{u1}

\subsection{Weierstrass  models from non-standard $\P^{2,3,1}$-fibered polytopes}
\label{non}

For the remaining eight Hodge pairs with large $h^{1,1}$ in the KS
database that were missed by our Tate construction (see table
\ref{restt}), the CY hypersurface equations (\ref{p}) with suitable
homogeneous coordinates cannot be in Tate form, although the $\nabla$
polytopes are still $\P^{2,3,1}$ fibered.  The failure to be in the
Tate form arises from the feature that there are lattice points in
$\Delta$ that give rise to non-trivial base dependence in the
coefficients of the monomials $x^3$ or $y^2$; i.e., these should be
sections of non-trivial line bundles over the base.

These $\nabla$ polytopes do not the form of standard $\P^{2,3,1}$-fibered
polytopes that we have defined in \S\ref{sp231}, although they still
have
$\P^{2,3,1}$ fibers.  We refer to  such polytopes
as \emph{non-standard} $\P^{2,3,1}$-fibered polytopes. In fact, the
feature of having base-dependent terms in $x^3$ or $y^2$ is equivalent
to being a non-standard $\P^{2,3,1}$-fibered polytope.  Geometrically
this feature corresponds to the condition that there is only a single
lattice point in $\ds$ that projects to each of the vertices
associated with these monomials.  We prove this equivalence as follows: Without loss of
generality, we choose a coordinate system such that the three vertices
of the $\P^{2,3,1}$ subpolytope $\dd_2$ are as given in equation
(\ref{fiberrays}), and such that the projection matrix to the base is
$\pi=\set{\set{1,0,0,0},\set{0,1,0,0},\set{0,0,0,0},\set{0,0,0,0}}$. Therefore, the set of the
vertices of the dual subpolytope $\ds_2$ is
$\set{(-2,1),(1,-1),(1,1)}$, and the lattice points in
$\ds$ are all in one of the forms in the set $\{(\_, \_, 1, -1), (\_, \_,
-2, 1), (\_,\_,0,0), (\_,\_,-1,1), (\_,\_,1,0), (\_,\_,0,1),$
$(\_,\_,1,1)\}$. Let us first show the forward direction: 
%Suppose we
%have a term in $x^3$ that is also base dependent.  If
%$m_{x^3}\equiv(0,0,-2,1)$ is the only lattice point in $\ds$ that
%contributes to a term with a $x^3$ factor, there must be a ray in the
%base $v^{(B)}$ that do not have a preimage of the form $v\equiv
%((v^{(B)})_{1,2},(pt_1)_{3,4})$, where $pt_{i=1,...,7}$ are defined in
%equation (\ref{pts}), as otherwise the exponent of the associated base
%coordinate $m_{x^3}\cdot v +1$ would be zero.  If there are more than
%one base dependent term that also have an $x^3$ factor, the associated
%lattice points must be of the form $(\_, \_, -2, 1)$. Let
%$m'_{x^3}\equiv(m_1,m_2,-2,1)$ be a second such lattice point, where
%$m_1, m_2$ are not simultaneously zero. Assume all the base rays
%$v^{(B)}_i$s have a preimage $v_i\equiv
%((v^{(B)}_i)_{1,2},(pt_1)_{3,4})$. The reflexivity condition gives
%$m'_{x^3}\cdot v_i\geq -1$; i.e., $v^{(B)}_i\cdot(m_1,m_2)\geq 0$ for
%all $i$, so all base rays lie within a half plane, which contradicts
%to the base being a compact toric variety.
We already showed  in \S\ref{sp231}
that the standard  $\P^{2,3,1}$-fibered polytope
construction in the coordinates given in (\ref{23}) gives a dual polytope $\ds$ that
contains at most the single points corresponding to ${\cal O} (0)$
at the vertices $(-2, 1), (1, -1)$ associated with the $y^2, x^3$
terms
  (assuming that the base is
compact),
and both of these points must be present for the polytope $\ds$ to contain
the origin as an interior point.  Thus, any
standard
$\P^{2,3,1}$-fibered polytope can be put in a coordinate system where it
has only the points $(0, 0, -2, 1)$ and $(0, 0, 1, -1)$ that project
to $(-2, 1)$ and $(1, -1)$ in $\ds_2$.
We can prove the backward direction as follows:
Assume there is only a single lattice point in $\ds$ taking each of
the forms  $(\_, \_, -2, 1)$ and $(\_,\_, 1, -1)$.
There is always a linear transformation that leaves the last two
coordinates fixed that moves these to the points
 $(0, 0, -2, 1)$ and $(0, 0, 1, -1)$; this linear transformation will
also leave the form of the fiber fixed as  (\ref{fiberrays}).
The presence of these two points in $\ds$ shows that every lattice
point
$(v_{i, 1}^{(B)}, v_{i, 2}^{(B)}, \xi, \eta)$ has coordinates $ \xi,
\eta$ that satisfy $\eta \leq \xi + 1, \eta \geq 2 \xi -1$.
For each ray in the base, however, the presence of any such lattice
point imposes conditions on the points in $\ds$ over each of
the points other than $(-2, 1)$ and $(1, -1)$ that are at least as
strong as those imposed by the ray $(v_{i, 1}^{(B)}, v_{i, 2}^{(B)},
2, 3)$; the conditions over these two points can be weaker, but as long as there is only the one point $(0,0,-2,1)$, $(0,0,1,-1)$ over these two points in the dual fiber, the ray $(v_{i, 1}^{(B)}, v_{i, 2}^{(B)}, 2,3)$ will be included in the polytope. Thus, for each ray in the base $(v_{i, 1}^{(B)}, v_{i, 2}^{(B)},
2, 3)\in\dd$ in this coordinate system.  This proves that the presence of a single lattice point of
each of the forms  $(\_, \_, -2, 1)$ and $(\_,\_, 1, -1)$ implies that
the polytope $\dd$ has the form of a standard
$\P^{2,3,1}$-fibered polytope.
% has a base ray $v^{(B)}$ that has a
%preimage of the form $((v^{(B)})_{1,2},z, w)$, where $(z, w) \neq (2,
%3)$.
%For any such ray, the exponents of the associated homogeneous
%coordinate 
%in the $y^2$ and $x^3$ coefficients associated with the lattice points
%$(0,0,-2,1)$ or $(0,0,1,-1)$ in $\ds$ are
%$-2z + y + 1$ and $z-y + 1$ respectively.
%\footnote{A ray has to be of the
%  form $v_1$ to have a power zero to the associated homogeneous
%  coordinate in both the monomials \set{0, 0, -2, 1} and \set{0, 0, 1,
%    -1}, which have factors $x^3$ and $y^2$, respectively:
%\begin{eqnarray}
%&&\set{0, 0, -2, 1}\cdot v_{i=1,2,3,4,5,6,7}+1=0, 1, 0, 2, 1, 3, 0,\\
%&&\set{0, 0, 1, -1}\cdot v_{i=1,2,3,4,5,6,7}+1=0, 0, 1, 0, 1, 0, 2.
%\end{eqnarray}} 
%Therefore, one of these two lattice points gives rise
%to a  monomial in either $x^3$ or $y^2$ with non-trivial base
%dependence, since $-2z +  w = -1, z- w = -1$ have no solutions other
%than $z = 2, w = 3$.

%We try to understand the fibration of non-standard $\P^{2,3,1}$-fibered polytope models by studying the Weierstrass models of the associated Jacobian fibration models so that we can use the F-theory technique to analyze the geometry. 

We would like to have a Weierstrass description of the non-standard
$\P^{2,3,1}$-fibered polytopes so that we can use the methodology of
F-theory to understand and analyze the geometry.  To this end, we
treat the $\P^{2,3,1}$ fiber as a twice blown up $\P^{1,1,2}$ fiber,
as depicted in figure \ref{fig:F10asF4}; following the procedure in appendix
A of \cite{Braun:2011ux} to obtain the Weierstrass model of the
associated Jacobian fibration model of a $\P^{1,1,2}$-fibered
polytope, we can obtain similarly that of the blownup
$\P^{1,1,2}$-fibered polytope.
Note that because  even non-standard $\P^{2,3,1}$-fibered polytopes
give elliptic Calabi-Yau hypersurfaces that  have a global section,
the Jacobian fibration should have the same geometry as the original
Calabi-Yau hypersurface; this would not be true for example if the
original elliptic fibration had no section \cite{Morrison:2014era}.
Explicitly in
coordinates, instead of treating the elliptic fiber as being embedded
in the $\P^{2,3,1}$ ambient fiber with $v_x=(0, 0, -1, 0), v_y=(0, 0,
0, -1)$, $v_z=(0, 0, 2, 3)$, we treat the $\P^{2,3,1}$ fiber as a
blownup $\P^{1,1,2}$, and embed the elliptic fiber in this blownup
$\P^{1,1,2}$ ambient fiber with $v_x=(0, 0, -1, 0), v_y=(0, 0, 0,
-1)$, $v_z=(0, 0, 1, 2)$.  The blowup rays of $\P^{1,1,2}$ reflect
the fact
that two of the nine sections of a $\P^{1,1,2}$-fibered polytope model
are completely tuned off (see figure \ref{fig:M10asM4}) --- the hypersurface
equation of a non-standard $\P^{2,3,1}$-fibered polytope is a
specialization of that of a generic $\P^{1,1,2}$-fibered polytope, and
the blowups of the $\P^{1,1,2}$ fiber resolve the singularities
of the tunings.

\begin{figure}
\centering
\begin{subfigure}{.45\textwidth}
  \centering
  \includegraphics[width=6.5cm]{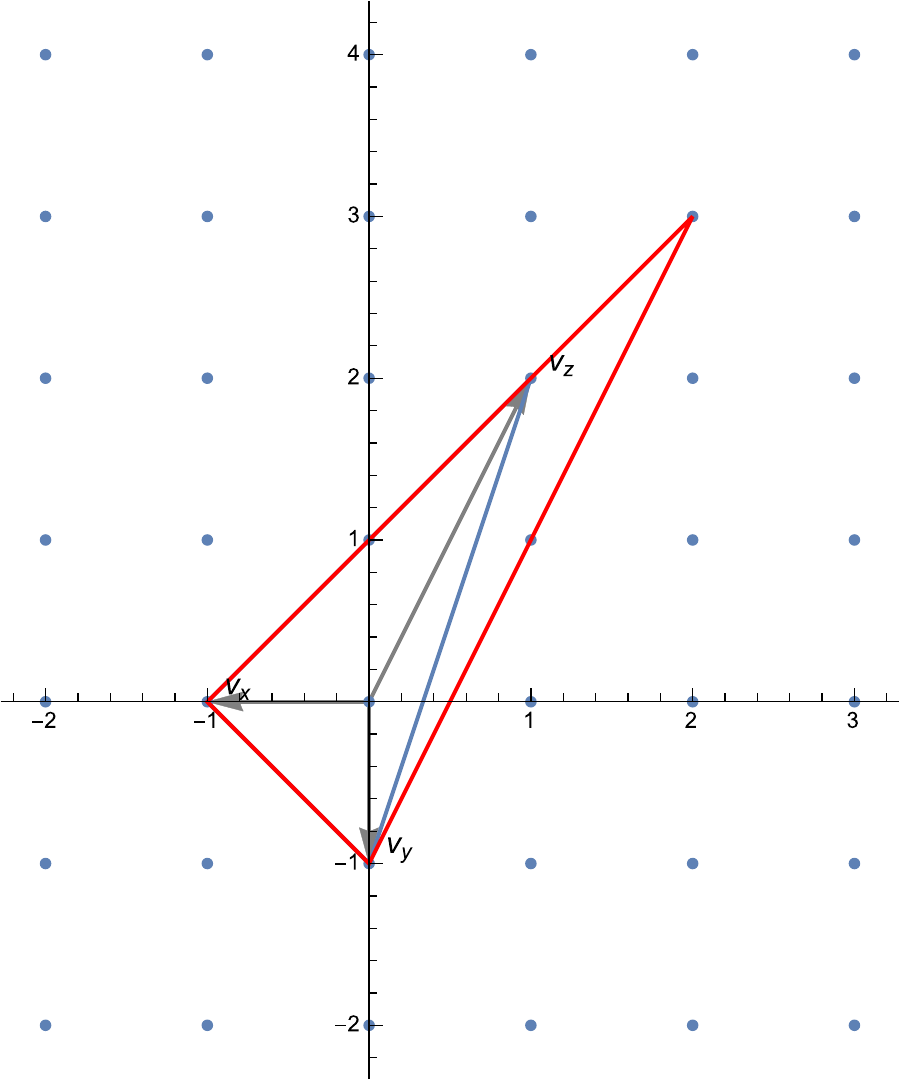}
  \caption{\footnotesize
Comparing the polytopes for $\P^{2,3,1}$
(red) and $\P^{1,1, 2}$ (blue). 
The red triangle
$\dd_{231}$ comes from blowing up the (fiber) fan of $\dd_{112}$
twice (cf. indicated rays in figure
    \ref{fig:sub1} for standard $\P^{2,3,1}$-fibered polytopes.)}
  \label{fig:F10asF4}
\end{subfigure}%
\hspace*{0.1in}
\begin{subfigure}{.45\textwidth}
  \centering
  \includegraphics[width=7.5cm]{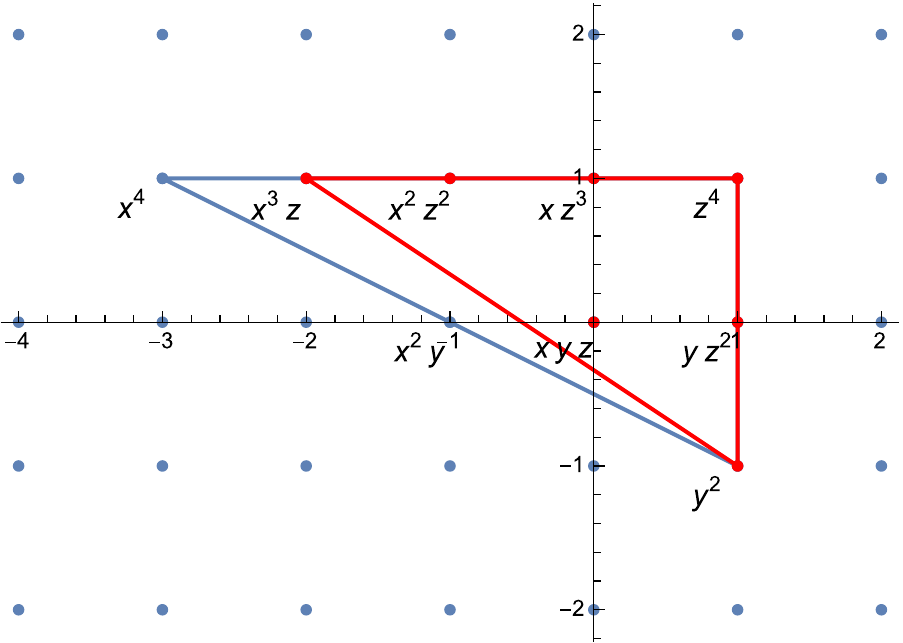}
  \caption{\footnotesize The blue triangle  $\ds_{112}$ is the dual of
    $\dd_{112}$ and the red triangle  $\ds_{231}$ is the dual
    of  $\dd_{231}$. The monomials in different sections
    are categorized by projection to the different lattice points in
    $\ds_{112}$, labeled in terms of the homogeneous coordinates
    $x, y, z$ in the fiber. The equations describing the hypersurface
    Calabi-Yau for a
non-standard $\P^{2,3,1}$ polytope
can be characterized as tunings for a $\P^{1,1,2}$-fibered polytope,
 in which there
are no nonzero monomials in
the sections labeled $x^{4}$ and $x^{2}y$. This interpretation
    allows the possibility of having sections in $y^2$ or $x^{3}$.}
  \label{fig:M10asM4}
\end{subfigure}
 \caption{The reflexive polytope pairs for the $\P^{1,1,2}$ ambient toric fiber (in blue) and the $\P^{2,3,1} = $  Bl$_2\P^{1,1,2}$ ambient toric fiber (in red).}
%  \caption{The reflexive polytope pairs for the $\P^{1,1,2}$ ambient toric fiber (in blue) and the Bl$_{[0,0,1]}\P^{1,1,2}$ ambient toric fiber (in red).}
\label{p112}
\end{figure}

The Weierstrass models obtained in this way from non-standard
$\P^{2,3,1}$-fibered polytopes have the novel feature that they can
have gauge groups tuned over non-toric curves in the base.  Moreover,
unlike the toric curves, which are always genus zero curves
(isomorphic to $\P^1$), non-toric curves can be of higher genus, and
this class of global Weierstrass models gives examples of tunings of
gauge groups over higher genus curves in the bases.
%Note that we do not analyze the CYs associated with the non-standard
%$\P^{2,3,1}$-fibered polytope directly, but we analyze the
%Weierstrass model of the associated Jacobian fibration model, which
%has the same discriminant locus as the original CY. 
As a check on this picture, we can verify that the Hodge numbers of
these Weierstrass models calculated from anomaly cancellation match
with those of the polytope data.

We give some examples of Weierstrass models from non-standard
$\P^{2,3,1}$-fibered polytopes in the following subsections. In \S
\ref{warmup}, we give a simple example that illustrates the non-toric curve
enhancement feature. In \S \ref{the8}, we analyze the eight remaining
polytope
data with large $h^{1,1}$ from the KS database
that were missing in the Tate-tuned construction. 
We also give some further examples of interesting geometries from the KS
database at smaller Hodge numbers to illustrate the unusual nature of
the non-standard $\P^{2,3,1}$-fibered polytope construction.
In \S \ref{g=1}, we give a model
with an $\gsu(2)$ tuning on a non-toric curve of genus one in the
base.

\subsubsection{A warmup example}
\label{warmup}
As an illustration of the two different types of $\P^{2,3,1}$-fibered
polytopes, we contrast the two polytopes in the KS database associated
with Calabi-Yau threefolds having Hodge numbers $\set{8,250}$: M:346 8
N:16 7 H:8,250 and M:345 8 N:17 7 H:8,250.

The second $\dd_{\text{2nd}}$ polytope is a standard
$\P^{2,3,1}$-fibered polytope, with vertices (in the standard
coordinates)
$\set{(0,0,-1,0),(-1,-4,2,3),(0,-2,1,2),(0,1,2,3),(1,0,1,2)$,
  $(1,0,2,3)$, $(0, 0,$ $0, -1)}$. This is a Tate-tuned model over the
base $\F_4$, with $\gso(9)\oplus\gsp(1)$ enhanced on the $-4$-curve
and the $0$-curve $\set{b_2=0}$. The base rays are $\set{(0,1),(1,0),
  (0,-1), (-1,-4)}$, and in particular, $\set{(0,1,2,3),(1,0,2,3),$
  $(0,-1,2,3), (-1,-4,2,3)}$ are lattice points. This polytope can be
obtained by tuning the $\ds_{\gso(8)}$ polytope of either one of the
$\gso(8)$ KS models (a generic elliptic fibration over $\F_4$) by
requiring the vanishing orders with respect to the coordinate
$b_2\leftrightsquigarrow (1,0,2,3)$ to be $\set{0, 0, 1, 1, 2}$ and
those for the coordinate $b_3\leftrightsquigarrow (0,-1,2,3)$ to be
$\set{1, 1, 2, 3, 4}$.  The $\dd_\text{2nd}$ polytope, which is then
the dual of the reduction of the $\ds_{\gso(8)}$ polytope, has an
$\gso(9)$ top over the base ray $(0,-1)$ and an $\gsp(1)$ top over
$(1,0)$.

%Tate tuning model over the $\F_4$ base with $\gso(9)\oplus\gsp(1)$ enhanced on the $-4$- and one of the  $0$-curve: in some coordinate, the base rays are $\{(0,1), (1,0), (0,-1), (-1,-4)\}$, the polytope contains  $\{(0, 1, 2,3), (1, 0,2,3), (0,-1,2,3), (-1,-4,2,3),\}$, 

The first $\dd_\text{1st}$ polytope is, on the other hand, of a
non-standard $\P^{2,3,1}$-fibered form. The data of this polytope 
can be obtained by removing the vertex $(1, 0, 2, 3)$ from
$\dd_\text{2nd}$ or equivalently by adding the lattice point $(-1, 0,
-2, 1)$ to $\ds_\text{2nd}$, which becomes a vertex of
$\ds_\text{1st}$. The one lattice point reduction of $\dd_\text{2nd}$
corresponds to the one lattice point enhancement of $\ds_\text{2nd}.$
Let us now show explicitly that $\dd_\text{1st}$ is a non-standard
$\P^{2,3,1}$-fibered polytope and check that it satisfies each of the
two equivalent conditions (i.e., the absence of an appropriate
preimage of the base in $\dd$ and the condition that $\ds$ has lattice
points associated with monomials in $x^3$ or $y^2$ that have base
dependence): The base rays of $\dd_\text{1st}$ are the same as those
of $\dd_\text{2nd}$, but the ray $(1,0)$ lacks the preimage $(1, 0, 2,
3)$ that we have removed; instead, the base ray $(1,0)$ comes from the
projection of the 4D ray $(1, 0, 1, 2)$; nonetheless $\dd_\text{1st}$
still has $\P^{2,3,1}$ as a
subpolytope, and therefore $\dd_\text{1st}$ is a non-standard
$\P^{2,3,1}$-fibered polytope. The equivalent condition for a
non-standard $\P^{2,3,1}$-fibered polytope is also satisfied from the
$\ds$ point of view: let us associate base coordinates
$\set{b_1,b_2,b_3,b_4}$ to the set of 4D rays $\set{ (0, 1, 2, 3),(1,
  0, 1, 2), (0, -1, 2, 3), (-1, -4, 2, 3)}$, and calculate the
set of monomials. The two lattice points $(-1, 0, -2, 1),(0, 0, -2,
1)$ give monomials of the form $x^3$ with base-dependent coefficients,
$b_4 x^3$ and $b_2 x^3$ respectively.

%The 1st version: Alternatively, associate base coordinates $\set{b_1,b_2,b_3,b_4}$ to the set of 4D rays $\set{ (0, 1, 2, 3),(1, 0, 1, 2), (0, -1, 2, 3), (-1, -4, 2, 3)}$, and calculate the set of monomials with the base coordinates and the fiber coordinates $\set{x,y}$ associated to $\set{(0,0,-1,0),(0,0,0,-1)}$. There is the $y^2$ monomial term as in the standard $\P^{2,3,1}$ fiber case, but the two monomial terms with the $x^3$ factor obtain also factors of base coordinates: $\set{b_4  x^3 , b_2  x^3}\leftrightsquigarrow \set{(-1, 0, -2, 1),(0, 0, -2, 1)}$, where $(-1, 0, -2, 1)$ was the new lattice point that we added to $\ds_\text{2nd}$.

%The 2nd version: Alternatively, we see that adding the lattice point $(-1, 0, -2, 1)$ to $\ds$,  gives a second monomial of the form $x^3$ with an independent base-dependent coefficient, so the equivalent condition for a non-standard $\P^{2,3,1}$-fibered polytope is also satisfied.

Although we do not have a Weierstrass model from a Tate form for this polytope,
we instead have a Weierstrass form for the hypersurface in
the $\text{Bl}_2 \P^{1,1,2}$-fibered polytope (where we have
substituted some
generic $\Z$ values in the complex structure moduli):
\begin{eqnarray}
\nonumber
f&=&1/48 b_3^2 (-1009274573279509056 + 34622237106205930350000 b_3\\\nonumber
&-&    274589065851262777907525390625 b_3^2  \cdots-528582381600 b_3^6 b_4^{22}\\ \nonumber
 &-& 22258660320 b_3^6 b_4^{23} + 388841808 b_3^6 b_4^{24}),\\ \nonumber
g&=&-(1/864) b_3^3 (344205633835899813888000  \\ \nonumber
&+&       1926547706542277636888364004147200 b_3  
  \cdots+6291082311776640 b_3^9 b_4^{35} \\ \nonumber
 &+&  27125536688271 b_3^9 b_4^{36}),\\\nonumber
\Delta&=&19683/2 b_3^7 (35 b_2 + 24 b_4)^2 (109370724968448 b_1^17 b_2^2 \\\nonumber
&+&    588208065199776 b_1^{16} b_2^6 b_3 +    1344055426083360 b_1^{15} b_2^{10} b_3^2 \\\nonumber
&+&\cdots+681083735457852 b_2 b_3^{17} b_4^{69} + 217077176379771 b_3^{17} b_4^{70}).
\end{eqnarray}
According to this analysis, there is an $\gso(9)$ enhancement on the
$-4$-curve ($b_3 = 0$) and an $\gsp(1)$ enhancement on the non-toric $0$-curve
$\set{35 b_2 + 24 b_4=0}$.
Note that
this non-toric curve is a (rational) $0$-curve
  because it is in the same class as the two toric 0-curves.
The curve supporting the $\gsp(1)$ algebra
intersects both the $-4$- and the $4$-curve at one point. This is
essentially the same configuration as the second model, so the Hodge
numbers from an  anomaly calculation also give the same
result,
$\set{8,250}$, in both cases.
While in this case, the non-toric curve supporting the $\gsp(1)$ can
be trivially transformed into a toric curve by a simple linear change
of variables, this is not the case in the more complicated examples
that we consider in the later subsections.

\subsubsection{The eight remaining
missing cases at large $h^{1,1}$}
\label{the8}

Now let us come back to the polytopes of the eight Hodge pairs in the
large $h^{1,1}$ region that we did not obtain through Tate tunings
and that have non-standard $\P^{2,3,1}$ fibration structure.
We go through one example in detail; the others have similar structure.
\vspace*{0.1in}

As a specific example,
we consider the polytope M:65 8 N:357 8 H:261,45. The vertex set of $\ds$ is
\begin{eqnarray}
\nonumber
 \set{(-3, -3, 1, 1), (0, 0, -2, 1), (1, -7, 1, 1), (-3, 1, 1, 1), (-1, 
  1, -1, 1), (0, 1, 1, 1),\\ (-1, 1, 1, -1), (0, 0, 1, -1)},
  \end{eqnarray}
where both the lattice points in the second line contribute to a $y^2$
term but with base dependence. Performing the projection we find that
 $\dd$ is a non-standard
$\P^{2,3,1}$-fibered polytope over the base
\begin{equation}
\{-12//-11//-12//-12//-12//-12//-12//-12//-9,-1,-2,-2,-1,0\}.
\label{expbase}
\end{equation}
There are  in total 102 base rays, and all rays  but
$v^{(B)}_{i=98,99,100,101}$ have a preimage of the form $(\_,\_,2,3)$. 
%The generic elliptic fibration model over this base is given by a standard $\P^{2,3,1}$-fibered polytope M:110 7 N:356 7 H:257,77.

The generic Weierstrass model over this base has the Hodge numbers
$\set{257,77}$, so the tunings must be such that the shifts are
$\set{4,-32}$.  We analyze the Weierstrass model of the non-standard
$\P^{2,3,1}$ polytope; as in the preceding example we
treat $\dd$ as a Bl${}_2$ $\P^{1,1,2}$-fibered
polytope (in particular, the fiber coordinates are associated to
$\set{v_x, v_y, v_z}=\set{(0,0,-1,0),(0,0,0,-1),$ $(0,0,1,2)}$), and find
the associated tuned Weierstrass model. The resulting computation of $\set{f,g,\Delta}$
shows that
\begin{itemize}
\item Over the toric curve $D_{100}\equiv\{b_{100}=0\}$ the vanishing
  order is enhanced to $\{0,0,2\}$, which corresponds to an $\gsu(2)$
  gauge symmetry on the $-2$-curve. 
\item Over the non-toric curve $D_\text{non-toric}\equiv\set{b_\text{non-toric}=0}$, where \begin{eqnarray}
\nonumber
b_\text{non-toric}&=&c_7 b_{100} b_{101} b_{98} b_{99}\\\nonumber
&&+c_6 b_{1} b_{10}^{22} b_{11}^{29} b_{12}^{36} b_{13}^7 b_{14}^{41} b_{15}^{34} b_{16}^{27} b_{17}^{20} b_{18}^{33} b_{19}^{13} b_{2}^{12} b_{20}^{32} b_{21}^{19} b_{22}^{25} b_{23}^{31} b_{24}^{37} b_{25}^6 b_{26}^{35} b_{27}^{29} b_{28}^{23} b_{29}^{17} b_{3}^{11} b_{30}^{28} b_{31}^{11}\\\nonumber&& b_{32}^{27} b_{33}^{16} b_{34}^{21}
 b_{35}^{26} b_{36}^{31} b_{37}^5 b_{38}^{29} b_{39}^{24} b_{4}^{10} b_{40}^{19} b_{41}^{14} b_{42}^{23} b_{43}^9 b_{44}^{22} b_{45}^{13} b_{46}^{17} b_{47}^{21} b_{48}^{25} b_{49}^4 b_{5}^9 b_{50}^{23} b_{51}^{19} b_{52}^{15} b_{53}^{11} b_{54}^{18} \\\nonumber
&&b_{55}^7 b_{56}^{17} b_{57}^{10} b_{58}^{13} b_{59}^{16} b_{6}^{17} b_{60}^{19} b_{61}^3 b_{62}^{17} b_{63}^{14} b_{64}^{11} b_{65}^8 b_{66}^{13} b_{67}^5 b_{68}^{12} b_{69}^7 b_{7}^8 b_{70}^9 b_{71}^{11} b_{72}^{13} b_{73}^2 b_{74}^{11} b_{75}^9 b_{76}^7 b_{77}^5\\
&& b_{78}^8 b_{79}^3 b_{8}^{23} b_{80}^7 b_{81}^4 b_{82}^5 b_{83}^6 b_{84}^7 b_{85} b_{86}^5 b_{87}^4 b_{88}^3 b_{89}^2 b_{9}^{15} b_{90}^3 b_{91} b_{92}^2 b_{93} b_{94} b_{95} b_{96},
\label{ntc}
\end{eqnarray}
the vanishing order is enhanced to $\{0,0,3\}$, which corresponds to an
$\gsu(3)$ gauge symmetry on the non-toric curve. 
In this expression,  $c_i$ are constant coefficients, while $b_k$  are
the variables associated with toric divisors $D_k$.
Note that the
non-toric curve intersects the two toric curves $\{b_{102}=0\}$ and
$\{b_{97}=0\}$ ($b_{102}$ and $b_{97 }$ are the only coordinates that
do not appear in equation (\ref{ntc}), and there are no intersections
between the divisors associated with the variables in the first and
second terms).
As in the preceding example, this non-toric curve is a 0-curve,
and is linearly equivalent to the combination of curves $D_{98} +
D_{99} + D_{100} + D_{101}$, as can be seen from the first term in
(\ref{ntc}).
The complicated combination of powers in the second term in
(\ref{ntc}) arise from the structure of the toric rays and the
sequence of blow-ups needed to build those rays from a fiber of a
minimal model Hirzebruch base.

\item 
Over the curve $D_{97}\equiv\set{b_{97}=0}$ (a $-9$-curve), there is a
two-dimensional resolved fiber; due, however, to the enhancement over
the non-toric curve intersecting the $-9$-curve, there are some
differences between the fiber structure over this $-9$-curve and the
one in an isolated $-9$-curve such as we discussed in
\S\ref{multiplicity} (see also Appendix \ref{2dfiber}): The top is the
same as that in equation (\ref{top}), so the 9 components that are the
boundary of the 3-dimensional face intersect with the CY over a
generic point in the $-9$-curve in a locus of $\P^1$s that compose the
extended $E_8$ Dynkin diagram, just as in equation
(\ref{boundary}). However, as opposed to having three distinct
$(4,6)$-points, as occur in the isolated $-9$-curve, there is only one
$(4,6)$ point.  Over this point the CY intersects the four irreducible
components interior to the 3-face (while in the previous case, there
is only one irreducible component that intersects the CY)
\begin{equation}
S=\set{(-3, -3, 1, 2), (-2, -2, 1, 2), (-1, -1, 1, 2), (-1, -1, 0, 1)}.
\label{4in}
\end{equation}
 Explicitly, 
\begin{eqnarray}
\nonumber
p\rvert{_I}&=&c_7 b_{100} b_{101} b_{98} b_{99}\\\nonumber
&+&c_6 b_{1} b_{10}^{22} b_{11}^{29} b_{12}^{36} b_{13}^7 b_{14}^{41} b_{15}^{34} b_{16}^{27} b_{17}^{20} b_{18}^{33} b_{19}^{13} b_{2}^{12} b_{20}^{32} b_{21}^{19} b_{22}^{25} b_{23}^{31} b_{24}^{37} b_{25}^6 b_{26}^{35} b_{27}^{29} b_{28}^{23} b_{29}^{17} b_{3}^{11} b_{30}^{28} b_{31}^{11}\\\nonumber&& b_{32}^{27} b_{33}^{16} b_{34}^{21}
 b_{35}^{26} b_{36}^{31} b_{37}^5 b_{38}^{29} b_{39}^{24} b_{4}^{10} b_{40}^{19} b_{41}^{14} b_{42}^{23} b_{43}^9 b_{44}^{22} b_{45}^{13} b_{46}^{17} b_{47}^{21} b_{48}^{25} b_{49}^4 b_{5}^9 b_{50}^{23} b_{51}^{19} b_{52}^{15} b_{53}^{11} b_{54}^{18} \\\nonumber
&&b_{55}^7 b_{56}^{17} b_{57}^{10} b_{58}^{13} b_{59}^{16} b_{6}^{17} b_{60}^{19} b_{61}^3 b_{62}^{17} b_{63}^{14} b_{64}^{11} b_{65}^8 b_{66}^{13} b_{67}^5 b_{68}^{12} b_{69}^7 b_{7}^8 b_{70}^9 b_{71}^{11} b_{72}^{13} b_{73}^2 b_{74}^{11} b_{75}^9 b_{76}^7 b_{77}^5\\
&& b_{78}^8 b_{79}^3 b_{8}^{23} b_{80}^7 b_{81}^4 b_{82}^5 b_{83}^6 b_{84}^7 b_{85} b_{86}^5 b_{87}^4 b_{88}^3 b_{89}^2 b_{9}^{15} b_{90}^3 b_{91} b_{92}^2 b_{93} b_{94} b_{95} b_{96}, \forall I\in S.
\label{prest}
\end{eqnarray}
Moreover, by comparing equations (\ref{ntc}) and (\ref{prest}), we
know that the $(4,6)$ point is exactly at the intersection of the
divisors $\{b_{97}=0\}$ and $\{b_\text{non-toric}=0\}$.

\end{itemize}

\begin{figure}
\centering
\begin{subfigure}{.5\textwidth}
  \centering
  \includegraphics[width=7cm]{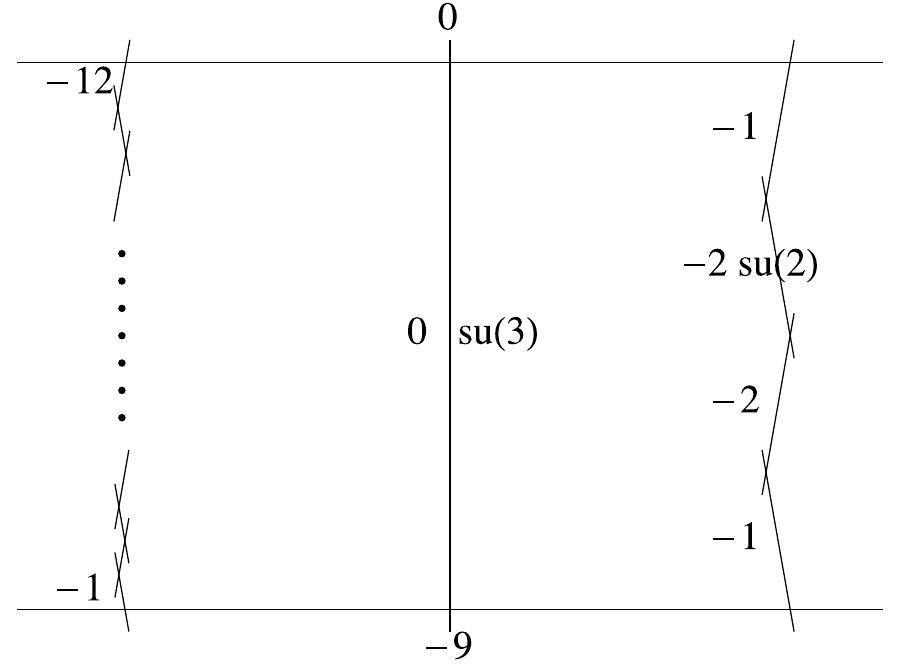}
  \label{fig:26145b}
\end{subfigure}%
\begin{subfigure}{.5\textwidth}
  \centering
  \includegraphics[width=7cm]{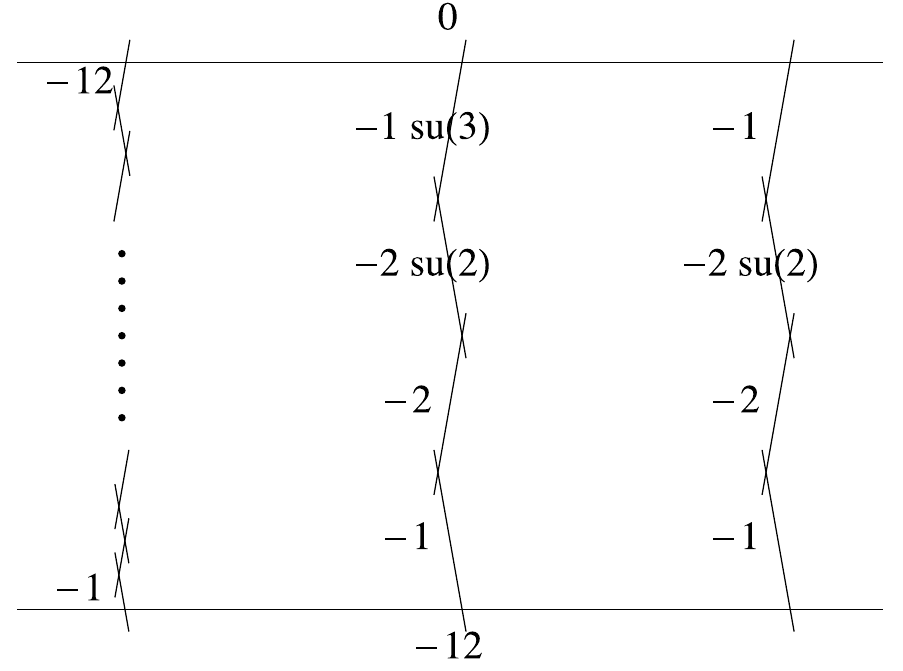}
  \label{fig:26145a}
\end{subfigure}
\caption{\footnotesize 
The base of the example with Hodge numbers $\{261, 45\}$.
Left: 
before resolution. Right: after resolving
  (4,6) points in the base. The top curve is $D_{102}$ and the bottom
  curve is $\tilde{D}_{97}$.  The curves in the left chain from top to
  down are in the order \{$D_1, D_2$,\ldots,$D_{96}$\}, in the middle
  chain \{$\tilde{D}_\text{non-toric}, \tilde{E_1},
  \tilde{E_2},E_3$\}, in the right chain \{$D_{101}, D_{100}, D_{99},
  D_{98}$\}.}
\label{fig:nontoric}
\end{figure}

%Back to that large $h^{1,1}$ example with 2-dimensional resolved fiber, we can analogously find 
%Resolve the $(4,6)$ points, if any, in the base, to obtain a flat elliptic fibration model over a semi-toric base as those we discussed in \S\ref{semitoric}

We now find the associated flat elliptic fibration model, so that we
can use F-theory techniques to compute the Hodge number shifts.  We
first identify the resolved  base, which is semi-toric, and then determine the
tunings.  Since there is the only one $(4,6)$ point, we blow up
successively three times at this point to turn the $-9$-curve into a
$-12$-curve, and the non-toric $0$-curve is replaced by a chain of
curves of self-intersection numbers $-1, -2, -2, -1$ (similar to the
last graph of the Hodge pair $\set{14,404}$ in Table \ref{t:f9}).
%its corresponding resolution in the base with gauge symmetries enhanced on the some divisors:  blow up three times at the non-toric point of the intersection of the $-9$-curve $D_{97}\equiv\{b_{97}=0\}$ and the non-toric curve $D_\text{non-toric}\equiv\{b_\text{non-toric}=0\}$, so 
The divisor classes of the curves after the blow-up process can be
determined in the usual fashion: The $-12$-curve
$\tilde{D}_{97}$ is the proper transform of the $-9$-curve after the
three blowups
\begin{equation}
\tilde{D}_{97}=D_{97}-E_1-E_2-E_3,
\end{equation}
%becomes a $-12$-curve $\tilde{D}_{97}=D_{97}-E_1-E_2-E_3$, 
where $E_1, E_2, E_3$ are the exceptional divisors associated with the
three blowups. The proper transform of the non-toric curve is
\begin{equation}
\tilde{D}_\text{non-toric}= D_\text{non-toric}-E_1,
\end{equation}
which is a $-1$-curve.  The three curves, $-2, -2, -1$, connecting
$\tilde{D}_{97}$ and $\tilde{D}_\text{non-toric}$ are respectively
\begin{equation}
\tilde{E_1}=E_1-E_2, \,\,\,\tilde{E_2}=E_2-E_3, \,\,\,\text{and }E_3.
\end{equation}
Now we figure out the gauge symmetries on these divisors. There was an
$\gsu(3)$ on the $0$-curve, $D_\text{non-toric}$, which is now on the
$-1$-curve, $\tilde{D}_\text{non-toric}$. This forces an $\gsu(2)$ on
the $-2$-curve, $\tilde{E_1}$, connecting to
$\tilde{D}_\text{non-toric}$. The configuration of the intersecting
curves and the symmetry enhancements are drawn in
Fig. \ref{fig:nontoric}.

Remarkably, the described configuration gives the correct counting of the shifts in Hodge numbers through the  anomaly calculation. The contributions to $h^{1,1}$ and $h^{2,1}$ from the tunings through equations (\ref{eq:dh11}) and (\ref{eq:dh21}) are
%The gauge symmetry configuration on the resolved base is then an $\gsu(2)$ on $D_{102}$ as before, $\gsu(3)$ on $\tilde{D}_\text{non-toric}$, and furthermore the $\gsu(2)$ on $\tilde{E_1}$ forced by the $\gsu(3)$ neighboring to it. 
% the original non-toric $0$-curve (which intersects also the toric $0$-curve $D_{102}\equiv\{b_{102}=0\}$) is then replaced by a chain of non-toric  curves $-1, -2, -2, -1$, where the first $-1$-curve  is the proper transformation of the original non-toric $0$-curve by the first blowup exceptional divisor,  $\tilde{D}_\text{non-toric}= D_\text{non-toric}-E_1$, and the following three curves are  the proper transformations of the exceptional divisors $\tilde{E_1}=E_1-E_2, \tilde{E_2}=E_2-E_3$, and the exceptional divisor $E_3$. The gauge symmetry configuration on the resolved base is then an $\gsu(2)$ on $D_{102}$ as before, $\gsu(3)$ on $\tilde{D}_\text{non-toric}$, and furthermore the $\gsu(2)$ on $\tilde{E_1}$ forced by the $\gsu(3)$ neighboring to it. The described structure (see fig. \ref{fig:nontoric})  gives the correct counting of the shifts in Hodge numbers:
\begin{itemize}
\item  $\gsu(2)$ on $D_{100}$: $\Delta h^{1,1}=\Delta r=+1$ and $\Delta h^{2,1}=\Delta V-\Delta H_\text{charge}=3- 4\times\bold{2}=-5$.

%$\Delta h^{1,1}=+1$ from the rank $\Delta r=1$ of the $\gsu(2)$ on $D_{102}$ and  $\Delta h^{2,1}=\Delta V-\Delta H_\text{charge}=3-8=-5$ from the dimension of $\gsu(2)$ and charge matter representations $4\times\bold{2}$ on the $-2$-curve.
\item $\gsu(2)\oplus\gsu(3)$ on $\set{\tilde{E_1},\tilde{D}_\text{non-toric}}$: $\Delta h^{1,1} =+1+2$ and  $\Delta h^{2,1}=(3+8)-(12\times\bold{3}+4\times\bold{2}-\bold{2}\times\bold{ 3} \text{ (shared)})=-27$.
%$\Delta h^{1,1} =+1+2$ from the rank of $\gsu(2)\oplus\gsu(3)$ enhancement on $\tilde{D}_\text{non-toric}$ and $\tilde{E_1}$ and $\Delta h^{2,1}=(3+8)-(10\times\bold{3}+\bold{2}\times\bold{ 3} (\text{shared})+1\times\bold{2})=-27$ from the dimensions of the guage algebras and the shared charge matter representations $12\times \bold{3}$ on $-3$-curve and $4\times\bold{2}$ on $-2$-curve.
\end{itemize}
The total gives the desired Hodge number shifts $\set{\Delta
  h^{1,1},\Delta h^{2,1}}=\set{4,-32}$. Note that the final Hodge
numbers correspond to the flat elliptic fibration over the non-toric
base, and correctly reflect the associated contribution of three
tensor multiplets from the three blow-ups of the base.\footnote{The
  Hodge numbers denoted in a generic model containing a
  $-9/-10/-11$-curve are understood to be those of the flat elliptic
  fibration over the base that has been resolved at the (4,6) points.}
The correspondence between the non-flat and the flat models may be
considered as that the four irreducible components of the
2-dimensional fiber over the $(4,6)$ point transform to the three
divisors resolving the $-9$-curve in the base and one divisor in the
fiber to resolve the forced $\gsu(2)$ on $\tilde{E}_1$.
%So we have in total $h^{1,1}=257+1+1+2=261$ and $h^{2,1}=77-5-27=45$, which agrees with the Hodge numbers from the polytope. 

We now consider the remaining non-standard $\P^{2,3,1}$
fibered ``extra'' cases at large $h^{1,1}$.  The
model associated with Hodge numbers \{$245, 57$\}  has a gauge symmetry that is enhanced on the non-toric
curve, but there are no $(4,6)$ singularities involved, which is 
similar to the example in \ref{warmup}; we must, however,  be
careful to properly include the shared matter contribution to the
matter multiplets, as a curve intersecting the non-toric
curve also carries a gauge symmetry. The models with Hodge numbers \{$261, 51$\},
\{$260, 62$\}, \{$260, 54$\}, \{$259, 55$\}, \{$258, 84$\}, and
$\set{254,56}$ are all similar to that of \{$261, 45$\} that we have
treated in detail here. 
\vspace*{0.1in}

We conclude the discussion of these cases by briefly summarizing the
details of the model M:82 10 N:351 10 H:254,56, where we need to
include one extra tensor multiplet in the Hodge number counting.

%have slightly fancier structures; and of \{$245, 57$\}, which is already a good elliptic fibration model but with gauge symmetry enhanced on some non-toric curve like the illustrative small $h^{1,1}$ example  \{8,250\} we gave at the beginning of this section.
\begin{itemize}
\item generic model (total 100 toric curves in the base)
\begin{eqnarray}
\{251,79,\{-12//-11//-12//-12//-12//-12//-12//-12,\\-1,-2,-2,-3,-1,-5,-1,-3,-2,-1,-7,-1,-2,-1,0\}\}
\nonumber
\end{eqnarray}
\item one $(4,6)$ point on  the $-7$-curve $D_{96}$.
%2D toric fiber component at one non-toric point $p$ on  the $-7$-curve $D_{96}$ 
\item gauge symmetry enhancements
\begin{enumerate}
\item $\gsu(2)$ on the $-2$-curve $D_{98}$
\item $\gsu(2)$ on the non-toric $0$-curve $D_\text{non-toric}$ intersecting the $0$-curve $D_{100}$ and the $-7$-curve $D_{96}$ at the $(4,6)$ point.
\end{enumerate}

\end{itemize}
The corresponding flat elliptic fibration model has
\begin{itemize}

\item base structure:
\newline The $(4,6)$ is blown up. $D_{96}$ is resolved into a $-8$-curve $\tilde{D}_{96}=D_{96}-E_1$ and $D_\text{non-toric}$ into $\tilde{D}_\text{non-toric}=D_\text{non-toric}-E_1$, where $E_1$ is the exceptional divisor of the blowup.
\item enhanced gauge symmetries:
\begin{enumerate}
\item  $\gsu(2)$ on the $-2$-curve $D_{98}$, which shifts $h^{1,1}$ by $r(\gsu(2))=1$ and $h^{2,1}$ by $V(\gsu(2))-H_\text{charged}(\gsu(2) \text{ on $-2$-curve})=3-4\times\bold{2}=-5$.
\item $\gsu(2)$ on the $-1$-curve $\tilde{D}_\text{non-toric}$, which shifts  $h^{1,1}$ by $r(\gsu(2))=1$ and $h^{2,1}$ by $V(\gsu(2))-H_\text{charged}(\gsu(2) \text{ on $-1$-curve})=3-10\times\bold{2}=-17$.
\end{enumerate}

\item Hodge numbers
\begin{enumerate}
\item contributions from the enhanced gauge symmetries $\Delta h^{1,1}= 1+1=2, \Delta h^{2,1}= -5-17=-22$
\item contribution from the tensor multiplet associated with the one extra blowup in the base $\Delta h^{1,1}= 1, \Delta h^{2,1}= -29$
\item compensation of $\Delta h^{2,1}= \frac1{2} \bold{56}=28$ due to
  the fact that there are half-hyper multiplets $\frac1{2} \bold{56}$
  on the NHC $-7$-curve, but there are no localized matter
fields on the NHC $-8$-curve.
\end{enumerate}
In total we have $h^{1,1}=251+2+1=254$  and $h^{2,1}=79-22-29+28=56$, which agree with the Hodge numbers of the polytope model.
\end{itemize}

\subsubsection{Example: a model with a tuned genus one curve in the base}
\label{g=1}

In the final part this section we consider an
additional non-standard $\P^{2,3,1}$-fibered models
that has the further interesting feature that a gauge group is tuned
on a non-toric curve that has nonzero genus.  While this phenomenon
does not occur in the ``extra'' models at large Hodge numbers that we
have focused on here, the fact that this non-toric tuning
structure can arise
even in the context of toric hypersurface Calabi-Yau constructions
seems sufficiently interesting and novel that we provide some details
for understanding the structure of models of this type.

We  study in particular a model with an $\gsu(2)$ tuning on a non-toric curve
of genus one in the base: M:223 7 N:10 6 H:3,165. The vertex set of
$\ds$ is
\begin{eqnarray}
\nonumber
 \set{((-1, -1, -2, 1), (2, -1, -2, 1), (-1, 2, -2, 1), &(-4, -4, 1, 
  1), (8, -4, 1, 1), \\&(-4, 8, 1, 1), (0, 0, 1, -1))},
  \end{eqnarray}
where the first three lattice points contribute to a $x^3$ term 
with base dependence. Then $\dd$ is a non-standard
$\P^{2,3,1}$-fibered polytope over $\P^2$. The base rays are
\begin{equation}
\set{(1,0),(0,1),(-1,-1)}\leftrightsquigarrow\set{b_1,b_2,b_3},
\label{p2base}
\end{equation}
 which come from the projection of the 4D rays $\set{(1, 0, 1, 2), (0,
   1, 1, 2),(-1, -1, 1, 2)}$ (in fact, these are the only three
 lattice points in $\dd$ that do not project to $(0,0)$, so none of
 the preimages are in the form $((v^{(B)})_{1,2},2,3)$.)

We analyze the Weierstrass model of the non-standard $\P^{2,3,1}$
polytope: Treating $\dd$again as the Bl${}_2$ $\P^{1,1,2}$-fibered polytope (in
particular, the fiber coordinates are associated to $\set{v_x, v_y,
  v_z}=\set{(0,0,-1,0),(0,0,0,-1),(0,0,1,2)}$)  we find the associated
tuned Weierstrass model. The orders of vanishing of $\set{f,g,\Delta}$
are enhanced to $\set{0,0,2}$ on the curve in the base
$D_\text{non-toric}\equiv \set{I_{\gsu(2)}=0}$, where

\begin{eqnarray}
\nonumber
I_{\gsu(2)}=& c_1 b_1^3 + c_{179} b_2 b_3^2  + c_{180} b_2^2 b_3 + c_{181}b_1 b_3^2  + c_{182}  b_1 b_2 b_3 \\
& + c_{183}b_1 b_2^2  + c_{184} b_1^2 b_3  + 
c_{185}  b_1^2 b_2  + c_2b_2^3  + c_3b_3^3.
\end{eqnarray}
In particular, the result for the discriminant $\Delta$ is
\begin{equation}
\Delta=I_{\gsu(2)}^2 I_1,
\end{equation}
where the $I_1$ component of $\Delta$ is a degree 30 polynomial in the homogeneous coordinates.
Note that $D_\text{non-toric}$ is a smooth curve of genus one, which can be calculated by the formula (\ref{eq:genus-relation})
\begin{equation}
3[b_1]\cdot (3[b_1]-([b_1]+[b_2]+[b_3]))=0=2g-2 \Rightarrow g=1.
\end{equation}
We calculate Hodge numbers from the anomaly conditions: the matter
representations of $\gsu(2)$ on a $g=1$ curve of self-intersection
$D_\text{non-toric}^2=9$ is $54\times \mathbf{2}+\mathbf{3}$
\cite{Johnson:2016qar}, but only two components of the adjoint
representation $\mathbf{3}$ are charged under the Cartan (see the
footnote in \S\ref{hns}). Therefore,
$H_\text{charged}=108+2=110.$ Then $h^{1,1}=\Delta r=1$ and
$h^{2,1}=\Delta V-H_\text{charged}=3-110=-107$, which agree with
$\set{3,165}-\set{2,272} =\{+1, -107\}$.

\section{Conclusions}
\label{sec:conclusions}

\subsection{Summary of results}

In this paper we have carried out a systematic comparison of elliptic
Calabi-Yau threefolds with large Hodge numbers that are realized by
tuning Tate-form Weierstrass models over toric bases and those that
are realized as hypersurfaces in toric varieties through the Batyrev
construction.  Specifically, we have considered a  class of
Tate-tuned models over toric bases that have nonabelian gauge groups
tuned over toric divisors.  These tunings give a specific class of 
``standard''
$\P^{2,3,1}$-fibered reflexive polytopes,
all of which give Calabi-Yau threefolds with Hodge numbers that appear
in the Kreuzer-Skarke database.
\vspace*{0.1in}

\noindent
$\bullet$  Almost all Hodge number pairs of known CY3's 
in the regime studied
come from
elliptically fibered Calabi-Yau threefolds associated with polytopes
constructed in this fashion that are associated with an explicit
Tate/Weierstrass construction of the restricted class that we
considered in our initial analysis.
\vspace*{0.1in}

\noindent
$\bullet$ We have explicitly analyzed the structure of the Calabi-Yau
threefolds in the Kreuzer-Skarke database for the 18 Hodge number
pairs not found in our initial analysis from Tate constructions.  All
of these admit elliptic fibrations of slightly more complicated forms.
\vspace*{0.1in}

\noindent
$\bullet$ Thus, we have found explicit realizations of elliptic
Calabi-Yau threefolds that produce all Hodge number pairs with
$h^{1,1}\geq 240$ or $h^{2, 1}\geq 240$ that are known to be possible
for Calabi-Yau threefolds.
This matches with the results of a companion paper
\cite{Huang-Taylor-fibers} showing that all polytopes in the KS
database giving Calabi-Yau threefolds with $h^{1,1}\geq 150$ or $h^{2,
  1}\geq 150$ have a genus one fibration, and have
an elliptic fibration whenever $h^{1,1}\geq 195$ or $h^{2, 1} \geq 228$.
These results provide additional evidence that
 virtually all known Calabi-Yau threefolds with large
Hodge numbers are elliptically fibered, building on a variety of other
recent work that has led to similar observations \cite{Candelas-cs,
  Gray-hl, Johnson-WT, Anderson-aggl,
Anderson-ggl,  Johnson:2016qar}.  Since the
number of elliptic Calabi-Yau threefolds is finite, this in turn
suggests that the number of distinct topological classes of Calabi-Yau
threefolds is in fact finite.
\vspace*{0.1in}

\noindent
$\bullet$ In the course of this analysis we have encountered some novel
structures in the Tate/Weierstrass tunings needed to reproduce certain
CY3's associated with polytopes in the KS database.
This has led to new insights into the Tate algorithm as well as in the
structure of fibrations that may occur through polytopes.

---  A novel Tate tuning of SU(6) gives rise to exotic 3-index
antisymmetric matter, of a form recently studied in
\cite{Morrison:2011mb, transitions}.

--- Some polytopes in the KS database correspond to tunings of very
large gauge algebras with components like $\gso(20)$.

--- Polytopes in the KS database include non-flat elliptic fibrations over
toric bases that resolve into flat elliptic fibrations over
more complicated non-toric bases including not only blow-ups of $-9, -10, -11$
curves, but also more exotic structure such as an $\ge_8$ over a $-8$
curve that must be blown up four times, or 
tunings of $\gso(n), n \geq -13$ on $-3$
curves that must be blown up to $-4$ curves to satisfy anomaly
conditions.  In some of the $\gso(n)$
cases the resolved geometry also gives rise to a nontrivial
Mordell-Weil group associated with a U(1) factor in the gauge group.

--- Some polytopes in the KS database have elliptic fibrations over
toric bases in which nonabelian gauge algebras are tuned over
non-toric curves in the base.

--- We worked out the tops associated with the gauge algebras
$\gso(n)$, $13\leq n\leq 25$, as well as the tops associated with
gauge algebras $\gsu(n)$, $7\leq n\leq 13$. 
For $\gso(n)$, these match the tops found in
\cite{Bouchard:2003bu} after an appropriate linear transformation; our
construction gives explicit realizations of these tops in reflexive
polytopes for the range of algebras listed, which is not guaranteed
from the construction of \cite{Bouchard:2003bu}.
The tops associated with
$I_n$ and $I_n^*$ types have the feature that they develop along the
fiber direction, and the projection to the fiber plane falls outside
the $\P^{2,3,1}$ fiber subpolytope.
Another interesting feature of the $\gso(n)$ tops is that there can be
multiple distinct tops for certain gauge algebras, corresponding to
monodromy conditions on the associated Tate tunings.
%, which is a novel feature not found
%in the previously known
%tops.

\subsection{Possible extensions of this work}

In the companion paper \cite{Huang-Taylor-fibers}, we carry out a
complementary analysis to that of this paper.
Here we have started from the Tate tuning picture and matched to data
in the Kreuzer-Skarke database.  One can instead start with the
polytopes in the database and try to derive the elliptic fibration
structure.  This is essentially the approach taken by Braun in
\cite{Braun:2011ux}, in which the database was scanned for elliptic
fibrations over the base $\P^2$.  
In \cite{Huang-Taylor-fibers}, we take that point of view and analyze
the fibration structure of the polytopes in the KS database directly.
The approach taken in this paper, however, shows that at large Hodge
numbers most Calabi-Yau threefolds have a standard elliptic fibration
structure; the ``sieve'' approach taken here enables us to identify
some of the most interesting cases that present novel features.

There are several closely related analyses that could be carried out
that we have not done here or in \cite{Huang-Taylor-fibers}; each of
these represents an opportunity for further work that would give
increased understanding of the set of Calabi-Yau threefolds, the
role of elliptic fibrations, and the landscape of 6D F-theory models.

First, we have started from the point of view of tuning Tate models
and used the output of that analysis to match Hodge numbers in the KS
database.  In principle, we could have tried to reproduce all the
polytopes in the database, i.e. included multiplicity information.
For reasons discussed in \S\ref{multiplicity}, this would be a more
complicated analysis.  In many cases there are multiple local Tate
tunings that give equivalent gauge groups, and we have in each case
systematically taken only the lowest possible choice for NHCs and the
lowest order choice with no further monodromy condition required for a
given gauge group tuning.  For bases with many toric divisors, the
number of combinatorial possibilities of local tunings can become
quite large.  There are also many equivalent models that correspond to
carrying out explicitly different subsets of toric blow-ups to partially
resolve (4, 6) singularities.  We have checked in some cases that the
multiplicity of Hodge numbers in the KS database is reproduced by
distinct Tate/Weierstrass tunings of elliptic fibrations, but we have
not approached this systematically.  This would be a natural next step
for this kind of analysis, and might reveal additional novel
structures for the elliptic fibrations found in the KS database.

Second, we have restricted to large Hodge numbers in part because we
have only focused on Tate models associated with the most generic
$\P^{2,3,1}$ fiber structure for the polytope.  There are 16 distinct
possible toric fibers, analyzed in detail in the F-theory context in
\cite{Braun:2011ux, Braun-16, Klevers-16}, each leading to a distinct
class of Weierstrass tuning types with characteristic nonabelian and
abelian gauge structure, and in principle we could systematically
analyze all tunings that correspond to each of the different fiber
types.  This would be necessary to extend the analysis of this paper
systematically to smaller Hodge numbers, where the other fiber types
become prevalent \cite{Huang-Taylor-fibers}.  We leave such an
endeavor for future work.
It would also be interesting to see whether the more general class of
fibers considered in
\cite{Braun:2014qka} may give further insights into other Weierstrass
tuning types that may be possible with complete intersection fibers.

%\footnote{\bf{YH: Tate tuning swampland (limitation of reflexive polytope models in Weierstrass model)}}

\acknowledgments{
We would like to thank Lara Anderson,
Andreas Braun,
James Gray,
Sam Johnson,
Nikhil Raghuram, 
David
  Morrison, 
Andrew Turner, and
Yinan Wang  for helpful discussions. The authors are supported by DOE grant
  DE-SC00012567.
}

\appendix

\section{Standard $\P^{2,3,1}$-fibered polytope tuning}
\label{polytopetuning}

In this Appendix, we go through the details of a standard
$\P^{2,3,1}$-fibered polytope tuning with an example of polytopes for
elliptic fibrations with tuned $\gsu_3, \gg_2$ over
the curve of self-intersection $-2$ in the Hirzebruch surface base $\F_2$.

Standard $\P^{2,3,1}$-fibered polytopes naturally correspond to Tate
(tuned) models. In principle, as long as the Tate tunings on adjacent
curves do not lead to $(4,6)$ singularities\footnote{Although in some
  cases such $(4, 6)$ singularities still lead to reflexive polytopes
  that can be associated with flat elliptic fibrations over blown-up
  bases, as encountered in the examples of \S\ref{semitoric}.}, and
are not merely further specialization of existing tunings that do not
change the gauge algebra\footnote{None of the lattice points
  corresponding to such further specialization are vertices of $\ds$,
  so removing those points does not affect the polytope.}, removing
the lattice points corresponding to a given tuning gives a different
reflexive polytope, associated with the resolved CY of the tuned
singular model.  The Hodge numbers of the new resolved polytope model
can be computed either directly from the polytopes or through F-theory
by anomaly cancellation.

As an example, consider tuning a type $I^s_3$ $\gsu(3)$ gauge algebra
on the $-2$ curve in the base $\F_2$. The polytope
model for the generic CY
is M:335 5 N:11 5 H:3,243, of which the set of vertices of $\dd$ is
\begin{equation}
\{(1,0,2,3), (0,1,2,3), (-1,-2,2,3), (0,0,-1,0), (0,0,0,-1)\},
\end{equation}
and the set of vertices of $\ds$ is
\begin{equation}
 \{(-6,-6,1,1), (0,0,-2,1), (18,-6,1,1), (0,0,1,-1), (-6,6,1,1)\}.
\end{equation}
The projection along the fiber gives the rays in the base
$\{v_{i}^{(B)}\}=\{(1,0), (0,1), (-1,-2), (0,-1)\}$ corresponding to
curves of self-intersection $\{0,2,0,-2\}$. We calculate the
hypersurface equation (\ref{p}) and take the set of homogeneous
coordinates $\{z_j\} =\{x, y, z, b_4\}$ associated respectively to
rays $v_x, v_y, v_z$ in the fiber plane and
$(v_{B4}^{(1)},v_{B4}^{(2)},2,3)$ in the base plane to get
\begin{equation}
y^2 + a_1xyz + a_3yz^3 = x^3 + a_2x^2z^2 + a_4xz^4 + a_6z^6,
\end{equation}
where the 5 sections $a_i$ in the coordinates $b_4$ and some second
coordinate $\zeta$ in the base have the forms 
\begin{eqnarray}
a_1(b_4,\zeta)&=& a_{1,0}(\zeta) + a_{1,1} (\zeta)b_4 + a_{1,2} (\zeta)b_4^2,\\
a_2(b_4,\zeta)&=& a_{2,0}(\zeta) + a_{2,1} (\zeta)b_4 + a_{2,2} (\zeta)b_4^2+ a_{2,3} (\zeta)b_4^3+ a_{2,4} (\zeta)b_4^4,\\
a_3(b_4,\zeta)&=& a_{3,0}(\zeta) + a_{3,1} (\zeta)b_4 + \cdots+a_{3,5} (\zeta)b_4^5+ a_{3,6} (\zeta)b_4^6,\\
a_4(b_4,\zeta)&=& a_{4,0}(\zeta) + a_{4,1} (\zeta)b_4 +\cdots+ a_{4,7} (\zeta)b_4^7+ a_{4,8} (\zeta)b_4^8,\\
a_6(b_4,\zeta)&=& a_{6,0}(\zeta) + a_{6,1} (\zeta)b_4 +\cdots+ a_{6,11} (\zeta)b_4^{11}+ a_{6,12} (\zeta)b_4^{12}.
\end{eqnarray}
The numbers of monomials (lattice points) in the sections $a_i$ are \{9,
25, 49, 81, 169\}; together with the two points associated with $x^3$
and $y^2$ these compose
the total set of 335 lattice points in the M polytope $\ds$. The
number of monomials in each section
can be further divided according to the power of the monomials in the
$b_4$ expansion. According to Tate Table \ref{tab:tatealg}, the
vanishing orders have to reach $\{0, 1, 1, 2, 3\}$ in $b_4$ to tune an
$I^s_3$ $\gsu(3)$ over $D_{B4}$, so all lattice points contributing to
$a_{2,0}, a_{3,0}, a_{4,0}, a_{4,1}, a_{6,0}, a_{6,1}, a_{6,2}$ should
be removed. As one can check those are
\begin{eqnarray}
&&a_{2,0}\leftrightarrow \{(-2, 2, -1, 1)\},\\
&&a_{3,0} \leftrightarrow \{(-3, 3, 1, 0)\},\\
&&a_{4,0} \leftrightarrow \{(-4, 4, 0, 1)\},\\
&&a_{4,1} \leftrightarrow\{(-4, 3, 0, 1), (-3, 3, 0, 1), (-2, 3, 0, 1)\},\\
&&a_{6,0} \leftrightarrow\{(-6, 6, 1, 1)\},\\
&&a_{6,1} \leftrightarrow\{(-6, 5, 1, 1), (-5, 5, 1, 1), (-4, 5, 1, 1)\},\\
&&a_{6,2}\leftrightarrow\{(-6, 4, 1, 1), (-5, 4, 1, 1), (-4, 4, 1, 1), (-3, 4, 1, 1), (-2, 4, 
  1, 1)\}.
\end{eqnarray}
After reduction, the vertex set of
the new dual polytope $\ds'$ for the tuned model
becomes
\begin{eqnarray}
&&  \{(-6,-6,1,1), (0,0,-2,1), (18,-6,1,1), (0,0,1,-1), (-6,3,1,1),\\&&(-3,2,1,0), (-1,1,0,0), (-1,2,1,0),
(0,3,1,1)\},
\end{eqnarray}
This new polytope is again reflexive, and corresponds to the example
M:320 9 N:13 7 H:5,233 in the KS database. Comparing the two sets of
data (for the generic and tuned models), the difference in the number
of lattice points of the monomial polytopes $\ds$ and $\ds'$, $335-320=15$,
is the number of the lattice points being removed. On the other hand,
the fan polytope is enlarged $\dd \rightarrow\dd'$, and the increased
number $N$, $13-11=2$, comes from lattice points \{(0, -1, 1, 1), (0,
-1, 1, 2)\}, which together with the affine node $(0,-1,2,3)$ form
exactly the $\gsu(3)$ top. The Hodge shifts
$\{5,233\}-\{3,243\}=\{2,-10\}$ match exactly with the calculation
from anomalies for tuning the algebra $\gsu(3)$ on an isolated
$-2$-curve.

There are two polytopes in the KS database with Hodge numbers
$\{5,233\}$. The other polytope M:316 6 N:14 6 H:5,233 is the polytope
arising from an enhancement to a $\gg_2$ gauge algebra by further removing from
the $\gsu(3)$ model
\begin{eqnarray}
&&a_{1,0}\leftrightarrow \{(-1, 1, 0, 0)\},\\
&&a_{3,1} \leftrightarrow \{(-3,2,1,0), (-1,2,1,0), (-2,2,1,0)\};
\end{eqnarray}
so that the vanishing orders along the $-2$-curve becomes
$\{1,1,2,2,3\}$, and the number of lattice points in
$\ds$ (M) decreases by
4. Comparing the fan polytope of the $\gg_2$ tuning model to that of the
generic model, there are three more lattice points $\{(0,-2,2,3),
(0,-1,1,1), (0,-1,1,2)\}$, which together with $(0,-1,2,3)$ form the
$\gg_2$ top. The Hodge numbers are the same as those of the $\gsu(3)$
model, since $\gsu(3)\rightarrow\gg_2$  is a rank-preserving tuning
(\S\ref{hns}).

\section{Elliptic fibrations over the bases $\F_9, \F_{10}$, and $\F_{11}$}
\label{2dfiber}

The ``standard stacking'' construction (\S\ref{polytune}) of a polytope
for a standard $\P^{2, 3, 1}$-fibered model over a base surface
containing $-9, -10, -11$-curves produces a flat toric fibration that
leads to a hypersurface that is a non-flat elliptic fibration.  There
are $(4,6)$-points in the $-9, -10, -11$-curves where the fiber
becomes two-dimensional; the singular fiber is resolved into an
irreducible component of the non-generic toric fiber, which is
two-dimensional, as the hypersurface CY equation restricting to the
component is trivially satisfied over these points.  For a flat
elliptic fibration, the $(4, 6)$-points in the base must be blown up,
which in general leads to a non-toric base.  Note that in the
Calabi-Yau hypersurface picture, some flops may be necessary before
the blow-ups can be done in the toric picture \cite{Braun:2011ux}.
Nonetheless, this provides a clear correspondence between the non-flat
elliptic fibrations associated with polytopes leading to $(4, 6)$
points in the base and flat elliptic fibrations over blown up bases,
which provide Calabi-Yau threefolds with the same Hodge numbers.  In
this Appendix we go through the details of these constructions for the
Hirzebruch surface bases $\F_m, m = 9, 10, 11$.

The flat toric fibration of  M:560 6 N:26 6 H:14,404 gives  a
non-flat elliptic fibration model over the toric base $\F_{m=9}$. The
vertices of the $\dd$ polytope are 
\begin{equation}
\set{((0, 0, -1, 0), (0, -6, 2, 3), (-1, -m, 2, 3),(1, 0, 2, 3), (0, 1, 2,
   3), (0, 0, 0, -1))}.
   \label{HBvert}
\end{equation}
 We associate the base coordinates $\set{b_1,b_2,b_3,b_4}$
 to the toric curves $\set{0,-m,0,m}$ whose corresponding rays in
 the base are $\set{(1,0),(0,-1),(-1,-m),$ $(0,1)}$. The set of
 lattice points in the top over the $-m$-curve is given by the set of
lattice points in $\dd$ of the form $(0, a, x, y)$,
   \begin{eqnarray}
\nonumber
&&\{(0,-6,2,3), (0,-5,2,3), (0,-4,1,2), (0,-4,2,3), (0,-3,1,1), (0,-3,1,2),\\\nonumber
&&(0,-3,2,3), (0,-2,0,1), (0,-2,1,1), (0,-2,1,2), (0,-2,2,3), (0,-1,0,0),\\
&&(0,-1,0,1), (0,-1,1,1), (0,-1,1,2), (0,-1,2,3)\}\,.
\label{top}
\end{eqnarray}
Each of these points represents an irreducible component of the
 2-dimensional non-generic toric fiber over the $-m$-curve $\{b_2=0\}$
 and projects to the corresponding base ray $(0,-1)$. Over a generic
 point on the $-m$
curve, the hypersurface CY, $p$
given by equation
 (\ref{p}), intersects with only the irreducible components on the
 boundary of the top giving a $\P^1$ for each, which combine to form the $E_8$ affine Dynkin diagram. 
 These nine components are
 \begin{eqnarray}
\nonumber
\set{((0,-6,2,3),(0,-5,2,3),(0,-4,2,3),(0,-3,2,3),(0,-2,2,3),
(0,-1,2,3),\\(0,-4,1,2),(0,-2,0,1),(0,-3,1,1)},
  \label{boundary}
\end{eqnarray}
where the set of components in first line forms the longest leg of the diagram, and the sets $\set{(0,-6,2,3),$ $(0,-4,1,2), (0,-2,0,1)}$ and $\set{(0,-6,2,3),(0,-3,1,1)}$ form the other two legs. ($(0,-6,2,3)$ is the node where three legs connect, and $(0,-1,2,3)$ is the affine node.)

  However,  $p$ also intersects the full irreducible component
  $(0,-1,0,0)$ over three points in the $-m$-curve, but does not meet the component over the other points:    $p$ restricted to the divisor  $I=(0,-1,0,0)$ is
 \begin{equation}
p\rvert_I=b_2^5 (c_4 b_1^3+c_{284} b_1^2 b_3+c_{285} b_1 b_3^2+c_3 b_3^3) b_4^7,
\end{equation}
where $c_3, c_4, c_{284}$, and $c_{285}$ are some complex structure moduli.   This  vanishes identically over the three points    $\set{c_4 b_1^3+c_{284} b_1^2 b_3+c_{285} b_1 b_3^2+c_3 b_3^3=0}$ in the $-m$-curve ($I$ projects to the ray of the $-m$-curve), and is otherwise a constant.
%Therefore, the CY intersects the full irreducible component $(0,-1,0,0)$ at the three points $\set{c_4 b_1^3+c_{284} b_1^2 b_3+c_{285} b_1 b_3^2+c_3 b_3^3=0}$, and do not meet the component over the other points in the $-m$-curve. 
It is these three points in the toric base that must be blown up to give
the $-12$-curve and the semi-toric base over which the elliptic
fibration model becomes flat and gives a good model for F-theory
compactification.

Similarly, the flat toric fibrations of M:600 6 N:26 6 H:13,433 and
M:640 6 N:26 6 H:12,462 give non-flat elliptic fibration models over
the toric bases $\F_{m=10}$ and $\F_{m=11}$, respectively. Both vertex
sets are given by equation (\ref{HBvert}), and the tops over the $-m$-curves
are the same as that over the $-9$ curve in equation (\ref{top}). We
know that a $-10$-curve (resp.\ a $-11$-curve) would need two blowups
(resp.\ one blowup) to become a $-12$-curve, so we expect there are two
$(4,6)$ points (resp. one point) in the $-m$-curve over which the
resolved fiber is two-dimensional. Indeed, we calculate the CY
hypersurface in equation (\ref{p}), and restrict it on each component
in (\ref{top}), and we find 
\begin{equation}
p\rvert_I=b_ 2^5 (c_ 4 b_ 1^2 + c_ {305} b_ 1 b_ 3 + c_ 3 b_ 3^2) b_ 3 b_ 4^7
\end{equation}
in the case of $m=10$, and
\begin{equation}
p\rvert_I=b_ 2^5 (c_ 4 b_ 1 + c_ 3 b_ 3) b_ 4^7
\end{equation}
in the case of $m=11$. Over a generic point in the $-m$-curve,
$p\rvert_I$ is non-vanishing, and $p$ intersects with the nine
components in (\ref{boundary}), each giving a $\P^{1}$
that corresponds
to a node in the extended $E_8$ Dynkin diagram.

The correspondence between the non-flat and the flat models may be
thought of as encoding the relationship between the irreducible
component of the 2-dimensional fiber over a $(4,6)$ point and divisors
that resolve the $-m$-curve to a $-12$-curve in the base.

\section{An example with a nonabelian
tuning that forces a U(1) factor.}
\label{26332cal}
\label{u1}

\begin{figure}
\centering
\begin{subfigure}{.5\textwidth}
  \centering
  \includegraphics[width=7cm]{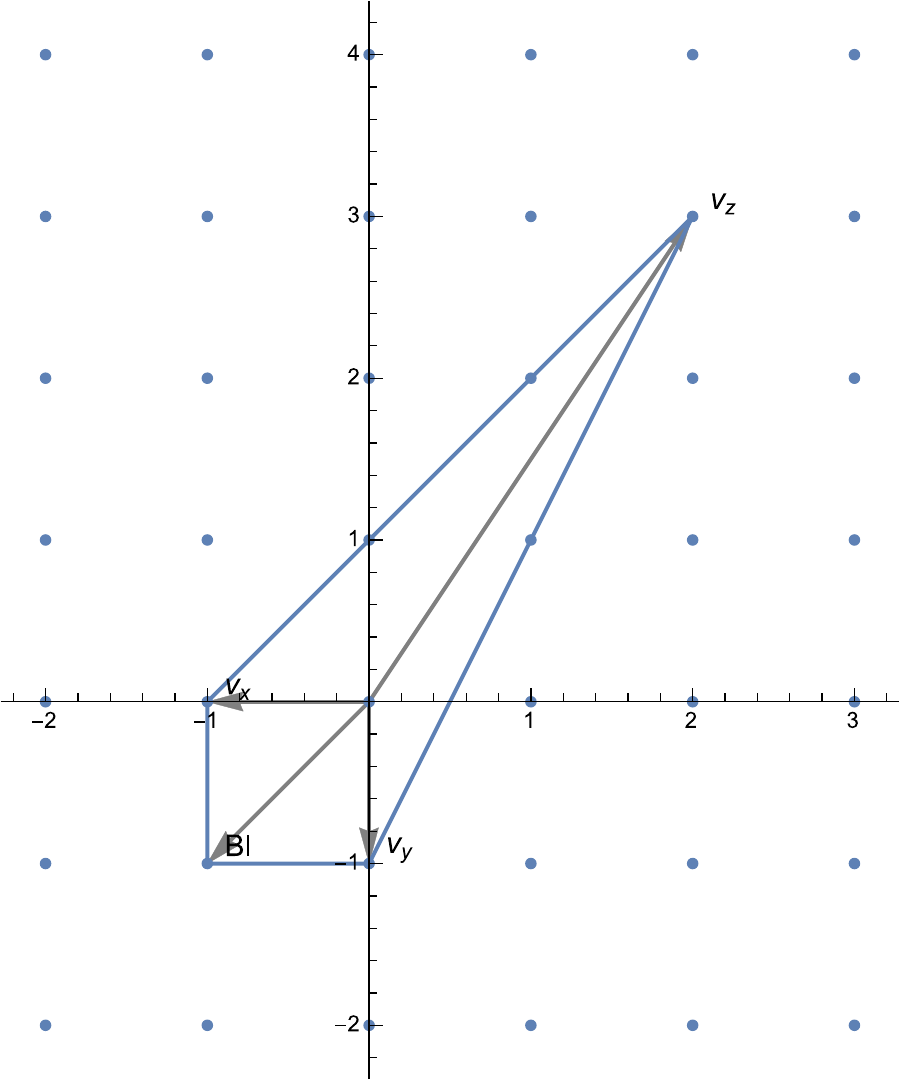}
  \caption{\footnotesize $\dd_2$: the additional ray blown up from $\P^{2,3,1}$ resolves $\gu(1)$-tuned models.}
  \label{fig:F11}
\end{subfigure}%
\begin{subfigure}{.5\textwidth}
  \centering
  \includegraphics[height=6.3cm]{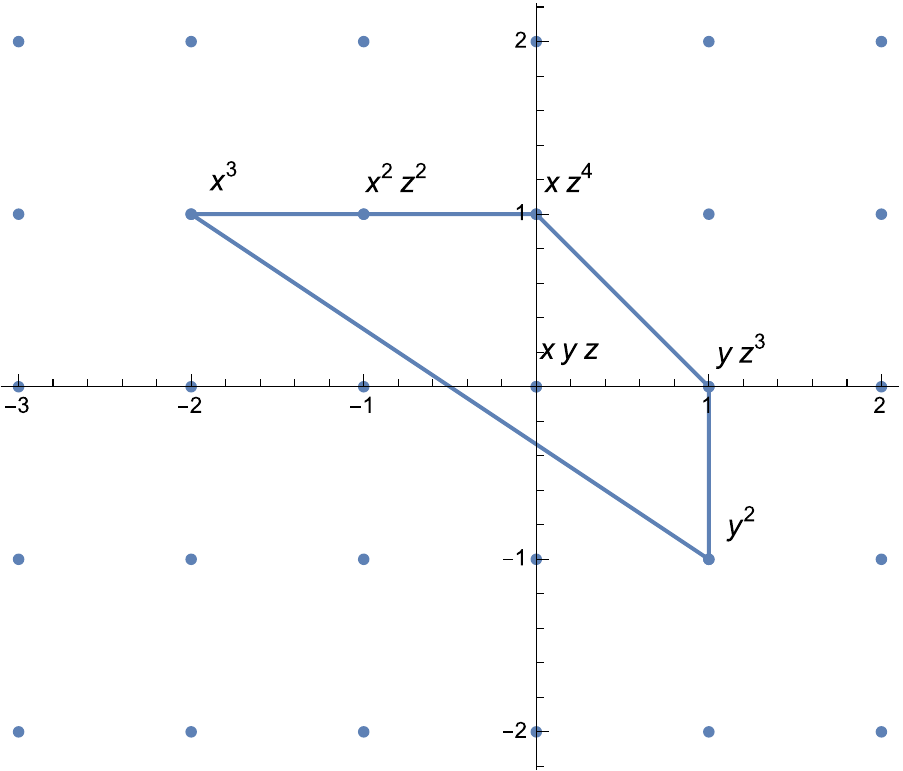}
  \caption{\footnotesize 
$\ds_2$: all monomials in the section $a_6$ are removed in the tuning, which leads to a global $\gu(1)$ factor in Bl$_{[0,0,1]}\P^{2,3,1}$-fibered polytopes.}
  \label{fig:M11}
\end{subfigure}
\caption{The reflexive polytope pair for the Bl$_{[0,0,1]}\P^{2,3,1}$ ambient toric fiber.}
\label{11}
\end{figure}

In this Appendix, we work through the details of an example of the
missing Hodge pairs in
the last part of Table \ref{restt}: M:47 11 N:362 11 H:263,32.  This
example involves huge tunings, a blow-up from an $\gso(n)$ tuning on a
$-3$ curve, and the further feature of a forced  nontrivial
Mordell-Weil group giving a U(1) factor.
After describing the geometry, we do a detailed
calculation of the Hodge numbers through the associated flat elliptic
fibration model.

The rational sections of an elliptic fibration form the Mordell-Weil
group, which is a finitely generated group of the form
$\Z^\text{rank}\times$ (torsion subgroup). If an elliptically fibered
Calabi-Yau has a non-trivial Mordell-Weil rank, the F-theory
compactification on it has an abelian sector $U(1)^\text{rank}$
\cite{Morrison:1996pp}.  The Weierstrass model of an elliptically
fibered Calabi-Yau automatically comes with a zero section $z=0$.
Additional sections can be produced through constraints in the toric
geometry
 \cite{Braun-16}.  For instance, an
abelian global $\gu(1)$ symmetry is forced when we set all the
monomials in the section $a_6$
to vanish (the
condition $a_6=0$ in \cite{Mayrhofer:2014opa}.)  
While this can be simply imposed as a constraint to tune a U(1)
factor, this condition can also be imposed when we tune a large enough
set of nonabelian gauge algebras on the toric curves.
The lack of the
monomials in $a_6$ occurs in this way in
the four missing Hodge pairs $\{263,
32\}, \{251, 35\}, \{247, 35\}, \{240, 37\}$ in Table \ref{restt},
which are therefore Bl$_{[0,0,1]}\P^{2,3,1}$-fibered polytope models
(see Figure \ref{11}).

The $\dd$ polytope of M:47 11 N:362 11 H:263,32 is
Bl$_{[0,0,1]}\P^{2,3,1}$-fibered over the base
\begin{equation}
\label{26332generic}
\{-4, -1, -3, -1, -4, -1, -4, -1, -4, 0, 2\}.
\end{equation}
The generic model over this base has Hodge numbers $\set{28,
  160}$. The polytope of interest can be obtained by Tate-tuning a polytope
model, for example M:225 6 N:31 6 H:28,160, associated with the
generic model over the base \eq{26332generic}.  Indeed,
$\dd$ is a standard $\P^{2,3,1}$-fibered polytope, where the tunings
are
\begin{eqnarray}
\label{26332before}
\{\{-4,\gso(38)\},\{-1,\gsp(29)\},\{-3,\gso(92)\},\{-1,\gsp(36)\},\{-4,\gso(68)\},\{-1,\gsp(24)\},\\\nonumber
\{-4,\gso(44)\},\{-1,\gsp(12)\},\{-4,\gso(20)\},\{0,\cdot\},\{2,\cdot\}\}.
\end{eqnarray}
The non-flat fiber results from the $\gso(92)$ on the $-3$-curve, as
it exceeds the upper bound $\gso(12)$ associated with anomaly
conditions. As the non-abelian tuning uses all of the monomials in
$a_6$, the dual fiber subpolytope $\dd_2$ becomes a blowup of
$\P^{2,3,1}$, Bl$_{[0,0,1]}\P^{2,3,1}$ (see Fig.~\ref{11}).

Now we compute the Hodge numbers from the associated flat elliptic fibration model over the resolved base
\begin{eqnarray}
\label{26332base}
&{-1}&\\\nonumber
-4, -1, &{-4},&  -1, -4, -1, -4, -1, -4, 0, 2,
\end{eqnarray}
with  tuned gauge symmetries
%Accurate speaking, the ``$-3$''-curve is a $-4$ curve intersecting a non-toric exceptional $-1$-curve  with forced $\gsp(19)$ gauge algebra:
\begin{eqnarray}
\label{26332}
&{\gsp(19)}&\\\nonumber
\gso(38),\gsp(29), & {\gso(92)},& \gsp(36), \gso(68), \gsp(24), \gso(44), \gsp(12), \gso(20),\cdot, \cdot.
\label{nab}
\end{eqnarray}
The $\gsp(19)$ on the exceptional $-1$-curve is forced by the
$\gso(92)$ on the intersecting $-4$-curve. Again, the tuned
non-abelian symmetries force
a global U(1).  The dimensions of the
non-abelian gauge group factors in equation (\ref{nab}) are
\begin{eqnarray}
\nonumber
&741&\\\nonumber
703, 1711, &4186&, 2628, 2278, 1176, 946, 300, 190, 0, 0,
\end{eqnarray}
which differ from the total dimension of the gauge groups in the NHCs in
(\ref{26332generic}) by $\Delta V_{\text{non-abelian}}=14859-(4\times
28 + 8)=14739.$ The representations of the gauge group factors on the
individual curves are \cite{Johnson:2016qar}
\begin{eqnarray}
\label{reps}
&46\times \mathbf{38}&\\\nonumber
30\times\mathbf{38}, 66\times\mathbf{58},&84\times\mathbf{92},&80\times\mathbf{72}, 60\times\mathbf{68}, 56\times\mathbf{48}, 36\times\mathbf{44}, 32\times\mathbf{24}, 12\times\mathbf{20}, \cdot, \cdot.
\end{eqnarray}
But some representations are shared between each pair of intersecting
curves. The representations that are charged under both of the two
corresponding group factors, $\gso(n)\oplus\gsp(m)$, are:
\begin{eqnarray}
\label{shared}
{\frac{1}{2}}&{\cdot}&{92\cdot38}
\\\nonumber
\frac{1}{2}\cdot38\cdot58,{\frac{1}{2}\cdot58\cdot92}&,&{\frac{1}{2}\cdot92\cdot72,}\frac{1}{2}\cdot72\cdot68,\frac{1}{2}\cdot68\cdot48,\frac{1}{2}\cdot48\cdot44,\frac{1}{2}\cdot44\cdot24,\frac{1}{2}\cdot24\cdot20, \cdot,\cdot,
\end{eqnarray}
where the $1/2$ factors come from the group theoretic normalization constant of $\gso(n)$. Hence,
$\Delta H_{\text{non-abelian charged}}=(\text{sum of all terms in (\ref{reps})}-\text{ sum of all terms in (\ref{shared})})=14830$. Note that all representations of a forced non-abelian gauge group are shared: $1/2(92)=46$ on the exceptional $-1$-curve are shared. All representations on the blown up $-4$-curve are also shared: $1/2(38+58+72)=84$, so the gauge symmetries can not be enhanced further on the three intersecting $-1$-curves.

The final piece needed is the U(1) charged matter. These fields are
not charged under the non-abelian group, and therefore have not yet
been
taken into account in our computations. These matter fields are
localized at codimension two
on the $I_1$ component (away from the non-abelian components) of the
discriminant locus (equation (\ref{delta})), and the number of the
U(1) charged matter fields corresponds to the number of the nodes,
over which the fiber is Kodaira $I_2$, on the $I_1$ component
\cite{Morrison-Park}. Concretely, as described for example in
\cite{Braun:2014oya}, we calculate the discriminant locus of the $I_1$
with respect to one of the two local coordinates, which we choose to
be $b_1$ associated with the $2$-curve and $b_2$ associated with the
$0$-curve; then the $I_1$ discriminant locus factors into
\begin{equation}
\Delta_{I_1}(b_2)=p_1(b_2)(p_2(b_2))^2(p_3(b_2))^3,
\end{equation}
where $p_1$ is a polynomial of degree 76 in $b_2$, $p_2$ is a polynomial of degree 9 in $b_2$, and $p_3$ is a polynomial of degree 63 in $b_2$. The degrees of the polynomials $p_2$, and  $p_3$ correspond to the number of nodes and cusps on the $I_1$, respectively. The hypermultiplets charged only under the U(1) are localized at the nodes, and therefore $\Delta H_{\text{abelian charged}}=9$ in this example.

Summing up all the pieces, we obtain $\Delta h^{1,1}= \Delta T+\Delta
r_{\text{non-abelian}}+\Delta r_{\text{abelian}}=1+(251-18)+1=235$ and
$\Delta h^{2,1}=(\Delta V_{\text{non-abelian}}+ \Delta
V_{\text{abelian}})-29\Delta T-(\Delta H_\text{non-abelian
  charged}+\Delta V_{\text{abelian
    charged}})=(14739+1)-29-(14830+9)=-128$, which agrees with the
differences in
Hodge numbers from the polytopes:
$\set{263,32}-\set{28,160}=\set{235,-128}.$


\begin{thebibliography}{99}


\bibitem{chsw}
  P.~Candelas, G.~T.~Horowitz, A.~Strominger and E.~Witten,
  ``Vacuum Configurations for Superstrings,''  Nucl.\ Phys.\ B {\bf 258}, 46 (1985).  %%CITATION = NUPHA,B258,46;

\bibitem{Batyrev}
V.\ Batyrev,
Duke Math.\ Journ.\ {\bf 69}, 349 (1993).

\bibitem{Kreuzer:2000xy} 
  M.~Kreuzer and H.~Skarke,
  ``Complete classification of reflexive polyhedra in four-dimensions,''
  Adv.\ Theor.\ Math.\ Phys.\  {\bf 4}, 1209 (2002)
  {\tt hep-th/0002240}.
  %%CITATION = HEP-TH/0002240;
  

\bibitem{database} 
M. Kreuzer and H. Skarke, 
\url{http://hep.itp.tuwien.ac.at/~kreuzer/CY.html}.

\bibitem{Vafa-F-theory}
  C.~Vafa,
  ``Evidence for F-Theory,''
  Nucl.\ Phys.\  B {\bf 469}, 403 (1996)
  {\tt arXiv:hep-th/9602022}.
  %%CITATION = NUPHA,B469,403;

\bibitem{Morrison-Vafa}
  D.~R.~Morrison and C.~Vafa,
  ``Compactifications of F-Theory on Calabi--Yau Threefolds -- I,''
  Nucl.\ Phys.\  B {\bf 473}, 74 (1996)
  {\tt arXiv:hep-th/9602114}.
  %%CITATION = NUPHA,B473,74;
  
  
  
\bibitem{Morrison:1996pp} 
  D.~R.~Morrison and C.~Vafa,
  ``Compactifications of F-Theory on Calabi--Yau Threefolds -- II,''
  Nucl.\ Phys.\  B {\bf 476}, 437 (1996)
  {\tt arXiv:hep-th/9603161}.
  %%CITATION = NUPHA,B476,437;

\bibitem{clusters} 
  D.~R.~Morrison and W.~Taylor,
  ``Classifying bases for 6D F-theory models,''
  Central Eur.\ J.\ Phys.\  {\bf 10}, 1072 (2012)
  {\tt arXiv:1201.1943 [hep-th]}.
  %%CITATION = 
  
 

\bibitem{Morrison:2012js} 
  D.~R.~Morrison and W.~Taylor,
  ``Toric bases for 6D F-theory models,''
  Fortsch.\ Phys.\  {\bf 60}, 1187 (2012)
  {\tt arXiv:1204.0283 [hep-th]}.
  %%CITATION = doi:10.1002/prop.201200086;
  

\bibitem{Martini-WT} 
  G.~Martini and W.~Taylor,
  ``6D F-theory models and elliptically fibered Calabi-Yau threefolds
  over semi-toric base surfaces,''
JHEP {\bf 1506}, 061 (2015) {\tt arXiv:1404.6300 [hep-th]}.

\bibitem{Wang-WT}
W.~Taylor, Y.~Wang,
``Non-toric bases for elliptic Calabi-Yau Threefolds and 6D F-theory Vacua,''
{\tt arXiv:1504.07689}. 

\bibitem{Johnson-WT}
S.~Johnson, W.~Taylor,
``Calabi-Yau Threefolds with Large $h^{2,1}$,''
JHEP {\bf 1410}, 23 (2014),
{\tt arXiv:1406.0514}.

\bibitem{Johnson:2016qar} 
  S.~B.~Johnson and W.~Taylor,
  ``Enhanced gauge symmetry in 6D F-theory models and tuned elliptic Calabi-Yau threefolds,''
Fortsch.\ Phys.\  {\bf 64}, 581 (2016)
{\tt arXiv:1605.08052 [hep-th]}.

\bibitem{Candelas-cs} 
  P.~Candelas, A.~Constantin and H.~Skarke,
  ``An Abundance of K3 Fibrations from Polyhedra with Interchangeable
  Parts,''  Commun.\  Math.\  Phys.\  {\bf 324}, 937 (2013)
{\tt arXiv:1207.4792 [hep-th]}.  
%%CITATION = ARXIV:1207.4792;

\bibitem{Gray-hl} 
  J.~Gray, A.~S.~Haupt and A.~Lukas,
  ``Topological Invariants and Fibration Structure of Complete
  Intersection Calabi-Yau Four-Folds,''
JHEP {\bf 1409}, 093 (2014)
{\tt arXiv:1405.2073 [hep-th]}.
%%CITATION = ARXIV:1405.2073;

%\cite{Anderson:2015iia}
\bibitem{Anderson:2015iia} 
  L.~B.~Anderson, F.~Apruzzi, X.~Gao, J.~Gray and S.~J.~Lee,
  ``A new construction of Calabi-Yau manifolds: Generalized CICYs,''
  Nucl.\ Phys.\ B {\bf 906}, 441 (2016)

  [arXiv:1507.03235 [hep-th]].
  %%CITATION = doi:10.1016/j.nuclphysb.2016.03.016;%%
  %30 citations counted in INSPIRE as of 06 Feb 2019



%\cite{Anderson:2016ler}
\bibitem{Anderson:2016ler} 
  L.~B.~Anderson, X.~Gao, J.~Gray and S.~J.~Lee,
  ``Tools for CICYs in F-theory,''
  JHEP {\bf 1611}, 004 (2016)

  [arXiv:1608.07554 [hep-th]].
  %%CITATION = doi:10.1007/JHEP11(2016)004;%%
  %12 citations counted in INSPIRE as of 06 Feb 2019


%\cite{Anderson:2016cdu}
\bibitem{Anderson:2016cdu} 
  L.~B.~Anderson, X.~Gao, J.~Gray and S.~J.~Lee,
  ``Multiple Fibrations in Calabi-Yau Geometry and String Dualities,''
  JHEP {\bf 1610}, 105 (2016)

  [arXiv:1608.07555 [hep-th]].
  %%CITATION = doi:10.1007/JHEP10(2016)105;%%
  %19 citations counted in INSPIRE as of 06 Feb 2019


%\cite{Anderson:2017aux}
\bibitem{Anderson:2017aux} 
  L.~B.~Anderson, X.~Gao, J.~Gray and S.~J.~Lee,
  ``Fibrations in CICY Threefolds,''
  JHEP {\bf 1710}, 077 (2017)

  [arXiv:1708.07907 [hep-th]].
  %%CITATION = doi:10.1007/JHEP10(2017)077;%%
  %15 citations counted in INSPIRE as of 06 Feb 2019


\bibitem{Huang-Taylor-fibers} 
%\cite{Huang:2018esr}

  Y.~C.~Huang and W.~Taylor,
  ``On the prevalence of elliptic and genus one fibrations among toric hypersurface Calabi-Yau threefolds,''
  arXiv:1809.05160 [hep-th].
  %%CITATION = ARXIV:1809.05160;%%
  %5 citations counted in INSPIRE as of 06 Feb 2019



\bibitem{Candelas:1996su} 
  P.~Candelas and A.~Font,
  ``Duality between the webs of heterotic and type II vacua,''
  Nucl.\ Phys.\ B {\bf 511}, 295 (1998)
  {\tt hep-th/9603170}.
  %%CITATION = 
  
  
  
  %\cite{Bouchard:2003bu}
\bibitem{Bouchard:2003bu} 
  V.~Bouchard and H.~Skarke,
  %``Affine Kac-Moody algebras, CHL strings and the classification of tops,''
  Adv.\ Theor.\ Math.\ Phys.\  {\bf 7}, no. 2, 205 (2003)

  [hep-th/0303218].
  %%CITATION = doi:10.4310/ATMP.2003.v7.n2.a1;%%
  %43 citations counted in INSPIRE as of 21 Dec 2018
  

\bibitem{Braun-16} 
  V.~Braun, T.~W.~Grimm and J.~Keitel,
  ``Geometric Engineering in Toric F-Theory and GUTs with U(1) Gauge Factors,''
JHEP {\bf 1312}, 069 (2013)
{\tt arXiv:1306.0577 [hep-th]}.

\bibitem{bmpw} 
  J.~Borchmann, C.~Mayrhofer, E.~Palti and T.~Weigand,
  ``Elliptic fibrations for $SU(5)\times U(1)\times U(1)$ F-theory vacua,''
Phys.\ Rev.\ D {\bf 88}, no. 4, 046005 (2013)
%doi:10.1103/PhysRevD.88.046005
{\tt arXiv:1303.5054 [hep-th]};


%\cite{Borchmann:2013hta}
\bibitem{Borchmann:2013hta} 
  J.~Borchmann, C.~Mayrhofer, E.~Palti and T.~Weigand,
  ``SU(5) Tops with Multiple U(1)s in F-theory,''
  Nucl.\ Phys.\ B {\bf 882}, 1 (2014)

  [arXiv:1307.2902 [hep-th]].
  %%CITATION = doi:10.1016/j.nuclphysb.2014.02.006;%%
  %83 citations counted in INSPIRE as of 06 Feb 2019


\bibitem{Morrison-TASI} 
  D.~R.~Morrison,
  ``TASI lectures on compactification and duality,''
{\tt hep-th/0411120}.
%%CITATION = HEP-TH/0411120;

\bibitem{Taylor:2011wt} 
  W.~Taylor,
  ``TASI Lectures on Supergravity and String Vacua in Various Dimensions,''
  arXiv:1104.2051 [hep-th].
  %%CITATION = ARXIV:1104.2051;
  

\bibitem{Katz:1996xe} 
  S.~H.~Katz and C.~Vafa,
  ``Matter from geometry,''
  Nucl.\ Phys.\ B {\bf 497}, 146 (1997)
  {\tt hep-th/9606086}.
  %%CITATION = 
  

\bibitem{Bershadsky:1996nh} 
  M.~Bershadsky, K.~A.~Intriligator, S.~Kachru, D.~R.~Morrison, V.~Sadov and C.~Vafa,
  ``Geometric singularities and enhanced gauge symmetries,''
  Nucl.\ Phys.\ B {\bf 481}, 215 (1996)
  {\tt hep-th/9605200}.
  %%CITATION = 
  

\bibitem{Morrison:2011mb} 
  D.~R.~Morrison and W.~Taylor,
  ``Matter and singularities,''
  JHEP {\bf 1201}, 022 (2012)
  {\tt arXiv:1106.3563 [hep-th]}.
  %%CITATION = 

\bibitem{Braun:2011ux} 
  V.~Braun,
  ``Toric Elliptic Fibrations and F-Theory Compactifications,''
  JHEP {\bf 1301}, 016 (2013)
  {\tt arXiv:1110.4883 [hep-th]}.

\bibitem{Braun:2014oya} 
  V.~Braun and D.~R.~Morrison,
he  ``F-theory on Genus-One Fibrations,''
  JHEP {\bf 1408}, 132 (2014)
  {\tt arXiv:1401.7844 [hep-th]}.
  %%CITATION = 

  
\newcommand{\etalchar}[1]{$^{#1}$}

\bibitem{Morrison:2014era} 
  D.~R.~Morrison and W.~Taylor,
  ``Sections, multisections, and U(1) fields in F-theory,''
  arXiv:1404.1527 [hep-th].
  %%CITATION = ARXIV:1404.1527;
  
\bibitem{Huang:2013yta} 
  M.~X.~Huang, A.~Klemm and M.~Poretschkin,
  ``Refined stable pair invariants for E-, M- and $[p, q]$-strings,''
JHEP {\bf 1311}, 112 (2013)
{\tt arXiv:1308.0619 [hep-th]}.

\bibitem{aggk} 
  L.~B.~Anderson, I.~Garcia-Etxebarria, T.~W.~Grimm and J.~Keitel,
  ``Physics of F-theory compactifications without section,''
  JHEP {\bf 1412}, 156 (2014)
  
  {\tt arXiv:1406.5180 [hep-th]}.
  %%CITATION = doi:10.1007/JHEP12(2014)156;
  

\bibitem{Mayrhofer:2014opa} 
  C.~Mayrhofer, D.~R.~Morrison, O.~Till and T.~Weigand,
  ``Mordell-Weil Torsion and the Global Structure of Gauge Groups in F-theory,''
  JHEP {\bf 1410}, 16 (2014)
  
  {\tt arXiv:1405.3656 [hep-th]}.
  %%CITATION = doi:10.1007/JHEP10(2014)016;

\bibitem{Cvetic:2015moa} 
  M.~Cvetic, R.~Donagi, D.~Klevers, H.~Piragua and M.~Poretschkin,
  ``F-theory vacua with $\mathbb Z_3$ gauge symmetry,''
  Nucl.\ Phys.\ B {\bf 898}, 736 (2015)
  {\tt arXiv:1502.06953 [hep-th]]}.
    %%CITATION = doi:10.1016/j.nuclphysb.2015.07.011;


  

\bibitem{Bonetti:2011mw} 
  F.~Bonetti and T.~W.~Grimm,
  ``Six-dimensional (1,0) effective action of F-theory via M-theory on Calabi-Yau threefolds,''
  JHEP {\bf 1205}, 019 (2012)
  {\tt arXiv:1112.1082 [hep-th]}.
  %%CITATION = 
  
  

\bibitem{Hodge} 
  W.~Taylor,
  ``On the Hodge structure of elliptically fibered Calabi-Yau threefolds,''
JHEP {\bf 1208}, 032 (2012)
{\tt arXiv:1205.0952 [hep-th]}.
%%CITATION = ARXIV:1205.0952;



\bibitem{Buchmuller:2017wpe} 
  W.~Buchmuller, M.~Dierigl, P.~K.~Oehlmann and F.~Ruehle,
  ``The Toric SO(10) F-Theory Landscape,''
JHEP {\bf 1712}, 035 (2017)
{\tt arXiv:1709.06609 [hep-th]}.
%%CITATION = doi:10.1007/JHEP12(2017)035;%%<br />  %11 citations counted in INSPIRE as of 27 Dec 2018

\bibitem{Dierigl:2018nlv} 
  M.~Dierigl, P.~K.~Oehlmann and F.~Ruehle,
  ``Global Tensor-Matter Transitions in F-Theory,''
Fortsch.\ Phys.\  {\bf 66}, no. 7, 1800037 (2018)
{\tt arXiv:1804.07386 [hep-th]}.
%%CITATION = doi:10.1002/prop.201800037;%%<br />  %1 citations counted in INSPIRE as of 27 Dec 2018



\bibitem{Bhardwaj:2018vuu} 
  L.~Bhardwaj and P.~Jefferson,
  ``Classifying 5d SCFTs via 6d SCFTs: Arbitrary rank,''
{\tt arXiv:1811.10616 [hep-th]}.

%\cite{Apruzzi:2018nre}
\bibitem{Apruzzi:2018nre} 
  F.~Apruzzi, L.~Lin and C.~Mayrhofer,
  ``Phases of 5d SCFTs from M-/F-theory on Non-Flat Fibrations,''
{\tt arXiv:1811.12400 [hep-th]}.
%%CITATION = ARXIV:1811.12400;%%<br />  %1 citations counted in INSPIRE as of 27 Dec 2018



\bibitem{Katz:2011qp} 
  S.~Katz, D.~R.~Morrison, S.~Schafer-Nameki and J.~Sully,
  ``Tate's algorithm and F-theory,''
  JHEP {\bf 1108}, 094 (2011)
  {\tt arXiv:1106.3854 [hep-th]}.
  %%CITATION = 
  

\bibitem{Grassi-Morrison}
  A.~Grassi and D.~R.~Morrison,
  ``Anomalies and the Euler characteristic of elliptic Calabi-Yau threefolds,''
Commun.\ Num.\ Theor.\ Phys.\  {\bf 6}, 51 (2012)
{\tt arXiv:1109.0042 [hep-th]}.
 

\bibitem{KMT-II}
  V.~Kumar, D.~R.~Morrison and W.~Taylor,
  ``Global aspects of the space of 6D ${\cal N} = 1$ supergravities,''
  JHEP {\bf 1011}, 118 (2010)
  {\tt arXiv:1008.1062 [hep-th]}.
  %%CITATION = JHEPA,1011,118;

\bibitem{Bertolini:2015bwa} 
  M.~Bertolini, P.~R.~Merkx and D.~R.~Morrison,
  ``On the global symmetries of 6D superconformal field theories,''
  JHEP {\bf 1607}, 005 (2016)
  {\tt arXiv:1510.08056 [hep-th]}.
  %%CITATION = 
  

\bibitem{Heckman:2013pva} 
  J.~J.~Heckman, D.~R.~Morrison and C.~Vafa,
  ``On the Classification of 6D SCFTs and Generalized ADE Orbifolds,''
  JHEP {\bf 1405}, 028 (2014)
  Erratum: [JHEP {\bf 1506}, 017 (2015)]
  {\tt arXiv:1312.5746 [hep-th]}.
  %%CITATION = 
  

\bibitem{Grassi}
A.~Grassi, ``On minimal models of elliptic threefolds,'' Math. Ann. {\bf 290}
  (1991) 287--301.

  

  

\bibitem{Fulton} 
William Fulton, ``Introduction to Toric Varieties,'' Annals of
Mathematics Study 131, Princeton University Press, Princeton, 1993.

\bibitem{mirror-symmetry} 
K.\ Hori {\it et al.}, ``Mirror Symmetry,'' American Mathematical
Society, 2003.

\bibitem{Batyrev:1994hm} 
  V.~V.~Batyrev,
  ``Dual polyhedra and mirror symmetry for Calabi-Yau hypersurfaces in toric varieties,''
  J.\ Alg.\ Geom.\  {\bf 3}, 493 (1994)
  [alg-geom/9310003].
  %%CITATION = ALG-GEOM/9310003;
  

\bibitem{palp} 
M. Kreuzer and H. Skarke, \emph{{PALP}: a {P}ackage for {A}nalyzing {L}attice {P}olytopes.}
\url{http://hep.itp.tuwien.ac.at/~kreuzer/CY/CYpalp.html}.

\bibitem{Avram:1996pj} 
  A.~C.~Avram, M.~Kreuzer, M.~Mandelberg and H.~Skarke,
  ``Searching for K3 fibrations,''
  Nucl.\ Phys.\ B {\bf 494}, 567 (1997)
  {\tt hep-th/9610154}.
  %%CITATION = 
  

\bibitem{Kreuzer:1997zg} 
  M.~Kreuzer and H.~Skarke,
  ``Calabi-Yau four folds and toric fibrations,''
  J.\ Geom.\ Phys.\  {\bf 26}, 272 (1998)
  {\tt hep-th/9701175}.
  %%CITATION = 
  

\bibitem{Skarke:1998yk} 
  H.~Skarke,
  ``String dualities and toric geometry: An Introduction,''
  Chaos Solitons Fractals {\bf 10}, 543 (1999)
  {\tt hep-th/9806059}.
  %%CITATION = 
  

\bibitem{Klevers-16} 
  D.~Klevers, D.~K.~Mayorga Pena, P.~K.~Oehlmann, H.~Piragua and J.~Reuter,
  ``F-Theory on all Toric Hypersurface Fibrations and its Higgs Branches,''
JHEP {\bf 1501}, 142 (2015)
{\tt arXiv:1408.4808 [hep-th]}.

\bibitem{cpr}
  P.~Candelas, E.~Perevalov and G.~Rajesh,
  ``Toric geometry and enhanced gauge symmetry of F theory / heterotic vacua,''
  Nucl.\ Phys.\ B {\bf 507}, 445 (1997)
%  doi:10.1016/S0550-3213(97)00563-4
  {\tt hep-th/9704097}

\bibitem{Perevalov:1997vw} 
  E.~Perevalov and H.~Skarke,
  ``Enhanced gauged symmetry in type II and F theory compactifications: Dynkin diagrams from polyhedra,''
  Nucl.\ Phys.\ B {\bf 505}, 679 (1997)
  {\tt hep-th/9704129}.
  %%CITATION = 
  
  

\bibitem{sage}
\emph{{S}ageMath, the {S}age {M}athematics {S}oftware {S}ystem ({V}ersion
  7.1)}, The Sage Developers, 2016. \url{http://sagemath.org/doc/reference/schemes/sage/schemes/toric/weierstrass.html}.

  


\bibitem{slansky} 
R.~Slansky,
  ``Group theory for unified model building,''
  Phys.\ Reports\   {\bf 79}, 1-128 (1981).

  


  

  

\bibitem{transitions}
	L.~Anderson, J.~Gray,N.~Raghuram and W.~Taylor,
    ``Matter in Transition,''
    JHEP {\bf1604}, 080 (2016).
    {\tt arXiv:1512.05791 [hep-th]}
    %%CITATION = ARXIV:1512.05791

\bibitem{Morrison-Park} 
  D.~R.~Morrison and D.~S.~Park,
  ``F-Theory and the Mordell-Weil Group of Elliptically-Fibered Calabi-Yau Threefolds,''
  JHEP {\bf 1210}, 128 (2012)
  {\tt arXiv:1208.2695 [hep-th]}.
  %%CITATION = ARXIV:1208.2695;

  
  
\bibitem{Braun:2014qka} 
  V.~Braun, T.~W.~Grimm and J.~Keitel,
  ``Complete Intersection Fibers in F-Theory,''
JHEP {\bf 1503}, 125 (2015)
{\tt arXiv:1411.2615 [hep-th]}.
% 

\end{thebibliography}
\end{document}